\documentclass[amsmath,aps,a4paper]{article}
\usepackage{color}
\usepackage{amssymb}
\usepackage{amsmath}
\usepackage{epsfig}
\usepackage{subfigure}
\usepackage{mathrsfs}
\usepackage{longtable}
\usepackage[usenames,dvipsnames]{xcolor}
\usepackage[linktoc=all]{hyperref}
\usepackage{titlesec}
\usepackage{lscape}
\hypersetup{
    colorlinks,
    citecolor=blue,
    filecolor=black,
    linkcolor=blue,
    urlcolor=black}

\usepackage[margin=1.4in]{geometry}
\usepackage{mathtools}
\usepackage{relsize}

\newcommand{\be}{\begin{equation}}
\newcommand{\ee}{\end{equation}}

\begin{document}

    \title{Axion Cosmology}
    \author{David J. E. Marsh${}^{1}$}
    \date{~}

\maketitle

    \thispagestyle{empty}
    \setcounter{page}{0}
    {\centerline{\it ${}^{1}$~Department of Physics,}}
     {\centerline{\it King's College London,}}
    {\centerline{\it Strand, London, WC2R 2LS,}}
     {\centerline{\it United Kingdom.}}
    \vskip 0.25in

%
\begin{abstract}
Axions comprise a broad class of particles that can play a major role in explaining the unknown aspects of cosmology. They are also well-motivated within high energy physics, appearing in theories related to $CP$-violation in the standard model, supersymmetric theories, and theories with extra-dimensions, including string theory, and so axion cosmology offers us a unique view onto these theories. I review the motivation and models for axions in particle physics and string theory. I then present a comprehensive and pedagogical view on the cosmology and astrophysics of axion-like particles, starting from inflation and progressing via BBN, the CMB, reionization and structure formation, up to the present-day Universe. Topics covered include: axion dark matter (DM); direct and indirect detection of axions, reviewing existing and future experiments; axions as dark radiation; axions and the cosmological constant problem; decays of heavy axions; axions and stellar astrophysics; black hole superradiance; axions and astrophysical magnetic fields; axion inflation, and axion DM as an indirect probe of inflation. A major focus is on the population of ultralight axions created via vacuum realignment, and its role as a DM candidate with distinctive phenomenology. Cosmological observations place robust constraints on the axion mass and relic density in this scenario, and I review where such constraints come from. I next cover aspects of galaxy formation with axion DM, and ways this can be used to further search for evidence of axions. An absolute lower bound on DM particle mass is established. It is $m_a>10^{-24}\text{ eV}$ from linear observables, extending to $m_a\gtrsim 10^{-22}\text{ eV}$ from non-linear observables, and has the potential to reach $m_a\gtrsim 10^{-18}\text{ eV}$ in the future. These bounds are weaker if the axion is not all of the DM, giving rise to limits on the relic density at low mass. This leads to the exciting possibility that the effects of axion DM on structure formation could one day be detected, and the axion mass and relic density measured from cosmological observables. 

\end{abstract}

\vskip 1.1in
\hfill KCL-PH-TH/2015-50

\newpage

\begingroup
\hypersetup{linkcolor=black}
\tableofcontents
\endgroup

\newpage

\section{Introduction}


As Weinberg said, ``physics thrives on crisis"~\cite{1989RvMP...61....1W}. In 1989 when Weinberg wrote that famous review, he said that physics was short on crises. Happily, these days, thanks in large part to the advent of precision cosmology, it is full of them. 

The standard cosmological model is described by just six numbers: two for initial conditions, one for dark matter (DM), one for the baryons, one for cosmic structure formation and reionization, and one for the cosmological constant (c.c.). Each of these numbers presents a problem for our understanding of fundamental physics. The initial conditions appear close to scale invariant: producing such initial conditions requires a period of rapid acceleration (or slow deceleration) in the early Universe, a state of affairs that cannot be realised in the usual hot big bang. Dark matter constitutes the vast majority of matter in the Universe, and no particle in the standard model of particle physics can fit the role of being stable, cold, and weakly coupled. The standard model also provides no obvious way to tip the matter-anti-matter asymmetry in favour of baryons instead of anti-baryons. Structure formation and reionization are sensitive to the initial conditions, matter content, and complex astrophysical processes in ways that we are only just learning. And then finally there is Weinberg's problem of the c.c..

In 1989 Weinberg selected just the c.c. as a major problem: even without precision cosmology, it was clear that the theoretical expectations about this number were wildly off the mark. All of the other problems were known at that time, but without the precision measurements we have today their importance could easily be debated and there was no need to call ``crisis.'' We are no longer in that position of blissful ignorance: all the numbers in the standard cosmological model need to be considered and their theoretical implications taken seriously.

\begin{landscape}
\begin{figure}\vspace{-0.8in}
    \includegraphics[width=1.5\textwidth]{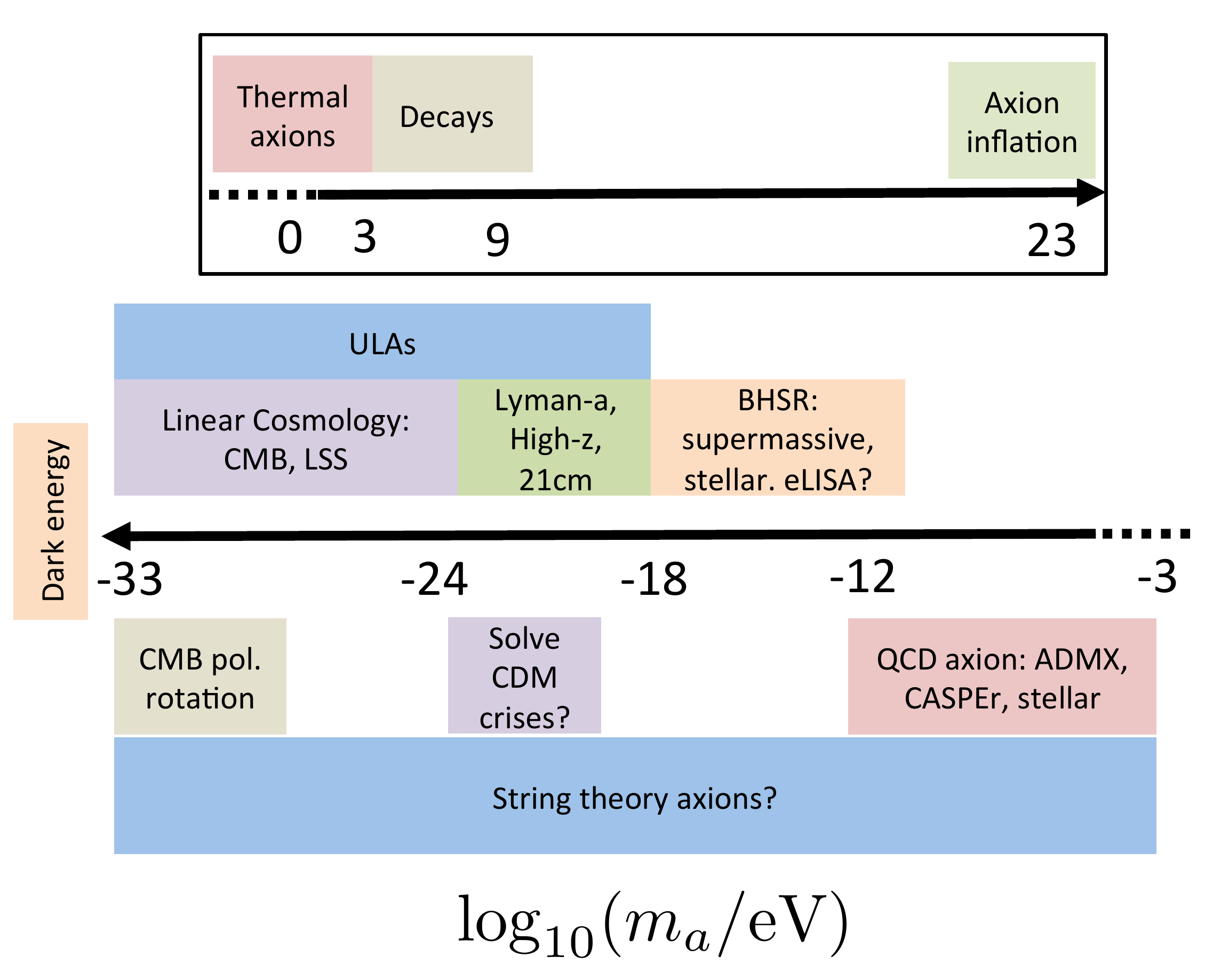}
    \caption{Summary of constraints and probes of axion cosmology.}
    \label{fig:massPlot}
\end{figure}
\end{landscape}

In seeking a unified view of the problems presented by precision cosmology, we will focus in this review on a class of particles known as axions. Ever since the earliest days of the QCD axion it has been realised that it offers an exceptionally good DM candidate. With the advent of string theory and the corresponding profusion of axion-like particles (ALPs), axions have come to play important roles in inflation and the generation of cosmological initial conditions, and in the solution of the c.c. problem. String axions also offer the posisbility to resolve problems of structure formation inherent in more vanilla models of DM. Axions can even assist in baryogenesis thanks to their role in $CP$-violation. A summary of constraints and probes of axion cosmology, as a function of axion mass, is shown in Fig.~\ref{fig:massPlot}.

A large portion of this review will focus on ALPs in the mass range
\be
10^{-33}\text{ eV}\lesssim m_a \lesssim 10^{-18}\text{ eV} \, .
\label{eqn:axion_mass_range}
\ee
I will refer to axions in this mass range as ultralight axions, or ULAs. The lower bound is of order the present day Hubble constant, $H_0/h=M_H=2.13\times 10^{-33}\text{ eV}=100\text{ km s}^{-1}\text{ Mpc}^{-1}$, and reflects constraints on axion dark energy (DE). The upper bound is related to the baryon Jeans scale, and reflects a distinctive role of ULAs in cosmological structure formation and reionization. This vast range of axion masses can be probed using the tools that led us to our crises in the first place, i.e. those of precision cosmology: the cosmic microwave background (CMB), large scale structure (LSS), galaxy formation in the local Universe and at high redshift, and by the epoch of reionization (EOR).

It is worth noting here, for clarity, that the word ``axion'' can take on a variety of meanings. It was first coined by Wilczek~\cite{wilczek1978} to name the particle associated to the axial anomaly in QCD and the Peccei-Quinn~\cite{pecceiquinn1977} solution to the strong-$CP$ problem. It is so named after the eponymous American laundry detergent, using the axial anomaly to clean up the mess of $CP$ symmetry in the strong interactions~\cite{citation_classics}. The QCD axion acquires mass from QCD chiral symmetry breaking, giving a one parameter model described by the axion decay constant, $f_a$. In quantum field theory, the term can apply generally to any pseudoscalar Goldstone bosons of spontaneously broken global chiral symmetries, typically giving a two parameter model with $(m_a,f_a)$. In string theory and supergravity, the term ``axion'' is more general and can refer either to such matter fields, or to pseudoscalar fields associated to the geometry of compact spatial dimensions~\cite{2006JHEP...06..051S}. In these theories there are typically many axion fields, each with a number of free parameters in their potentials and kinetic terms. In this review, we will use the term in its most general sense for a light pseudoscalar field (indeed in some cosmological cases, apart from naturalness considerations, even the distinction between scalar and pseudoscalar will be irrelevant).

Since the QCD axion was first proposed in 1977-1978, there have been many reviews written on axion physics. Many such reviews and published lecture notes focus on the QCD axion and its role in solving the strong-$CP$ problem~\cite{2008LNP...741....3P,2010RvMP...82..557K}, as well as its important cosmological role~\cite{2008LNP...741...19S}. Of ALPs, there are technical reviews of axions in field theory and string theory~\cite{Kim:1986ax,2006JHEP...06..051S}, as well as reviews of axions in astrophysics~\cite{2008LNP...741...51R}, and of axion inflation~\cite{2013CQGra..30u4002P}. There is also a vast number of reviews in the field of axion direct detection~\cite{2012PDU.....1..116R,2011ConPh..52..211R,2012arXiv1209.2299R,2015arXiv150608364S,2013arXiv1309.7035C}. It is the purpose of this review firstly to focus on ULAs, the cosmology of which has not been reviewed before, and with a particular emphasis on methods of modern precision cosmology, including computational aspects both analytic and numerical, and with an eye to data. Secondly, it is to bring together the disparate topics of other axion reviews into one place, expressing the unity of axion particle physics and cosmology: a task, which, to my knowledge, has not been fully addressed since the review of Ref.~\cite{Kim:1986ax}, more than 30 years ago in this very journal. 

\subsection*{Notes}

Useful notation and equations for cosmology are defined in the Appendix. I (mostly) use units where $c=\hbar=k_B=1$ and express everything in terms of either electronvolts, eV, solar masses, $M_\odot$, parsecs, pc, or Kelvin, K, depending on the context. The Fourier conjugate variable to $x$ is $k$ and my Fourier convention puts the $2\pi$'s under the $dk$'s. I use the reduced Planck mass, $M_{pl}=1/\sqrt{8\pi G}=2.435\times 10^{27}\text{ eV}$, and a ``mostly positive'' metric signature.

\newpage

\section{Models}

A classic review of models for axions in particle physics and string theory is Ref.~\cite{Kim:1986ax}, where many more details are given. A modern review of axions in string theory is Ref.~\cite{2006JHEP...06..051S}, and for pedagogical introductions and phenomenology see e.g. Refs.~\cite{axiverse,2012arXiv1209.2299R}. This section is intended only as an overview: we will wave our hands through the particle physics computations, and wave them even more wildly through the string theory. This section is also self-contained, and can be skipped for those interested only in cosmology and astrophysics. The salient points for cosmology are repeated in Section~\ref{sec:non-pert}.

\subsection{The QCD Axion}
\label{sec:qcd_axion}

\subsubsection{The Strong-$CP$ Problem and the PQ Solution}

QCD suffers from the ``strong-$CP$ problem.'' A topological (total derivative) term is allowed in the Lagrangian:
\be
\mathcal{L}_{\theta{\rm QCD}}=\frac{\theta_{\rm QCD}}{32\pi^2}\text{Tr } G_{\mu\nu}\tilde{G}^{\mu\nu}\, ,
\label{eqn:qcd_topological}
\ee
where $G_{\mu\nu}$ is the gluon field strength tensor, $\tilde{G}^{\mu\nu}=\epsilon^{\mu\nu\alpha\beta}G_{\alpha\beta}/2$ is its dual, and the trace is over the adjoint representation of $SU(3)$ (a notation I drop from now on).\footnote{I have chosen the normalization for the gluon field, $A_\mu$, appropriate for the vacuum topological term, which takes $\theta_{\rm QCD}\in [0,2\pi]$. In this normalization the gluon kinetic term is $-G_{\mu\nu}G^{\mu\nu}/4g_3^2$, where $g_3$ is the $SU(3)$ gauge coupling constant.} This term arises due to the so-called ``$\theta$-vacua'' of QCD~\cite{1988assy.book.....C}, which are discussed in Appendix~\ref{appendix:theta_vacua}. 

The $\theta$ term is $CP$ violating and gives rise to an electric dipole moment (EDM) for the neutron~\cite{1979PhLB...88..123C}:
\be
d_n\approx 3.6 \times 10^{-16}\theta_{\rm QCD} \, e\text{ cm}\, ,
\ee
where $e$ is the charge on the electron. The (permanent, static) dipole moment is constrained to $|d_n|<2.9\times 10^{-26}\,e\,\text{cm}$ (90\% C.L.)~\cite{2006PhRvL..97m1801B}, implying $\theta_{\rm QCD}\lesssim 10^{-10}$. 

This is a true fine tuning problem, since $\theta_{\rm QCD}$ could obtain an $\mathcal{O}(1)$ contribution from the observed $CP$-violation in the electroweak (EW) sector~\cite{2014ChPhC..38i0001O}, which must be cancelled to high precision by the (unrelated) gluon term. Specifically, the measurable quantity is
\be
\theta_{\rm QCD}=\tilde{\theta}_{\rm QCD}+\text{arg det}M_uM_d\, ,
\ee
where $\tilde{\theta}$ is the bare quantity and $M_u$, $M_d$ are the quark mass matrices.\footnote{The phase of the quark mass matrix is not measured, but could be $\mathcal{O}(1)$. $CP$-violation in the standard model leads to a calculable minimum value for $\theta_{\rm QCD}$ even in the axion model (e.g. Ref.~\cite{2005AnPhy.318..119P}).}

The QCD axion is the dynamical pseudoscalar field coupling to $G\tilde{G}$, proposed by Peccei and Quinnn (PQ)~\cite{pecceiquinn1977}, which dynamically sets $\theta_{\rm QCD}=0$ via QCD non-perturbative effects (instantons)~\cite{Vafa:1984xg}. The simple idea is that there is a field, $\phi$, which enjoys a shift symmetry, with only derivatives of $\phi$ appearing in the action. Taking $\theta_{\rm QCD}=\mathcal{C}\phi/f_a$, where $\phi$ is the canonically normalized axion field, $f_a$ is the axion decay constant and $\mathcal{C}$ is the ``colour anomaly'' (discussed in Section~\ref{sec:instantons}), this is a symmetry under $\phi\rightarrow\phi+{\rm const}$. Then, as long as shift symmetry violation is induced only by quantum effects as $(\mathcal{C}\phi/f_a)G\tilde{G}$, any contribution to $\theta_{\rm QCD}$ can be absorbed in a shift of $\phi$. The action, and thus the potential induced by QCD non-perturbative effects, only depends on the overall field multiplying $G\tilde{G}$. If the potential for the shifted field is minimized at $\mathcal{C}\phi/f_a=0\text{ mod }2\pi$, then the strong $CP$ problem is solved.  In fact, a theorem of Vafa and Witten~\cite{Vafa:1984xg} guarantees that the instanton potential is minimized at the $CP$ conserving value. We will discuss the instanton potential in more detail in Section~\ref{sec:instantons}.

The axion mass, $m_a$, induced by QCD instantons can be calculated in chiral perturbation theory~\cite{weinberg1978,wilczek1978}. It is given by
\be
m_{a,{\rm QCD}}\approx 6\times 10^{-6}\text{ eV}\left(\frac{10^{12}\text{ GeV}}{f_a/\mathcal{C}} \right)\, .
\label{eqn:qcd_mass_zero_T}
\ee
This is a (largely) model-independent statement, and the approximate symbol, ``$\approx$,'' takes model and QCD uncertainties into account. If $f_a$ is large, the QCD axion can be extremely light and stable, and is thus an excellent DM candidate~\cite{1983PhLB..120..127P,1983PhLB..120..133A,1983PhLB..120..137D}.

We will consider three general types of QCD axion model:\footnote{One can also construct more general particle physics models along these lines with multiple ALPs as well as the QCD axion, but we will not discuss such models in detail. We consider all ALPs within a string theory context in Section~\ref{sec:string_models}.}
\begin{itemize}
\item The Peccei-Quinn-Weinberg-Wilczek (PQWW)~\cite{pecceiquinn1977,weinberg1978,wilczek1978} axion, which introduces one additional complex scalar field only, tied to the EW Higgs sector. It is excluded by experiment.
\item The Kim-Shifman-Vainshtein-Zakharov (KSVZ)~\cite{1979PhRvL..43..103K,1980NuPhB.166..493S} axion, which introduces heavy quarks as well as the PQ scalar.
\item The Dine-Fischler-Srednicki-Zhitnitsky (DFSZ)~\cite{1981PhLB..104..199D,Zhitnitsky:1980tq} axion, which introduces an additional Higgs field as well as the PQ scalar.
\end{itemize}

\subsubsection{PQWW axion}

The PQWW model introduces a single additional complex scalar field, $\varphi$, to the standard model as a second Higgs doublet. One Higgs field gives mass to the $u$-type quarks, while the other gives mass to the $d$-type quarks (a freedom of the model is the choice of which doublet, if not a third field, gives mass to the leptons). This fixes the representation of $\varphi$ in $SU(2)\times U(1)$. The whole Lagrangian is then taken to be invariant under a global $U(1)_{\rm PQ}$ symmetry, which acts with \emph{chiral} rotations, i.e. with a factor of $\gamma_5$. These chiral rotations shift the angular part of $\varphi$ by a constant. The PQ field couples to the standard model via the Yukawa interactions which give mass to the fermions as in the usual Higgs model. The invariance of these terms under global $U(1)_{\rm PQ}$ rotations fixes the PQ charges of the fermions.
\begin{figure}
\begin{center}
\includegraphics[scale=0.45]{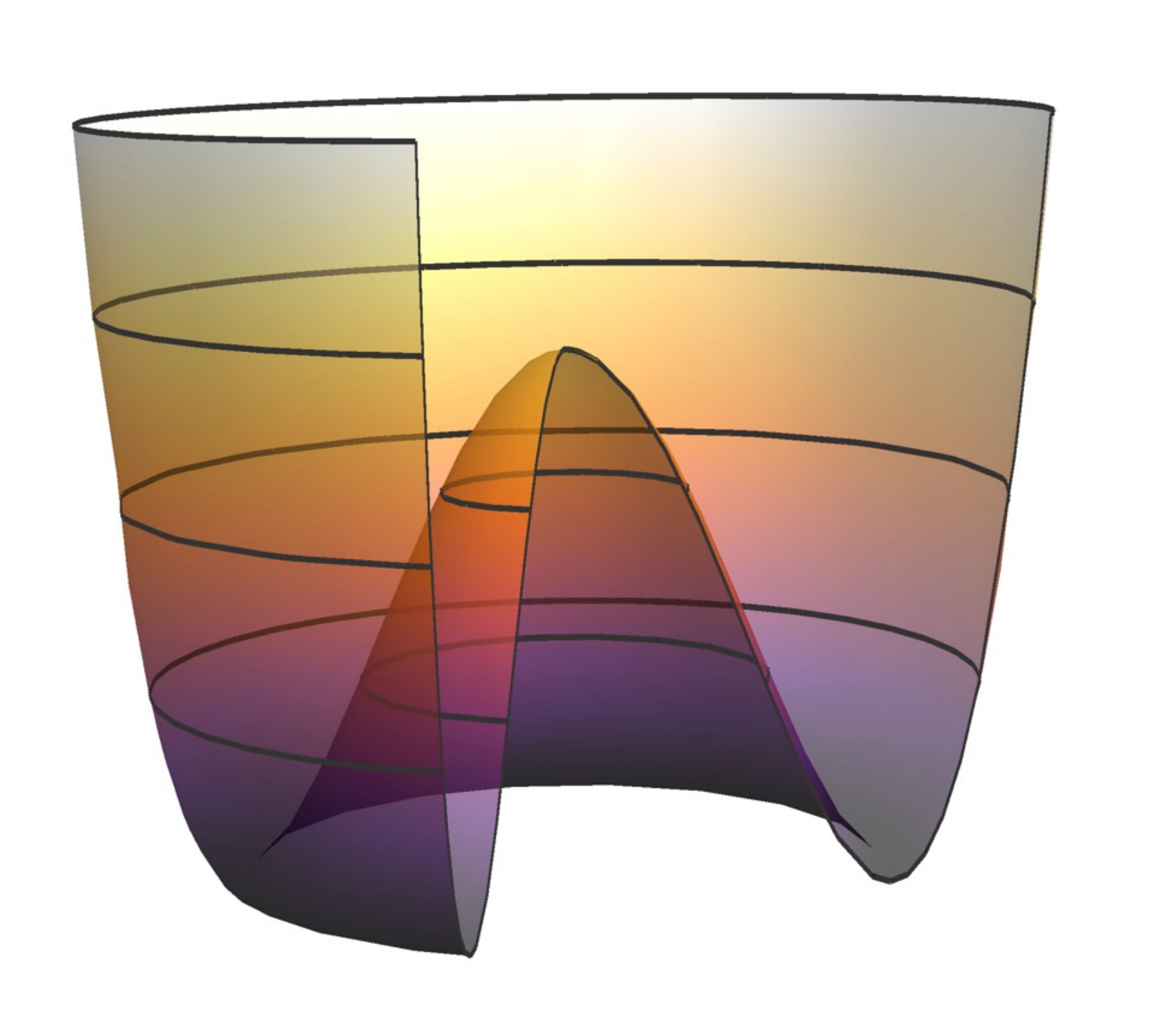}
\caption{A symmetry breaking potential in the complex $\varphi$ plane. The vev of the radial mode is $f_a/\sqrt{2}$ and the axion is the massless angular degree of freedom at the potential minimum.}
\label{fig:symm_break}
\end{center}
\end{figure}

Just like the Higgs, $\varphi$ has a symmetry breaking potential (see Fig.~\ref{fig:symm_break}):
\be
V(\varphi)=\lambda \left(|\varphi|^2-\frac{f_a^2}{2}\right)^2 \, ,
\label{eqn:pq_symmetry_breaking}
\ee
and takes a vacuum expectation value (vev), $\langle\varphi\rangle=f_a/\sqrt{2}$ \emph{at the EW phase transition}. Just as for the Higgs, this fixes the scale of the vev $f_a\approx 250\text{ GeV}$. 

There are four real, electromagnetically (EM) neutral scalars left after EW symmetry breaking: one gives the $Z$-boson mass, one is the standard model Higgs~\cite{Aad:2012tfa,Chatrchyan:2012ufa}, one is the heavy radial $\varphi$ field, and one is the angular $\varphi$ field. The angular degree of freedom appears as $\langle\varphi\rangle e^{i\phi/f_a}$ after canonically normlaizing the kinetic term. The field $\phi$ is the axion and is the Goldstone boson of the spontaneously broken $U(1)_{\rm PQ}$ symmetry.

The axion couples to the standard model via the chiral rotations and the PQ charges of the standard model fermions, e.g. expanding in powers of $1/f_a$ the quark coupling is $m_q(\phi/f_a)i\bar{q}\gamma_5 q$. The chiral anomaly~\cite{tHooft:1976fv} then induces couplings to gauge bosons via fermion loops\footnote{See Appendix~\ref{appendix:eft} for a heuristic description of effective field theory (EFT).} $\propto\phi G\tilde{G}/f_a$ and $\propto \phi F\tilde{F}/f_a$, where $F$ is the EM field strength. The gluon term is the desired term and leads to the PQ solution of the strong-$CP$ problem. Notice that \emph{all axion couplings come suppressed by the scale $f_a$}, which in the PQWW model is fixed to be the EW vev. \emph{In the PQWW model $f_a$ is too small, the axion couplings are too large, and it is excluded, e.g. by beam-dump experiments}~\cite{Kim:1986ax}. The PQWW axion is also excluded by collider experiments such as LEP (see the recent compilation of collider constraints in Ref.~\cite{2015JHEP...06..173M}, and Section~\ref{sec:decays}).

In the KSVZ and DFSZ models, which we now turn to, the PQ field, $\varphi$, is introduced independently of the EW scale. The decay constant is thus a free parameter in these models, and can be made large enough such that they are not excluded. For this reason, both the KSVZ and the DFSZ axions are known as \emph{invisible axions}. On the plus side, in these models the axion is stable and is an excellent DM candidate with its own phenomenology. 

\subsubsection{KSVZ axion}

The KSVZ axion model introduces a heavy quark doublet, $Q_L,Q_R$, each of which is an $SU(3)$ triplet, and the subscripts represent the charge under chiral rotations. The PQ scalar field, $\varphi$, has charge 2 under chiral rotations, but is now a standard model singlet. The PQ field and the heavy quarks interact via the PQ-invariant Yukawa term, which provides the heavy quark mass:
\be
\mathcal{L}_{\rm Y}=-\lambda_Q\varphi\bar{Q}_LQ_R+h.c. \, ,
\ee
where the Yukawa coupling $\lambda_Q$ is a free parameter of the model. As in the PQWW model, there is a global $U(1)_{\rm PQ}$ symmetry which acts as a chiral rotation with angle $\alpha=\phi/f_a$, shifting the axion field. Global $U(1)_{\rm PQ}$ symmetry is spontaneously broken by the potential, Eq.~\ref{eqn:pq_symmetry_breaking}. 

At the classical level, the Lagrangian is unaffected by chiral rotations, and $\varphi$ is not coupled to the standard model. However at the quantum level, chiral rotations on $Q$ affect the $\tilde{G}G$ term via the chiral anomaly~\cite{tHooft:1976fv}:
\be
\mathcal{L}\rightarrow \mathcal{L}+\frac{\alpha}{32\pi^2}G\tilde{G}\, ,
\ee 
where I have used that in the KSVZ model the colour anomaly is equal to unity (see Section~\ref{sec:instantons}).

At low energies, after PQ symmetry breaking, $\varphi$ takes a vev and the $Q$ fields obtain a large mass, $m_Q\sim \lambda_Qf_a$. The $Q$ fields can then be integrated out. The chiral anomaly induces the axion coupling to $\tilde{G}G$ as a ``memory'' of the chiral rotation applied at high energy. At the level of EFT, the induced topological term is the only modification to the standard model Lagrangian: \emph{the KSVZ axion has no unsuppressed tree-level couplings to standard model matter fields.}

There is an axion-photon coupling in this model that can be calculated via loops giving the EM anomaly. It's value depends on the electromagnetic charges assigned to the $Q$ fields. The canonical choice is that they are uncharged and the axion-photon coupling is induced solely by the longitudinal mode of the $Z$-boson (see e.g. Ref.~\cite{1998PhRvD..58e5006K}). Other couplings can also be induced by loops and mixing, since $Q$ must be charged under $SU(3)$. Couplings will be listed and discussed further in Section~\ref{sec:sm_couplings}.

\subsubsection{DFSZ axion}

The DFSZ axion couples to the standard model via the Higgs sector. It contains two Higgs doublets, $H_u$, $H_d$, like in the PQWW model, however the complex scalar, $\varphi$, which contains the axion as its angular degree of freedom, is introduced as a standard model singlet. Again, global $U(1)_{\rm PQ}$ symmetry is imposed and spontaneously broken by the potential, Eq.~\eqref{eqn:pq_symmetry_breaking}.  

The PQ and Higgs fields interact via the scalar potential:
\be
V=\lambda_H\varphi^2H_uH_d \, .
\ee
This term is PQ invariant for $\varphi$ with $U(1)_{\rm PQ}$ charge +1, and the Higgs fields each with charge -1. As in the KSVZ model, PQ rotations act by shifting the axion by $\phi/f_a\rightarrow \phi/f_a+\alpha$. When the PQ symmetry is broken and $\varphi$ obtains a vev, the parameters in the Higgs potential, and the coupling constant, $\lambda_H$, must be chosen such that the Higgs fields remain light, consistent with the observed 125 GeV standard model Higgs~\cite{Aad:2012tfa,Chatrchyan:2012ufa}, and the EW vev, $v_{\rm EW}=\sqrt{\langle H_u\rangle^2+\langle H_d\rangle^2}$.

The Higgs must also couple to all the standard model fermions, providing their mass through Yukawa terms as usual, e.g.
\be
\mathcal{L}_Y\supset \lambda_u\bar{q}_Lu_RH_u \, .
\ee
In order for this to be PQ invariant the standard model fermions must be charged under $U(1)_{\rm PQ}$. After EW symmetry breaking, $H$ is replaced by its vev, inducing axial current couplings between the axion and standard model fermions from the chiral term in the fermion mass matrix: $m_u(\phi/f_a)i\bar{u}\gamma_5u$. This axial current in turn induces the coupling between the axion and $G\tilde{G}$ via the colour anomaly. The difference between KSVZ and DFSZ is that for DFSZ this term is induced by light quark loops calculated at low energy, rather than via the integrating out of a heavy quark. In the DFSZ model all of the standard model quarks are charged under the PQ symmetry, giving rise to a larger colour anomaly, $\mathcal{C}=6$.

The same fermion loops induce the axion-photon coupling, $\phi F\tilde{F}$, which is computed via the electromagnetic anomaly. Freedom in this model appears through the lepton charges: we are free to choose whether it is $H_u$ or $H_d$ that gives mass to the electron via $H_{u,d}\bar{\ell}_Le_R$. The axion-photon coupling is the sum of quark and lepton loops, and the different lepton PQ charges give different values for the anomaly, and thus the coupling (see Section~\ref{sec:sm_couplings}).

The use of the Higgs in DFSZ leads to a number of important consequences that differentiate it from KSVZ. Firstly, \emph{in the DFSZ model there are tree-level couplings between the axion and standard model fermions}, via the chiral terms in the mass matrix. Secondly, the EW sector is modified by the \emph{addition of an extra axial Higgs field}, $A$, with mass of order the EW scale. This is constrained by collider data, and could potentially be discovered at the LHC, just like the additional Higgs fields of supersymmetry (SUSY, see e.g. Refs.~\cite{Djouadi:2015jea,Aad:2015pla}).

\subsection{Anomalies, Instantons, and the Axion Potential}
\label{sec:instantons}

A PQ rotation on a field $x_i$ with PQ charge $\mathcal{Q}_{{\rm PQ},i}$ acts as
\be
x_i\rightarrow e^{i\mathcal{Q}_{{\rm PQ},i}\phi/f_a}x_i\, .
\ee 
The rotation is chiral, meaning that, if $x_i$ is a spinor, left and right handed components of $x_i$ have opposite charges (for the two-component spinor $\psi=(\psi_L,\psi_R)$ one introduces a factor of $\gamma_5$ to achieve this). 

The axion model is set up so that at the classical level the Lagrangian is invariant under such transformations, which leads to the shift symmetry of the axion field, $\phi\rightarrow\phi+{\rm const}$. At the quantum level, however, PQ rotations of quarks are anomalous, meaning that the quantum theory violates the classical symmetry. This affects the QCD topological term, and shifts it by an amount $\propto (\phi/f_a)G\tilde{G}$. The question we now wish to answer is: what is the constant of proportionality?

The constant of proportionality is called the \emph{colour anomaly} of the PQ symmetry, and is given by (e.g. Ref.~\cite{Srednicki:1985xd}):
\be
\mathcal{C}\delta_{ab}=2\text{Tr }\mathcal{Q}_{\rm PQ}T_aT_b \, ,
\ee
where the trace is over all the fermions in the theory, and $T_a$ are the generators of the $SU(3)$ representations of the fermions (e.g. for the triplet these are the Gell-Mann matrices). A PQ rotation now shows up in the action as
\be
S\rightarrow S+\int d^4x\frac{\mathcal{C}}{32\pi^2}\frac{\phi}{f_a}{\rm Tr}\,G_{\mu\nu}\tilde{G}^{\mu\nu}\, .
\ee

Although the topological term in the QCD action, Eq.~\eqref{eqn:qcd_topological}, does not affect the classical equations of motion, it does affect the vacuum structure, and the vacuum energy depends on $\theta_{\rm QCD}$. This is because of the existence of \emph{instantons} and the so-called $\theta$-vacua of QCD (for more details, see Ref.~\cite{1988assy.book.....C} and Appendix~\ref{appendix:theta_vacua}). These emerge because the non-Abelian gauge group, $SU(3)$, can be mapped onto the symmetry group of the space-time boundary, allowing for topologically-distinct field configurations~\cite{1988assy.book.....C}. The different vacua of QCD are labelled by the value of $\theta_{\rm QCD}$. The vacuum energy is~\cite{1978PhRvD..17.2717C,1981RvMP...53...43G}:
\be
E_{\rm vac}\propto \cos \theta_{\rm QCD}\sim \theta_{\rm QCD}^2 \, .
\ee
However, because the $\theta$-vacua are topologically distinct, no process allows for transitions between them, and the energy cannot be minimized.\footnote{There is a ``superselection rule'' such that $\langle \theta|\text{Anything}|\theta'\rangle=\delta_{\theta\theta'}$.} Introducing a field that couples to $G\tilde{G}$, as the axion does, means that the vacuum energy now depends on the linear combination $E_{\rm vac}(\theta_{\rm QCD}+\mathcal{C}\phi/f_a)$.

Using the shift symmetry on $\phi$ to absorb any contribution to $\theta_{\rm QCD}$, the vacuum energy is 
\be
E_{\rm vac}\propto \cos \left( \frac{\mathcal{C}\phi}{f_a} \right) \, .
\ee
The vacuum energy now depends on a dynamical field, and so can be minimized by the equations of motion.

The colour anomaly sets the number of vacua that $\phi$ has in the range $[0,2\pi f_a]$. Because $\phi$ is an angular variable, we must have a symmetry under $\phi\rightarrow \phi+2\pi f_a$. This implies that the colour anomaly must be an integer (this can always be achieved by normalization~\cite{Srednicki:1985xd}). Because it sets the number of vacua, the colour anomaly is also known as the \emph{domain wall number}, $\mathcal{C}=N_{\rm DW}$ (see Section~\ref{sec:defects}). Dynamics of $\phi$ send it to one of these vacua, which is the essence of the PQ mechanism.

In this way, the instantons are said to induce a mass for the axion. Let's investigate this in the DFSZ model, though the argument is more general. The relevant terms in the Lagrangian are:
\be
m_q\bar{q}q+\frac{N_{\rm DW}\phi}{32 \pi^2 f_a}G\tilde{G}\, .
\ee
Applying a chiral rotation to the quarks by an angle $\alpha=N_{\rm DW}\phi/f_a$ shows up as an interaction between the axion and the quarks:
\be
\cos (N_{\rm DW}\phi/f_a) m_*(\bar{u}u+\bar{d}d)+\sin(N_{\rm DW}\phi/f_a) m_*(\bar{u}i\gamma_5u+\bar{d}i\gamma_5d)\, ,
\ee
where $m_*=m_um_d/(m_u+m_d).$

After the QCD confinement transition at $T\sim\Lambda_{\rm QCD}$ we can replace the quark bilinears with their condensates, $\langle q\bar{q}\rangle$. Expanding for large $f_a$ we see that the cosine term introduces a mass (i.e. $\phi^2$ term) for the axion proportional to $-(m_u+m_d)\langle q\bar{q}\rangle/f_a^2=m_\pi^2f_\pi^2/f_a^2$, where $m_\pi$ is the pion mass and $f_\pi$ is the pion decay constant. 

At lowest order the sine term introduces a Yukawa-like interaction between axions and quarks, and renormalizes the axion mass. The interaction allows for the quark condensate to appear in the axion two-point function. The structure of the interaction is such that the $\eta'$ meson dominates this effect and the axion mass is renormalized to
\be
m_a^2=\frac{m_\pi^2f_\pi^2}{(f_a/N_{\rm DW})^2}\frac{m_um_d}{(m_u+m_d)^2}\left\{ 1+\frac{m_\pi^2}{m_\eta^2}\left[-1+\mathcal{O}\left(1-\frac{m_\pi}{m_\eta}\right) \right] \right\}\, .
\ee

The masses of the mesons are known~\cite{pdg}, and the $\eta'$ is substantially heavier than the $\pi$. If the masses were the same, the quantum effects would cancel, and the axion would be massless. QCD non-perturbative effects are responsible for lifting the $\eta'$ above the $\pi$. Any non-perturbative physics will do the job, but it happens that the lifting is due to the same instantons that are responsible for the $\theta$-vacua. This is why we say that \emph{QCD instantons give mass to the axion for $T<\Lambda_{\rm QCD}$}. The non-perturbative effects break the axion shift symmetry down to the discrete shift symmetry, $\phi\rightarrow\phi+2\pi f_a/N_{\rm DW}$, and the axion is a \emph{pseudo Nambu-Goldstone boson} (pNGB).

The axion potential generated by QCD instantons is
\be
V(\phi)=m_u\Lambda_{\rm QCD}^3\left[1-\cos \left(\frac{N_{\rm DW}\phi}{f_a}\right)\right] \, .
\label{eqn:qcd_instanton_potential}
\ee
The cosine form comes from the dependence of the vacuum energy on $\theta_{\rm QCD}$ in the lowest order instanton calculation~\cite{1978PhRvD..17.2717C}, and I have applied a constant shift such that $V$ is minimized at zero, i.e. I have assumed a solution to the cosmological constant problem. The instanton potential given here is the zero temperature potential: we will discuss temperature dependence in Section~\ref{sec:qcd_misalignment}, as it is important when computing the axion relic abundance.

QCD is not the only non-abelian gauge theory in the standard model, there is also $SU(2)$ in the EW sector, and $SU(2)$ instantons also contribute to the axion potential. The weak force breaks $CP$, and the $SU(2)$ instantons lead to a shift in the minimum of the axion potential away from the $CP$-conserving value. The instanton action for a gauge group with coupling $g_i$ is~(this is typical of non-perturbative effects, and can be seen e.g. via dimensional transmutation~\cite{1978PhRvD..17.2717C})
\be
S_{\rm inst.}=\frac{8\pi^2}{g_i^2}\, .
\label{eqn:gauge_inst}
\ee
This action sets the co-efficient in front the axion potential from a given sector as $V_i(\theta)\propto \cos\theta e^{-S_{\rm inst.}(g_i)}$. Taking $g=g_{EW}\ll g_3$ we see that the potential from W-bosons only weakly breaks $CP$ compared to the QCD term. For more details, see Ref.~\cite{Kim:1986ax}.

We have so far discussed instantons and non-perturbative physics in the standard model, but the story can be extended to encompass general pNGBs, including ALPs. The steps are:
\begin{itemize}
\item There is a global $U(1)$ symmetry respected by the classical action.
\item Spontaneous breaking at scale $f_a$ leads to an angular degree of freedom, $\phi/f_a$, with a shift symmetry.
\item The $U(1)$ symmetry is anomalous and explicit breaking is generated by quantum effects (instantons etc.), which emerge with some particular scale, $\Lambda_a$. Because of the classical shift symmetry, these effects must be non-perturbative.
\item Since $\phi$ is an angular degree of freedom, the quantum effects must respect the residual shift symmetry $\phi\rightarrow \phi+2n\pi f_a$.
\end{itemize}
In this picture a pNGB or ALP obtains a periodic potential $U(\phi/f_a)$ when the non-perturbative quantum effects ``switch on.'' The mass induced by these effects is $m_a\sim \Lambda_a^2/f_a$.

\subsection{Couplings to the Standard Model}
\label{sec:sm_couplings}

The couplings of the QCD axion are computed in Ref.~\cite{Srednicki:1985xd}. Other references include Refs.~\cite{Kim:1986ax,1998PhRvD..58e5006K,1990eaun.book.....K}. 

The QCD axion is defined to have coupling strength unity to $G\tilde{G}$, via the term in Eq.~\eqref{eqn:qcd_topological}, replacing $\theta_{\rm QCD}\rightarrow \phi/(f_a/N_{\rm DW})$. Any ALP must couple more weakly to QCD (e.g. Ref~\cite{2014PhLB..737...30B}), and in any case a field redefinition can often define the QCD axion to be the linear combination that couples to QCD, leaving ALPs free of the QCD anomaly. 

Axion couplings to the rest of the standard model are defined by symmetry, and in specific models can be computed in EFT. The axion is a pseudoscalar Goldstone boson with a shift symmetry, so all couplings to fermions must be of the form
\be
\partial_\mu (\phi/f_a) (\bar{\psi}\gamma^\mu\gamma_5\psi) \, .
\label{eqn:fermion_coupling}
\ee 
The form of this coupling, as an axial current, means that the force mediated by axions is \emph{spin-dependent} and only acts between spin-polarised sources  (see Section~\ref{sec:axion_forces}). Thus \emph{no matter how light the axion, it transmits no long-range scalar forces between macroscopic bodies}. This has the important implication that, in an astrophysical setting, ULAs are not subject to the simplest fifth-force constraints like light scalars such as (non-axion) quintessence are. 

For example, in the DFSZ model, a coupling of the form Eq.~\eqref{eqn:fermion_coupling} is obtained from the $H\bar{\psi}\psi$ term after symmetry breaking and a PQ rotation, with the value of the co-efficient set by the PQ charge of the fermions. Such a term is generated at one loop in the KSVZ model. 

A coupling to EM of the form: 
\be
\phi \vec{E}\cdot\vec{B}=-\phi F_{\mu\nu}\tilde{F}^{\mu\nu}/4
\ee
is generated if there is an EM anomaly (see below). 

On symmetry grounds we can write a general interaction Lagrangian, applicable at low energies (after PQ symmetry breaking and non-perturbative effects have switched on):
\be
\mathcal{L}_{\rm int}=-\frac{g_{\phi\gamma}}{4}\phi F_{\mu\nu}\tilde{F}^{\mu\nu}+\frac{g_{\phi N}}{2m_N}\partial_\mu \phi (\bar{N}\gamma^\mu\gamma_5 N)+\frac{g_{\phi e}}{2m_e}\partial_\mu \phi (\bar{e}\gamma^\mu\gamma_5 e) -\frac{i}{2} g_d \phi \bar{N}\sigma_{\mu\nu}\gamma_5 NF^{\mu\nu}\, ,
\label{eqn:interaction_lagrangian}
\ee
where $\sigma^{\mu\nu}=\frac{i}{2}[\gamma^\mu,\gamma^\nu]$, and here $N$ is a nucleon (proton or neutron). The coupling $g_{\phi \gamma}$ has mass-dimension $-1$ and is proportional to $1/f_a$; the coupling $g_d$ has mass dimension $-2$ and is also proportional to $1/f_a$. The couplings $g_{\phi e}$ and $g_{\phi N}$ are dimensionless in the above conventions, but are related to commonly-used dimensionful couplings $\tilde{g}_{\phi e,N}=g_{\phi e,N}/(2 m_{e,N})\propto 1/f_a$. Notice how all dimensionful couplings are suppressed by $1/f_a$, which is a large energy scale. This is why axions are weakly coupled, and evade detection. Note the similarity to the suppression of quantum-gravitational effects by $1/M_{pl}$. 

In generic ALP models the couplings to the standard model are taken as free parameters that and can be very much less than they are in the QCD case if, e.g., they are loop suppressed, or forbidden on symmetry grounds. In specific models, the couplings of ALPs can be computed~(e.g. Refs.~\cite{2014JHEP...06..037D,2015arXiv151001701K}).

Expressions for all standard model couplings of the QCD axion can be found in, e.g. Ref.~\cite{1990eaun.book.....K} (though the notation differs slightly). The EDM coupling, $g_d$, is discussed in Ref.~\cite{2013PhRvD..88c5023G}. In this section, we will only discuss the two-photon coupling in detail, following Ref.~\cite{1998PhRvD..58e5006K}. We define:
\be
g_{\phi\gamma}=\frac{\alpha_{\rm EM}}{2\pi (f_a/\mathcal{C})}c_{\phi\gamma}\, ,
\ee
where $\alpha_{\rm EM}\approx 1/137$ is the EM coupling constant and $c_{\phi\gamma}$ is dimensionless. The dimensionless coupling obtains contributions from above the chiral symmetry breaking scale, via the EM anomaly, and below the chiral-symmetry breaking scale, by mixing with the longitudinal component of the $Z$-boson~\cite{Srednicki:1985xd}:
\be
c_{\phi\gamma}=\frac{\mathcal{E}}{\mathcal{C}}-\frac{2}{3}\cdot\frac{4+m_u/m_d}{1+m_u/m_d}\, ,
\label{eqn:c_phi_gamma}
\ee
where $\mathcal{E}$ is the EM anomaly:
\be
\mathcal{E}=2\text{Tr } \mathcal{Q}_{\rm PQ}\mathcal{Q}_{\rm EM}^2\, ,
\ee
and $\mathcal{Q}_{\rm EM}$ are the EM charges

We see clearly here how the KSVZ and DFSZ models differ. In KSVZ we only have the heavy $Q$ fields with PQ charge, and so the value of $c_{\phi\gamma}$ is fixed by the EM charge assigned to this field. Model dependence in KSVZ occurs if we introduce additional heavy quarks with PQ and EM charges. In the DFSZ model, all the standard model fermions carry PQ charges. Model dependence in DFSZ occurs because the coupling depends on the lepton PQ charges, i.e. whether $H_u$ or $H_d$ gives mass to the leptons.  If $H_u$ gives mass to the leptons, $c_{\phi\gamma}$ also depends on the ratio of Higgs vevs, $\tan \beta = \langle H_u\rangle/\langle H_d\rangle$. 

The QCD axion has certain canonical choices for the model dependence. For KSVZ one takes a single EM neutral $Q$ field. For DFSZ the $H_d$ gives mass to the leptons, allowing for $SU(5)$ unification. For $m_u/m_d=0.6$, the couplings are then:
\be
c_{\phi\gamma}=-1.92\,\, \text{(KSVZ)};\quad c_{\phi\gamma}=0.75 \,\,\text{(DFSZ)}.
\label{eqn:photon_dfsz_ksvz}
\ee

\subsection{Axions in String Theory}
\label{sec:string_models}

As is well known, string theory requires the existence of more spacetime dimensions than our usual four: 10 in the case of the critical superstring, and 11 in the case of M-theory~\cite{1987cup..bookQ....G,1987cup..bookR....G,2007stmt.book.....B}. The additional spacetime dimensions must be ``compactified," that is, rolled up and made compact, with a small size. Typically, for appropriate phenomenology containing some unbroken SUSY and chiral matter, the compact manifold must be ``Calabi-Yau"~\cite{1985NuPhB.258...46C}. The supergravity description of string theory contains antisymmetric tensor fields: for example, the antisymmetric partner of the metric, $B_{MN}$, is present in all string theories. 

\begin{figure}
\begin{center}
\includegraphics[scale=0.45]{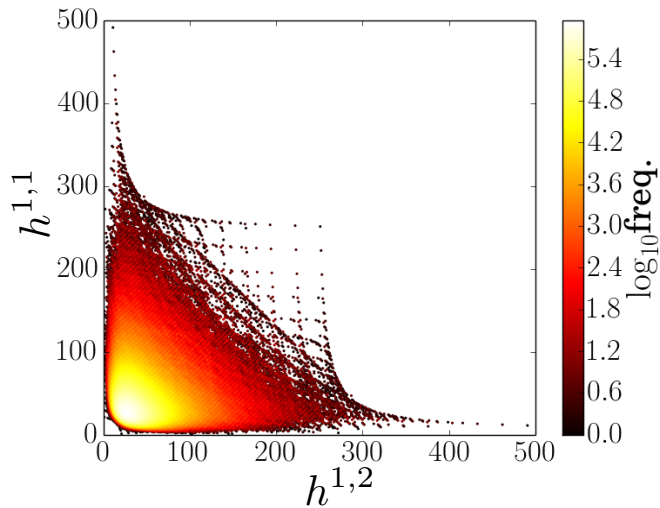}
\caption{The distribution of Hodge numbers $h^{1,1}$ and $h^{1,2}$ for the known Calabi-Yau manifolds in the Kreuzer-Skarke~\cite{2000math......1106K} list. Note that the frequency (=number of occurrences) color scale is logarithmic. There is a huge peak in the distribution at $h^{1,1}\approx h^{1,2}\approx 30$, which implies that a compactification picked at random from this list is most likely to contain of the order of 30 axions.}
\label{fig:kreuzer_skarke}
\end{center}
\end{figure}

Axions arise as the Kaluza-Klein (KK) zero modes of the antisymmetric tensors on the Calabi-Yau~\cite{1984PhLB..149..351W}. The number of axions present depends on the topology of the compact manifold, and in particular is determined by its Hodge numbers. Many Calabi-Yau manifolds are known to exist, and the distribution peaks for Hodge numbers in the dozens~\cite{2013IJMPA..2830032H}, as shown in Fig.~\ref{fig:kreuzer_skarke} for the Kreuzer-Skarke~\cite{2000math......1106K} list. Furthermore, axions arising in this way are massless to all orders in perturbation theory thanks to the higher-dimensional gauge invariance. The axions then obtain mass by non-perturbative effects, such as instantons. Thus axions, with symmetry properties similar to those axions in field theory that we have already discussed, are an extremely generic prediction of string theory, in the low-energy four-dimensional limit~\cite{2006JHEP...06..051S}. This scenario has come to be known as the \emph{string axiverse}~\cite{axiverse}.\footnote{Of course, there are many subtleties, and not all the axions present in the spectrum may survive to low energies. I defer to the references for discussion of this topic.}

Let's flesh out the discussion above with some simple examples and observations. I will use notation for forms, which can be found in e.g. Ref.~\cite{2004sgig.book.....C}. 

A $(p+1)$-form field strength $F_{p+1}$ appears in the action as:
\be
S\supset-\frac{1}{2}\int F_{p+1}\wedge\star F_{p+1}= -\frac{1}{2 (p+1)!}\int d^Dx\sqrt{-g_D} F_{\mu_1\cdots\mu_{p+1}}F^{\mu_1\cdots\mu_{p+1}}  \, ,
\label{eqn:form_action}
\ee
where $D$ is the number of spacetime dimensions, and $g_D$ is the $D$-dimensional metric determinant. The equation of motion is $\mathrm{d}F=0$, implying $F_{p+1}$ can be written as $F_{p+1}=\mathrm{d}A_{p}$, since $\mathrm{d}^2=0$ (this is just like the EM field strength and the usual vector potential). A general solution which is homogeneous and isotropic in the large dimensions is found by decomposing the potential $A$ into the basis of harmonic $p$-forms, $\omega_{p,i}$, on the compact manifold:
\be
A_p=\frac{1}{2\pi}\sum a_i(x)\omega_{p,i}(y)\, \Rightarrow a_i = \int_{C_{p,i}} A_p \, ,
\label{eqn:form_decomposition}
\ee
where $C_{p,i}$ are $p$-cycles in the compact space, $x$ are co-ordinates in the large $3+1$ dimensions, $y$ are co-ordinates in the compact space, and for symmetry under $CP$, $a_i(x)$ is a pseudoscalar.

The sum in Eq.~\eqref{eqn:form_decomposition} runs over the number of harmonic forms, and expresses the topologically distinct ways that $F$ can be ``wrapped'' on the compact space. The number of basis $p$-forms is determined by the number of homologically non-equivalent $p$-cycles, i.e. by the $p^{\rm th}$ Betty number, $b_p$. For example, taking the decomposition Eq.~\eqref{eqn:form_decomposition} for the two-form $B$ mentioned above, we would count the number of two-cycles, and for the $C_{4}$ four-form of Type IIB theory, we would count the number of four-cycles.\footnote{Take a simple example in non-string theory jargon. Imagine a vector field, $A_\mu$ with field strength $F_{\mu\nu}$ in 3+1 large dimensions, and a two dimensional compact space in the shape of a doughnut (or two-torus). There are two distinct ways the vector field can wrap the doughnut: along the tube, or all the way around. These are the distinct one-cycles of the torus. The vector field has co-ordinates in the large dimensions also, but if these are to be homogeneous and isotropic, the only dependence can be as a (pseudo)scalar expressing how wrapping varies from place to place. The two fields necessary are the axions: the KK zero-modes of the $A$ field wrapped on the one-cycles.} For a Calabi-Yau three-fold (three complex dimensions, six real dimensions), all the $b_p$ are determined by the two Hodge numbers $h^{1,1}$ and $h^{1,2}$ (see, e.g., Chapter 9 of Ref.~\cite{2007stmt.book.....B}, and Fig.~\ref{fig:kreuzer_skarke} above).

The axions of Eq.~\eqref{eqn:form_decomposition} are closed string axions. Each closed string axion is partnered into a complex field $z_i=\sigma_i+ia_i$ where $\sigma_i$ is a scalar modulus (saxion) field controlling the size of the corresponding $p$-cycle. The moduli come from KK reduction of the Ricci scalar as usual, and their pairing with axions is a consequence of SUSY, which demands the existence of the appropriate form fields in supergravity. Open string axions also exist in string theory, and are more like the field theory axions we discussed previously. Open string axions live on spacetime filling branes supporting gauge theories and are the phases of matter fields, $\varphi$, which break global PQ symmetries. Open string axions might be related to closed string axions by gauge/gravity duality~\cite{ads_cft,1998AdTMP...2..253W}.

We have just seen the basics of how string theory gives rise to axions and moduli, the number of which is determined by the topology of the compact space. Next we must ask what determines the spectrum of axion masses and decay constants.

After KK reduction of Eq.~\eqref{eqn:form_action} the $a_i(x)$ fields are found to be massless, i.e. there are only kinetic terms for them in the action, implying a shift symmetry. The shift symmetry descends from the higher-dimensional gauge invariance of $F$, and so is protected to all orders in perturbation theory. 

In Type IIB theory, the axion kinetic term resulting from KK reduction of the action for the $C_4$ four-form potential is (for the full axion action in Type IIB theory, see e.g. Ref.~\cite{2012arXiv1209.2299R})
\be
S\supset -\frac{1}{8}\int \mathrm{d}a_i \mathcal{K}_{ij}\wedge\star\mathrm{d}a_j \, , 
\ee
where $\mathcal{K}_{ij}$ is the K\"{a}hler metric,
\be
\mathcal{K}_{ij} = \frac{\partial^2 K}{\partial\sigma_i\partial\sigma_j}\, ,
\ee
and $K$ is the K\"{a}hler potential, which depends on the moduli. KK reduction kinetically mixes the axions and couples them to the moduli via the K\"{a}hler metric. Canonically normalizing the kinetic terms and diagonalizing the K\"{a}hler metric, we see that it is the moduli that determine the axion decay constants, since the canonical kinetic term is $\mathcal{L}_{\rm kin.}=-f_{a,i}^2(\partial a_i)^2/2 $. In particular we have that, parametrically,\footnote{I have assumed that the size of the cycle is of order the size of the manifold. See Refs.~\cite{2006JHEP...06..051S,2006hep.th....7086S} for more details.}
\be
f_{a,i}\sim \frac{M_{pl}}{\sigma_i}\lesssim M_{pl}\, ,
\label{eqn:string_f_scaling}
\ee
where the dimensionless modulus $\sigma_i$ measures the volume of the corresponding $p$-cycle in string units, i.e. $\sigma_i={\rm Vol}_i/l_s^p$, for string length $l_s$. The volume should be larger than the string scale in order for the effective field theory description to be valid, giving the inequality. This may be related to be a general feature, known as the ``weak gravity conjecture,'' following from properties of black holes~\cite{2007JHEP...06..060A}.\footnote{The relation of the conjecture to axion decay constants is only well formulated in the case of a single axion. Consider, for example, the two axion model of Ref.~\cite{2016arXiv160100647C} has a decay constant $\sim 3.25 M_{pl}$. Our simplistic description here has ignored the phenomenon of alignment~\cite{2005JCAP...01..005K,Bachlechner:2014hsa}.} We return to this question in the context of inflation in Section~\ref{sec:axion_inflation}.

Axions in string theory can obtain potentials from a variety of non-perturbative effects~(see e.g. Refs.~\cite{2006JHEP...06..051S,axiverse,2006hep.th....7086S,1988NuPhB.306..890G}). In general, instantons provide a contribution to the superpotential, $W$ for the axion field $a=\phi/f_a$:
\be
W=M^3 e^{-S_{\rm inst.}+ia}\, ,
\ee
where $S_{\rm inst.}$ is the instanton action and $M$ is the scale of instanton physics, which in string theory may be the Planck scale. If SUSY is broken at a scale $m_{\rm SUSY}$ then the axion potential at low energies is 
\be
V(\phi)=\Lambda_a^4 [1-\cos (\phi/f_a)] \, \text{ with } \Lambda_a^4=m_{\rm SUSY}^2 M_{pl}^2e^{-S_{\rm inst.}}\, .
\label{eqn:v_sinst_string}
\ee

A non-Abelian gauge group has instantons with action given by Eq.~\eqref{eqn:gauge_inst}. In string theory, the moduli couple to the gauge kinetic term for a non-Abelian group realized by a stack of $D$-branes wrapping the corresponding cycle, and the gauge coupling $g^2\propto 1/\sigma$ (this occurs e.g. in Type IIB theory for gauge theory on a stack of $D7$ branes filling 3+1 spacetime and wrapped on the same four-cycles as $C_4$). Thus, if an axion obtains mass from these instantons as above, we find that \emph{the axion mass scales exponentially with the cycle volumes}: 
\be
m_a^2 \sim \frac{\mu^4}{f_a^2}e^{-\#\sigma_i} \, ,
\label{eqn:string_m_scaling}
\ee
where $\mu$ is a hard scale. In general, from the above, we expect $\mu=\sqrt{m_{\rm SUSY}M_{pl}}$. If the moduli are stabilised by perturbative SUSY breaking effects giving $m_\sigma\sim m_{\rm SUSY}\gg m_a$ then the moduli can be set to constant values at late times in cosmology and the axion mass will be a constant (for dynamical moduli as dark energy, see Refs.~\cite{2011PhRvD..83l3526M,2012PhRvD..86b3508M}).

The two observations, Eqs.~(\ref{eqn:string_f_scaling},\ref{eqn:string_m_scaling}), form the key basis for the phenomenology of the axiverse. Thanks to the exponential scaling of the potential energy scale with respect to the moduli, string axions will have masses spanning many orders of magnitude. The axion decay constants will (generally) be parametrically smaller than the Planck scale, and are expected to span only a small range of scales due to the power-law scaling with the moduli. 

Let's end this discussion with a few examples of explicit string theory constructions displaying the above properties. The so-called ``model independent axion'' in heterotic string theory emerges from compactification of $B_{MN}$ on two-cycles. It has decay constant $f_a=\alpha_{\rm GUT}M_{pl}/2\sqrt{2}\pi$ and the shift symmetry of the axions is broken by wrapped NS-5 branes with $S_{\rm inst.}=2\pi/\alpha_{\rm GUT}$~\cite{2006JHEP...06..051S}. Gauge coupling unification at $\alpha_{\rm GUT}=1/25$ gives $f_a\sim 1.1\times 10^{16}\text{ GeV}$. 

The \emph{M-theory axiverse}~\cite{acharya2010a} is realized as a compactification of M-theory on a $G_2$ manifold, with axions arising from the number of three-cycles. The $G_2$ volume is small, fixing one heavy string-scale axion by leading non-perturbative effects, and giving $f_a\approx 10^{16}\text{ GeV}$. The remaining axions obtain potentials from higher order effects, and are hierarchically lighter. Fixing the GUT coupling requires that an additional axion take a mass $m_{a,{\rm GUT}}\approx 10^{-15}\text{ eV}$. The other axions in the theory will be distributed around these characteristic values according to the scalings we have discussed.

The \emph{Type IIB axiverse}~\cite{2012JHEP...10..146C} is a LARGE volume Calabi-Yau compactification~\cite{2005JHEP...03..007B,2005JHEP...08..007C}, with axions arising from $C_4$ as discussed above. At least two axions are required in this scenario, one of which is the almost-massless volume-axion associated to the exponentially large volume-modulus, and the other is again associated to the GUT coupling. The volume, $\mathcal{V}$, is exponentially large in string units and gives the decay constant of the volume-axion as $f_a\approx 10^{10}\text{ GeV}$. Other light axions are associated to perturbatively fixed moduli, since they must obtain masses only from higher order effects. Larger values of the effective decay constant for very light axions with $m_a\sim H_0$ can be achieved in this scenario by alignment~\cite{2014JCAP...08..012C}.

\section{Production and Initial Conditions}

\subsection{Symmetry Breaking and Non-Perturbative Physics}
\label{sec:non-pert}

Let's briefly review the general picture for axions given in the previous section, highlighting how this is relevant to axion cosmology in the very early Universe. Two important physical processes determine this behaviour. Symmetry breaking occurs at some high scale, $f_a$, and establishes the axion as a Goldstone boson. Next, non-perturbative physics becomes relevant, at some temperature $T_{\rm NP}\ll f_a$, and provides a potential for the axion.

Giving substance to this chain of events: the axion field, $\phi$, is related to the angular degree of freedom of a complex scalar, $\varphi=\chi e^{i\phi/f_a}$. The radial field, $\chi$, obtains the vev $\langle\chi\rangle=f_a/\sqrt{2}$ when a global $U(1)$ symmetry is broken (see Fig.~\ref{fig:symm_break}). The field $\chi$ is heavy, and $f_a$ is the PQ symmetry breaking scale. The axion is the Goldstone boson of this broken symmetry , and possesses a shift symmetry, $\phi\rightarrow \phi+\text{const.}$, making it massless to all orders in perturbation theory. Non-perturbative effects, for example instantons, ``switch on'' at some particular energy scale and break this shift symmetry, inducing a potential for the axion, $V(\phi)$. The potential must, however, respect the residual discrete shift symmetry, $\phi\rightarrow \phi +2n\pi f_a/N_{\rm DW}$, for some integer $n$, which remains because the axion is still the angular degree of freedom of a complex field. The potential is therefore \emph{periodic}.

The scale of non-perturbative physics is $\Lambda_a$ and the potential can be written as $V(\phi)=\Lambda_a^4 U(\phi/f_a)$, where $U(x)$ is periodic, and therefore possesses at least one minimum and one maximum on the interval $x\in [-\pi,\pi]$. We can choose the origin in field space such that $U(x)$ has its minimum at $x=0$.\footnote{When $x\neq 0$ is associated to the breaking of $CP$ symmetry, as is the case for the QCD axion, a theorem of Vafa and Witten \cite{Vafa:1984xg} guarantees that the induced potential has a minimum at the $CP$-conserving value $x=0$.} It is common practice to assume a solution to the cosmological constant problem such that the minimum is also obtained at $U(0)=0$ (see Section~\ref{sec:dark_energy} for further discussion). A particularly simple choice for the potential is then
\be
V(\phi)=\Lambda_a^4\left[1-\cos \left(\frac{N_{\rm DW}\phi}{f_a}\right)\right] \, ,
\label{eqn:cosine_potential}
\ee
where $N_{\rm DW}$ is an integer, which unless otherwise stated I will set equal to unity. I stress that the potential Eq.~\eqref{eqn:cosine_potential} is not unique and without detailed knowledge of the non-perturbative physics it cannot be predicted. For example, so-called ``higher order instanton corrections" might appear, as $\cos^n\phi/f_a$ (see e.g. Ref.~\cite{2008JCAP...08..003D}). The form of the potential given by Eq.~\eqref{eqn:cosine_potential} is, however, a useful benchmark for considering the form of axion self-interactions. 

We can study axions in a model-independent way if we consider only small, $\phi<f_a$, displacements from the potential minimum. In this case, the potential can be expanded as a Taylor series. The dominant term is the mass term:
\be
V(\phi)\approx \frac{1}{2}m_a^2\phi^2 \, ,
\label{eqn:mass_potential}
\ee
where $m_a^2=\Lambda_a^4/f_a^2$. The symmetry breaking scale is typically rather high, while the non-perturbative scale is lower. The axion mass is thus parametrically small. 

Let's consider some possible values for these scales. The QCD axion (see Section~\ref{sec:qcd_axion}) is the canonical example, where we have that $\Lambda_a^4\approx\Lambda_{\rm QCD}^3m_u$ with $\Lambda_{\rm QCD}\approx 200\text{ MeV}$ and $m_u$ the u-quark mass, and $10^9\text{ Gev}\lesssim f_a \lesssim 10^{17}\text{ GeV}$. The lower limit on $f_a$ comes from supernova cooling~\cite{1996PhRvL..77.2372G,1996PhLB..383..439B} (see Section~\ref{sec:stellar_astro}), while the upper limit comes from black hole superradiance~\cite{2015PhRvD..91h4011A} (BHSR, see Section~\ref{sec:superradiance}). This leads to an axion mass in the range $4\times 10^{-10}\text{ eV}\lesssim m_{a,{\rm QCD}}\lesssim 4\times 10^{-2}\text{ eV}$. 

In string theory models (see Section~\ref{sec:string_models}), things are much more uncertain. The decay constant typically takes values near the GUT scale, $f_a\sim 10^{16}\text{ GeV}$~\cite{2006JHEP...06..051S}, though lower values of $f_a\sim 10^{10-12}\text{ GeV}$ are possible~\cite{2012JHEP...10..146C}. In specific, controlled, examples one always finds $f_a\lesssim M_{pl}$ for individual axion fields. The ``weak gravity conjecture" places some constraints on realising super-Planckian decay constants within quantum gravity~\cite{2007JHEP...06..060A}.\footnote{Collective behaviour of multiple axion fields further complicates matters. We will return to this topic in Section~\ref{sec:axion_inflation}. A large literature surrounds the question of super-Planckian axions in string theory, see e.g. Refs.~\cite{1998PhRvD..58f1301L,2008JCAP...08..003D,2003JCAP...06..001B,2015arXiv150907049K,2015arXiv150400659B,2015arXiv150603447H,2015arXiv150304783B}, and references therein.} The potential energy scale in string models depends exponentially on details of the compactification, and large hierarchies between the non-perturbative scale and the string scale can easily be achieved. Explicitly, $\Lambda_a\sim\mu e^{-\sigma}$, where $\mu$ is the hard non-perturbative scale (e.g. SUSY breaking), and $\sigma$ is a modulus field describing the size of the compact dimensions in string units: small changes in $\sigma$ produce large changes in $\Lambda_a$ for fixed $\mu$. String models are expected to possess a large number of axions, with each axion associated to a different modulus. String axions thus have a mass spectrum spanning a vast number of orders of magnitude from the string scale down to zero. In particular, string models can realise a spectrum such as Eq.~\eqref{eqn:axion_mass_range}.

The axion mass is protected from quantum corrections, since these all break the underlying shift symmetry and must come suppressed by powers of $f_a$. For the same reason, self-interactions and interactions with standard model fields are also suppressed by powers of $f_a$ (for the self-interactions, we can see this easily by expanding the cosine potential to higher orders). This renders the axion a light, weakly interacting, long-lived particle. These properties are protected by a symmetry and as such the axion provides a \emph{natural} candidate to address cosmological problems that can be solved using a light scalar field. Axions can be used to drive inflation, to provide DM, and to provide DE. 

Taking only the mass term from the potential for simplicity, the homogeneous component of the axion field obeys the equation
\be
\ddot{\phi}+3H\dot{\phi}+m_a^2\phi=0 \, .
\label{eqn:axion_background}
\ee
This is the equation of a simple harmonic oscillator with time dependent friction determined by the Friedmann equations, Eqs.~\eqref{eqn:friedmann}. In general, the axion mass will be temperature dependent, as the non-perturbative effects switch on. We will study this equation in detail in Section~\ref{sec:cosmo_field}. An important stage in the evolution of the axion field is the transition form over-damped to under-damped motion, which occurs when $H\sim m_a$, and the axion field begins oscillating. 

\subsection{The Axion Field During Inflation}
\label{sec:axion_during_inflation}

This section refers explicitly to DM axions as a spectator fields during inflation.\footnote{I assume a standard, single-field, slow-roll inflationary model throughout these notes, as it gives us a concrete setting for performing calculations and comparing to data. I further assume (for the most part) that the Universe is radiation dominated from the end of inflation, and in particular when $V(\phi)$ switches on. The general principles, however, can be used as a guide for computing in non-standard cosmologies. The important aspects to consider are: when does symmetry breaking occur with respect to the epoch when initial conditions are set; what is the energy scale at which initial conditions are set; what dominates the energy density when the non-perturbative physics giving rise to $V(\phi)$ becomes relevant?} Inflation driven by an axion field is discussed in Sec.~\ref{sec:axion_inflation}.

The temperature of the Universe during inflation is given by the Gibbons-Hawking~\cite{1977PhRvD..15.2738G} temperature (Hawking radiation emitted from the de-Sitter horizon):
\be
T_I=\frac{H_I}{2\pi} \, ,
\ee
where $H_I$ is the inflationary Hubble scale. This temperature determines whether the PQ symmetry is broken or unbroken during inflation, with each scenario giving rise to a different cosmology. 

The inflationary Hubble scale is tied to the value of the tensor-to-scalar ratio, $r_T$:
\be
\frac{H_I}{2\pi}= M_{pl} \sqrt{A_sr_T/8} \, .
\ee
where $A_s$ is the scalar amplitude. Ever since the observation of the first acoustic peak in the CMB~\cite{1999ApJ...524L...1M,2000Natur.404..955D,2000ApJ...545L...5H}, we have known that $r_T<1$ and that cosmological fluctuations are dominantly scalar and adiabatic, with $\sqrt{A_s}\sim 10^{-5}$ first measured by COBE~\cite{1992ApJ...396L...1S}. This sets, very roughly, $H_I\lesssim 10^{14}\text{ GeV}$. The most up-to-date constraints come from the combined analysis of \emph{Planck} and BICEP2~\cite{2015PhRvL.114j1301B}, which give $A_s=2.20\times 10^{-9}$, $r_T<0.12$ and thus
\be
\frac{H_I}{2\pi}<1.4\times 10^{13}\text{ GeV} \, .
\ee 
High scale single-field slow-roll inflation has observably large tensor modes, $r_T\gtrsim 10^{-3}$, and requires super-Planckian motion of the inflaton~\cite{1997PhRvL..78.1861L}. We will discuss the importance of CMB tensor modes to axion phenomenology in more detail in Section~\ref{sec:tensors}. 

\subsubsection{PQ symmetry unbroken during inflation, $f_a<H_I/2\pi$}
\label{sec:unbroken_PQ}

This scenario occurs when $f_a<H_I/2\pi$. A large misalignment population of ULA DM (our main focus in these notes) requires $f_a\sim 10^{16}\text{ GeV}$, and so this scenario is irrelevant to that model. This is an important scenario for the QCD axion, however, since it applies to the ADMX~\cite{2010PhRvL.104d1301A} sensitivity range of $f_a\sim 10^{12}\text{ GeV}$ in the case of high scale standard inflation. 

During inflation, fluctuations induced by the Gibbons-Hawking temperature are large enough that the $U(1)$ symmetry is unbroken and $\varphi$ has zero vev. After inflation, the symmetry breaks when the radiation temperature drops below $f_a$. At this point, $\chi$ obtains a vev and each causally disconnected patch picks a different value for $\phi/f_a=\theta_{\rm PQ}$. Since the decay constant is larger than the scale of non-perturbative physics, the axion has no potential at this time, and $\theta_{\rm PQ}$ thus has no preferred value. Therefore, in each Hubble patch $\theta_{\rm PQ}$ is drawn at random from a uniform distribution on $[-\pi,\pi]$. The horizon size $R\sim 1/H$ when the PQ symmetry is broken. The symmetry is broken in the early Universe, and the present day Universe is made up of many patches that had different initial values of $\theta_{\rm PQ}$.

Given the $\theta_{\rm PQ}$ distribution, it is possible to compute the average value of the square of the axion field, $\langle\phi^2\rangle$. As we will see later, this value fixes the axion relic density produced by vacuum realignment in this scenario (see Sections~\ref{sec:populations} and \ref{sec:misalignment}). However, it is clear that there are $\mathcal{O}(1)$ fluctuations in the axion field from place to place on scales of order the horizon size when non-perturbative effects switch on ($R\sim 10\text{ pc}$ today for the QCD axion). These large fluctuations have been conjectured to give rise to so-called ``axion miniclusters" \cite{1988PhLB..205..228H}. Fluctuations of this type are non-adiabatic, but are \emph{not scale invariant} and give rise to additional power only on scales sub-horizon at PQ symmetry breaking. 

The breaking of global symmetries gives rise to topological defects. A broken $U(1)$ creates axion strings, while having $N_{\rm DW}>1$ in Eq.~\eqref{eqn:cosine_potential}, as in the DFSZ QCD axion model, gives rise to domain walls. When the PQ symmetry breaks after inflation, a number of such defects will remain in the present Universe. Domain walls, if stable, are phenomenologically disastrous, since their energy density scales like $1/a^2$ and they can quickly dominate the energy density of the Universe~\cite{Sikivie:1982qv}. They can be avoided if $N_{\rm DW}=1$ in Eq.~\eqref{eqn:cosine_potential}, which is possible in the KSVZ axion model, although other mechanisms to avoid their disastrous consequences exist (e.g. Ref.~\cite{2014PhRvL.113x1301B}). Cosmic strings have a host of additional phenomenology. Perturbations seeded by strings and the decay of domain walls may lead to the existence of heavy axion clumps~\cite{1983PhRvL..50..928S}. For our purposes, the most important impact of axion strings is that their decay can source a population of relic axions, which is discussed below.

The important phenomenological aspects of the unbroken PQ scenario are:
\begin{itemize}
\item \emph{The average (background) initial misalignment angle is not a free parameter:} $\langle\theta_{a,i}^2\rangle=\pi^2/3$.
\item \emph{Phase transition relics are present. Their consequences must be dealt with.}
\item \emph{Existence of axion miniclusters?}
\end{itemize}

\subsubsection{PQ symmetry broken during inflation, $f_a>H_I/2\pi$}
\label{sec:broken_PQ}

This scenario occurs when $f_a>H_I/2\pi$. It is particularly relevant for GUT scale axions, and all axion DM models combined with low-scale inflation.

As in the previous scenario, PQ symmetry breaking establishes causally disconnected patches with different values of $\theta_{\rm PQ}$, and produces topological defects. However, the rapid expansion during inflation dilutes all the phase transition relics away.\footnote{Recall that one of the original motivations for inflation was as a solution to the monopoloe problem of GUT theories~\cite{1981PhRvD..23..347G,1982PhLB..108..389L,1982PhRvL..48.1220A}.} It also stretches out each patch of $\theta_{\rm PQ}$, so that our current Hubble volume began life at the end of inflation with a single uniform value of $\theta_{\rm PQ}$ everywhere. This initial value of $\theta_{\rm PQ}$ is completely random. It is again drawn from a uniform distribution, but the existence of many different Hubble patches means that values of $\theta_{\rm PQ}$ arbitrarily close to zero or $\pi$ cannot be excluded, except on grounds of taste or anthropics.

Fluctuations in $\theta_{\rm PQ}$, which later seed structure formation with axion DM, are generated in two different ways in this scenario. Firstly, as we will show in Section~\ref{sec:perturbation_theory}, the axion field has a gravitational Jeans instability. Axion DM will fall into the potential wells established by photons in the radiation era (which were in turn established by quantum fluctuations during inflation). This leads to adiabatic fluctuations.

The second source of axion fluctuations are inflationary isocurvature modes. When the PQ symmetry is broken during inflation, the axion exists as a massless field (or in any case, one with $m_a\ll H_I$). All massless fields in de Sitter space undergo quantum fluctuations with amplitude
\be
\delta\phi=\frac{H_I}{2\pi} \, .
\label{eqn:axion_field_iso}
\ee
The amplitude of the power spectrum of these perturbations is proportional to $r_T$. In de Sitter space, the power spectrum would be scale invariant. Slow roll inflation imparts a red tilt. The isocurvature spectral index is the same as the tensor spectral index, and is also fixed by $H_I$ via inflationary consistency conditions.

Just like tensor modes, DM isocurvature perturbations of this type do not give rise to a large first acoustic peak in the CMB, and are thus constrained to be sub-dominant. The latest \emph{Planck} constraints give $A_I/A_s<0.038$~\cite{2015arXiv150202114P}. As we will discuss in detail in Section~\ref{sec:tensors}, this typically forbids the compatibility of $f_a\gtrsim 10^{11}\text{ GeV}$ axion DM and an observably large $r_T$. 

Isocurvature perturbations also give rise to a backreaction contribution to the homogeneous field displacement (see e.g. Ref.~\cite{1992PhRvD..45.3394L})
\begin{align}
\langle \phi^2_i \rangle &= f_a^2\theta_{a,i}^2+\langle\delta\phi^2\rangle \, , \nonumber \\
&=f_a^2\theta_{a,i}^2+(H_I/2\pi)^2 \, .
\end{align}
The backreaction sets a minimum value to the misalignment population of axions that can be significant in high scale inflation for heavier ALPs, $m_a\gtrsim 10^{-12}\text{ eV}$, and the QCD axion.

The important phenomenological aspects of the broken PQ scenario are:
\begin{itemize}
\item \emph{The average (background) initial misalignment angle is a free parameter, with a minimum value fixed by backreaction.} 
\item \emph{Isocurvature perturbations are produced. Their consequences must be dealt with.}
\item \emph{Use as a probe of inflation?}
\end{itemize}

\subsection{Cosmological Populations of Axions}
\label{sec:populations}

The relic density of axions is $\rho_a=\Omega_a\rho_{\rm crit}$. In cosmology we often discuss the physical density, $\Omega_a h^2$, by factoring out the dimensionless Hubble parameter, $h$, from the critical density. This gives $\rho_a=\Omega_a h^2 \times (3.0\times 10^{-3}\text{ eV})^4$.

A relic axion population can be produced in a number of different ways. The four principle mechanisms are:
\begin{itemize}
\item Decay product of parent particle.
\item Decay product of topological defect.
\item Thermal population from the radiation bath.
\item Vacuum Realignment.
\end{itemize}
I will discuss the first three briefly here, but leave most of the details to the references. Vacuum realignment is discussed in detail in Section~\ref{sec:misalignment}.

\subsubsection{Decay Product of Parent Particle}
\label{sec:direct_decay}

A massive particle, $X$, with $m_X>m_a$, is coupled to the axion field, and decays, producing a population of relativistic axions. If the decay occurs after the axions have decoupled from the standard model then they remain relativistic throughout the history of the Universe. In this case, axions are dark radiation (DR). In cosmology, DR is parameterised via the ``effective number of relativistic neutrinos,'' $N_{\rm eff}$, defined as:
\be
\rho_r=\rho_\gamma \left[ 1+\frac{7}{8}\left(\frac{4}{11}\right)^{4/3}N_{\rm eff}\right] \, .
\label{eqn:neff_def}
\ee
Recall that three species of massless neutrinos in the standard model of particle physics contribute $N_{\rm eff}=3.04$, the additional 0.04 being contributed by heating after $e^+e^-$ annihilation~\cite{2002PhLB..534....8M}.

Assuming instantaneous decay of the parent particle when it dominates the energy density of the Universe gives:\footnote{If the parent particle does not dominate the energy density of the Universe when it decays, then under certain circumstances it may act as a curvaton~\cite{2001PhLB..522..215M,2002NuPhB.626..395E,2002PhLB..524....5L} and sources correlated isocurvature perturbations, which are also constrained by the CMB. See, e.g., Ref.~\cite{2014PhRvD..89j3513I}.}
\be
\Delta N_{\rm eff}=\frac{43}{7}\left( \frac{10.75}{g_{\star S}(T_r)} \right)^{1/3}\frac{B_a}{1-B_a}\, ,
\ee
where $T_r$ is the reheating temperature of the decay of the parent particle, $B_a$ is the branching ratio to axions, and $g_{\star S}(T)$ is the entropic degrees of freedom. The evolution of  $g_{\star,S}(T)$ in the standard model can be computed or can be looked up, e.g. in the Review of Particle Physics \cite{2014ChPhC..38i0001O}.

DR can affect the CMB in a number of ways; for a concise description, see Ref.~\cite{2013PhRvD..87h3008H}. If we hold the angular size of the sound horizon fixed (compensating the change in matter radiation equality with a different Hubble constant or DE density), the main effect of DR is to cause additional damping of the high-multipole acoustic peaks in the CMB.\footnote{Recent constraints on $N_{\rm eff}$ in Ref.~\cite{2015arXiv150307863F} have separated the damping tail effect from the neutrino anisotropic stress, which changes the angular scale of the higher acoustic peaks (see also constraints on neutrino viscosity in Ref.~\cite{planck_2015_params}).} This damping tail is well measured by \emph{Planck}, ACT and SPT, giving $N_{\rm eff}=3.15\pm 0.23$ from a representative combination of CMB data~\cite{planck_2015_params}. $N_{\rm eff}$ is also constrained by big bang nucleosynthesis (BBN, again see Ref.~\cite{planck_2015_params}). Whether this should be combined with the CMB constraint depends on whether the decay producing the axions occurred before or after BBN. An important point to note about neutrino constraints form the CMB is that they do not care whether the DR is a boson or a fermion. We discuss more consequences of axionic dark radiation in Section~\ref{sec:axion_dr}.

A scenario in which axions are produced in this way arises in models with SUSY and extra dimensions. The DR ``cosmic axion background'' is thus considered a generic prediction of many string and M-theory compactifications, and it has a rich phenomenology (see e.g. Refs.~\cite{acharya2010a,2013JHEP...07..005H,2013PhRvD..87d3520C,2013JHEP...10..214C} and Sections~\ref{sec:axion_dr} and \ref{sec:x-ray_bg} of this review). In these models, a K\"{a}hler modulus, $\sigma$, of the compact space comes to dominate the energy density of the Universe after inflation, leading to an additional matter dominated era and a non-thermal history. The modulus must decay and reheat the Universe to a temperature above $T_{\rm BBN}\sim 3\text{ MeV}$, since BBN does not occur successfully in a matter dominated universe.\footnote{This is the ``cosmological moduli problem,'' see e.g. Refs.~\cite{1983PhLB..131...59C,2008JHEP...06..064A}.} Moduli are gravitationally coupled and are therfore expected to have comparable branching ratios to hidden and visible sectors, and in particular have a large branching ratio to axions, since axions are partnered to moduli by SUSY. The modulus decay rate is given by its mass, $\Gamma_\sigma\sim m_\sigma^3/M_{pl}^2$ and it decays when $H\sim \Gamma_\sigma$. Decay before BBN requires $m_\sigma\gtrsim 10\text{ TeV}$. Moduli are thus much heavier than axions, and their decay produces a sizeable relativistic axion population, surviving from before BBN until today. 

\subsubsection{Decay Product of Topological Defect}
\label{sec:defects}

The breaking of global symmetries leads to the formation of topological defects. In the case of a global $U(1)$ symmetry, like the PQ symmetry, this means global (axionic) strings and (if $N_{\rm DW}>1$) domain walls. In the broken PQ scenario, topological defects and their decay products are inflated away, and can be ignored, so here we focus on the unbroken PQ scenario. Axion strings decay, producing a population of cold axions, which we discuss below. The energy density in domain walls scales like $\rho_{\rm DW}\sim a^{-2}$ and can quickly dominate the energy density of the Universe, with phenomenologically disastrous results. Thus $N_{\rm DW}>1$ models (like the DFSZ model) typically require the broken PQ scenario, or some other mechanism to remove the domain walls (see e.g. Ref.~\cite{2014PhRvL.113x1301B} and references therein). In this Section I give only the briefest overview of axion production from topological defects: see e.g. Refs.~\cite{1990eaun.book.....K,2008LNP...741...19S,2012PhRvD..85j5020H} for more details.

Let's focus on strings. Strings are formed by the ``winding'' of the $\theta$ angle. The value of the $\theta$ angle is set independently at each point in space when the PQ symmetry breaks. The Goldstone nature of $\theta$ homogenizes this value in each horizon volume. As the horizon grows, the homogenized area grows. However, in different horizon volumes, $\theta$ will be different. Then, if the $\theta$ angle undergoes a winding around any given point in space, the mapping between $\theta$ and the spatial co-ordinates does not allow a continuous unwinding, leading to a string-like topological defect along the length of the region enclosed by the winding. Formation of topological defects in cosmology in this manner is known as the Kibble mechanism~\cite{1976JPhA....9.1387K}.

Strings in cosmology enter into a ``scaling solution,'' caused by strings within any horizon volume cutting themselves into loops. During the radiation dominated epoch, this requires the string energy density to scale as:
\be
\rho_{\rm string}\propto \mu_{\rm string}/t^2 \, , \quad \mu_{\rm string} \sim f_a^2 \ln (f_a d)\, ,
\ee
where $\mu_{\rm string}$ is the energy per unit length of the axion string, and $d$ the characteristic distance between strings. For global strings, this scaling symmetry is maintained by the continuous emission of axions. The change in the number density of axions, $n_a$, per entropy density, $s$, per Hubble time, required for this is~\cite{1990eaun.book.....K}:
\be
\Delta (n_a/s)\sim \frac{\mu_{\rm string} t^2}{\omega T^3}\Delta (H t)
\label{eqn:string_decay_differential}
\ee
where $\omega$ is the average energy of the radiated axion.

Recall from Eq.~\eqref{eqn:axion_background} that the axion field begins oscillating when $m_a\sim H$, which occurs at a temperature $T_{\rm osc.}$, and depends on the temperature evolution of the axion mass (we discuss this in more detail for the misalignment population of axions in Section~\ref{sec:cosmo_field}). When oscillations commence, axion strings become the boundaries of domain walls connected by strings. For $N_{\rm DW}=1$, these walls can be ``unzipped'' by the strings (as explained in Ref.~\cite{2008LNP...741...19S}), and the decay of the topological defects is complete. Therefore, the total number of axions produced by string decay in a comoving volume is given by the integral of Eq.~\eqref{eqn:string_decay_differential} from the time of the PQ phase transition at $T=f_a$ up to $T_{\rm osc}$:
\be
\frac{n_a}{s}\sim\int_{T_{\rm osc}}^{f_a} \frac{ \mu_{\rm string}d T}{\omega (T) M_{pl}^2}\, .
\ee

Axions produced by string decay are dominated by the low-frequency modes, making them non-relativistic and contributing as CDM to the cosmic energy budget. Accurate computation of the relic density requires numerical simulation of the PQ phase transition and decay of axion strings in order to determine the energy spectrum, $\omega (T)$. Results of such simulations are commonly expressed as the ratio of axion energy density produced by topological defect decay compared to that produced by misalignment:
\be
\Omega_a h^2=\Omega_{a,{\rm mis}} h^2 (1+\alpha_{\rm dec.}) \, .
\label{eqn:defect_decay}
\ee 

For the specific case of the QCD axion, with known temperature dependence of the mass, the value of $\alpha_{\rm dec}$ is calculated.\footnote{As we will show shortly, the contribution from misalignment, $\Omega_{a,{\rm mis}} h^2$, has a particular scaling with $f_a$ for the QCD axion. Quoting a constant value for $\alpha_{\rm dec.}$ in the parameterisation Eq.~\eqref{eqn:defect_decay} assumes the same scaling with $f_a$ for the population produced by topological defect decay. Ref.~\cite{Wantz:2009it} show slightly different scalings, but argue that the uncertainty due to mass-dependence is sub-dominant to other uncertainties in the string calculation.} There is a long-standing controversy over what the value of $\alpha_{\rm dec.}$ should be, with quoted values ranging from 0.16 to 186~\cite{1985PhRvD..32.3172D,1987PhLB..195..361H,1994PhRvL..73.2954B,1996PhRvL..76.2203B}, with the true value possibly lying somewhere in between~\cite{2012PhRvD..85j5020H}. 

The uncertainty arises from the form of the spectrum $\omega$. If the radiated axions have the longest wavelengths possible, of order the horizon, then $\omega(t)\sim t^{-1}$~\cite{1985PhRvD..32.3172D}, while if the spectrum $\sim 1/k$ (cut off at the horizon and the string size) then $\omega(t)\sim \ln (f_a t)t^{-1}$~\cite{1987PhLB..195..361H}. These stem from different assumptions about simulating strings. For the QCD axion mass-temperature relation, this factor of $\ln (f_a t_{\rm osc})\sim 70$, with the enhancement occurring for the case where $\omega\sim t^{-1}$ (accounting for the $t$ dependence of $\mu$ with $d\sim t$). The modern direct simulation of the PQ field yields the somewhat intermediate result of Ref~\cite{2012PhRvD..85j5020H}.

This is clearly a very important area of uncertainty in models of high scale inflation and intermediate scale axions that could have consequences for direct detection of the QCD axion. If decay products from topological defects can produce a relic density larger than misalignment ($\alpha_{\rm dec.}\gg 1$), then axions with $f_a$ as low as $10^9\text{ GeV}$ could be relevant DM candidates (see Section~\ref{sec:qcd_misalignment} for quantitative details). Ultimately, if $\alpha_{\rm dec.}$ were too large, then QCD axion DM would be excluded by stellar astrophysics (see Section~\ref{sec:stellar_astro}). Direct detection of low-$f_a$ axions is outside the reach of ADMX, but may be possible with e.g. open resonator searches (see Section~\ref{sec:admx}).

Topological defects also source CMB fluctuations (e.g. Ref.~\cite{1996ApJ...473L...5T}). A cosmic string network generates power on all sub-horizon scales~\cite{1984Natur.310..391K}. Therefore, axion strings only generate power on scales of order the horizon size at string decay. This scale is small, and is not constrained by the CMB power spectrum, but axion strings may source additional power on minicluster scales.

\subsubsection{Thermal Production}

If axions are in thermal contact with the standard model radiation, then mutual production and annihilation can lead to a thermal relic population of axions, just as for massive standard model neutrinos and WIMPs. The couplings of an axion to the standard model are only specified in the case of the QCD axion. Furthermore, generic ALPs are often more weakly coupled to the standard model, or at least to QCD, than the QCD axion. For these reasons, we will consider only the thermal population of the QCD axion. 

Axions are produced from the standard model plasma by pion scattering, and decouple when the rate for the $\pi+\pi\rightarrow \pi+a$ process drops below the Hubble rate. The thermal axion abundance is fixed by freeze-out at the decoupling temperature (see, e.g. Ref.~\cite{1990eaun.book.....K}), with a larger relic density for lower decoupling temperatures. The number density in thermal axions, $n_a$, relative to the photon number density, $n_\gamma$ is given by
\be
n_a=\frac{n_\gamma}{2}\frac{g_{\star,S}(T_0)}{g_{\star,S}(T_D)} \, ,
\ee
with $T_D$ the decoupling temperature, and $T_0$ the CMB temperature today. See Ref.~\cite{2014JCAP...01..011S} for a more complete formula and a computation involving all relevant standard model production channels. Thermal axions contribute to the effective number of neutrinos as $\Delta N_{\rm eff}\approx 0.0264 n_a/n_{a,{\rm eq}}\approx 10 n_a$, with $n_{a,{\rm eq}}$ the thermal equilibrium number density.

Since axion couplings scale inversely with $f_a$, only low $f_a$ (higher mass) thermally produced axions can contribute a significant amount to the energy budget of the Universe. Thermal populations are significant for $m_a\gtrsim 0.15\text{ eV}$, when decoupling occurs after the QCD phase transition (recall that $g_{\star,S}$ reduces dramatically after the QCD phase transition, diluting the abundance of particles produced before it). For the QCD axion respecting $f_a\gtrsim 10^9\text{ GeV}$, as suggested by stellar cooling constraints (ses Section~\ref{sec:stellar_astro}), the thermal population is negligible.

Thermal axions produced in this way are relativistic as long as $T_D>m_a$. Once decoupled the axion temperature, $T_a$, redshifts independently from the standard model temperature, and the axions become non-relativistic when $T_a<m_a$. Thermal axions behave cosmologically in a manner similar to massive neutrinos, and contribute as hot DM, suppressing cosmological structure formation below the free-streaming scale (see Section~\ref{sec:transfers}). Assuming a standard thermal history, current CMB limits from \emph{Planck} on axion hot DM constrain $m_a<0.529 \rightarrow 0.67\text{ eV}$ at 95\% confidence~\cite{2013JCAP...10..020A,2015PhRvD..91l3505D,2015arXiv150708665D} (for older limits from different datasets including large scale structure and WMAP, see Refs.~\cite{2005JCAP...07..002H,2007JCAP...08..015H,2008JCAP...04..019H,2010JCAP...08..001H}). AFuture galaxy redshift surveys will be sensitive enough to detect a thermal axion population for all $m_a\geq 0.15\text{ eV}$~\cite{2015JCAP...05..050A}. Relaxing the assumption of a standard thermal history and introducing an early matter-dominated phase and low temperature reheating relaxes the bound on thermal axions, allowing masses as large as a keV~\cite{2008PhRvD..77h5020G}.

\subsubsection{Vacuum Realignment}

The process of vacuum realignment is a model independent production mode for axions, also known as the \emph{misalignment mechanism}. It relies only on their defining properties (being associated to spontaneous symmetry breaking, and being a pNGB), and depends only on gravitational (and to some extent self-) interactions. This production mode is our primary focus, and is discussed in detail in Section~\ref{sec:misalignment}.

\section{The Cosmological Axion Field}
\label{sec:cosmo_field}

If axions are to have observable effects on cosmology, they must contribute an appreciable amount to the energy density of the Universe. Since the axion mass is so small, this implies large occupation numbers. In this case, axions can be modelled by solving the classical field equations of a condensate. This condensate can have excited states carrying energy and momentum, and indeed it will. There is nothing more mysterious here than using Maxwell's equations to describe the behaviour of electric and magnetic fields. It is also the standard way that scalar field models of inflation and DE are treated. 

It is a separate question to ask whether axions form a Bose-Einstein condensate (BEC), and even to define a ``BEC'' in a cosmological context, where we are certainly not in the ground state. I comment briefly on this in Section~\ref{sec:two_cents}. The results I present below are valid whenever the classical field equations hold, and \emph{do not assume BEC occurs} (except to the extent that it is captured by the classical field equations). Many of the results below also apply to other models of scalar field DM at late times (when oscillations about a quadratic minimum are the only important aspect), though the early time cosmology can be markedly different (e.g. complex fields in Ref.~\cite{2014PhRvD..89h3536L}, which have equation of state $w=1$ at early times).

\subsection{Action and Energy Momentum Tensor}

The action for a minimally coupled real scalar field in General Relativity is:
\be
S_\phi=\int d^4 x \sqrt{-g}\left[ -\frac{1}{2}(\partial\phi)^2-V(\phi) \right] \, .
\ee
For an axion, this action is only valid after symmetry breaking, and after non-perturbative effects have switched on. Before non-perturbative effects have switched on, the axion is massless. Non-perturbative effects do not switch on instantaneously, either, and time (temperature) dependence of the potential can be important. We discuss this shortly, in Section~\ref{sec:misalignment}.

Varying the action with respect to $\phi$ gives the equation of motion
\be
\Box\phi-\frac{\partial V}{\partial \phi}=0 \, ,
\ee
where the D'Alembertian is
\be
\Box=\frac{1}{\sqrt{-g}}\partial_\mu(\sqrt{-g}g^{\mu\nu}\partial_\nu) \, .
\ee
Varying the action with respect to the metric gives the energy momentum tensor
\be
T^\mu_{\,\,\,\nu}=g^{\mu\alpha}\partial_\alpha\phi\partial_\nu\phi-\frac{\delta^\mu_{\,\,\,\nu}}{2}[g^{\alpha\beta}\partial_\alpha\phi\partial_\beta\phi+2V(\phi)]\, .
\label{eqn:axion_em_tensor}
\ee

As we will show below, there are certain limits in which the axion field behaves as a fluid. See Appendix~\ref{appendix:fluid} for useful definitions for the components of the energy momentum tensor in the fluid case.

\subsection{Background Evolution}

The background cosmology is defined in Appendix~\ref{appendix:frw}. Computing the D'Alembertian for the FRW metric and taking $V=m_a^2\phi^2/2$, the axion equation of motion is:
\be
\ddot{\phi}+3H\dot{\phi}+m_a^2\phi=0 \, .
\label{eqn:axion_background_later}
\ee
The background energy density and pressure of the axion field are:
\begin{align}
\bar{\rho}_a&=\frac{1}{2}\dot{\phi}^2+\frac{1}{2}m_a^2\phi^2 \, ,\\
\bar{P}_a&=\frac{1}{2}\dot{\phi}^2-\frac{1}{2}m_a^2\phi^2 \, .
\end{align}

When the universe is matter or radiation dominated the scale factor evolves as a power law, $a\propto t^p$. In this case, Eq.~\eqref{eqn:axion_background_later} has an exact solution:
\be
\phi=a^{-3/2}(t/t_i)^{1/2}[C_1J_n(m_at)+C_2Y_n(m_at)] \, ,
\label{eqn:exact_background}
\ee
where $n=(3p-1)/2$, $J_n(x)$, $Y_n(x)$ are Bessel functions of the first and second kind, and $t_i$ is the initial time. The dimensionful coefficients $C_1$ and $C_2$ are determined by the initial conditions. For axions in the vacuum realigment mode, the initial conditions are well defined when $H(t_i)\gg m_a$:
\be
\phi (t_i) = f_a \theta_{a,i} \, , \quad \dot{\phi}(t_i) = 0 \, .
\label{eqn:misalignment_initial}
\ee

When matter and radiation are both important, such as near matter-radiation equality,\footnote{Recall that in $\Lambda$CDM equality occurs at $z_{\rm eq}\approx 3000$, while the CMB is formed at decoupling, $z_{\rm dec}\approx 1020$. The contribution of radiation to the expansion rate at decoupling cannot be neglected.} or when the axion field can itself dominate the energy density, Eq.~\eqref{eqn:axion_background_later} must be solved either by approximation or numerically. In the case of axion DM produced by the misalignment mechanism, the most useful approximation to solve Eq.~\eqref{eqn:axion_background_later} is the WKB approximation.

\subsection{Misalignment Production of DM Axions}
\label{sec:misalignment}

The misalignment production of DM axions can be computed given the initial conditions of Eq.~\eqref{eqn:misalignment_initial}. At symmetry breaking the Hubble rate is much larger than the axion mass, and the field is overdamped. This sets $\dot{\phi}=0$ initially. The homogeneous value of the field is specified by the scenario for when symmetry breaking occurs with respect to inflation. The term ``misalignment'' refers to this scenario where there is a coherent initial displacement of the axion field, and ``vacuum realignment'' to the process by which this value relaxes to the potential minimum.

An important buzz-word to remember about the misalignment production of DM axions is that it is \emph{non-thermal}.

\subsubsection{Axion-Like Particles}
\label{sec:alp_misalignment}

Let's begin with the simple case of an ALP. Given ignorance of the non-perturbative physics, I will describe such an axion only by its mass, which I take to be constant in time. The general picture described here applies to the QCD axion also. The validity of the constant mass assumption will be discussed later in this subsection.

\begin{figure}
\begin{center}
\includegraphics[scale=0.85]{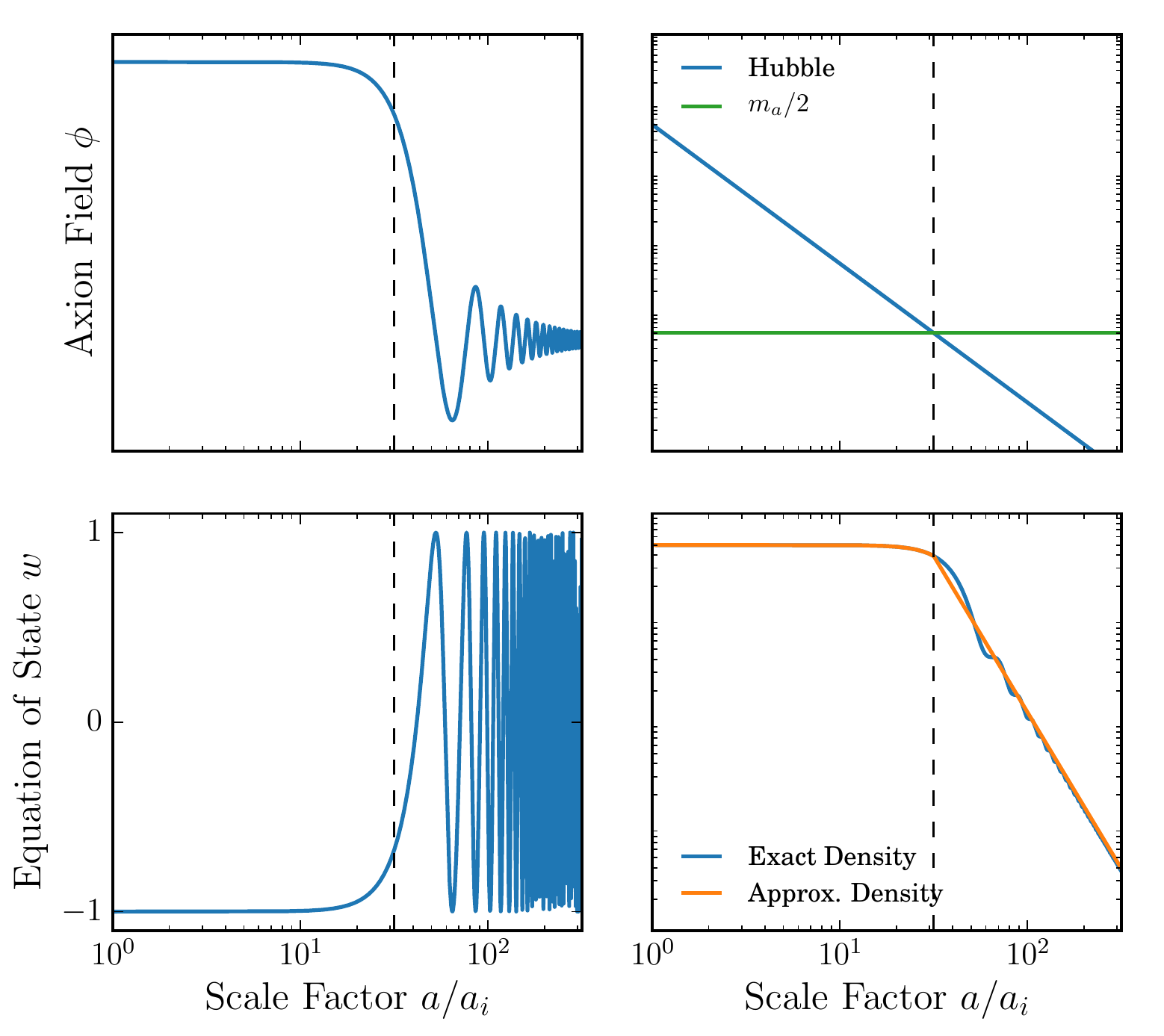}
\caption{Evolution  of various quantities in the exact solution to the background evolution of an ALP, Eq.~\eqref{eqn:exact_background}, for a radiation-dominated universe ($p=1/2$). Dimensionful quantities have arbitrary normalization. Vertical dashed lines show the condition defining $a_{\rm osc.}$. Further discussion of this choice, and the approximate solution for the energy density, is given in the text. }
\label{fig:exact_background}
\end{center}
\end{figure}

The initial condition $\dot{\phi}=0$ fixes the relative values of $C_1$ and $C_2$ in the exact solution to the background evolution, Eq.~\eqref{eqn:exact_background}. The equation of motion is linear, and so the initial field value can be scaled out. Fig.~\ref{fig:exact_background} shows the evolution of the axion field, Hubble rate, axion equation of state, and the axion energy density for the solution Eq.~\eqref{eqn:exact_background} in a radiation-dominated universe ($p=1/2$), with arbitrary normalization of all dimensionful parameters. The scale factor is shown relative to the initial value, $a_i$.

At early times when $H>m_a$, the axion field is overdamped and is frozen at its initial value by Hubble friction. The equation of state at early times is $w_a=-1$, and the axion behaves as a contribution to the vacuum energy. This is why axions can serve as models for DE and inflation. All other components of the Universe scale as $a$ to some negative power. If the axion can come to dominate the energy density while it is still overdamped with $w_a<-1/3$, it can drive a period of accelerated expansion. The length of this period depends on the ratio $H/m_a$ when the axion comes to dominate the energy density, which is in turn fixed by the initial displacement of the field (in inflation, this fixes the values of the slow-roll parameters).

Later, when $H<m_a$, the axion field is underdamped and oscillations begin. The equation of state oscillates around $w_a=0$, and the energy density scales as $\rho_a\propto a^{-3}$. This is the same behaviour as ordinary matter, and is why \emph{misalignment axions are a valid DM candidate.} The Hubble rate at matter-radiation equality in $\Lambda$CDM is approximately $H(a_{\rm eq})\sim 10^{-28}\text{ eV}$. \emph{Axions heavier than this begin oscillations in the radiation dominated era and are suitable candidates to compose all the DM.}

The transition in the axion equation of state can be approximated if we define a fixed value of the scale factor, $a_{\rm osc}$, and simply fix the behaviour of $\rho_a(a)$ at late times to be
\be
\rho_a(a)\approx\rho_a(a_{\rm osc})(a_{\rm osc}/a)^3\, ; \quad (a>a_{\rm osc})\, .
\label{eqn:background_rho_approx}
\ee
Furthermore, the energy density is approximately constant up until $a_{\rm osc}$ and so we can further approximate $\rho_a(a_{\rm osc})\approx m_a^2\phi_i^2/2$. This gives the usual approximation used to calculate axion DM energy density. \emph{The energy density in the misalignment population is fixed by the initial field displacement and the mass alone.}

How shall we define $a_{\rm osc}$? Roughly, it is when $m_a\gtrsim H$, so we can let $AH(a_{\rm osc})=m_a$ for some constant $A>1$. The larger we set $A$ to be, the better the approximation (assuming we compute $\rho_a(a_{\rm osc})$ from the exact solution). However, we must also play this off against the expense of following oscillations in a numerical solution. The equation of motion, Eq.~\eqref{eqn:axion_background}, suggests $A=3$ is as a sensible-looking choice. In the example with a radiation dominated universe, I found $A=3$ leads to a 40\% error in the energy density at late times, with $A=2$ giving a better approximation.\footnote{As already stated, the approximation in general improves as $A$ gets larger. The poor performance at $A=3$ is just because the energy density is falling rapidly at this point and errors are amplified. In this case, 3 is not a lucky number. In numerical solutions including perturbations, taking a larger $A$ will always be better, as the improvement shown here for $A=2$ applies only to the exact background solution.} The approximation Eq.~\eqref{eqn:background_rho_approx} and the location of $a_{\rm osc}$ for $A=2$ are also shown in Fig.~\ref{fig:exact_background}.

In real-Universe examples with a matter-to-radiation transition and late time $\Lambda$ domination, we found in Ref.~\cite{Hlozek:2014lca} that $A=3$ works well in most cases. Using the known solutions in matter and radiation domination for $H(t)$ to fix $a_{\rm osc}$ in terms of other cosmological parameters, this gives the following useful approximation to the ULA fractional energy density as a function of the initial displacement \cite{2010PhRvD..82j3528M}:
\begin{eqnarray}
\Omega_a \approx \left\{ 
\begin{array}{ll} 
\frac{1}{6}(9 \Omega_r)^{3/4} \left( \frac{m_a}{H_0} \right)^{1/2} \left\langle\left( \frac{\phi_{i}}{M_{pl}} \right)^2\right\rangle\mbox{if $a_{\rm osc}< a_{\rm eq}$}\, ,\\
\frac{9}{6}\Omega_m \left\langle\left( \frac{\phi_{i}}{M_{pl}} \right)^2\right\rangle\mbox{if $a_{\rm eq}<a_{\rm osc}\lesssim 1$} \, , \\
\frac{1}{6}\left( \frac{m_a}{H_0} \right)^2 \left\langle\left( \frac{\phi_{i}}{M_{pl}} \right)^2\right\rangle\mbox{if $a_{\rm osc}\gtrsim 1$} \, ,\label{eqn:simpledens}
\end{array}
\right. ,
\label{eqn:full_ula_omega}
\end{eqnarray}
where I have used angle brackets to denote the average homogeneous value, to remind us of the consequences when the PQ symmetry is broken or unbroken during inflation. 

Let's use the WKB approximation to understand the background evolution further. The WKB approximation for $H\ll m_a$ consists of the ansatz solution
\be
\phi(t)=\mathcal{A}(t)\cos (m_at+\vartheta) \, ,
\ee 
where $\vartheta$ is an arbitrary phase, and $\mathcal{A}$ is slowly varying such that $\dot{\mathcal{A}}/m_a\sim H/m_a\sim \epsilon\ll 1$. Plugging this into Eq.~\eqref{eqn:axion_background} and working to leading order in $\epsilon$ gives the solution $\mathcal{A}(a)\propto a^{-3/2}$. Using this solution we find that the energy density simply scales as $\rho_a\propto \mathcal{A}^2\propto a^{-3}$, while $w_a$ has rapid oscillations with frequency $2 m_a$. Consequently, the average equation of state on time scales $t\gg 1/m_a$ is $\langle w_a\rangle_t=0$. This gives a general proof as to why $w_a$ oscillates around zero and $\rho_a\propto a^{-3}$ at late times when $H\ll m_a$, independent of any assumptions about the background evolution being matter or radiation dominated.\footnote{This applies to fields oscillating in a harmonic potential, $V(\phi)\sim \phi^2$. Turner \cite{1983PhRvD..28.1243T} proved the more general result for fields oscillating in an anharmonic potential, $V(\phi)\sim \phi^\alpha$, giving $\rho \propto a^{-6\alpha/(\alpha+2)}$.}

The solution for $\phi$ and $\rho_a$ in the WKB approximation sheds light on the constant-mass assumption we made at the beginning of this section. The magnitude of non-perturbative effects generally varies with temperature, and so the axion mass varies with cosmological time, approaching an asymptotic value for $T\ll T_{\rm NP}$. If the asymptotic value of the mass has been reached before the axion becomes relevant in the energy density and when $a<a_{\rm osc}$ then cosmology will proceed as if we simply take $m_a=m_a(T=0)$ everywhere. Only the quantities evaluated at $a=a_{\rm osc}$ matter. In string models, non-perturbative effects stabilise moduli and break SUSY at high energies, while ULAs oscillate in the post-BBN Universe, with $T_{\rm BBN}\ll T_{\rm SUSY}$. In that context, i.e. ULAs from string theory, constant mass is an excellent approximation.
\begin{figure}
\begin{center}
\includegraphics[width=0.75\textwidth]{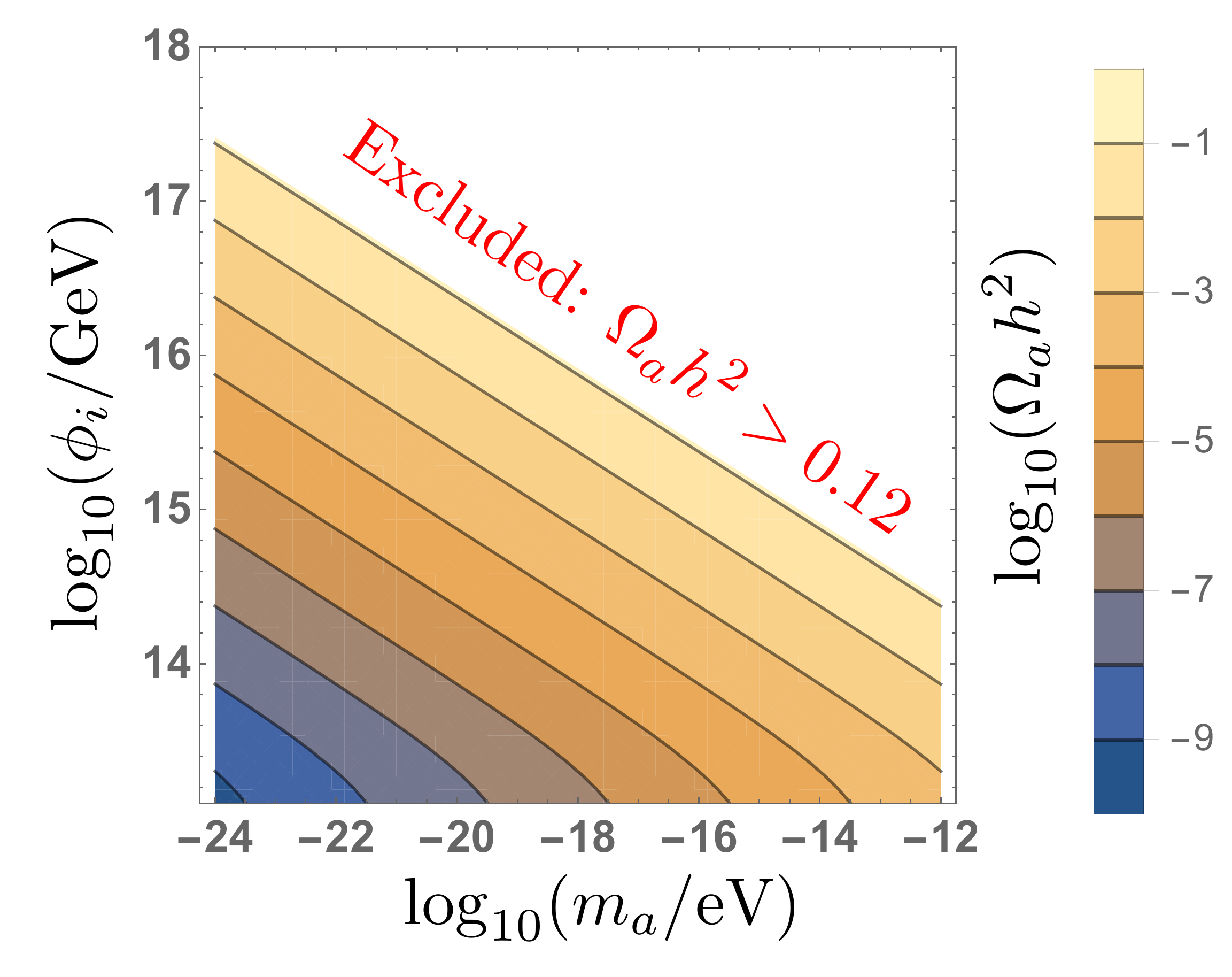}
\caption{ULA relic density from vacuum realignment in the broken PQ scenario with high scale inflation, $H_I\approx 10^{14}\text{ GeV}$. ULAs require $\phi_i>10^{14}\text{ GeV}$ in order to contribute more than a few percent to the DM density. Even with high scale inflation, the contribution of isocurvature backreaction is less than a percent of the total DM across the entire ULA parameter space. See Fig.~\ref{fig:contours_combinedLinear} for more details on the allowed region at lower mass.}
\label{fig:alp_misalignment}
\end{center}
\end{figure}

Fig.~\ref{fig:alp_misalignment} shows $\Omega_a h^2$ in the broken PQ scenario, for ULAs in the range $10^{-24}\text{ eV}\leq m_a\leq 10^{-12}\text{ eV}$ (where $a_{\rm osc}<a_{\rm eq}$ and ULAs are safe from linear cosmological constraints, see Section~\ref{sec:cosmological_constraints}), with $H_I=7.8\times 10^{13}\text{ GeV}$ (the maximum allowed value with $r_T=0.1$) for varying $\phi_i=f_a\theta_{a,i}$. The contribution from $H_I$ backreaction to $\Omega_a h^2$ is less than $10^{-4}$ across the entire range of masses shown: \emph{backreaction of isocurvature perturbations can safely be neglected for all ULAs} and $\langle \phi_i^2\rangle\approx \phi_i^2$ can be taken as a completely free parameter. All ULAs require $\phi_i>10^{14}\text{ GeV}$ in order to contribute more than a few percent to the DM density. Since $\phi_i\lesssim f_a$ and $H_{I,{\rm max}}<10^{14}\text{ GeV}$ this implies that \emph{ULAs should always be considered in the broken PQ scenario.}

The ``anthropic boundary'' for ULAs in string theory is defined as the minimum mass where $\Omega_a h^2=0.12$~\cite{planck_2015_params} can be obtained with $f_a\leq 10^{16}\text{ GeV}$~\cite{axiverse}. Plugging $\phi_i=10^{16}\text{ GeV}$ into Eq.~\ref{eqn:simpledens} gives:
\be
m_a= 5.3\times 10^{-19}\text{ eV}\left(\frac{f_a}{10^{16}\text{ GeV}}\right)^{-4}\, \quad \text{(string anthropic boundary)}\, ,
\label{eqn:anthopic_boundary_DM}
\ee
where I have used $z_{\rm eq}=3400$, $\Omega_c h^2=0.12$, $\Omega_b h^2=0.022$ and $h=0.67$ to fix the radiation density. ULAs heavier than this require (anthropic) tuning of $\phi_i$ if $f_a\sim 10^{16}\text{ GeV}$. ULAs lighter than this require larger decay constants, a large number of individual axions, or some other production mechanism, to contribute a significant amount to the DM density. Since $f_a\leq 10^{16}\text{ GeV}$ is by no means a hard prediction of string theory, it is worth considering the limit of the anthropic boundary for DM-like axions with $m_a= 10^{-24}\text{ eV}$. This is visible in Fig.~\ref{fig:alp_misalignment}, and from the $f_a$ scaling of Eq.~\eqref{eqn:anthopic_boundary_DM}. We find $f_a\leq 4\times 10^{17}\text{ GeV}$: ULA DM is natural for comfortably sub-Planckian values of the decay constant.

\subsubsection{The QCD Axion}
\label{sec:qcd_misalignment}

QCD non-perturbative effects switch on at $T\sim \Lambda_{\rm QCD}\sim 200\text{ MeV}$, precisely when the QCD axion with intermediate $f_a$ begins oscillations. The temperature dependence of the axion mass in QCD is given by:
\be
m_a^2(T)f_a^2=\chi_{\rm top.}(T) \, ,
\ee 
where $\chi_{\rm top.}(T)$ is the QCD topological susceptibility, which must be calculated. The original calculation is due to Ref.~\cite{1981RvMP...53...43G} and is reviewed in e.g. Ref.~\cite{2004hep.th....9059F}, while a modern calculation in the `interacting instanton liquid model' (IILM) is given in Ref.~\cite{Wantz:2009it}.  A simple power-law dependence of the axion mass on temperature applies at high temperatures, $T>1\text{ GeV}$:
\be
m_a^2(T)=\alpha_a\frac{\Lambda_{\rm QCD}^3m_u}{f_a^2}\left(\frac{T}{\Lambda_{\rm QCD}}\right)^{-n}\, .
\ee
This should be matched to the zero temperature value, Eq.~\eqref{eqn:qcd_mass_zero_T}, at low $T\lesssim \Lambda_{\rm QCD}$.

The standard \cite{1981RvMP...53...43G} value for the power-law from the dilute instanton gas model (DIGM) is $n=7+n_f/3+\cdots\approx 8$ (where $n_f$ is the number of fermions active at a given temperature). The fits of Ref.~\cite{Wantz:2009it} from the IILM give $n=6.68$ and $\alpha_a=1.68\times 10^{-7}$ (which also agrees with Ref.~\cite{2008JCAP...09..005B}). The temperature dependence can also be computed non-perturbatively on the lattice in the pure Yang-Mills limit~(e.g. Refs.~\cite{2015PhRvD..92c4507B,2015arXiv150902976B,2015JHEP...10..136K,2016PhLB..752..175B}), and at low temperatures from chiral perturbation theory (for a recent calculation, see Ref.~\cite{2015arXiv151102867G} and references therein). The lattice calculations of Ref.~\cite{2015PhRvD..92c4507B} find $n=5.64$ (compare to the pure Yang-Mills, $n_f=0$, DIGM). Ref.~\cite{2015arXiv151102867G} consider a range between $n=2$ and $n=8$ from lattice and instanton calculations respectively.

The temperature of the Universe in the radiation dominated era is determined by the Friedmann equation in the form
\be
3H^2M_{pl}^2=\frac{\pi^2}{30}g_{\star}T^4 \, .
\ee
Taking the standard $n=8$ result, using that $g_\star=61.75$ for tempertaures just above the QCD phase transition, and defining $3H(T_{\rm osc})=m_a$, the QCD axion with $f_a< 2\times 10^{15}\text{ GeV}$ begins oscillating when $T>1\text{ GeV}$~\cite{2004hep.th....9059F}. From this point on, axion energy density scales as $a^{-3}$ independently of the behaviour of $m_a(T)$. The relic density can thus be reliably computed from the high-temperature power-law behaviour of $m_a(T)$, scaled as $a^{-3}$ from $T_{\rm osc}$. The relic density is fixed by the initial misalignment angle and $f_a$. For $f_a< 2\times 10^{15}\text{ GeV}$ it is given by~\cite{2004hep.th....9059F}
\be
\Omega_a h^2 \sim 2\times 10^4 \left(\frac{f_a}{10^{16}\text{ GeV}} \right)^{7/6} \langle \theta_{a,i}^2\rangle \, .
\label{eqn:relic_qcd}
\ee
For $f_a\gtrsim 2\times 10^{17}\text{ GeV}$ oscillations begin when $T<\Lambda_{\rm QCD}$, such that the mass has reached its zero-temperature value. In this case the relic density is
\be
\Omega_a h^2 \approx 5\times 10^3 \left(\frac{f_a}{10^{16}\text{ GeV}} \right)^{3/2} \langle \theta_{a,i}^2\rangle \, .
\label{eqn:relic_qcd_large_f}
\ee
Note that there is not an overlapping region of validity for Eqs.~\eqref{eqn:relic_qcd} and \eqref{eqn:relic_qcd_large_f}. For $2\times 10^{15}\text{ GeV}\lesssim f_a\lesssim 2\times 10^{17}\text{ GeV}$ oscillations begin during the QCD epoch, the dilute instanton gas approximation breaks down and the relic density calculation is more complicated (see e.g. Refs.~\cite{2004hep.th....9059F,Wantz:2009it,Visinelli:2014twa}). However, it is argued in Ref.~\cite{2004hep.th....9059F} that Eq.~\eqref{eqn:relic_qcd} is a good approximation for $f_a<6\times 10^{17}\text{ GeV}$. For our simple purposes of illustration, we use Eq.~\eqref{eqn:relic_qcd} for all $f_a<M_{pl}$. 

So far, we have computed the relic density using the harmonic potential, $V(\phi)=m_a^2\phi^2/2$. For large initial displacements, $\theta_i\gtrsim 1$, anharmonic corrections caused by axion self-interactions become important. The potential becomes flatter at increased $\theta_{\rm PQ}$, causing the axion field to spend more time with $w_a\approx -1$, thus delaying $a_{\rm osc}$ and increasing the relic abundance relative to the harmonic approximation. Anharmonic effects can be taken into account with a correction factor by replacing
\be
\langle \theta_{a,i}^2\rangle \rightarrow \langle \theta_{a,i}^2 \mathcal{F}_{\rm anh.}(\theta_{a,i})\rangle \, ,
\label{eqn:anharmonic1}
\ee
where $\mathcal{F}_{\rm anh.}(x)\rightarrow 1$ for small $x$ and monotonically increases as $x\rightarrow \pi$. An analytic approximation to $\mathcal{F}_{\rm anh.}(x)$ for the cosine potential is \cite{2009PhRvD..80c5024V}
\be
\mathcal{F}_{\rm anh.}(x)=\left[\ln\left(\frac{e}{1-x^2/\pi^2}\right)\right]^{7/6} \, .
\label{eqn:anharmonic2}
\ee
Note that the use of Eqs.~\eqref{eqn:anharmonic1} and \eqref{eqn:anharmonic2} breaks down if the axion field comes to dominate the energy density, driving a period of inflation, since they rely on the assumption that oscillations begin in a radiation-dominated background.

A full numerical computation of the relic abundance valid for all $f_a$ in the IILM, taking into account the temperature dependence of $g_\star$ in the standard model and all anharmonic effects, is given in Ref.~\cite{Wantz:2009it}.

Axions produced by misalignment behave as DM, and we know that the DM density is $\Omega_c h^2\approx 0.12$. Axions may not be all the DM, but they had better not produce too much of it, so we must have $\Omega_ah^2<0.12$.\footnote{Violating this constraint is sometimes, misleadingly, called ``overclosing the Universe,'' a phrase which dates from before the precision cosmology era, when one simply demanded $\rho_a<\rho_{\rm crit}$ for some approximate value of $H_0$.} Eq.~\eqref{eqn:relic_qcd}, and its anharmonic corrections Eqs.~\eqref{eqn:anharmonic1} and \eqref{eqn:anharmonic2}, inform the classic discussions on the QCD axion and ``natural'' values for $f_a$~\cite{1983PhLB..120..127P,1983PhLB..120..137D,1983PhLB..120..133A,1983PhLB..129...51S}. 

First, let's just take $\langle \theta_{a,i}^2 \rangle$ to be a free parameter, and work out the consequences. High $f_a$ axions produce too much DM unless $\theta_{a,i}\ll 1$. On the other hand, low $f_a$ axions can only produce a fraction of the DM unless $\theta_{a,i}$ is tuned very close to $\pi$ such that anharmonic corrections can boost the relic density. The ``sweet spot'' where $\Omega_a h^2=0.12$ is achieved for $\theta_{a,i}\approx 1$ is at $f_a\approx 3 \times 10^{11}\text{ GeV}$. The range of $f_a$ where $\Omega_a h^2\approx 0.12$ can be achieved with minimal tuning of $\theta_{a,i}$ towards zero or $\pi$ is the region where broken PQ axions are ``natural." It's boundaries clearly depend on taste, but allowing for tuning at the level $10^{-2}$ it is:
\be
8\times 10^9\text{ GeV}\lesssim f_a\lesssim 1\times 10^{15}\text{ GeV} \quad (\text{no tuning, broken PQ}) \, .
\ee
In the unbroken PQ scenario the relic abundance is fixed by taking $\langle \theta_{a,i}^2 \rangle=\pi^2/3$. Keeping $\Omega_a h^2<0.12$ and satisfying bounds from stellar cooling and supernovae defines the \emph{classic axion window}:
\be
1\times 10^9\text{ GeV}\lesssim f_a\lesssim 8.5\times 10^{10}\text{ GeV} \quad (\text{classic axion window, unbroken PQ}) \, .
\ee

Axions with $f_a\gtrsim 10^{15}\text{ GeV}$ are sometimes referred to as living in the \emph{anthropic axion window} \cite{2004hep.ph....8167W,2006PhRvD..73b3505T,hertzberg2008}. It is so-called because although $\theta_{a,i}$ must be tuned small, if it was not small and the DM density was too large, the Universe would not be conducive to the formation of galaxies and life.\footnote{Refs.~\cite{2011JCAP...07..021M,2011JCAP...05..001M} discuss the interesting case of anthropic selection with multiple axion fields. An additional fine-tuning measure is also applied based on isocurvature perturbations (see Section~\ref{sec:tensors}). However, when applied to iscourvature, the measure used in Refs.~\cite{2011JCAP...07..021M,2011JCAP...05..001M} assumes that the inflationary parameter $\epsilon_{\rm inf}$ has a flat prior. A least information (Jeffreys) prior on the unknown physical scale $H_I$ would yield very different conclusions.} Note that the anthropic window is automatically open to high $f_a$ axions, since for $r_T<1$, $f_a\gtrsim 10^{15}\text{ GeV}$ is always in the broken PQ scenario where $\theta_{a,i}$ is a free parameter, although the backreaction contribution may be important depending on the value of $H_I$.
\begin{figure}
\begin{center}
$\begin{array}{@{\hspace{-0.6in}}l@{\hspace{+0.1in}}l}
\includegraphics[scale=0.33]{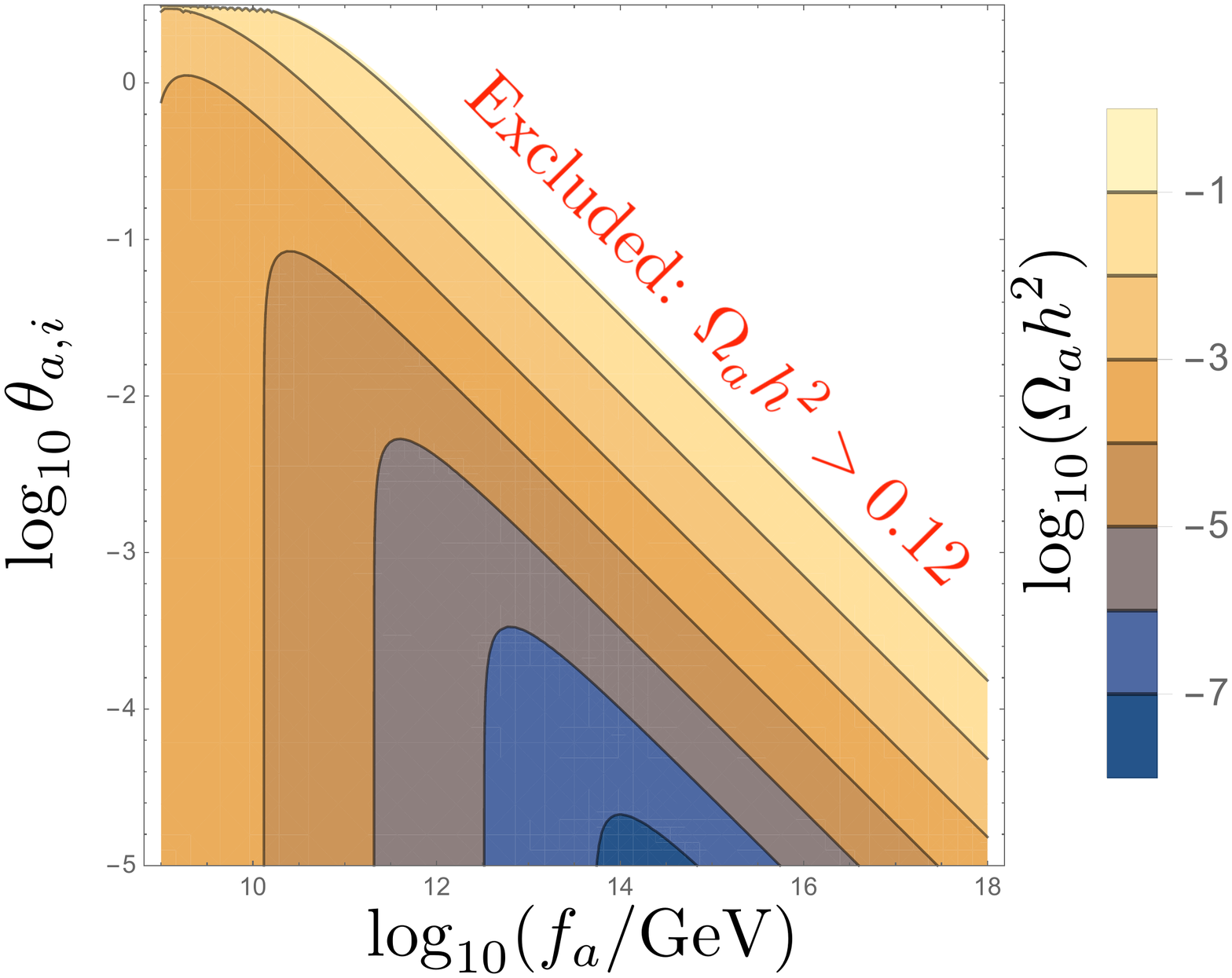}&
\includegraphics[scale=0.33]{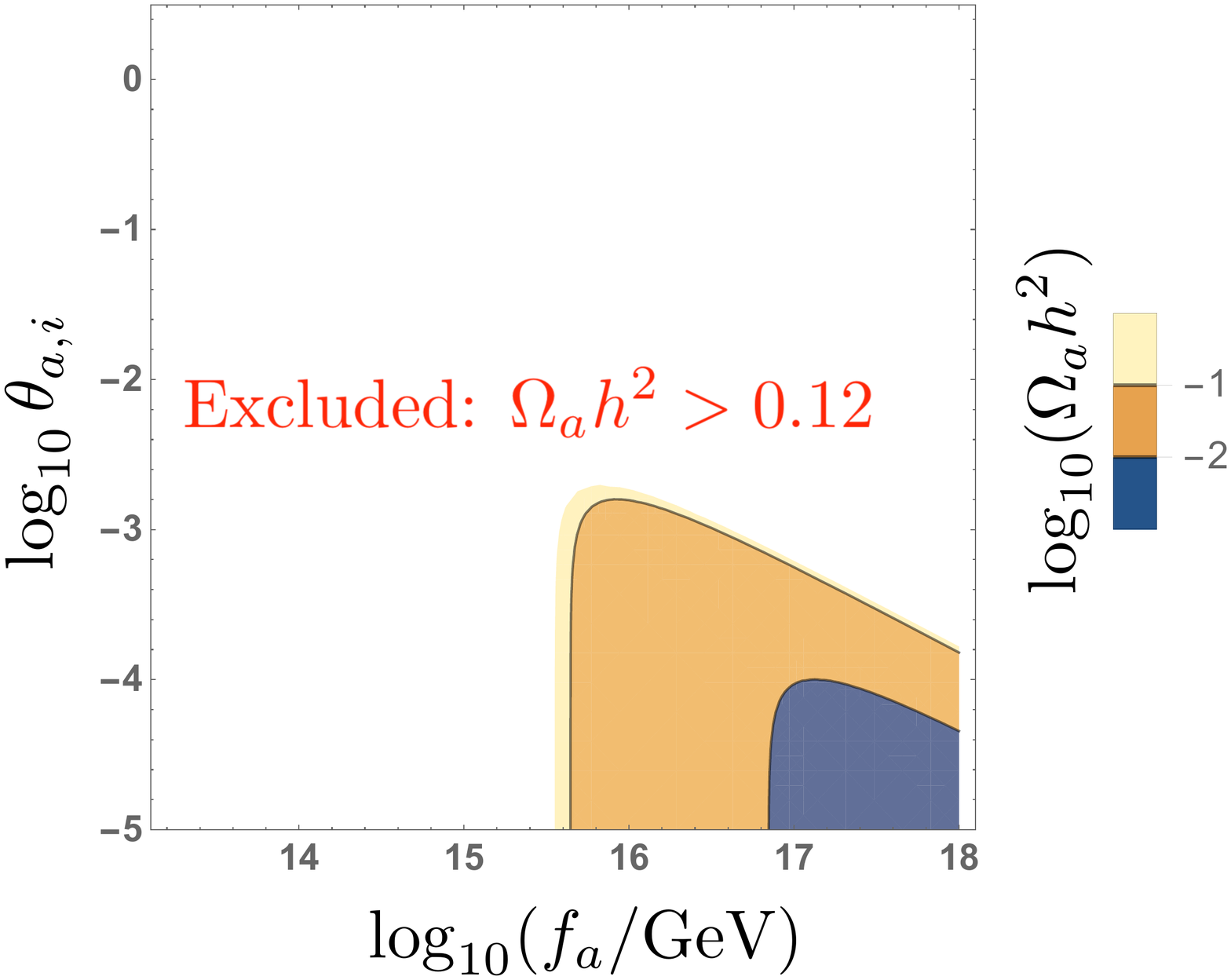}
 \end{array}$
 \end{center}
\caption{QCD axion DM relic density from vacuum realignment in the broken PQ scenario. Isocurvature constraints are ignored, see Fig.~\ref{fig:qcd_iso}. \emph{Left panel}: Low scale inflation, $H_I=2\pi \times 10^9\text{ GeV}$. All of the allowed range of $f_a$ has PQ symmetry unbroken during inflation. Large $f_a$ requires tuning $\theta_{a,i}$ in order not to produce too much DM. \emph{Right Panel:} High scale inflation, $H_I=10^{14}\text{ GeV}$. Backreaction produces too much DM for all $f_a\lesssim 3\times 10^{15}\text{ GeV}$.}
\label{fig:qcd_broken_contour}
\end{figure}
\begin{figure}
\begin{center}
\includegraphics[width=0.75\textwidth]{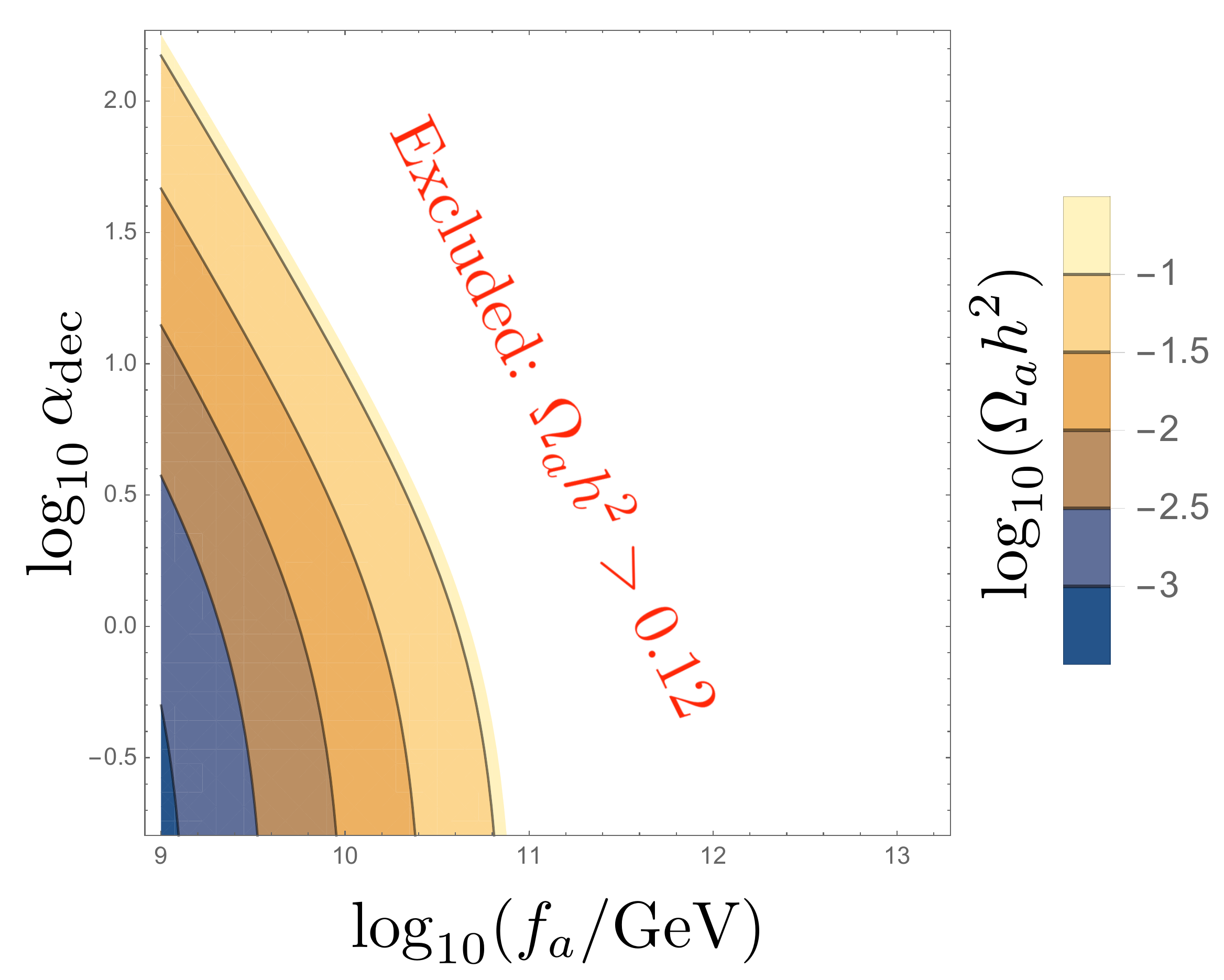}
\caption{QCD axion DM relic density from vacuum realignment in the unbroken PQ scenario. The fixed value of $\langle \theta_{a,i}^2\rangle=\pi^2/3$ excludes all axions with $f_a\gtrsim 9\times 10^{10}\text{ GeV}$ for producing too much DM. The uncertainty in axion production from string decay, reflected in the range for $\alpha_{\rm dec}$, means that all axions with lower $f_a$ can produce a significant contribution to the DM.}
\label{fig:qcd_unbroken_contour}
\end{center}
\end{figure}

Let's bring together everything we know about the QCD axion DM relic density from vacuum realignment into two equations:
\begin{eqnarray}
\Omega_a h^2 \approx \left\{ 
\begin{array}{ll} 
2\times 10^4 \left(\frac{f_a}{10^{16}\text{ GeV}} \right)^{7/6} \frac{\pi^2}{3} \mathcal{F}_{\rm anh.}(\pi/\sqrt{3}) (1+\alpha_{\rm dec})\quad\mbox{(unbroken PQ)}\, ,\\
2\times 10^4 \left(\frac{f_a}{10^{16}\text{ GeV}} \right)^{7/6} (\theta_{a,i}^2+H_I^2/(2\pi f_a)^2) \mathcal{F}_{\rm anh.}\left(\sqrt{\theta_{a,i}^2+H_I^2/(2\pi f_a)^2}\right) \quad\mbox{(broken PQ)} \, . 
\end{array}
\right. 
\label{eqn:qcd_relic}
\end{eqnarray}
For simplicity, as stated above, I am going to assume that Eq.~\eqref{eqn:relic_qcd} holds for all $f_a$ (see the discussion below Eq.~\ref{eqn:relic_qcd_large_f}). See Section~\ref{sec:defects} and \ref{sec:two_cents} for discussion on the difference between the misalignment and topological defect populations.

Fig.~\ref{fig:qcd_broken_contour} is a contour plot of $\Omega_a h^2$ as a function of $f_a$ and $\theta_{a,i}$ for the broken PQ scenario in two different inflation models. The first takes $H_I=2\pi \times 10^9\text{ GeV}$, so that all of the allowed range of $f_a$ has the PQ symmetry broken during inflation. The second scenario takes $H_I=10^{14}\text{ GeV}$, i.e. about as large as it can be without violating current tensor constraints. In the case of low scale inflation, the entire allowed range of $f_a$ can produce the required DM density by vacuum realignment. Large $f_a$ requires tuning of $\theta_{a,i}$ in order to satisfy $\Omega_a h^2<0.12$. In the high scale inflation case, backreaction of isocurvature perturbations leads to too much DM production for $f_a\lesssim 5\times 10^{16}\text{ GeV}$. Large $f_a\gtrsim 5\times 10^{16}\text{ GeV}$ anthropic axions appear compatible with high scale inflation if we allow $\theta_{a,i}$ to be tuned, however we have so far only considered constraints from the relic density, and not from the isocurvature amplitude. We will see in Section~\ref{sec:tensors} that isocurvature constraints imply that high-$f_a$ axions are essentially incompatible with high-scale inflation. 

Fig.~\ref{fig:qcd_unbroken_contour} is a contour plot of $\Omega_a h^2$ as a function of $f_a$ and $\alpha_{\rm dec.}$ in the unbroken PQ scenario. Based on constraints from $r_T$, the largest possible value of $f_a$ in this scenario is $f_a\approx 10^{14}\text{ GeV}/2\pi$, and I allow $\alpha_{\rm dec}\in [0.16,186]$. In the unbroken PQ scenario, the fixed value of $\langle \theta_{a,i}^2\rangle=\pi^2/3$ excludes all axions with $f_a\gtrsim 9\times 10^{10}\text{ GeV}$ for producing too much DM. The possible range of $\alpha_{\rm dec}$ values means that all axions with lower $f_a$ than this have the possibility of providing the correct DM abundance. This defines the classic axion window. Note that if $\alpha_{\rm dec}\gtrsim 200$ then the QCD axion in the unbroken PQ scenario, satisfying astrophysical constraints, would be completely excluded unless the excess DM abundance could be diluted (e.g. by late-time entropy production). This possibility is the source of the controversy over the axion abundance by string decay discussed in Section~\ref{sec:defects}.

\subsection{Cosmological Perturbation Theory}
\label{sec:perturbation_theory}

All specific results here assume that cosmological history begins in the radiation dominated universe after reheating. I work in two gauges: the synchronous gauge and the Newtonian gauge. These gauges, the gauge transformations between them and the equations of motion for matter and radiation, are given in the classic, and endlessly useful, Ref.~\cite{bertschinger1995} (see also Ref.~\cite{2004astro.ph..2060H}).\footnote{As usual in cosmology, note that the adage ``the Russians did it first'' holds very well here. If you are so inclined, you can find everything you need in Landau and Lifschitz~\cite{1971ctf..book.....L_cosmoPT}. Another useful early reference is Ref.~\cite{1984PThPS..78....1K}. I refer explicitly to Ref.~\cite{bertschinger1995} as it addresses specifically the CMB computation.} The Newtonian gauge is useful (obviously) for the Newtonian limit (discussed in more detail in the following subsection). The Newtonian potentials $\Psi$ and $\Phi$ are also transparently related to the gauge invariant curvature perturbation, and to the integrated Sachs-Wolfe (ISW) source terms for the CMB. The synchronous gauge, with potentials $h$ (not to be confused with the reduced Hubble rate, also denoted $h$) and $\eta$, on the other hand, makes the CDM evolution particularly simple, as $\theta_c\equiv 0$. The synchronous gauge is also used by the popular CMB Boltzmann solver \textsc{camb} \cite{camb}. The full treatment of ULAs in the synchronous gauge has been implemented in \textsc{axionCAMB}, described in Ref.~\cite{Hlozek:2014lca}, and soon to be publicly released. Another popular Boltzmann solver is \textsc{class} \cite{2011arXiv1104.2932L,2011JCAP...07..034B}, with a ULA model implemented in Ref.~\cite{2015arXiv151108195U}.

In this section I work primarily in the fluid treatment of axion perturbations. This can be derived from the perturbed field equation. In Fourier space in synchronous gauge this is
\be
\delta\phi''+2\mathcal{H}\delta\phi'+(k^2+m_a^2a^2)\delta\phi=-\frac{1}{2}\phi'h' \, ,
\label{eqn:dphi_sync}
\ee
while in Newtonian gauge it is
\be
\delta\phi''+2\mathcal{H}\delta\phi'+(k^2+m_a^2a^2)\delta\phi=(\Psi'+3\Phi')\phi'-2m_a^2a^2\phi\Psi \, ,
\label{eqn:dphi_newt}
\ee
where primes denote derivatives with respect to conformal time, $ad\tau=dt$ (not to be confused with the optical depth, also denoted $\tau$), and the conformal Hubble rate is $\mathcal{H}=aH$. The perturbed axion field is $\delta\phi$; the background field is $\phi$.

\subsubsection{Initial Conditions}

Initial conditions are set for all modes, $k$, when they are super-horizon $k\ll aH$ and at early times during the radiation era. I present the simplest, zeroth order initial conditions. Corrections to these results can be derived order-by-order in the super-horizon/early-time limit. The computation is described in Ref.~\cite{bucher2000}, with results specific to axions given in Ref.~\cite{Hlozek:2014lca}.

If all cosmological perturbations are seeded by single field inflation, the initial conditions are \emph{adiabatic}. Radiation is the dominant component at early times, and carries the inflationary curvature perturbation. The adiabatic condition relates the overdensity in photons to the overdensity in any other fluid component, $i$:
\be
\delta_i=\frac{3}{4}(1+w_i)\delta_\gamma \, .
\label{eqn:adiabatic_def}
\ee
At early times, the axion equation of state is $w_a\approx -1$ and so $\delta_a=\delta\phi=0$ in the adiabatic mode in the early-time, super-horizon perturbative-expansion.  

This adiabatic initial condition seems very different from the standard CDM adiabatic initial condition where $\delta_c=3\delta_\gamma/4$. That is because we are beginning when axions are not behaving as CDM. As the axion field rolls and begins oscillating around $w_a=0$, the axions begin to cluster and fall into the potential wells set up by the photons. At late times, $a>a_{\rm osc}$, this evolution ``locks on'' to the standard CDM behaviour on large scales, as we will show from numerical results shortly.

Isocurvature initial conditions can be thought of in a number of ways. Commonly, they are thought of as entropy perturbations: i.e. perturbations in relative number density of particles of different species that leave the total curvature unperturbed. An isocurvature perturbation between two species, $i$ and $j$, can be written in a gauge invariant way as (e.g. Ref.~\cite{2009PhR...475....1M} and references therein)
\be
S_{ij}=3(\zeta_i-\zeta_j)
\ee
where $\zeta_i$ is the curvature perturbation due to a single species:
\be
\zeta_i=-\Psi-H\frac{\delta\rho_i}{\dot{\rho}_i} \, .
\ee
The total curvature perturbation is 
\be
\zeta = \frac{\sum_i (\rho_i+P_i)\zeta_i}{\sum_i (\rho_i+P_i)}\, .
\ee

The most useful practical definition for all cosmological initial conditions is to think of them as simply the different normal (eigen) modes of the energy momentum tensor \cite{bucher2000}. One then finds the early time, $\tau\ll 1$, super horizon, $k\tau\ll 1$, expansion for each mode. In the synchronous gauge each mode can be identified by the leading, zeroth order, behaviour of the fluid variables and the metric potentials:
\begin{align}
\eta&=1 \quad \text{(adiabatic mode)} \, , \\
\delta_i&=1 \quad \text{(density isocurvature in species }i)\, , \\
\theta_i &=k \quad \text{(velocity isocurvature in species }i) \, ,
\end{align}
with all other components unperturbed. At higher orders one then selects the growing mode for each component. The correct selection of this is crucial. For example the adiabatic mode has (e.g. Refs.~\cite{bertschinger1995,bucher2000})
\be
\delta_\gamma=-\frac{1}{3}(k\tau)^2 \, ,
\ee
and from the equations of motion one finds the condition Eq.~\eqref{eqn:adiabatic_def} relates this to the other species at each order in the perturbative-expansion, and also accounts for possible evolution of $w_i$ (as is the case for the slowly rolling axion field at early times \cite{Hlozek:2014lca}).

In the axion iscocurvature mode, relevant for the broken PQ scenario, the initial condition is $\delta_a=1$, with all other species unperturbed at zeroth order. The normalization and spectrum can be multiplied afterwards since the equations are linear. The spectrum is a power law, with spectral index $(1-n_I)=2\epsilon_{\rm inf}$ (for inflationary slow-roll parameter $\epsilon_{\rm inf}$, see Section~\ref{sec:axion_inflation}). 

\subsubsection{Early Time Treatment}

At early times, the background equation of motion should be solved numerically to find the evolution of the axion equation of state, $w_a(\tau)$. With this in hand, the background energy density evolves as
\be
\rho'_a=-3\mathcal{H}\rho_a(1+w_a) \, .
\ee
The equation of state also specifies the evolution of the adiabatic background sound speed:
\be
c_{\rm ad}^2 = w_a-\frac{w_a'}{3\mathcal{H}(1+w_a)}\, .
\ee

The second order perturbed equations of motion can be rewritten as two first order equations for the axion overdensity, $\delta_a$ and dimensionless perturbed heat flux, $u_a=(1+w_a)v_a$. The equation of state and adiabatic sound speed specify the background evolution-dependent co-efficients in the equations of motion for the fluid components. Using the result that the sound speed in perturbations, $c_s^2=\delta P_a/\delta \rho_a=1$ in the $\delta\phi=0$ axion comoving gauge, the transformation to fluid variables can be performed exactly \cite{hu1998b}. Performing a gauge transformation to the synchronous gauge, the equations of motion read \cite{Hlozek:2014lca}:
\begin{align}
\delta_a'&=-ku_a-(1+w_a)h'/2-3\mathcal{H}(1-w_a)\delta_a-9\mathcal{H}^2(1-c_{\rm ad}^2)u_a/k \, , \\
u_a'&=2\mathcal{H}u_a+k\delta_a+3\mathcal{H}(w_a-c_{\rm ad}^2)u_a \, .
\end{align} 
I stress that at this stage no approximations have been made. Given the evolution of $w_a(\tau)$ (or equivalently $\phi(\tau)$) the evolution of $\delta_a$ and $u_a$ specify the evolution of $\delta\phi$ (with metric potentials sourced by all species). 

Note that if $\phi'=0$ then $w_a=-1$ and $w_a'=0$. In this case, an adiabatic fluctuation with $\delta\phi=\delta\phi'=0$ in Eq.~\eqref{eqn:dphi_sync} has no source and will not grow. The same holds in the fluid variables: $w_a=-1$ leads to vanishing metric source in the fluid equations, and so if $\delta_a=u_a=0$ initially then this remains so, and no growth occurs.

In this picture, the axions source the Einstein equations with density, pressure and velocity perturbations as
\begin{align}
\delta\rho_a &=\rho_a\delta_a \, ,  \\
\delta P_a&=\rho_a[\delta_a+3\mathcal{H}(1-c_{\rm ad}^2)(1+w_a)u_a/k]\, , \\
\rho_a(1+w_a)v_a&=\rho_au_a \, .
\end{align}

\subsubsection{The Axion Effective Sound Speed}
\label{sec:eff_sound_speed}

When $a>a_{\rm osc}$, $w_a$ and $c_{\rm ad}^2$ oscillate rapidly in time compared to the Hubble scale and all other quantities of interest (e.g. the curvature perturbation evolves on time scales of order $H$). The exact fluid equations now become numerically expensive to solve, and an approximation for the perturbed fluid equations, akin to the $w_a=0$ approximation in the background equations of motion, is necessary.

Consider the general equation of motion for fluids in synchronous gauge \cite{bertschinger1995}:
\begin{align}
\delta'&=-(1+w)(\theta+h'/2)-3\mathcal{H}(c_s^2-w)\delta \, , \nonumber\\
\theta'&=-\mathcal{H}(1-3w)\theta-\frac{w'}{1+w}\theta+\frac{c_s^2}{1+w}k^2\delta\, ,
\label{eqn:general_perturbed_fluid}
\end{align}
where I have only assumed the vanishing of anisotropic stress, which is valid at first order in perturbation theory for a scalar field. The evolution is specified by two quantites: the equation of state, $w$, and the sound speed in perturbations:\footnote{See Appendix~\ref{appendix:fluid} for discussion of different definitions of the scalar field sound speed and the relations between them.}
\be
c_s^2=\frac{\delta P}{\delta \rho} \, .
\ee

For an axion at late times, $a>a_{\rm osc}$, we know how to approximate the time averaged equation of state: $\langle w_a\rangle_t=\langle w_a'\rangle_t=0$ (see Section~\ref{sec:alp_misalignment}). If we can simply find a similar expression for $\langle c_s^2\rangle_t$ \emph{evaluated in the appropriate gauge}, then we can use Eqs.~\eqref{eqn:general_perturbed_fluid} to specify the evolution of the axion overdensity. The pressure source of the Einstein equations due to axions will then be given by $\delta P_a=\langle c_s^2\rangle_t\rho_a\delta_a$.

Just as for the background, we can use the WKB approximation by writing the background field and field perturbation as
\begin{align}
\phi&=a^{-3/2}[\phi_+\cos mt+\phi_-\sin mt] \, , \\
\delta\phi&=\delta\phi_+ \cos mt + \delta\phi_- \sin mt \, ,
\end{align}
where the functions $\delta\phi_{\pm}$ depend on wavenumber $k$ as well as time. It is now possible to find the effective sound sound speed in the gauge comoving with the time-averaged axion fluid (see e.g. Refs.~\cite{2009PhLB..680....1H,2012PhRvD..86h3535P} for the derivation):
\be
\langle c_s^2\rangle_t=c_{s,{\rm eff}}^2=\frac{k^2/4m_a^2a^2}{1+k^2/4m_a^2a^2} \, .
\ee
This effective sound speed is the key to understanding the difference between ULAs and CDM in terms of structure formation.

The metric potentials in the axion comoving gauge are defined in the same way as the synchronous gauge. The gauge transformation between the two gauges induces additional terms to Eqs.~\eqref{eqn:general_perturbed_fluid} that decay on sub-horizon scales~\cite{Hlozek:2014lca}. The axion fluid equations of motion in the synchronous gauge are:
\begin{align}
\delta'_{a}&=-ku_{a}-\frac{h'}{2}-3\mathcal{H}c_{s,{\rm eff}}^{2}\delta_{a}-9\mathcal{H}^{2}c_{s,{\rm eff}}^{2}u_{a}/k,\label{dens_late}\\
u'_{a}&=-\mathcal{H}u_{a}+c_{s,{\rm eff}}^{2}k\delta_{a}+3c_{s,{\rm eff}}^{2}\mathcal{H}^{2}u_{a}.\label{v_late}
\end{align}

\subsubsection{Growth of Perturbations and the Axion Jeans Scale}

So far, we've been very precise and set up the equations of motion and initial conditions as they would be used in numerical Boltzmann equation solver to compute cosmological observables in the real Universe. 

Let's take a step back for a moment to a simplified situation, and consider a Universe dominated by axion DM, and work in the Newtonian gauge. Let's take the sub-horizon limit, so that we can use the Poisson equation in its usual form:
\be
k^2\Psi^2=-4\pi G a^2 \rho \delta
\ee
Gauge transformations on the effective sound speed between the synchronous and Newtonian gauge also vanish in this limit. Combining the equations for $\dot{\delta}_a$ and $\dot{\theta}_a$ into a single second order equation for $\delta_a$, and using the Poisson equation to eliminate the Netwonian potential, gives the equation of motion for $\delta_a$ in physical time:
\be
\ddot{\delta}_a+2H\dot{\delta}_a+(k^2c_{s,{\rm eff}}^2/a^2-4\pi G \rho_a)\delta_a=0 \, .
\label{eqn:delta_newtonian}
\ee
This is the equation for an oscillator with time-dependent mass and friction. The mass term in this equation expresses the competition between density and pressure during gravitational collapse. The origin of the effective sound speed and pressure in the axion equation of motion is scalar field gradient energy.

On large scales, $k^2c_s^2\rightarrow 0$, density wins and \emph{axion DM has a Jeans instability}~\cite{khlopov_scalar}.\footnote{The growth of perturbations for small $k$, despite positive mass-squared for the perturbations in Eqs.~\eqref{eqn:dphi_sync} and \eqref{eqn:dphi_newt}, can be understood from the rapid oscillations in $\phi'$ causing the system to act as a driven oscillator~\cite{hu1998b,2015arXiv150106918A}.} The equation of motion is exactly the same as for CDM, with the usual growing, $\delta_a\propto a$, and decaying, $\delta_a\propto a^{-3/2}$, modes. On small scales, the pressure term dominates over the density, and $\delta_a$ oscillates without growing. 

The scale where density and pressure are in equilibrium and $4\pi G\rho_a=k^2c_s^2$ is known as a the \emph{axion Jeans scale}, and it defines a particular wavenumber, $k_J$. Modes with $k<k_J$ grow, while modes with $k>k_J$ oscillate. The buzz-phrase to remember referring to axion perturbations is that there is \emph{scale-dependent growth}, and that \emph{axion DM differs from CDM on scales below the axion Jeans scale}.
\begin{figure}
\begin{center}
$\begin{array}{@{\hspace{-0.6in}}l@{\hspace{-0.1in}}l}
\includegraphics[scale=0.4]{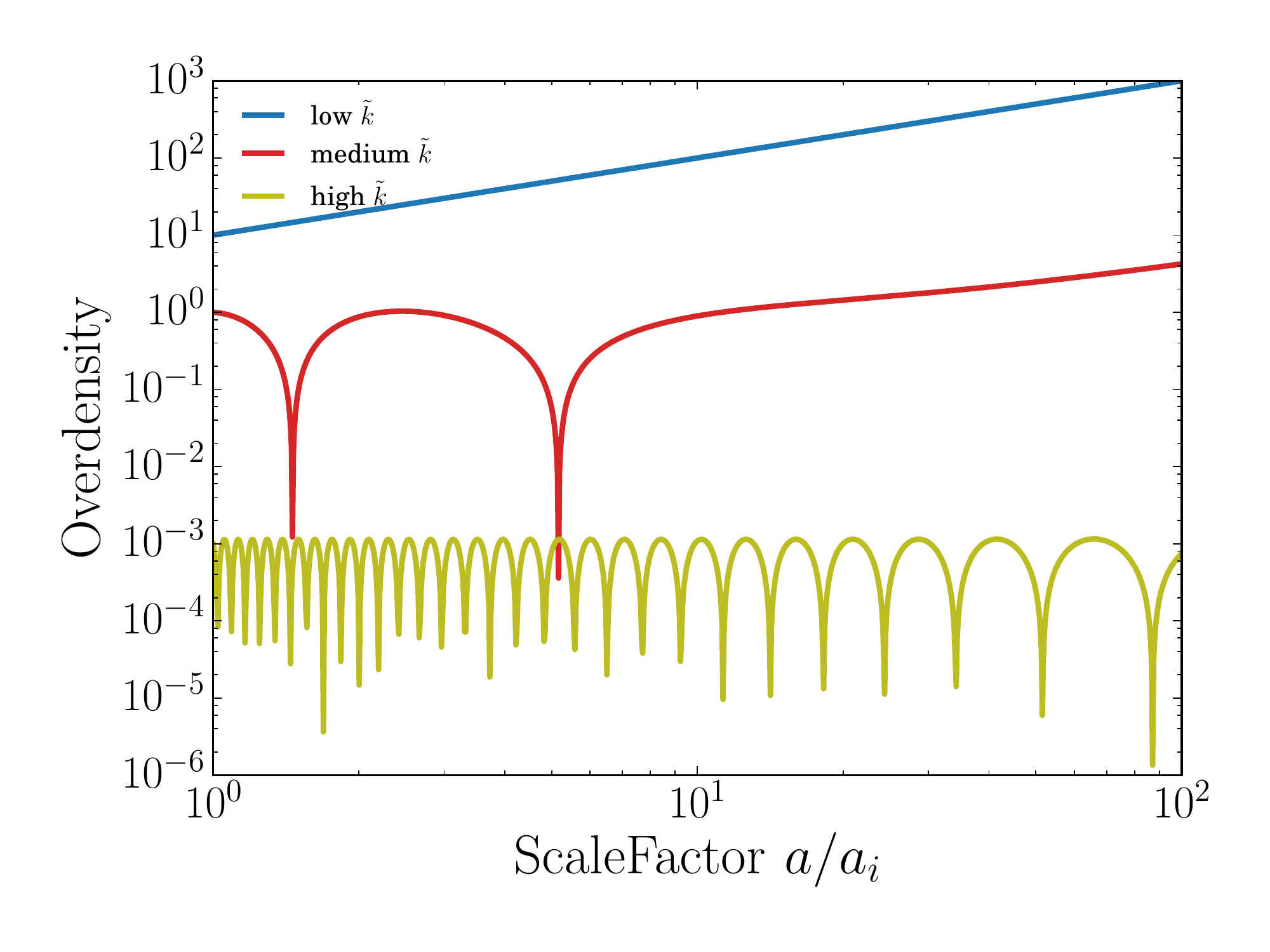}&
\includegraphics[scale=0.4]{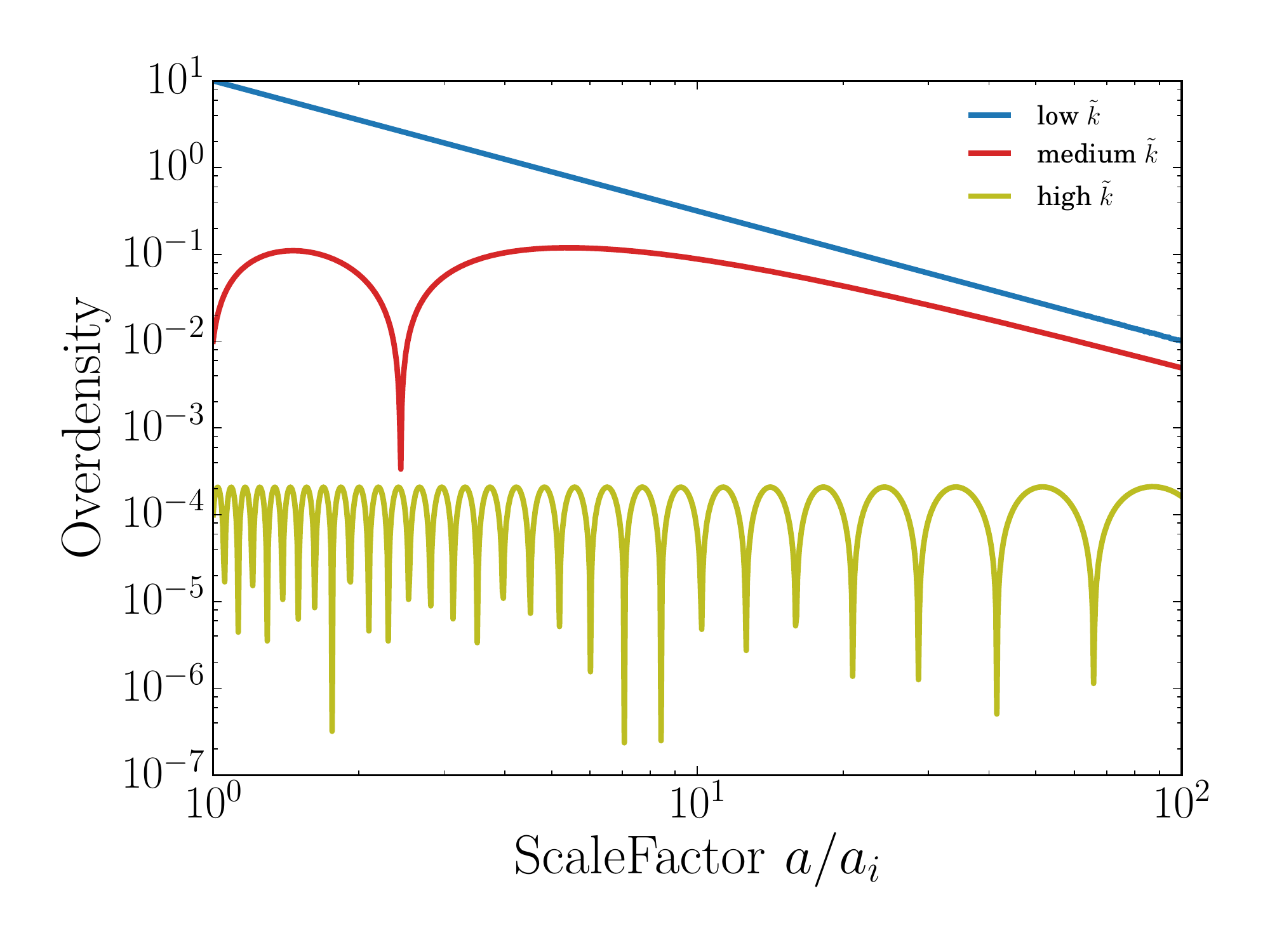}
 \end{array}$
 \end{center}
\caption{The exact scale-dependent linear growth for an axion DM dominated universe, Eq.~\eqref{eqn:delta_exact}, at three values of $\tilde{k}=k/\sqrt{m_a H_0}$, as a function of $a/a_i$. Normalization is arbitrary. Note that the initial scale factor in this case must obey $a_i>a_{\rm osc}$ for the solutions to hold. \emph{Left panel}: The growing mode, $D_+(k,a)$, Eq.~\eqref{eqn:delta_exact2}. \emph{Right Panel:} The decaying mode, $D_-(k,a)$, Eq.~\eqref{eqn:delta_exact3}.}
\label{fig:linear_growth_exact}
\end{figure}

In the limit $k/m_a a<1$ the sound speed has the approximate form:
\be
c_{s,{\rm eff}}^2\approx \frac{k^2}{4 m_a^2 a^2} \, .
\label{eqn:approx_sound_speed}
\ee
The Jeans scale is given by
\be
k_J=(16 \pi Ga \rho_{a,0})^{1/4}m_a^{1/2}=66.5 a^{1/4}\left(\frac{\Omega_a h^2}{0.12}\right)^{1/4}\left( \frac{m_a}{10^{-22}\text{ eV}}\right)^{1/2}\text{ Mpc}^{-1}\, .
\label{eqn:axion_jeans}
\ee

With $\rho_a=\rho_{\rm crit}a^{-3}$ giving the matter-dominated solution for $H$, and using the approximation Eq.~\eqref{eqn:approx_sound_speed} for the sound speed,  there is an exact solution to Eq.~\eqref{eqn:delta_newtonian} given by:
\be
\delta_a=C_1 D_+(k,a)+C_2D_-(k,a) \, .
\label{eqn:delta_exact}
\ee
The closed-form expressions for $D_\pm(k,a)$ are:
\begin{align}
D_+(k,a)&= \frac{3\sqrt{a}}{\tilde{k}^2}\sin \left( \frac{\tilde{k}^2}{\sqrt{a}} \right) +\left[ \frac{3a}{\tilde{k}^4}-1 \right]\cos \left( \frac{\tilde{k}^2}{\sqrt{a}} \right)		\, , \label{eqn:delta_exact2}\\
D_-(k,a)&=   \left[ \frac{3a}{\tilde{k}^4}-1\right] \sin \left( \frac{\tilde{k}^2}{\sqrt{a}} \right) -\frac{3\sqrt{a}}{\tilde{k}^2}\cos \left( \frac{\tilde{k}^2}{\sqrt{a}} \right)   \, , 
\label{eqn:delta_exact3}
\end{align}
where $\tilde{k}=k/\sqrt{m_aH_0}\propto k/k_J$. The solutions $D_\pm(k,a)$ are plotted in Fig.~\ref{fig:linear_growth_exact} at three different values of $\tilde{k}$. For low $\tilde{k}$, $D_+(k,a)\sim a$ is the usual growing mode, and $D_-(k,a)\sim a^{-3/2}$ is the usual decaying mode. For intermediate $\tilde{k}$ there are some oscillations at early times while the mode is below the Jeans scale. At late times, it moves above the Jeans scale and picks up the same growing/decaying behaviour as the low $\tilde{k}$ mode. For high $\tilde{k}$ the mode is always below the Jeans scale, and both $D_+$ and $D_-$ oscillate, retaining constant amplitude.
\begin{figure}
\begin{center}
\includegraphics[width=0.75\textwidth]{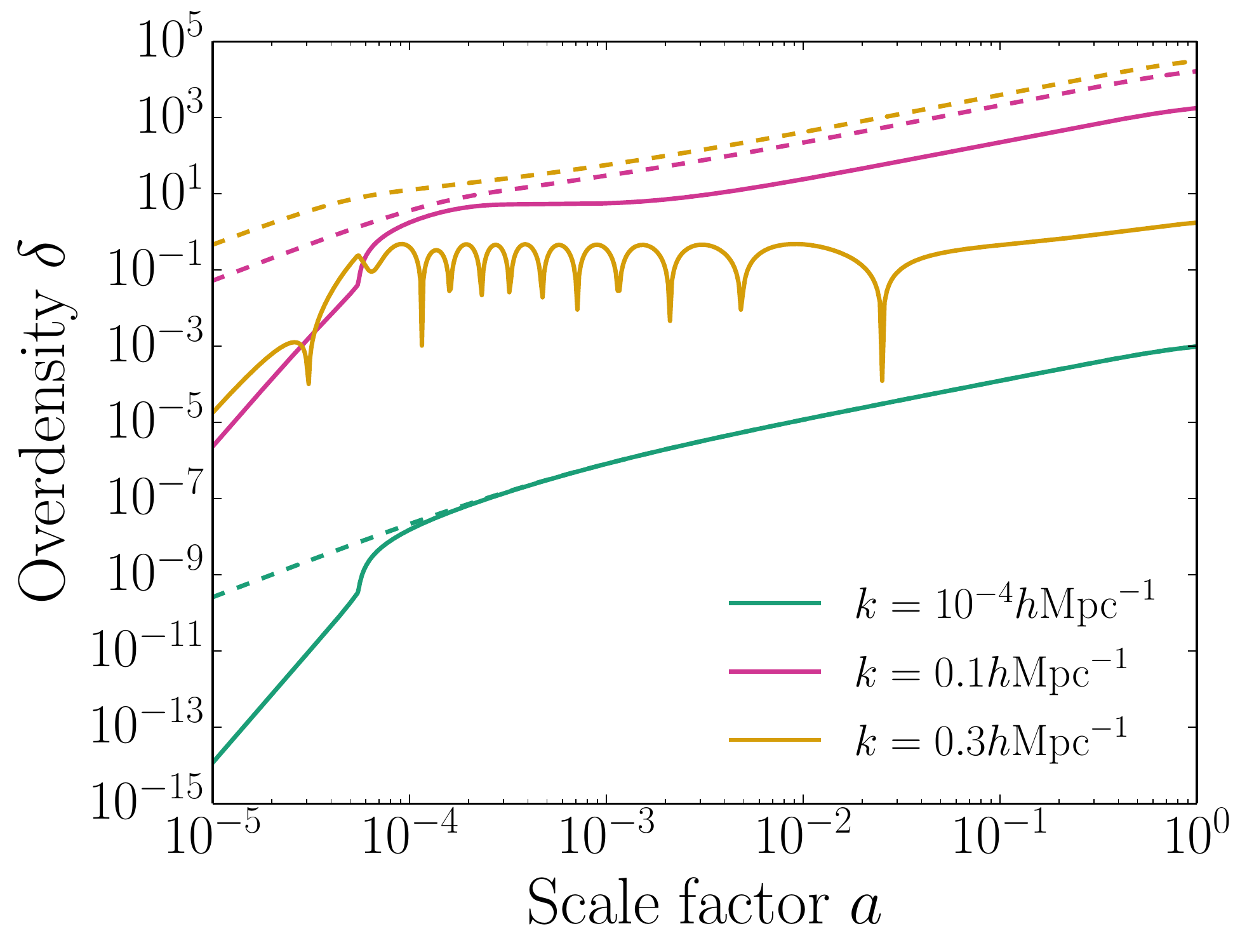}
\caption{Evolution of the axion overdensity, for a ULA mass of $m_{a}=10^{-26}~{\rm eV}$ and a series of wave-numbers $k$ (as shown in the figure), compared to standard CDM (dashed). Axions compose all the DM in this model. Normalization is arbitrary. All cosmological parameters take realistic values. Reproduced (with permission) from Ref.~\cite{Hlozek:2014lca}. Copyright (2015) by The American Physical Society.}
\label{fig:overdensity_plot}
\end{center}
\end{figure} 

Finally, let's return to the real Universe. Fig.~\ref{fig:overdensity_plot} shows the evolution of the axion overdensity computed using \textsc{axionCAMB}, in a realistic model. The axion mass is $m_a=10^{-26}\text{ eV}$, and axions compose all the DM (we will see shortly that this combination of mass and energy density contribution are actually ruled out precisely because of the effects shown here). During the radiation era, before $a_{\rm osc}$, the adiabatic axion perturbation is small. As the axion field begins to roll, the overdensity grows, approaching the CDM value. At low $k$ (large scales), the overdensity locks on to the standard CDM adiabatic evolution, despite the different initial conditions between axions and CDM. This occurs before matter-radiation-equality ($a\sim 10^{-3}$), and today ($a=1$) the CDM and axion models have the same amplitude of density perturbations on large scales. At intermediate $k$, growth is suppressed relative to CDM for some time after equality, and at $a=1$ the axion amplitude is slightly suppressed relative to CDM. The highest $k$ mode has $k>k_J$ initially, and oscillates for some time, leading to a greatly suppressed axion amplitude relative to CDM on small scales.

\subsubsection{Transfer Functions: Relation to WDM and Neutrinos}
\label{sec:transfers}

Thermal DM that was relativistic at freeze-out leads to suppressed clustering power compared to CDM on scales that were sub-horizon while the particles were still relativistic. This gives rise to the free-streaming scale, $k_{\rm fs}$~\cite{1990eaun.book.....K}, which is of cosmological size in models of hot dark matter (HDM, including $m_\nu\lesssim 1\text{ eV}$ standard model neutrinos, see e.g. Refs.~\cite{1980PhRvL..45.1980B,1989ApJ...347..575S,1996ApJ...467...10D}) and warm dark matter (WDM, including sterile neutrinos and thermal gravitinos with $m_X \sim 1\text{ keV}$, see e.g. Refs.~\cite{1982PhRvL..48.1636B,1994PhRvL..72...17D,1996ApJ...458....1C,bode2001}). Suppression of clustering power below the axion Jeans scale (large wavenumbers, $k>k_J$) bears a qualitative similarity to the effects of these low-mass thermal DM models~\cite{amendola2005,marsh2011b}. 

In linear theory, modifications to the power spectrum relative to $\Lambda$CDM can always be expressed by the use of a transfer function:
\be
P_X(k,z)=T_X^2(k,z)P_{\Lambda {\rm CDM}}(k,z) \, .
\ee
The function $T_X(k,z)$ accounts for both scale and redshift dependence. In $\Lambda$CDM, growth is scale-independent for $z\lesssim \mathcal{O}(100)$, after the baryon acoustic oscillations (BAO) have frozen-in, and radiation ceases to be relevant in the expansion rate. Therefore, the linear-theory $\Lambda$CDM power spectrum at any redshift $z\lesssim 100$ can be obtained from the one at $z=0$ by use of the linear growth factor, $D(z)$:\footnote{The $z=0$ power spectrum must in general be computed numerically. It is itself a product of the primordial power spectrum with some transfer function. Some useful fits for this transfer function can be found in Refs.~\cite{1986ApJ...304...15B,1998ApJ...496..605E}.}
\be
P_{\Lambda {\rm CDM}}(k,z)=\left(\frac{D(z)}{D(0)}\right)^2P_{\Lambda {\rm CDM}}(k) \, .
\ee
The linear growth factor is~\cite{1993ppc..book.....P}:
\be
D(z)=\frac{5 \Omega_m}{2H(z)}\int_0^{a(z)}\frac{da'}{(a'H(a')/H_0)^3} \, .
\ee

Axions and thermal DM induce scale-dependent growth, which causes the suppression of power relative to $\Lambda$CDM. However, if this can be neglected on the scales and redshifts of interest, then a redshift-independent transfer function, $T(k)$, can be used to describe the effects of the alternative DM model on structure formation. 

Over a range of scales, the redshift-independent transfer function is a useful description of WDM, for $m_X\gtrsim 0.1\text{ keV}$, and for ULAs with $m_a\gtrsim 10^{-24}\text{ eV}$. For lighter ULAs and for HDM, scale-dependent growth remains relevant at late times and the transfer function is redshift-dependent. These lightest ULAs and HDM require their own detailed treatment, and physics other than the power suppression currently drives constraints. We will discuss them independently when the time comes.

WDM and ULAs with $m_a\gtrsim 10^{-24}\text{ eV}$ can be described by the transfer functions~\cite{bode2001,hu2000}:\footnote{The WDM transfer function can be computed exactly in the Boltzmann code \textsc{class}~\cite{2011JCAP...07..034B}.}
\begin{align}
T_{\rm W}(k) &=(1+(\alpha k)^{2\mu})^{-5/\mu} \, , \label{eqn:wdm_transfer}\\
T_{\rm F}(k) &= \frac{\cos x_J^3(k)}{1+x_J^8(k)} \, , \label{eqn:fcdm_transfer}
\end{align}
where I have used ``F'' standing for ``Fuzzy CDM'' for ULAs described by this transfer function. These transfer functions assume that all of the DM is composed of ULAs or WDM, and cannot be used for mixed DM models. The fitting parameters are
\begin{align}
\mu &= 1.12 \, , \\
\alpha &= 0.074 \left(\frac{m_X}{\text{keV}}\right)^{-1.15}\left(\frac{0.7}{h}\right) \text{ Mpc}\, , \\
x_J(k)&=1.61 \left(\frac{m_a}{10^{-22}\text{ eV}} \right)^{1/18}\frac{k}{k_{J,{\rm eq}}} \, , \\
k_{J,{\rm eq}}&=9 \left(\frac{m_a}{10^{-22}\text{ eV}} \right)^{1/2}\text{ Mpc}^{-1} \, .
\end{align}

The WDM transfer function falls off as a power-law in wavenumber. Intuitively, this is because it is caused by thermal velocities, with temperature scaling as $T\sim 1/a$, and is related to the comoving wavenumber of order the horizon size when $T\sim m_X$. This wavenumber, and the scale factor, evolve as power laws in cosmic time during matter or radiation domination. The ULA transfer function falls off more rapidly, as a cosine. Intuitively, this can be understood from the Jeans scale: solutions to a harmonic equation transition from exponential growth to harmonic oscillations when the growth exponent changes from real to imaginary.

Note that the WDM mass used here, and throughout this review, $m_X$, is the ``thermal relic mass,'' which can be mapped to the larger mass of a sterile neutrino with the same free streaming scale~\cite{1996ApJ...458....1C,2005PhRvD..71f3534V}:
\be
m_{\nu,\rm sterile}=4.43\text{ keV}\left(\frac{m_X}{\text{ keV}}\right)^{4/3}\left(\frac{0.12}{\Omega_W}\right)^{1/3} \, .
\ee

The characteristic scale in the WDM transfer function is fixed by $\alpha^{-1}$, while in the axion transfer function it is fixed by the Jeans scale at equality, $k_{J,{\rm eq}}$. Note that for axions scale-dependent growth is still important on scales $k>k_J(z)$, and the transfer function Eq.~\eqref{eqn:fcdm_transfer} is only valid for smaller wavenumbers. The mild redshift dependence of $k_J\propto a^{1/4}$ means that the current Jeans scale is not so far separated from $k_{J,{\rm eq}}$ (see Eq.~\ref{eqn:axion_jeans}).

A very rough estimate for when structure suppression is relevant on the same scales for WDM and ULAs can be obtained in the following way. For ULAs, assume that structure is suppressed for modes inside the horizon at $a_{\rm osc}$, while for WDM assume the same for the temperature at which particles became non-relativistic, $T_{\rm non.\,rel.}$. Furthermore, assume for both that this happened during the radiation dominated epoch. If these transitions happened at the same time for WDM and ULAs, they will each suppress structure on the same scale relative to CDM. Taking $T_{\rm non.\,rel.}\sim m_X$ and $H(a_{\rm osc.})\sim m_a$, and using that during the radiation dominated epoch $T\sim \sqrt{HM_{pl}}$ gives that WDM suppresses structure on the same scales as a ULA if:
\be
m_X\sim \sqrt{m_a M_{pl}}= 0.5\left(\frac{m_a}{10^{-22}\text{ eV}}\right)^{0.5}\text{ keV} \,  \quad \text{(approximate match)}\, .
\label{eqn:mw_ma_approx}
\ee
We see that it is the large value of $M_{pl}$ that generates the huge separation of mass scales between ULAs and WDM in their effects on structure formation.

A more precise relation between $m_X$ and $m_a$ can be obtained using the transfer functions Eqs.~\eqref{eqn:wdm_transfer} and \eqref{eqn:fcdm_transfer}. The FCDM transfer function falls off more rapidly than the WDM transfer function, so first we must define a scale at which to match them. We can take this to be the half-mode, $k_{1/2}$, defined by $T(k_{1/2})=0.5$. For the FCDM transfer function the half-mode is~\cite{hu2000}:\footnote{We define the half mode using $T(k)$ rather than $T^2(k)$ as Ref.~\cite{hu2000} does, which explains the different co-efficient. Thanks to H.~Y. Schive for noticing this.}
\be
k_{1/2}\approx  5.1\left(\frac{m_a}{10^{-22}\text{ eV}}\right)^{4/9}\text{ Mpc}^{-1} \, .
\ee
Matching this to the WDM half-mode gives:
\be
m_X=0.84\left(\frac{m_a}{10^{-22}\text{ eV}}\right)^{0.39}\text{ keV}\, \quad \text{(half-mode matching)} \, .
\label{eqn:mw_ma_half_mode}
\ee
This agrees with the fit found using ULA transfer functions computed from \textsc{axionCAMB} in Ref.~\cite{2014MNRAS.437.2652M}, and also agrees surprisingly well with the simple estimate of Eq.~\eqref{eqn:mw_ma_approx}. 

\begin{figure}
\begin{center}
\includegraphics[scale=0.5]{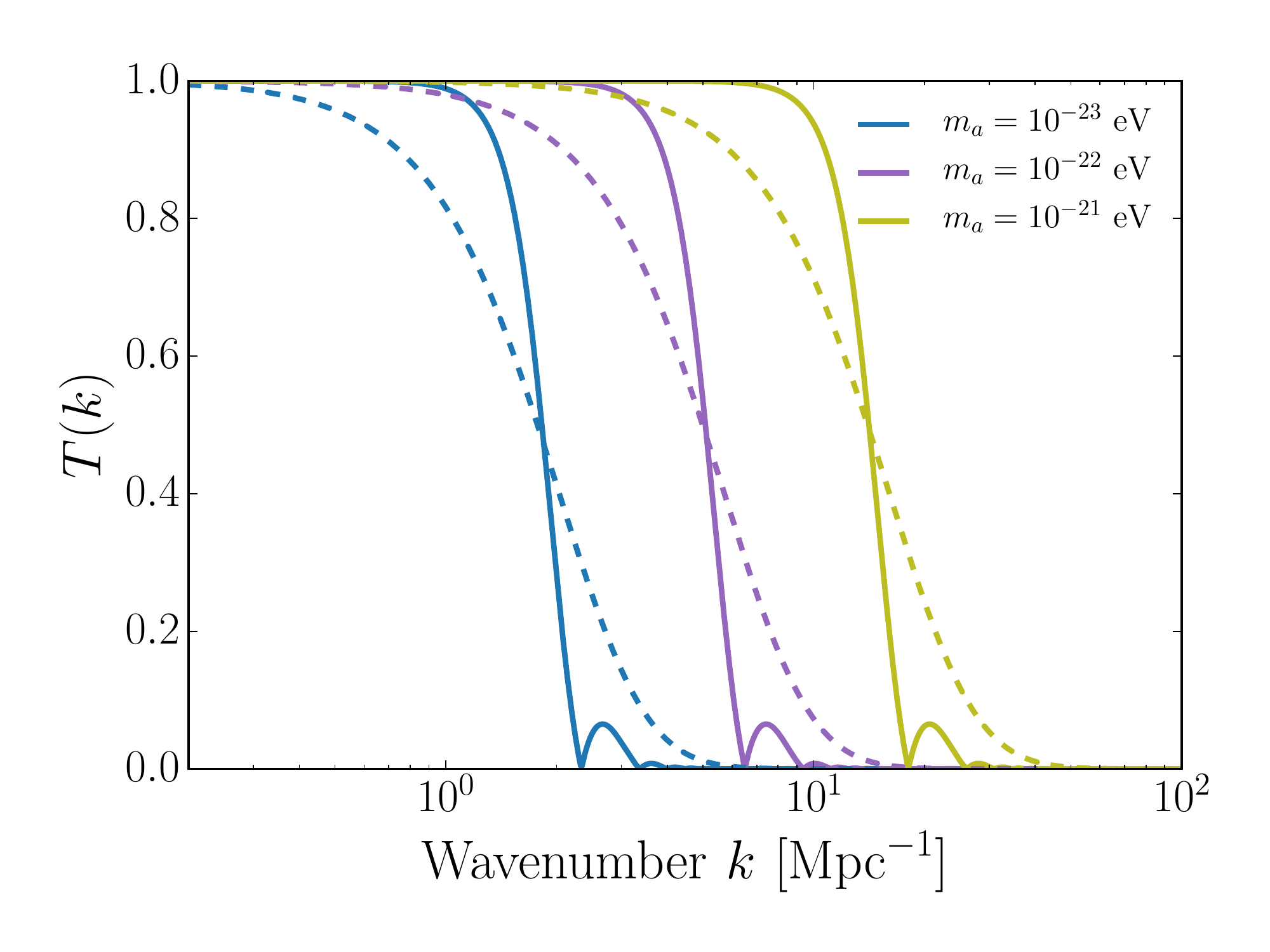}
\caption{ULA (solid) and WDM (dashed) transfer functions, Eqs.~\eqref{eqn:wdm_transfer} and \eqref{eqn:fcdm_transfer}. WDM mass is computed to give the same value of $k_{1/2}$, using Eq.~\eqref{eqn:mw_ma_half_mode}.}
\label{fig:wdm_fcdm_transfers}
\end{center}
\end{figure}

Transfer functions for WDM and ULAs, with the WDM mass computed using Eq.~\eqref{eqn:mw_ma_half_mode}, are shown in Fig.~\ref{fig:wdm_fcdm_transfers}. The lowest mass shown is $m_a=10^{-23}\text {eV}\rightarrow m_X=0.34\text{ keV}$, and has $k_{1/2}=1.6\text{ Mpc}^{-1}$. The non-linear scale is $k_{\rm nl}\sim 0.1\rightarrow 1\text{ Mpc}^{-1}$, and so we see that power suppression by $m_a\geq 10^{-23}\text{ eV}$ cannot be constrained by linear LSS observables.

As a common reference, CDM composed of a neutralino with mass $m_X=100\text{ GeV}$ and decoupling temperature $T=33\text{ MeV}$ cuts off power due to free streaming at $k\approx 10^6\text{ Mpc}^{-1}$ (e.g. Ref.~\cite{2004MNRAS.353L..23G}). Using Eq.~\eqref{eqn:mw_ma_half_mode} this is approximately the same scale as an axion with $m_a\approx 10^{-10}\text{ eV}$, and the QCD axion with $f_a\approx 10^{16}\text{ GeV}$. Thus, low-$f_a$ QCD axions in the classic window suppress structure formation on scales \emph{smaller} than standard WIMPs. 

\subsection{Non-linearities and the Schr\"{o}dinger Picture}
\label{sec:schrodinger}

To study the clustering of axions on non-linear scales, we need to make some approximations. Axions that cluster on galactic scales began oscillating in the very early Universe, with $a_{\rm osc}\ll 1$, so we can take the WKB approximation. The virial velocity in a typical galaxy is $v_{\rm vir}\sim 100\text{ km s}^{-1}\ll c$, and galaxies are much smaller than the horizon, so we can take the non-relativistic approximation. Overdensities in galaxies are $\delta\gtrsim \mathcal{O}(10^5)$, so perturbation theory on $\delta_a$ or $\phi$ is no good. However, except in the vicinity of black holes, the Newtonian potential is small, $\Psi\ll 1$. Thus the Newtonian limit is appropriate,  and $\Psi$ obeys the Poisson equation. We will also only be concerned with scales above the axion Compton wavelength (which is on relativistic scales in the Klein-Gordon equation).

To leading order in $\Psi$ the D'Alembertian is
\be
\Box=-(1-2\Psi)(\partial_t^2+3H\partial_t)+a^{-2}(1+2\Psi)\nabla^2-4\dot{\Psi}\partial_t \, ,
\ee
and the axion energy density is
\be
\rho_a = \frac{1}{2}[(1-2\Psi)\dot{\phi}^2+m_a^2\phi^2+a^{-2}(1+2\Psi)\partial^i\phi\partial_i\phi] \, .
\ee
We take the WKB approximation in the form
\be
\phi=(m_a\sqrt{2})^{-1} (\psi e^{-im_at}+\psi^*e^{im_at}) \, ,
\ee
where $\psi$ is a complex field, which can be written in polar co-ordinates as
\be
\psi=Re^{iS}\, .
\ee

We take our limits as $\Psi\sim \epsilon_{\rm NR}^2$, and $k/m_a\sim \epsilon_{\rm NR}$ and $H/m_a\sim \epsilon_{\rm WKB}$, and work to quadratic order in $\epsilon\sim \epsilon_{\rm NR}\sim\epsilon_{\rm WKB}$. In this limit, the energy density contains the leading order piece:
\be
\rho_a=|\psi|^2=R^2\, ,
\ee
and the equation of motion for $\psi$ is the Schr\"{o}dinger equation:
\be
i\dot{\psi}-3iH\psi/2+(2m_a a^2)^{-1}\nabla^2\psi-m_a\Psi\psi=0 \, .
\label{eqn:schrodinger_hubble}
\ee
This is a non-linear Schr\"{o}dinger equation, with $\Psi$ sourced by $|\psi|^2$ via the Poisson equation. The form shown here, including the Hubble friction explicitly, can be found from the usual form by going to comoving coordinates.

While the Schr\"{o}dinger equation is interesting and can provide insight into structure formation with axion DM, wave equations don't fit the bill as standard cosmologist's tools. We can make contact with standard perturbation theory \cite{2002PhR...367....1B} and non-linear simulation tools such as smoothed-particle hydrodynamics (SPH) using, as before, a fluid description. Substituting the polar form of the wavefunction, we can find conservation and Euler equations for an effective fluid described by $\psi$. The fluid velocity is
\be
\vec{v}_a\equiv (m_a a)^{-1}\nabla S \, .
\ee
We can now perform a background-fluctuation split and find the equations of motion in terms of the overdensity, $\delta_a$ (e.g. Refs.~\cite{2011PhRvD..84f3518C,2015PhRvD..91l3520M}):
\begin{align}
\dot{\delta}_a +a^{-1}\vec{v}_a\cdot\nabla\delta_a&=-a^{-1}(1+\delta_a)\nabla\cdot\vec{v}_a \, , \label{eqn:axion_conservation}\\
\dot{\vec{v}}_a+a^{-1}\left( \vec{v}_a\cdot \nabla\right)\vec{v}_a&= -a^{-1}\nabla (\Psi+Q)-H\vec{v}\, \label{eqn:axion_euler} \, , \\
Q\equiv -\frac{1}{2m_a^2 a^2}&\frac{\nabla^2\sqrt{1+\delta_a}}{\sqrt{1+\delta_a}} \, ,\label{eqn:q_def_for_eoms}
\end{align}
where I have defined the ``quantum potential'' $Q$.\footnote{We have used the Schr\"{o}dinger equation as an intermediate step to get a fluid form for the axion equations without needing to perform the background-fluctuation split on $\phi$ first. We were thus able to retain canonical equations of motion for $\rho$ and $\vec{v}$ beyond linear perturbation theory. For discussion on the use of hydrodynamics to describe quantum mechanics in the ``synthetic'' view of Bohmian mechanics, see Ref.~\cite{wyatt_trajectories}.} The quantum potential is all we need to model the axion gradient energy and Jeans scale in the full non-linear dynamics as a simple modification to the force on a fluid element~\cite{2015PhRvD..91l3520M,Mocz:2015sda}:
\be
F=-a^{-1}\nabla (\Psi+Q) \, .
\label{eqn:quantum_force_mod}
\ee

Eqs.~\eqref{eqn:axion_conservation} and \eqref{eqn:axion_euler} can also be used as the basis for a modified perturbation theory of axion DM, which takes into account the differences to CDM near the Jeans scale. Expanding Eq.~\eqref{eqn:q_def_for_eoms} to first order in $\delta_a$ and going to Fourier space provides a simple derivation of the asymptotic form of the effective sound speed, Eq.~\eqref{eqn:approx_sound_speed}. 

The Schr\"{o}dinger form of the field equations is useful and interesting in and of itself. It is a fundamental (though approximate) equation governing axion DM on non-linear scales. We will use the Schr\"{o}dinger equation to discuss axion halo density profiles in Section~\ref{sec:halo_profiles}. Above the de-Broglie wavelength Schr\"{o}dinger equation also accurately models CDM, and is an alternative to standard N-body simulation techniques~\cite{Widrow&Kaiser1993}. The wave properties below the de Broglie scale and the introduction of the quantum force in the fluid equations are a particular regularization and softening of the Vlasov equation \cite{Uhlemann:2014npa}. They also provide a setting to study modifications to the Zel'dovich approximation~\cite{Coles&Spencer2003,2011PhRvD..84f3518C}, which is the basis of Lagrangian perturbation theory. Perhaps most importantly, however, the Schr\"{o}dinger equation provides the best method currently available to accurately simulate axion and scalar field DM on small scales, which we will now discuss.

\subsection{Simulating axion DM}
\label{sec:schive_simulations}

A full description of DM clustering in any model can only truly be provided by non-pertrubative numerical simulations. Since the earliest days of computational cosmology, this been studied in $N$-body simulations, which simulate the dynamics of collisionless point particles interacting via Newton's gravitational law. The ``particles'' are not fundamental particles, but simulations particles, the mass of which is fixed by the simulations resolution. Newton's law is ``softened'' on small scales to prevent unphysical two-body pairs of these particles dominating the dynamics. These classic $N$-body simulations are the perfect picture of CDM, and their conceptual simplicity provides some explanation for the popularity of its study. 

A simulation of CDM is defined by two properties: initial conditions, and dynamics. The initial conditions are provided by the matter power spectrum from linear theory, with higher order effects to deal with transients~\cite{2002PhR...367....1B}; the dynamics is that of collisionless particles. Axions, particularly ULAs, modify both of these properties:
\begin{itemize}
\item \emph{Modified initial conditions}: The initial power spectrum is suppressed relative to CDM. Modes below the Jeans scale at matter-radiation equality have the power erased.
\item \emph{Modified dynamics}: On scales of order the axion de Broglie wavelength, wavelike effects must be included. The dynamics is not that of collisionless point particles.
\end{itemize}

Modified initial conditions are easily implemented in an $N$-body simulation, as long as the correct power spectrum is provided from a Boltzmann code. Such simulations provide an accurate description of axions above the de Broglie wavelength, and have been performed in Refs.~\cite{2016ApJ...818...89S,2016JCAP...04..012S}. These simulations are very similar to those of WDM in the case that streaming velocities are irrelevant (e.g. Refs.~\cite{bode2001,2014MNRAS.439..300L}). Special care must be taken, however, due to the appearance of ``spurious structures'' caused by discreteness effects~\cite{2007MNRAS.380...93W}. Such spurious structures can be removed based on the shape of the protohalos~\cite{2014MNRAS.439..300L} or on the functional shape of the halo mass function~\cite{2013MNRAS.433.1573S}. Removal of spurious structure for ULAs was carried out using the protohalo shape condition in Ref.~\cite{2016ApJ...818...89S}. We will discuss the halo mass function in more detail in Section~\ref{sec:hmf}.

Modified dynamics are somewhat less trivial to implement, in particular those relevant to ULAs. Modern simulations add new dynamics to the simplest CDM model such as hydrodynamics of the baryons (e.g. Ref.~\cite{2005MNRAS.364.1105S}), parameterised force law modifications for variants of SIDM~\cite{2015arXiv151205344C}, neutrino models with streaming velocities~\cite{2012MNRAS.420.2551B,2015PhRvD..92b3502I}, and even general relativistic effects~\cite{2016arXiv160406065A} or modified gravity~\cite{2015MNRAS.454.4208W}. At their core, all these methods are based, to some degree, on the $N$-body paradigm.

As long as the objects to be simulated are non-relativistic (as galactic halos are), the Schr\"{o}dinger equation provides the correct model of axion DM on small scales. A cosmological simulation of the Schr\"{o}dinger equation is a fundamental departure from $N$-body simulations. The first high-resolution cosmological simulations of the Schr\"{o}dinger form were recently performed in Ref.~\cite{2014NatPh..10..496S}. The modified dynamics caused by wavelike effects for $m_a\approx 10^{-22}\text{ eV}$ appear in dwarf galaxy-sized objects on scales of order 1~kpc. The modified dynamics can be seen to introduce effects including smooth halo density profiles and interference fringes (see Section~\ref{sec:halo_profiles} and Fig.~\ref{fig:schive}), which would be completely absent in a CDM-like $N$-body simulation. Resolving these features accurately in a cosmological simulation involves many computational technicalities, including e.g. the use of adaptive mesh refinement to solve the scalar field equation over a wide range of length scales. 

An alternative way to model the modified dynamics of ULAs and other scalar fields in cosmological simulations, which fits more easily into the $N$-body paradigm, is suggested by the modified force law in the fluid description, Eq.~\ref{eqn:quantum_force_mod}. This modified force law could be implemented in a hydrodynamic model (as suggested in Refs~\cite{2015PhRvD..91l3520M,Mocz:2015sda}), or indeed in any method where the local density and its derivatives can be accurately determined. This method was employed in toy models in Ref.~\cite{Mocz:2015sda}, but has yet to be applied to a cosmological simulation.

The Schr\"{o}dinger equation in this context models more than just axions. It is applicable to any model of scalar field DM, real or complex-valued, so long as the field is oscillating about a quadratic potential minimum, and self-interactions can be neglected. The simulations of Ref.~\cite{2014NatPh..10..496S} represent the state-of-the-art for simulations of these models. There is still much to be done in this area, however. For example, some of the many things not covered in Ref.~\cite{2014NatPh..10..496S}:
\begin{itemize}
\item Initial conditions. Use of full Boltzmann equation power spectra. Modified perturbation theory and Zel'dovich approximation.
\item Hydrodynamics. Modelling of baryonic effects in tandem with scalar field dynamics to assess complementary roles.
\item Zoom-in simulations. Dwarf galaxies and sub-structure modelled in Milky-Way and Local Group analogs from larger $N$-body simulations.
\end{itemize} 
This shopping list is not meant to detract from the achievements of Ref.~\cite{2014NatPh..10..496S}: the field of study of such simulations is simply young compared to that of CDM $N$-body simulations.

\subsection{My Two Cents on BEC}
\label{sec:two_cents}

In this section we discuss \emph{only} DM axions. There is some debate in the literature as to whether axion DM forms a Bose-Einstein condensate (BEC), and over what scales such a BEC differs from CDM.  For more discussion on this topic, see Refs.~\cite{2009PhRvL.103k1301S,2013JCAP...12..034D,2015APh....65..101D,2015PhRvD..92j3513G}. The original discussions of the link between quantum theory and classical fields for the axion are in Refs.~\cite{1990PhRvD..42.3918N,1991PhRvD..44..352R}.\footnote{There is a vast literature on so-called ``BEC dark matter'': as far as I can tell, for all practical purposes this simply maps to general scalar field models. Since the early Universe physics is often less well defined than in the case of axions, questions of condensate formation are also less clear. For a good source of references and history, see Ref.~\cite{2011PhRvD..84d3531C}.}

Davidson~\cite{2015APh....65..101D} defines a BEC as
\be
\text{BEC}\,\, =\,\, \text{condensed regime}\,\,=\,\, \text{classical field} \, .
\ee
This chimes with our usual notion form undergraduate statistical mechanics: the macroscopically-occupied ground-state obeys the classical equations of motion. The important characteristic, however, is not the ground-state, which is only accessible to a homogeneous system (which cosmology certainly is not), but it is that the Fourier modes are concentrated at a particular value and that the particles in this state are coherent.

Let's define some of these notions: we will not use these formal definitions, but it helps to be precise. QFT decomposes a field operator into modes of creation, $\hat{a}$, and annihilation, $\hat{a}^\dagger$, operators as
\be
\hat{\phi} (x) = \int \frac{d^3 p}{(2\pi)^3}\frac{1}{\sqrt{2E_p}}\left( \hat{a}_pe^{-\vec p\cdot\vec{x}}+\hat{a}_p^\dagger e^{-i\vec{p}\cdot\vec{x}}\right) \, ,
\ee
where $\vec{p}$ is the three-momentum, and $E_p$ is the energy. The ground state is defined by $\hat{a}_p|0\rangle=0$. The classical field is defined by the coherent state~\cite{1986qmv1.book.....C}
\be
|\phi\rangle = \frac{1}{N}\exp\left[ \int \frac{d^3 q}{(2\pi)^3}\tilde{\phi}(\vec{q})\hat{a}^\dagger_q \right]|0\rangle \, ,
\ee
where $\tilde{\phi}(\vec{q})$ is the Fourier transform of the classical field, and $N$ is a normalisation such that $\langle\phi|\phi\rangle=1$. The expectation value of the field operator in this state is the classical field:
\be
\langle\phi|\hat{\phi}(x)|\phi\rangle = \int \frac{d^3 p}{(2\pi)^3}\frac{1}{\sqrt{2E_p}}\tilde{\phi}(\vec{p}) e^{-i\vec{p}\cdot\vec{x}} = \phi (x) \, ,
\ee
i.e. this expectation value obeys the classical equations of motion as we have been discussing in the preceding subsections, and will continue to discuss throughout this review.

The questions now are: over what timescales do axions enter the state $|\phi\rangle$, how does this state evolve, and, crucially, what is its coherence length? The ``controversy'' of axion BEC is over what role gravity plays in this process, particularly at late times, and over the coherence length this induces for structures with vorticity, $\vec{\nabla}\times\vec{v}\neq 0$, within galaxies. 

Recall that there are two populations of DM axions: those formed from vacuum realignment, and those formed from decay of topological defects. The vacuum realignment population begins life already in the state $|\phi\rangle$. In the broken PQ scenario, the state $|\phi\rangle$ is formed by inflation, which super-cools and homgenises the axion field over the entire visible Universe. In the unbroken PQ scenario, the parent PQ field, $\varphi$, is in it is classical field state, $|\varphi\rangle$, and thus the axion field created after SSB is also coherent in the state $|\phi\rangle$ over the horizon size at SSB (leading to the classical field configurations of strings, domain walls, and miniclusters, as discussed above). 

Thus, for either the broken or unbroken PQ scenario, \emph{axions from the vacuum realignment mechanism are described entirely by the classical field equations}, as presented in the preceding parts of this section. \emph{Thermalisation at early times is irrelevant, as coherence is established by initial conditions}. The gravitational interactions lead to the usual structure formation on large scales: as perturbations grow, the field effectively loses some coherence. The Jeans scale supports the field against gravitational collapse and maintains total coherence on smaller scales. The characteristic size of collapsed objects is given by the soliton solutions to the Schro\"{o}dinger-Possion equation (see Section~\ref{sec:halo_profiles}).

For the population of cold axion particles produced by topological defect decay in the unbroken PQ scenario, axions can enter the state $|\phi\rangle$ via thermalisation. The condition for thermalisation due to any interactions is that the relaxation rate, $\Gamma$, is of order the Hubble rate. 

Consider the QCD axion for concreteness. The self interactions are computed by Taylor expanding the cosine potential, giving:
\be
V_{\rm int} = \frac{\lambda}{4!}\phi^4\, , \quad \lambda=\frac{m_a^2}{f_a^2 }\frac{m_d^3+m_u^3}{(m_d+m_u)^3}\approx 0.35\frac{m_a^2}{f_a^2 } \, .
\ee
Note that these interactions are \emph{attractive}.  The relaxation rate is~\cite{2009PhRvL.103k1301S}
\be
\Gamma_{\lambda}\sim n\sigma_0\delta v \mathcal{N}\, .
\ee
where $n$ is the number density of particles, $\sigma_0$ is the cross section for two-to-two axion scattering in vacuum, $\sigma_0=\lambda^2/(64\pi m_a^2)$, $\delta v$ is the velocity dispersion, and $\mathcal{N}$ is the average state occupation number. The number density is computed from the relic density, the velocity dispersion at time $t$ is computed by redshifting the initial momentum, $p(t_{\rm osc})\approx H(t_{\rm osc})$ (recall that topological defects decay when the classical field begins oscillating), and the occupation number is given by
\be
\mathcal{N}=\frac{(2\pi)^3 n}{V_{\rm coh.}}\, ,
\ee 
where $V_{\rm coh.}$ is the spherical volume of a coherence patch: $V_{\rm coh.}=4\pi(m\delta v)^3/3$.

By taking $m_at_{\rm osc}\sim 1$ we find that $\Gamma_{\lambda}(t_{\rm osc})/H(t_{\rm osc})\sim \mathcal{O}(1)$: \emph{self interactions thermalise the cold population of axions, with an initial coherence length of order $1/H(t_{\rm osc})$}.\footnote{The general scalings of these arguments hold also for generic ALPs with $\lambda\sim m_a^2/f_a^2$.} Thus, for the cold population of axions produced by topological defect decay, on all times later that $t_{\rm osc}$ we can also describe the axions as being in the state $|\phi\rangle$ obeying the classical equations of motion. This is as we expect: occupation numbers for axion DM from any production mechanism are so huge that classical field equations ought to be adequate. So far, so uncontroversial.

The question now arises as to whether axions can ``re-thermalise'' at later times. The two-to-two rate, $\Gamma_\lambda$, redshifts faster than $H$, such that at times after $t_{\rm osc}$ self-interactions are not sufficient for this purpose~\cite{2009PhRvL.103k1301S}. Now the controversial part: \emph{can gravitational interactions re-thermalise the axion condensate}? If re-thermalisation at times $t>t_{\rm osc}$ occurs, then a larger coherence length will be established, and axion DM will differ from CDM on scales larger than those set by the Jeans scale and quantum pressure in the classical equations of motion.

Sikivie and Yang~\cite{2009PhRvL.103k1301S} propose that gravitational scattering of axions can lead to re-thermalisation of the QCD axion at a temperature $T_{re.}\sim 100\text{ eV}(f_a/10^{12}\text{ GeV})^{1/2}$. This is argued based on the gravitational relaxation rate:
\be
\Gamma_G = \frac{G_N n m_a^2 }{(m\delta v)^2}\, .
\ee
If Sikivie and Yang are correct, this effect will induce a larger coherence length for the axion field, absent in the classical equations of motion. In particular, the claim is that re-thermalization due to $\Gamma_G$ is \emph{not} captured by the classical equations of motion.

However, this claim has been countered by Davidson and Elmer~\cite{2013JCAP...12..034D}, Davidson~\cite{2015APh....65..101D}, and Guth, Hertzberg and Prescod-Weinstein~\cite{2015PhRvD..92j3513G}, who show that the effects of the relaxation rate $\Gamma_G$ are already present in the classical equations of motion (the relevant case being the Schr\"{o}dinger-Poisson equation), and thus by solving them alone on times $t>t_{\rm osc}$ we miss nothing: there is no coherence on scales larger than the Jeans scale. The rate $\Gamma_G$ is the interaction rate between axions already in the condensate with one another, hence being linear in $G_N$. Davidson~\cite{2015APh....65..101D} also estimated the quadratic in $G_N$ scattering between cold axion particles and the condensate, concluding that this interaction is negligible for $f_a\lesssim M_{pl}$.\footnote{Recall that it is folk-wisdom that super-Planckian $f_a$ violates ``gravity as the weakest force''} In the end, Davidson notes, all such questions can ultimately be answered by the Path Integral, using the Closed Time Path 2PI action in curved space. Further treatment of this is far beyond the scope of this review.

A final note here is on the possible formation of vortices in the axion field (a well-known phenomenon in BEC in the laboratory~\cite{bec_book}), and their possible phenomenological role in galactic haloes. A net overall rotation of the axion field caused by tidal torques leading to $\nabla\times\vec{v}\neq 0$ would augment our system of classical equations due to anomalous stresses, and could lead to vortex formation. Sikivie and Yang (see also Ref.~\cite{2013PhRvD..88l3517B}) argued that this could be a distinctive feature of axion DM, and may explain the structure of caustics in DM haloes. This was explored in more detail by Rindler-Daller and Shapiro~\cite{2012MNRAS.422..135R}, who found that the axion self-interactions are of the wrong type (attractive rather than repulsive) to support vortex formation. Vortex formation depends on having self-interactions, and so goes beyond  the $m_a^2\phi^2$ simplified model we study mostly in this review. In any case, it is clearly a model-dependent effect, and one that appears not to occur for the QCD axion.

\section{Constraints from the CMB and LSS}
\label{sec:cosmological_constraints}

This section reviews work presented in Refs.~\cite{Hlozek:2014lca,marsh2011b,marsh2013,Marsh:2014qoa}. Bayes theorem is briefly reviewed in Appendix~\ref{appendix:bayes}. Issues related to sampling the axion parameter space are discussed in Appendix~\ref{appendix:degeneracies}.

\subsection{The Primary CMB}

The CMB temperature auto-power, $C_\ell^{TT}$, is the data product at the disposal of the precision cosmologist. We use CMB data from \emph{Planck} (2013 release)~\cite{2014A&A...571A...1P,2014A&A...571A..15P} and WMAP~\cite{2013ApJS..208...20B}, ACT~\cite{2014JCAP...04..014D} and SPT~\cite{2011ApJ...743...28K}.

ULAs affect the primary (adiabatic, unlensed, no secondaries) CMB primarily via the expansion rate. The first acoustic peak of the CMB temperature power occurs at $\ell\approx 200$ and is fixed by the angular size of the BAO at recombination, $z_{\rm rec}\approx 1100$. ULAs with $z_{\rm osc}\gtrsim 1100$ affect higher acoustic peaks, while those with $z_{\rm osc}\lesssim 1100$ affect the Sachs-Wolfe (SW) plateau. 

\begin{figure}
\begin{center}
\includegraphics[width=0.75\textwidth]{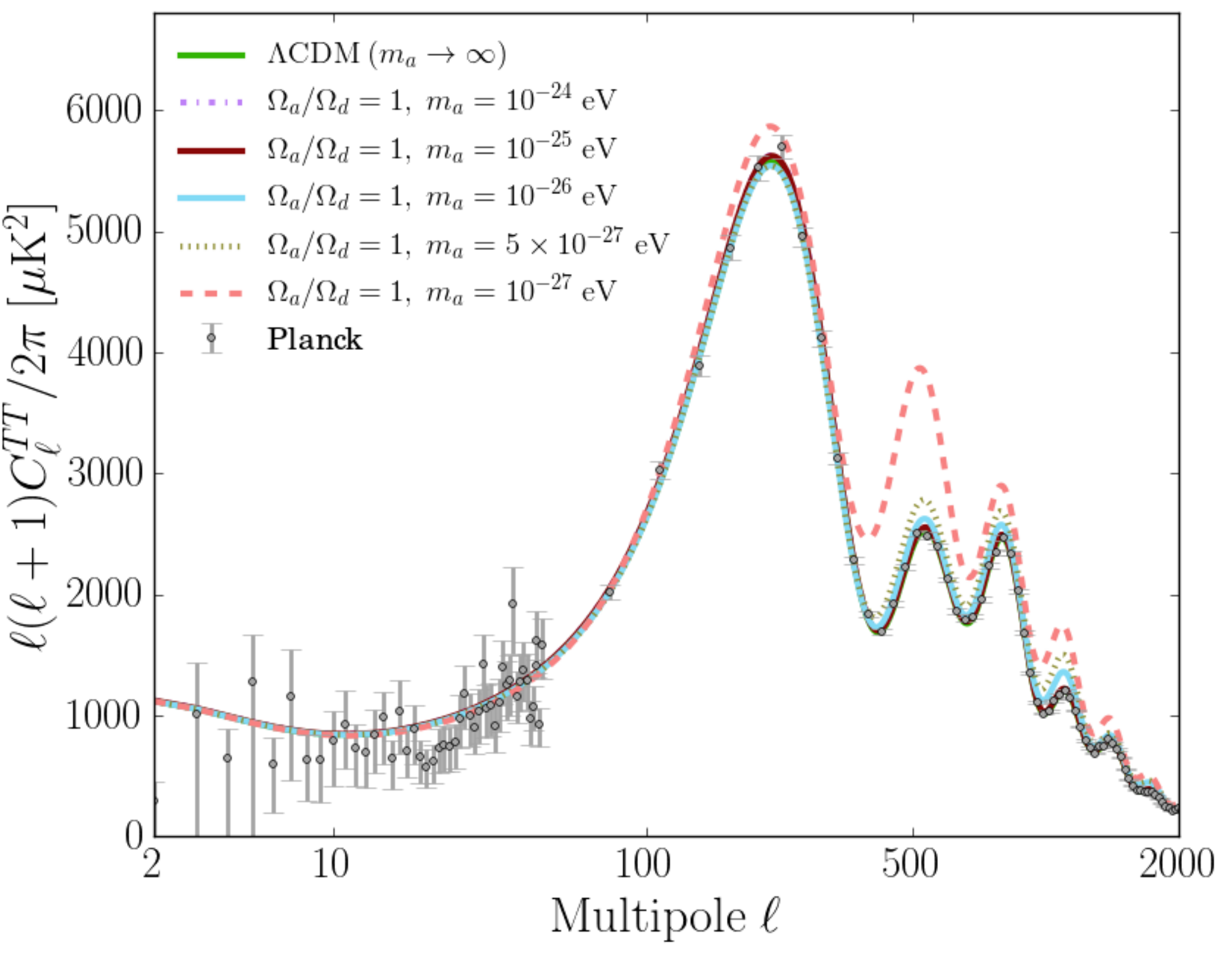}
\caption{Effect of ULAs on the CMB as a function of ULA mass. Here we demand that ULAs compose all the DM, with no CDM. The early-time expansion rate is altered, changing the relative heights of the higher acoustic peaks. Reproduced (with permission) from Ref.~\cite{Hlozek:2014lca}. Copyright (2015) by The American Physical Society.}
\label{fig:TT_models_DM_mass}
\end{center}
\end{figure}

The CMB acoustic peaks constrain the relative matter-to-radiation density at different epochs, fixing the DM to baryon ratio and the redshift of matter-radiation equality. Axions with $w_a\approx -1$ at any particular epoch alter the expansion rate relative to that in a pure CDM cosmology.  The higher acoustic peaks probe successively higher order effects on the expansion at earlier times, however radiation is increasingly dominant at early times, and the higher acoustic peaks also Silk-damp away. Thus, there is some maximum $z_{\rm osc}$ for heavy ULAs beyond which the effects on the higher acoustic peaks vanish and ULAs become indistinguishable from CDM. If we demand that ULAs compose all the DM, the effects on the CMB are more dramatic for low mass ULAs, where the expansion rate is significantly altered near matter-radiation equality. These effects are illustrated in Fig.~\ref{fig:TT_models_DM_mass}. The lightest ULA model shown has $m_a=10^{-27}\text{ eV}$. The mass is just large enough that matter-radiation equality and recombination are barely changed, leaving the first peak at the same location, and the SW plateau unchanged. Higher acoustic peaks depart significantly from the CDM case. Increasingly higher masses lead to increasingly smaller effects away from CDM, with the effects moving to higher acoustic peaks. By eye, it is impossible to distinguish $m_a=10^{-25}\text{ eV}$ from CDM.

\begin{figure}
\begin{center}
$\begin{array}{@{\hspace{-0.9in}}l@{\hspace{-0.1in}}l}
\includegraphics[scale=0.4]{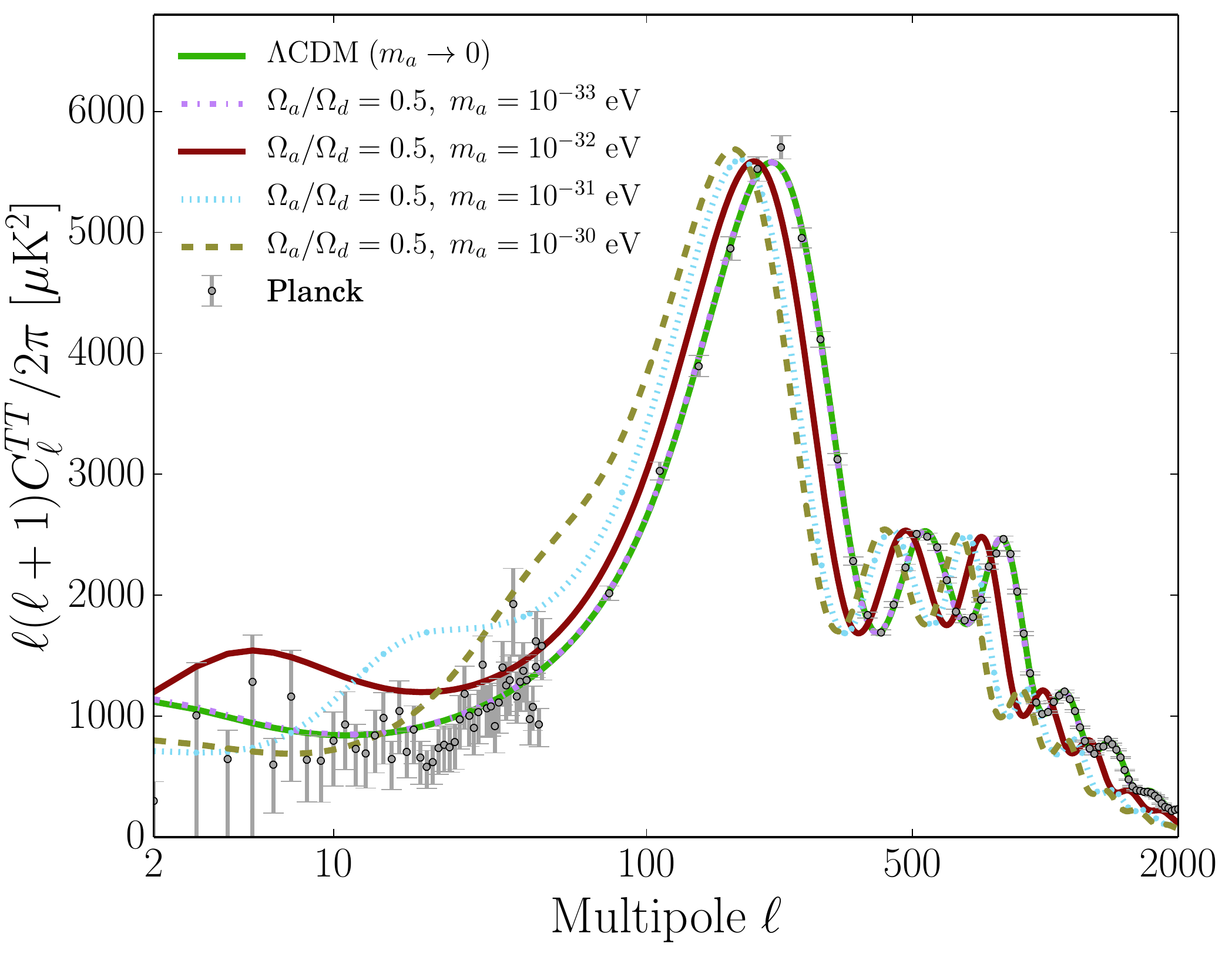}
\includegraphics[scale=0.4]{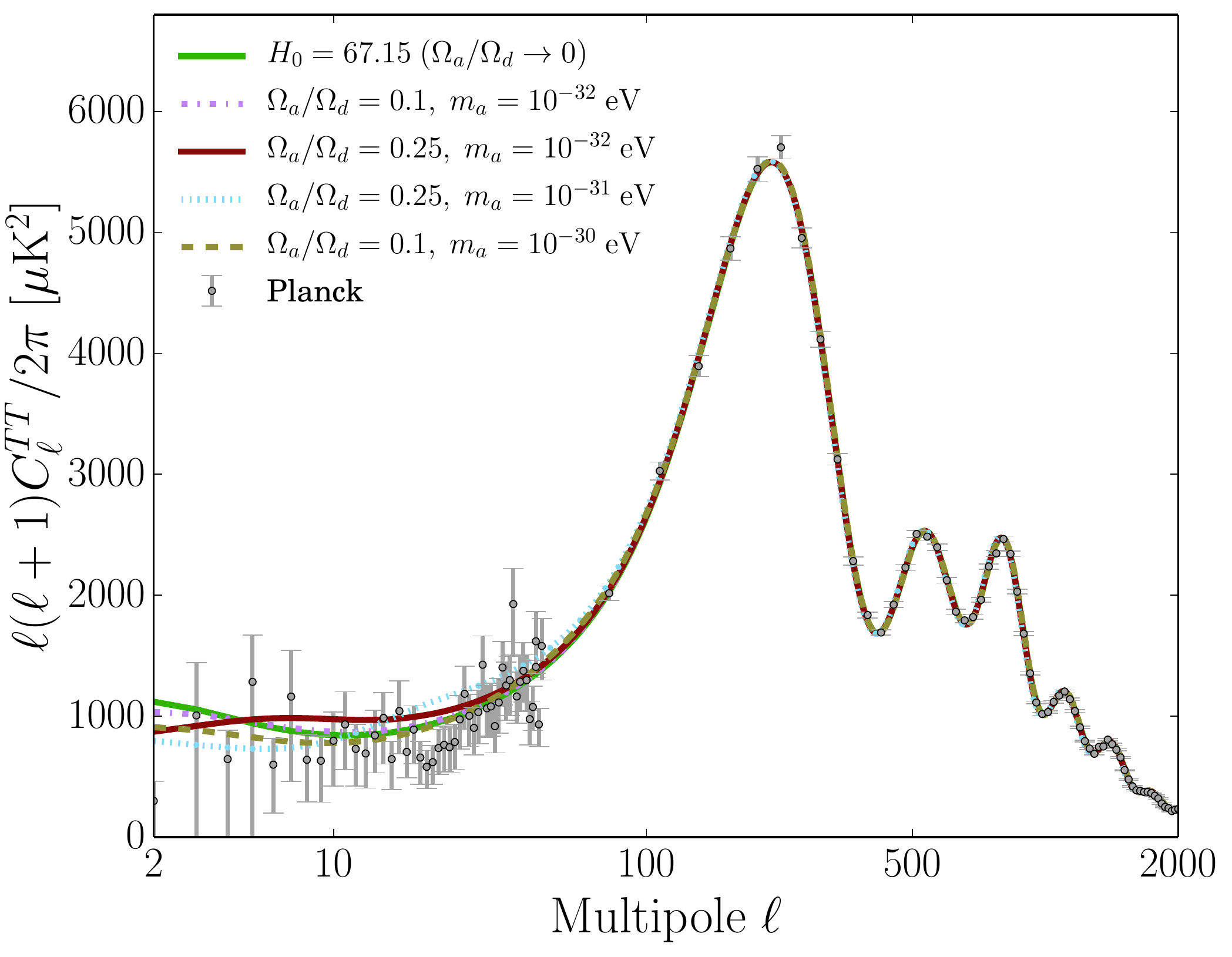}
 \end{array}$
 \end{center}
\caption{Effect of the lightest ULAs on the CMB. \emph{Left Panel:} I hold $\Omega_ch^2=0.12$ fixed and introduce successively heavier axions as a fraction of the DE at fixed $H_0$. The first acoustic peak moves and the ISW effect more pronounced compared to $\Lambda$CDM. \emph{Right Panel:} Here we demand that the location of the first peak remains fixed, which requires reducing $H_0$ compared to $\Lambda$CDM, isolating the ISW effect. Reproduced (with permission) from Ref.~\cite{Hlozek:2014lca}. Copyright (2015) by The American Physical Society.}
\label{fig:TT_models_DE}
\end{figure}

Lighter ULAs differ significantly from CDM in the post-recombination Universe. Getting matter-radiation equality right requires us to keep the CDM density at $\Omega_c h^2=0.12$. Introducing light ULAs at fixed $H_0$ thus reduces $\Omega_\Lambda$. The Universe is now younger, with reduced distance to the CMB. This moves the first acoustic peak to lower $\ell$. The ULAs have $w_a=-1$ transitioning to $w_a=0$ in the late Universe, and imprint this on the low $\ell$ CMB via the integrated (I)SW effect. Both of these effects are shown for varying ULA masses in Fig.~\ref{fig:TT_models_DE} (Left Panel). Notice that $m_a=10^{-33}\text{ eV}$ is indistinguishable from $\Lambda$CDM: axions this light have $w_a\approx-1$ today, and contribute to the effective cosmological constant and DE. 

The low $\ell$ CMB measurement is cosmic variance limited, leading to large uncertainties, while the first acoustic peak is measured exquisitely well. We can isolate the ISW effect of ULAs by changing the value of $H_0$ to leave the location of the first peak unchanged. Such a cosmology is shown in Fig.~\ref{fig:TT_models_DE} (Right Panel). With $\Omega_a/\Omega_d=0.1$ and $m_a=10^{-32}\text{ eV}$ the ULA model is indistinguishable from $\Lambda$CDM (except in the quadrupole, $\ell=2$, which is poorly measured).

\subsection{The Matter Power Spectrum}

The matter power spectrum, $P(k,z)$, contains a wealth of cosmological information. The BAO imprint a fixed physical scale on the power spectrum, and this is used as a measurement of the expansion rate (e.g. Ref.~\cite{2010deot.book..246B}). The BAO measure a single number, the angular size of the sound horizon, as a function of redshift. The full shape of the matter power spectrum contains more information than just the BAO, and is our focus here. The matter power spectrum can be measured from the two-point correlation function of some tracer of the DM. Here we focus on the galaxy power spectrum, $P_{\rm gal}(k,z)=b^2 P(k,z)$, where $b$ is the galaxy bias. It is measured by a number of surveys, of which we choose to use the WiggleZ survey~\cite{Parkinson:2012vd}, which measures the galaxy power spectrum in four redshift bins centred on $z=0.22,0.41,0.60$ and $0.78$. We further restrict to only linear scales, $k \lesssim 0.2 \,h\mathrm{Mpc}^{-1}$.

\begin{figure}
\begin{center}
$\begin{array}{@{\hspace{-0.7in}}l@{\hspace{-0.1in}}l}
\includegraphics[scale=0.4]{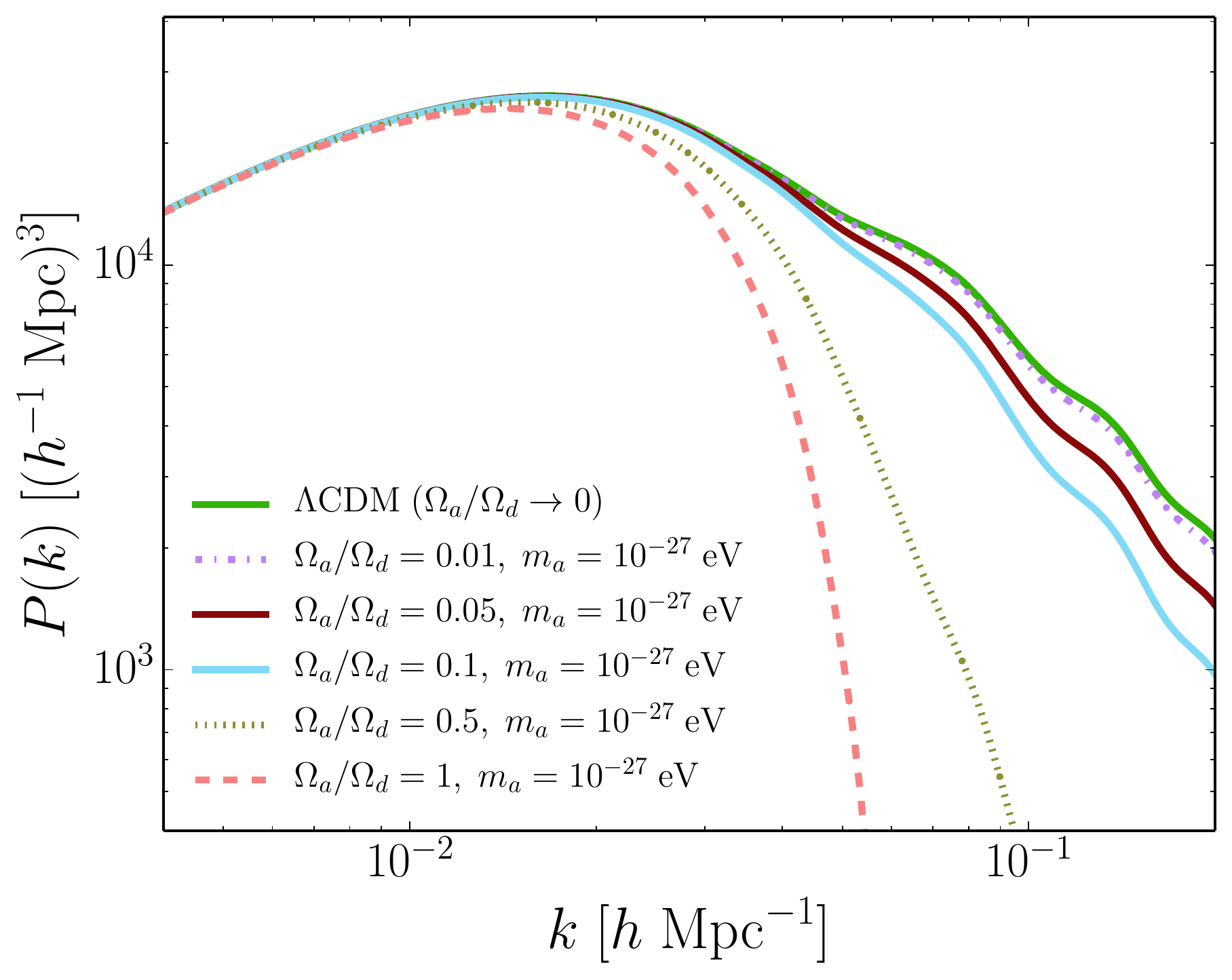}
\includegraphics[scale=0.4]{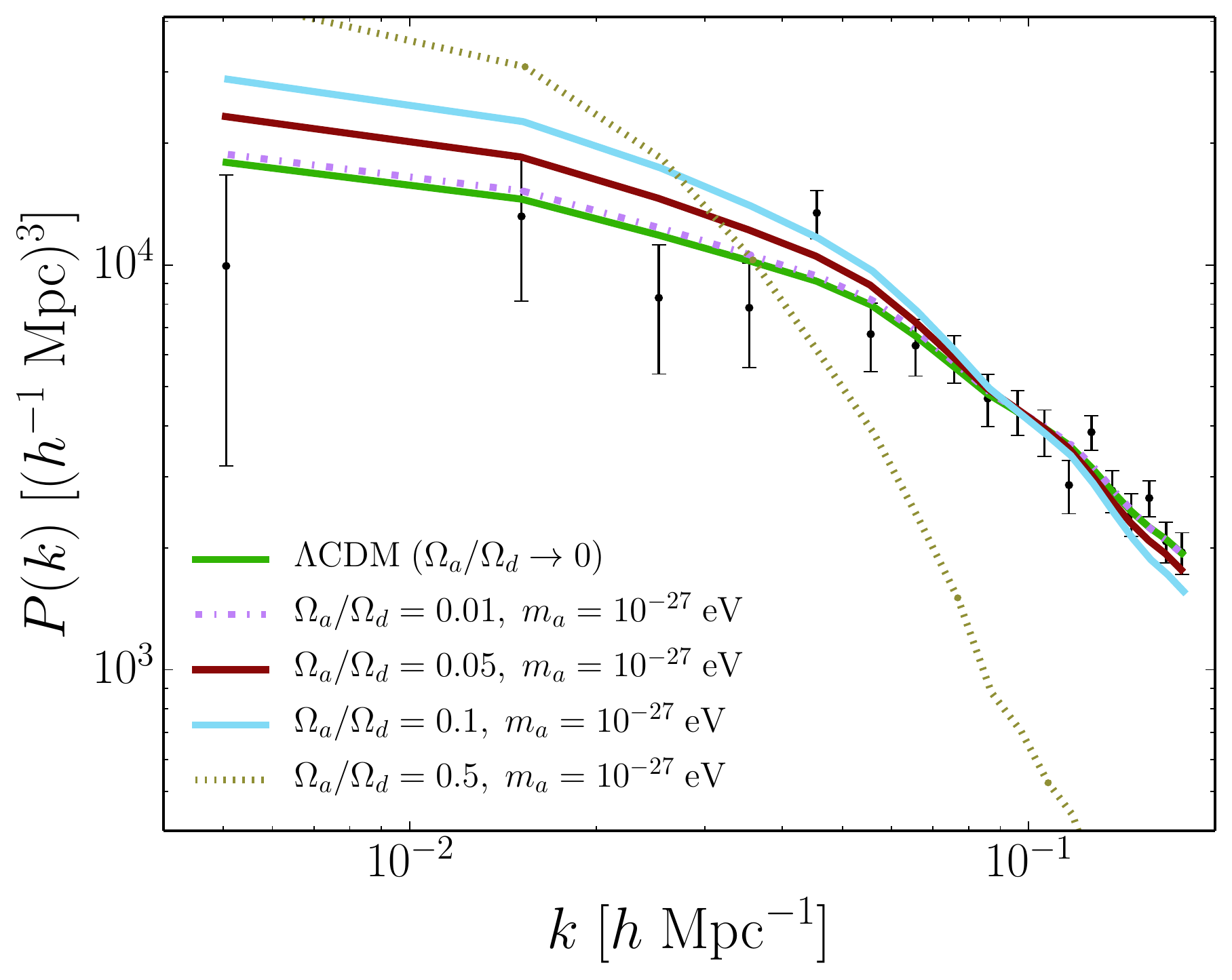}
 \end{array}$
 \end{center}
\caption{Effect of the lightest ULAs on the matter power spectrum, with fixed mass and varying contribution to the DM density. \emph{Left Panel:} The matter power spectrum. \emph{Right Panel:} After convolution with the WiggleZ survey window function and marginalization over galaxy bias at $z=0.60$. Reproduced (with permission) from Ref.~\cite{Hlozek:2014lca}. Copyright (2015) by The American Physical Society.}
\label{fig:pk_frac_models}
\end{figure}

The effect of axions on the matter power spectrum probes both the expansion rate (via the BAO) and the growth of structure, via the transfer and growth functions. The most well-known effect that we have already discussed is the suppression of power caused by the existence of the axion Jeans scale. This effect is shown in Fig.~\ref{fig:pk_frac_models}, where the left panel shows the idealized scenario with $P(k)$, and the right panel the effect convolved with the WiggleZ survey window function and marginalized over galaxy bias. 

In the idealized case, we see how reducing the axion contribution to the DM density reduces the amount of structure suppression compared to CDM~\cite{2010PhRvD..82j3528M,amendola2005}. For $m_a=10^{-27}\text{ eV}$ structure suppression kicks in at $k\approx 0.02\,h\,\text{Mpc}^{-1}$, and has a sub-percent effect on the power relative to CDM for $\Omega_a/\Omega_d=0.01$ (ULAs contributing $\sim 1\%$ to the total DM). The galaxy bias, $b$, changes the character of the effect. Galaxy bias is measured by the survey by allowing $b$ to float as a free parameter. When it varies, it can compensate, in a scale-independent manner, for suppression of power. The preferred value of $b$, and so the normalization of the power spectrum, is thus different for the ULA cosmologies than for $\Lambda$CDM, and this partial degeneracy reduces the constraining power of the galaxy survey.

The scale-dependent clustering of ULAs tells us that a full treatment of bias in these cosmologies should involve computing a \emph{scale-dependent bias}, $b(k)$, and its dependence on the ULA transfer function and growth rate. Scale-dependent bias in mixed DM cosmologies is a poorly understood problem, and it has particular relevance to studies of massive neutrinos (see e.g. Ref.~\cite{2014PhRvD..90h3530L}). Scale-dependent bias can be studied through numerical simulation, or semi-analytically via the halo model~\cite{2002PhR...372....1C}. Ref.~\cite{Hlozek:2014lca} proposed an approximate treatment of scale-dependent bias for ULAs, motivated by treatments of DE and neutrinos, and by the data, which we now outline.

Bias relates the galaxy power spectrum to the matter distribution. On scales where ULAs do not cluster (below the Jeans scale), we do not expect any correlation between the galaxies and the ULAs. Galaxy surveys only observe out to some smallest wavenumber (largest scale), $k_{\rm obs}$. The scale of the observations defines an epoch, $k_{\rm obs}= a_{\rm bias}H(a_{\rm bias})$: ULAs which only begin to behave like matter after this epoch will not be correlated with the galaxy distribution on observable scales. We can approximate the scale of structure suppression for ULAs as $k_{\rm osc}=a_{\rm osc}H(a_{\rm osc})$ and impose scale-dependent bias as a hard cut by excluding ULAs from the matter density if $a_{\rm osc}>a_{\rm bias}$:
\be
\delta\rho_m = \Theta (a_{\rm osc}-a_{\rm bias})(\delta\rho_c+\delta\rho_b)+\Theta (a_{\rm bias}-a_{\rm osc})(\delta\rho_c+\delta\rho_b+\delta\rho_a)\, ,
\ee
where $\Theta (x)$ is the Heaviside function, and $\rho_m$ in the overdensity is defined in the same manner.  Because no current galaxy surveys observe on scales larger than the horizon size at equality, Ref.~\cite{Hlozek:2014lca} made the simplification $a_{\rm bias}=a_{\rm eq}$, which effectively removes ULAs from the matter distribution used to compute the galaxy power spectrum for $m_a\lesssim 10^{-27}\text{ eV}$.

\begin{figure}
\begin{center}
\includegraphics[width=0.75\textwidth]{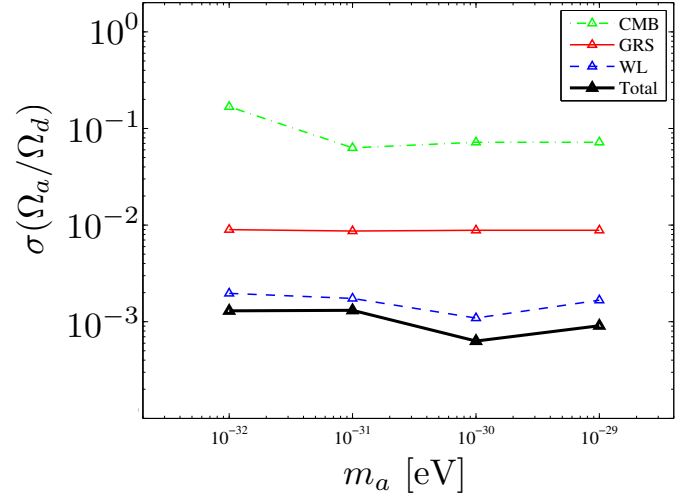}
\caption{Forecasted sensitivity of a \emph{Euclid}-like galaxy redshift (GRS) and weak lensing (WL) survey to axion DM fraction, $\Omega_a/\Omega_d$, as a function of mass. WL increases sensitivity to $\Omega_a$ by a factor of around ten compared to GRS alone. Reproduced and modified (with permission) from Ref.~\cite{marsh2011b}. Copyright (2012) by The American Physical Society.}
\label{fig:modified_euclid}
\end{center}
\end{figure}

An unbiased tracer of the matter distribution is provided by gravitational lensing. Upcoming surveys such as \emph{Euclid} propose to measure the galaxy shear power spectrum \cite{Laureijs:2011gra}, and could improve constraints on DM models considerably~\cite{marsh2011b,Smith:2011ev,Amendola:2012ys} if systematics can be controlled. The forecasted sensitivity to $\Omega_a$ of the lightest ULAs for a \emph{Euclid}-like survey is shown in Fig.~\ref{fig:modified_euclid}.\footnote{In this figure, neutrino parameters are included and marginalized over, lowering the CMB sensitivity compared to that found in Ref.~\cite{Hlozek:2014lca} (see next section, and Appendix~\ref{appendix:degeneracies}).} These optimistic forecasts for weak lensing show an increase in sensitivity of around a factor of ten compared to the galaxy redshift survey alone.

The effect of axions on the expansion rate is also seen in the power spectrum, and is particularly evident if axions replace $\Lambda$ (although now the issue of bias becomes more complicated~\cite{Hlozek:2014lca}). This changes the age of the Universe relative to $\Lambda$CDM, with a younger Universe having less time to grow structures, reducing the amplitude of $P(k)$. In the CMB the effect of a younger Universe could be largely compensated by reducing $H_0$; in $P(k)$ it can be compensated by changing the amplitude of primordial fluctuations, $A_s$. However, as both the CMB and $P(k)$ share common parameters, no choice of $A_s$ and $H_0$  can completely remove the effects of this change, demonstrating the complementarity of CMB and LSS measurements. See Ref.~\cite{Hlozek:2014lca} for further discussion.

\subsection{Combined Constraints}

\begin{figure}
\begin{center}
$\begin{array}{@{\hspace{-0.2in}}l@{\hspace{+0.1in}}l}
\includegraphics[scale=0.4]{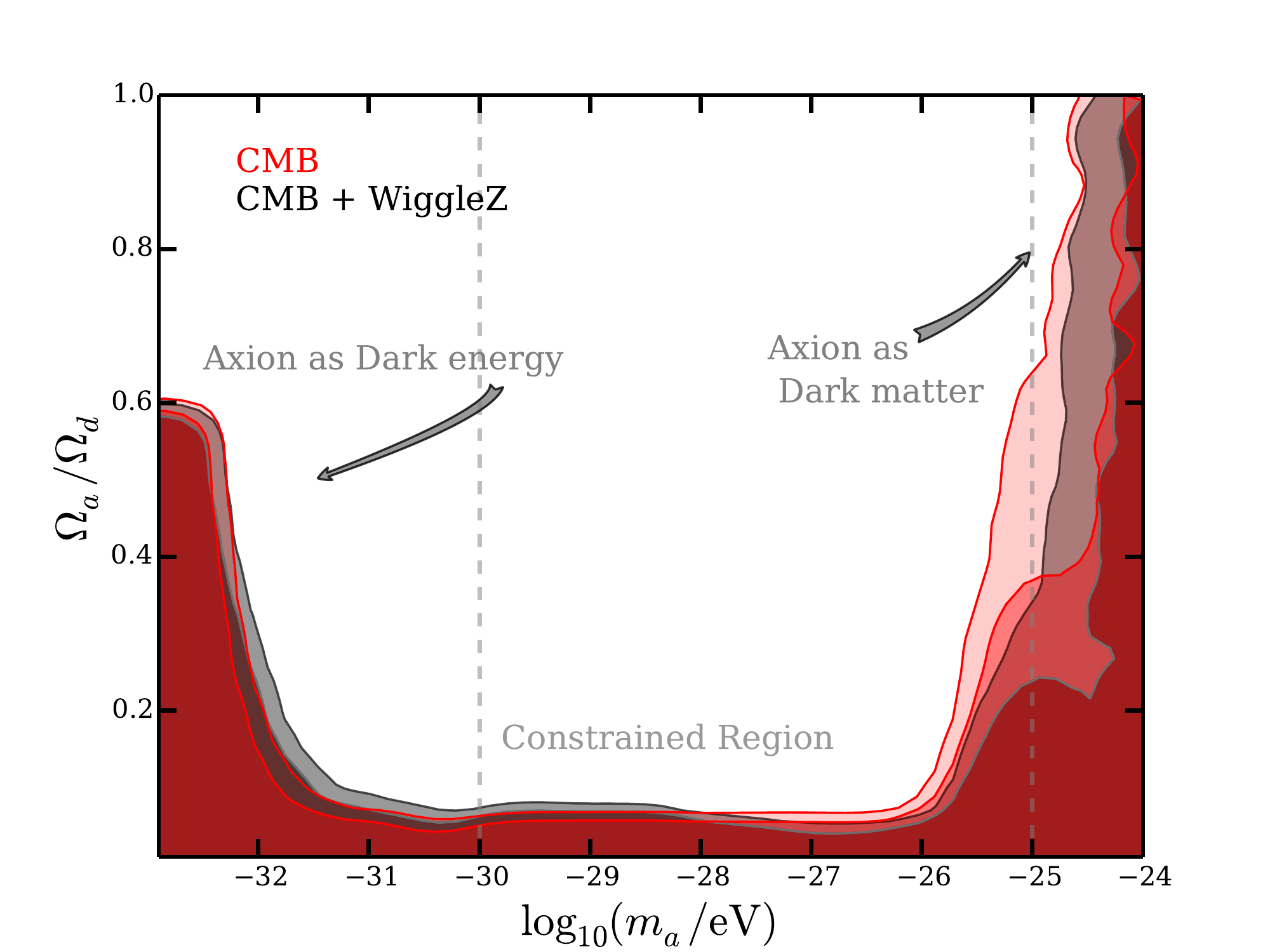}
\includegraphics[scale=0.4]{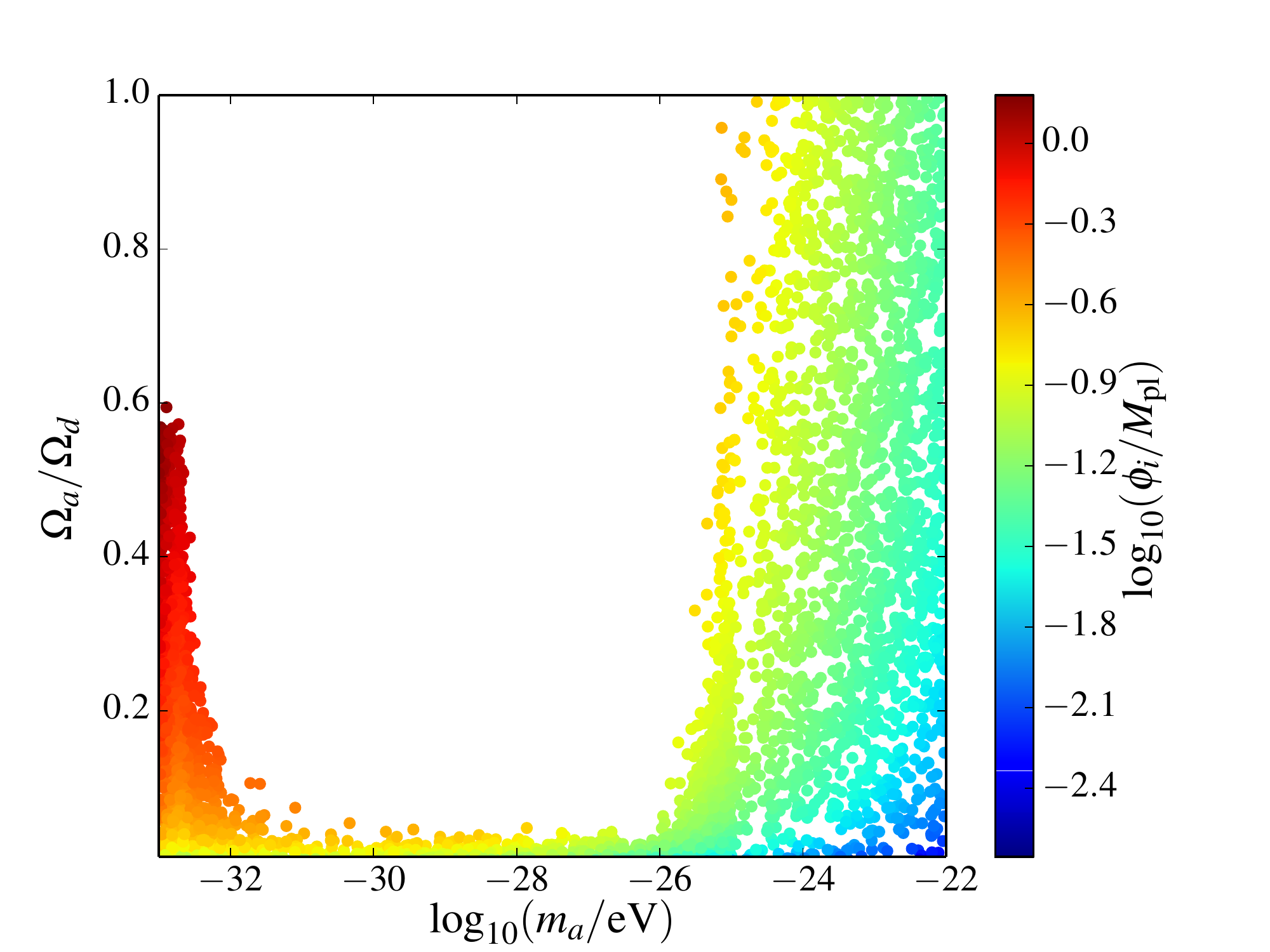}
 \end{array}$
 \end{center}
 \caption{Constraints to the axion dark sector energy fraction, $\Omega_a/\Omega_d$, as a function of axion mass from linear cosmological probes. \emph{Left Panel:} Contours show 2 and 3 $\sigma$ allowed regions comparing CMB and CMB+WiggleZ. \emph{Right Panel:} CMB constraints, with sample points from chains colour-coded by axion initial displacement in Planck units. Reproduced (with permission) from Ref.~\cite{Hlozek:2014lca}. Copyright (2015) by The American Physical Society.}
\label{fig:contours_combinedLinear}
\end{figure} 

Fig.~\ref{fig:contours_combinedLinear} (left panel) shows the constraints on the axion dark sector density fraction, $\Omega_a/\Omega_d$, as a function of axion mass for CMB and CMB+WiggleZ data set combinations, taken from Ref.~\cite{Hlozek:2014lca}. Including LSS data from WiggleZ as well as the CMB loosens constraints slightly at low mass, and tightens them slightly at high mass. The looser constraint at low mass is possibly being driven by the CMB/LSS tension in measurements of the power spectrum amplitude (commonly expressed as the ``$\sigma_8$ tension''). The tighter constraint at high mass is due to the WiggleZ data points with small error bars at $k\sim 0.1\,h\,\text{Mpc}^{-1}$.

The normalization is $\Omega_d=\Omega_a+\Omega_c$, i.e. we consider a mixed DM model with CDM and ULAs. The allowed value at the lowest ULA masses, $m_a\approx 10^{-33}\text{ eV}$, is $\Omega_a/\Omega_d=0.6$ implying $\Omega_a\approx 0.6$, with the CDM density held fixed at close to its usual value. These ULAs are DE and drive the current period of accelerated expansion. At high mass, we see that in order for axions to be all the DM, with $\Omega_a/\Omega_d=1$, requires $m_a\geq 10^{-24}\text{eV}$ at 95\% C.L. This is \emph{the lower bound on DM particle mass from linear cosmological probes}, as promised in the abstract. The constraint in the central, intermediate mass, region of $10^{-32}~{\rm eV}\leq m_{a}\leq 10^{-25.5}~{\rm eV}$ is $\Omega_{a}/\Omega_{d} \leq 0.05$ and $\Omega_{a}h^{2}\leq 0.006$ at $95\%$-confidence. That is, \emph{intermediate mass axions must make up less then 5\% of the total DM.} 

It is important to note that the constraints of Ref.~\cite{Hlozek:2014lca} apply to a cosmology with CDM plus a single light axion, and not to CDM plus multiple axions. It might be a good guess to assume that the constraint on the energy density in the intermediate mass regime applies to the sum total energy density for all such axions (because the constraint is independent of mass).  A dedicated study is necessary, but degeneracies will be even more problematic and a prudent choice of priors and sampling will be required (see Appendix~\ref{appendix:degeneracies}).

Fig.~\ref{fig:contours_combinedLinear} (right panel) shows the CMB only constraints, with sample points from \textsc{Multinest}~\cite{multinest} chains colour-coded by the initial axion field displacement in Planck units (and re-sampled such that point density is proportional to probability as in a Markov chain Monte Carlo, MCMC).\footnote{The field displacement is found by using Eq.~\eqref{eqn:simpledens} as the initial guess in a shooting method to obtain the desired $\Omega_a$. We solve the Klein-Gordon equation at early times, switching to $\rho_a\propto a^{-3}$ when $3H=m_a$.} The field displacement is always $\phi_i<\pi M_{pl}$, and is thus consistent with a quadratic potential and sub-Planckian $f_a$. Axion DE requires $f_a\sim M_{pl}$. For $m_a=10^{-22}\text{ eV}$ to be all the DM requires $\phi_i\sim \mathcal{O}(\text{few})\times 10^{16}\text{ GeV}$. This shows that a ULA with $f_a\leq 10^{16}\text{ GeV}$ will satisfy all current constraints on $\Omega_a$ without fine tuning. These conclusions from numerical computation and full comparison with CMB data agree with the discussion in Section~\ref{sec:alp_misalignment} based on Eq.~\ref{eqn:simpledens}. 

\subsection{Isocurvature and Axions as a Probe of Inflation}
\label{sec:tensors}

Axions in the broken PQ scenario pick up isocurvature perturbations. The amplitude of these perturbations is proportional to the energy scale of inflation. The CMB places strong constraints on the allowed amplitude of such perturbations. Therefore, if axions compose the DM, constraints on isocurvature constrain the energy scale of inflation, and a detection of both would uniquely probe inflation. An independent measurement of the energy scale of inflation can be used to place strong constraints on axion cosmology. 

Let's flesh these ideas out and quantify the possibilities. All of this Section assumes standard, single-field, slow-roll inflation. We'll focus on the QCD axion, which is also covered in detail in Refs.~\cite{2004hep.th....9059F,2009ApJS..180..330K,Visinelli:2014twa}. The case of ALPs is slightly more complicated than for the QCD axion, as the parameter space has more dimensions. ALPs are covered by Refs.~\cite{marsh2013,Marsh:2014qoa,acharya2010a}.
\begin{figure}
\begin{center}
\includegraphics[width=0.75\textwidth]{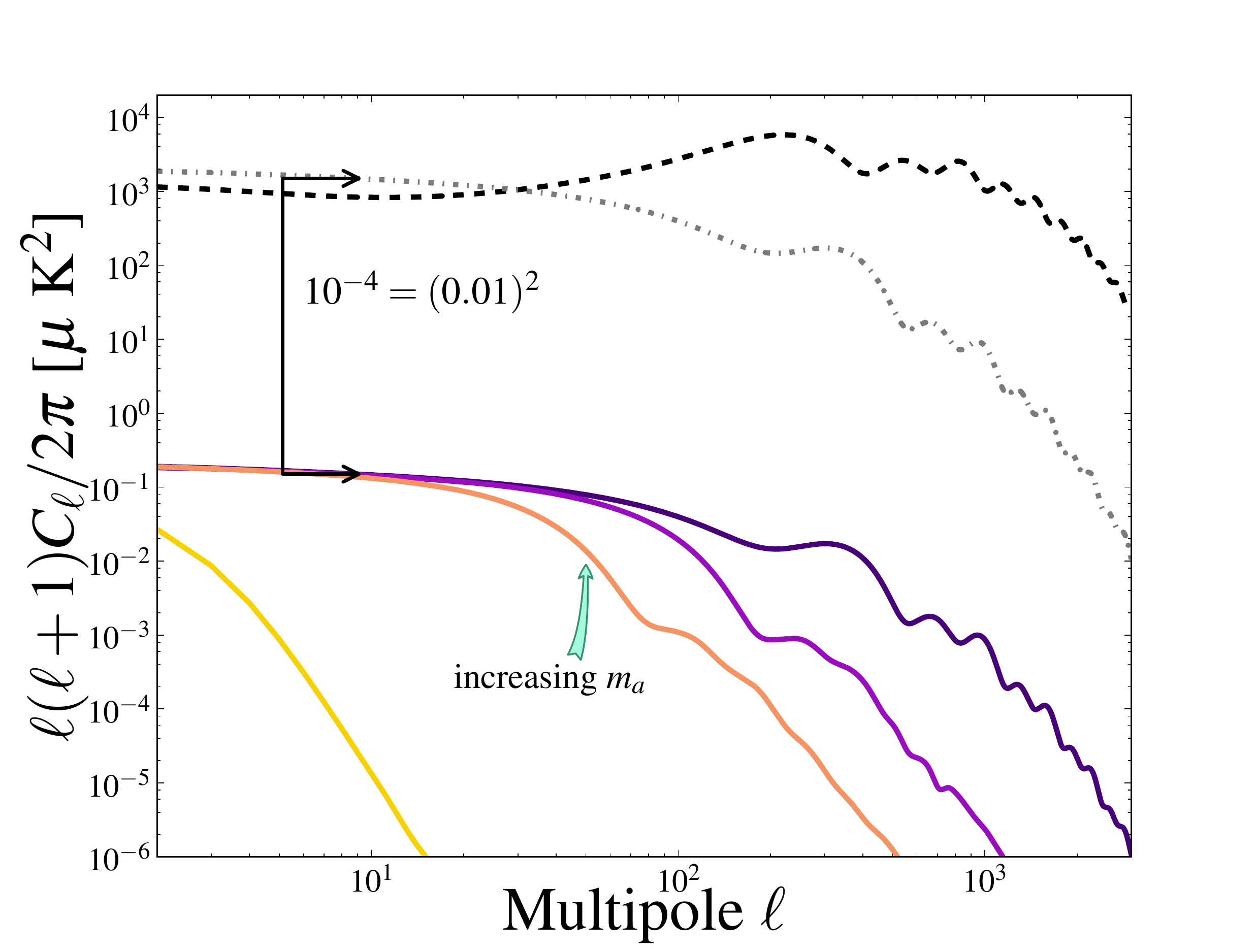}
\caption{CMB adiabatic and isocurvature spectra. $\Lambda$CDM adiabatic (dashed), CDM isocurvature with $\Omega_a/\Omega_d=A_I/A_s=1$ (dot dashed), ULA isocurvature with $\Omega_a/\Omega_d=0.01$ and increaing $m_a$ from left to right (solid, colour). Reproduced (with permission) from Ref.~\cite{marsh2013}. Copyright (2013) by The American Physical Society.}
\label{fig:schematic_isocurvature}
\end{center}
\end{figure} 

Axion isocurvature density perturbations are of \emph{uncorrelated CDM} type, as long as the Jeans scale can be neglected, which is the case for the QCD axion. The isocurvature CMB spectrum is shown in Fig.~\ref{fig:schematic_isocurvature}, where the effect of non-negligible ULA Jeans scale is also shown.  The isocurvature power spectrum generated by Eq.~\eqref{eqn:axion_field_iso} is:
\be
P_I=A_I\left(\frac{k}{k_0}\right)^{1-n_I}\, ,
\ee
with amplitude
\be
A_I=\left( \frac{\Omega_a}{\Omega_d}\right)^2\frac{(H_I/M_{pl})^2}{\pi^2(\phi_i/M_{pl})^2} \, .
\ee
The scalar power is: 
\be
P_\zeta=A_s\left(\frac{k}{k_0}\right)^{1-n_s}\, ,
\ee
with amplitude 
\be
A_s=\frac{1}{2\epsilon_{\rm inf}}\left(\frac{H_I}{2\pi M_{pl}} \right)^2=2.20\times 10^{-9}\, .
\ee
The measured value of $A_s$ is taken from \emph{Planck} (2015), and the scalar spectral index is measured to be $n_s=0.96$~\cite{planck_2015_params}. Uncorrelated CDM isocurvature is constrained to\footnote{This assumes scale invariance of the isocurvature power, $\epsilon\ll 1$, which is consistent with the implied value of $H_I$ and $r_T$. Compare this to the isocurvature power generated in the unbroken PQ scenario. In this case the amplitude is huge, $A_I\sim \langle(\delta\theta/\theta)^2\rangle\sim\mathcal{O}(1)\gg A_s$, but power is only generated on very small scales, $k\gg k_0$, that are not constrained by the CMB power spectrum. Spectral distortions and miniclusters may impose interesting additional constraints~\cite{2012PhRvD..85l5027B,2014arXiv1405.6938C}.} 
\be
\frac{A_I}{A_s}<0.038 \, .
\ee

The tensor-to-scalar ratio, $r_T=16\epsilon_{\rm inf}$, provides an independent constraint on the energy scale of inflation. \emph{Planck} and BICEP2~\cite{2015PhRvL.114j1301B} provide the limit $r_T<0.12$. The projected sensitivity of CMB-S4 experiments is $r_T\sim 10^{-3}$~\cite{2013arXiv1309.5381A}, while futuristic sensitivity from 21cm lensing could be as low as $r_T\sim 10^{-9}$~\cite{2005PhRvL..95u1303S,2012PhRvL.108u1301B}.
\begin{figure}
\begin{center}
\includegraphics[width=0.75\textwidth]{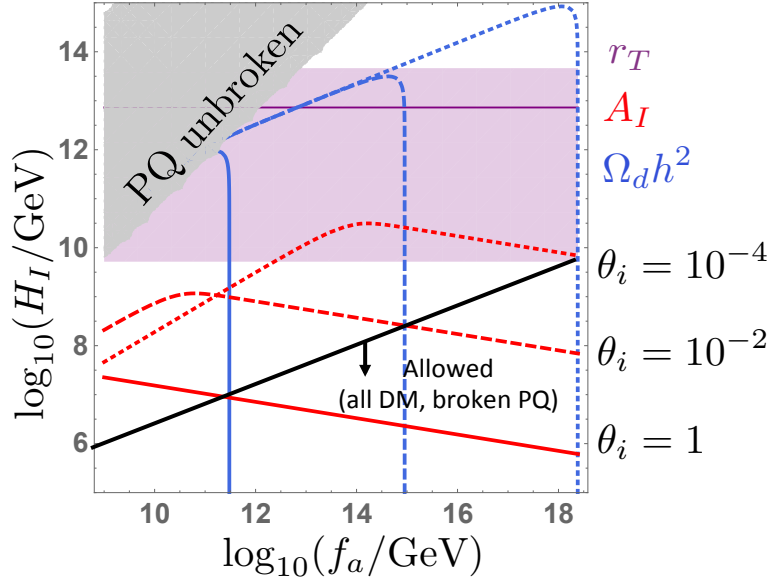}
\caption{The QCD axion and CMB tensor modes. The unbroken PQ scenario produces no isocurvature, and is allowed as long as the limits on $r_T$ and relic density (see Fig.~\ref{fig:qcd_unbroken_contour}) are satisfied, requiring low $f_a$. In the broken PQ scenario, we show various levels of tuning: $\theta_{a,i}=1$ (solid lines), $\theta_{a,i}=10^{-2}$ (dashed lines), $\theta_{a,i}=10^{-4}$ (dotted lines). Constraints are shown for relic density $\Omega_d h^2<0.12$ (blue, lie below-left) and isocurvature amplitude $A_I/A_s<0.04$ (red, lie below). The observable range of $10^{-9}<r_T<0.1$ is shown in purple, with a realistic near-future limit of $r_T=10^{-3}$ given by the solid line at $H_I\sim 10^{13}\text{ GeV}$. The allowed regime if the QCD axion in the broken PQ scenario is to be all the DM is given by the intersections of the red and blue lines (black), which always lies below a detectable tensor mode.}
\label{fig:qcd_iso}
\end{center}
\end{figure} 

All of these results are collected together for the QCD axion in Fig.~\ref{fig:qcd_iso}. I plot contours for $A_I/A_s=0.04$ and $\Omega_a h^2=0.12$ as functions of $(f_a,H_I)$ at fixed levels of fine tuning on $\theta_{a,i}$. Satisfying relic density and isocurvature constraints requires being below the intersections of these curves. For example, having $\theta_{a,i}=1$ and $\Omega_a h^2=0.12$, requires $f_a\approx 3\times 10^{11}\text{ GeV}$. The isocurvature constraint then enforces $H_I\lesssim 10^7\text{ GeV}$. The intersection of the $A_I$ and $\Omega_a$ constraints traces out, approximately, $H_I\lesssim 10^{10}(f_a/M_{pl})^{1/2}\text{ GeV}$ if axions are to be all the DM. The constraint quoted by \emph{Planck}~\cite{2015arXiv150202114P} in this scenario is $H_I<0.86\times 10^7\text{ GeV}(f_a/10^{11}\text{ GeV})^{0.408}$ (95\% C.L.), consistent with our rough estimates.

A range of measurably-large values of $r_T$ are shown shaded purple, corresponding to a range $10^{10}\text{ GeV}\lesssim H_I\lesssim 10^{14}\text{ GeV}$. There is nowhere on the $(f_a,H_I)$ plane where the QCD axion in the broken PQ scenario can be all of the DM, satisfy iscourvature bounds, and produce $r_T>10^{-9}$ (a realistically observable value, shown by the dark purple line). Note that such small values of $r$ can be obtained, consistent with $A_s$ and $n_s$ observations, in string inflation scenarios such as KKLT~\cite{2003PhRvD..68d6005K} or brane inflation (see Ref.~\cite{2014PDU.....5...75M} for details).

Relaxing the assumption that the QCD axion is all the DM, Fig.~\ref{fig:qcd_iso} shows that with $\theta_{a,i}\lesssim 10^{-4}$ a range of large $f_a$ starts to become consistent with $r_T>10^{-9}$. By trial and error, we find the maximum value of $r_T$ consistent with isocurvature constraints and $f_a<M_{pl}$ occurs for $\theta_{a,i}\approx 10^{-7}$ where we have $\Omega_a h^2<10^{-6}$ and $r_T\approx 10^{-4}$. There is no amount of tuning that can make the QCD axion in the broken PQ scenario consistent with tensor modes as large as $r_T=10^{-3}$, the CMB-S4 target.

CDM-type isocurvature modes are avoided completely in the unbroken PQ scenario. Thus, if tensor modes are observed, the QCD axion must live in the parameter space of Fig.~\ref{fig:qcd_unbroken_contour} contained within the grey shaded region of Fig.~\ref{fig:qcd_iso}, implying $f_a<10^{11}\text{ GeV}$. 

These conclusions can be avoided if some of our cosmological assumptions are relaxed. An example non-minimal inflation model producing $r_T>10^{-3}$ consistent with the broken PQ scenario and high $f_a$, uses the radial PQ field, $\chi$, as the inflaton, non-minimally coupled to gravity (similarly to Higgs inflation)~\cite{2015PhRvD..91b3509F}. Such a scenario can allow for simultaneous detection of DM axions by CASPEr~\cite{2014PhRvX...4b1030B} (see Section~\ref{sec:casper}), and detection of $r_T$ by, e.g., \textsc{spider}~\cite{2014SPIE.9153E..13R}. There are many other possibilities to avoid the isocurvature problem of high-scale axions by modifying inflation, particle physics, or the thermal history, for example Ref.~\cite{2014PhLB..734...21H}, and related works.  

We conclude our discussion of the QCD axion and isocurvature in summary:
\begin{itemize}
\item \emph{The QCD axion in the broken PQ scenario is incompatible with observably-large tensor modes from standard inflation.\footnote{It is, in fact, possible to make the QCD axion in the broken PQ scenario compatible with observable tensors if we allow $f_a\gtrsim 10^{10}M_{pl}$ and tune the initial misalignment angle at a level $\theta_{a,i}\ll 10^{-10}$. I exclude such a scenario as unreasonable. The tuning is worse than the strong-$CP$ problem, and the existence of a scale so much larger than the Planck scale is considered highly problematic in theories of quantum gravity.}  }
\item \emph{In the broken PQ scenario with standard inflation, axion isocurvature modes could probe $H_I$ as low as $10^7\text{ GeV}$, offering a unique probe of low-scale inflation.}
\item \emph{Simultaneously detecting a high $f_a\gtrsim 10^{13}\text{ GeV}$ QCD axion and tensor modes at $r_T=10^{-3}$ would falsify minimally coupled, single-field, slow-roll inflation with a standard thermal history.}
\end{itemize}

\section{Galaxy Formation}
\label{sec:galaxy_formation}

This section reviews work presented in Refs.~\cite{2014MNRAS.437.2652M,2015MNRAS.450..209B,2015MNRAS.451.2479M}.

\subsection{The Halo Mass Function}
\label{sec:hmf}

The halo mass function (HMF) gives the expected number of halos per logarithmic mass bin, per unit volume, for a given cosmology. It depends fundamentally on two quantities, both of which can depend on halo mass and redshift: the variance of fluctuations, $\sigma^2 (M,z)$, and the linearly extrapolated critical density required for such fluctuations to collapse, $\delta_{\rm crit}(M,z)$. The relevant standard formulae are given in Appendix~\ref{appendix:hmf_def}. 

We can compute $\sigma(M,z)$ given the linear power spectrum, $P(k,z)$. The cut-off in power caused by the axion Jeans scale leads to a suppression of $\sigma(M,z)$ compared to CDM at low halo mass, with $\sigma(M,z)$ going to a constant as $M\rightarrow 0$. The reduced value of $\sigma(M,z)$ reduces the abundance of low mass halos.

In an Einstein-de Sitter universe (CDM with $\Lambda=0$), spherical collapse can be solved exactly. Scale-independent growth gives a constant, mass-independent, value for $\delta_{\rm crit}$, which can be scaled to any redshift using the linear growth factor (the result also works well for $\Lambda$CDM on not-too-large scales):
\be
\delta_{\rm crit,EdS}(z)=\frac{1.686D_0}{D(z)}\, .
\label{eqn:delta_crit_cdm}
\ee

The collapse barrier is mass-independent for CDM because the growth equation is scale-invariant. In DM models with an effective pressure, the Jeans scale introduces scale-dependence into the collapse threshold. In spherical collapse simulations with WDM, where free-streaming was modelled by an effective pressure~\cite{2001ApJ...558..482B}, a mass-dependent critical barrier is found, with $\delta_{\rm crit}$ increasing below the WDM Jeans scale. This barrier can then be used in a full excursion set model of WDM halo formation, dramatically suppressing halo formation below the effective Jeans mass~\cite{2013MNRAS.428.1774B}. Spherical collapse and the excursion set have not been studied for axion DM. Instead, Ref.~\cite{2014MNRAS.437.2652M} proposed a simple model where $D(z)$ in Eq.~\eqref{eqn:delta_crit_cdm} is simply replaced by an appropriately normalized (in both scale and redshift relative to $\Lambda$CDM) scale-dependent growth factor, $\mathcal{G}$. The mass can be assigned from the wavenumber using the enclosed mean density in a sphere of radius $R=\pi/k$ giving:
\be
\delta_{\rm crit}(M,z)=1.686 \mathcal{G}(M,z) \, .
\label{eqn:delta_crit_ula}
\ee

We define $\mathcal{G}$ as the relative amount of growth between axion DM and CDM, normalized to unity on large scales, $k_0$, and at early times, $z_{\rm early}$:
\be
\mathcal{G}(k,z)=\frac{\delta_a(k_0,z)\delta_a(k,z_{\rm early})}{\delta_a(k,z)\delta_a(k_0,z_{\rm early})}\frac{\delta_c(k,z)\delta_c(k_0,z_{\rm early})}{\delta_c(k_0,z)\delta_c(k,z_{\rm early})} \, ,
\ee
where $\delta_a$ is computed in the axion cosmology, and $\delta_c$ is computed in the CDM cosmology, with $\Omega_ah^2=\Omega_ch^2$. In practice, $k_0$ should be chosen such that $k_0<k_J(z_{\rm early})$, but not so small such that scale dependent growth in $\Lambda$CDM due to $\Lambda$ domination becomes relevant. Similarly, $z_{\rm early}$ should be chosen such that the power spectrum shape in $\Lambda$CDM has frozen in, i.e. after BAO formation. For DM axions in a close-to-$\Lambda$CDM cosmology, reasonable choices are $k_0=0.002\,h\text{Mpc}^{-1}$ and $z_{\rm early}\approx 300$.\footnote{An interesting recent discussion of the relative importance of scale dependent growth to LSS simulations of axion DM is given in Ref.~\cite{2016ApJ...818...89S}, where a similar quantity to $\mathcal{G}$ is used to measure this.}

\begin{figure}
\begin{center}
$\begin{array}{@{\hspace{-0.7in}}l@{\hspace{-0.5in}}l}
\includegraphics[scale=0.4]{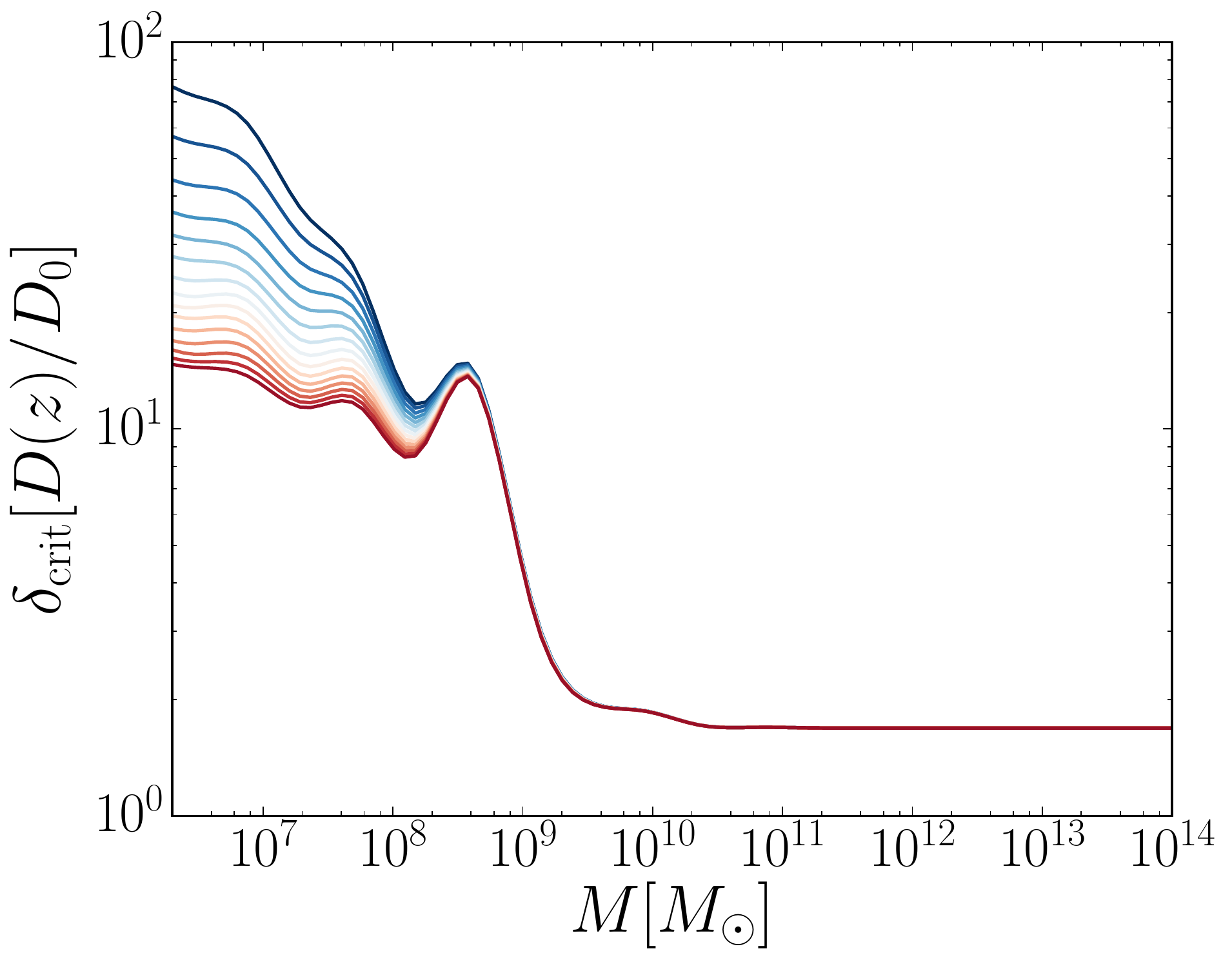}
\includegraphics[scale=0.4]{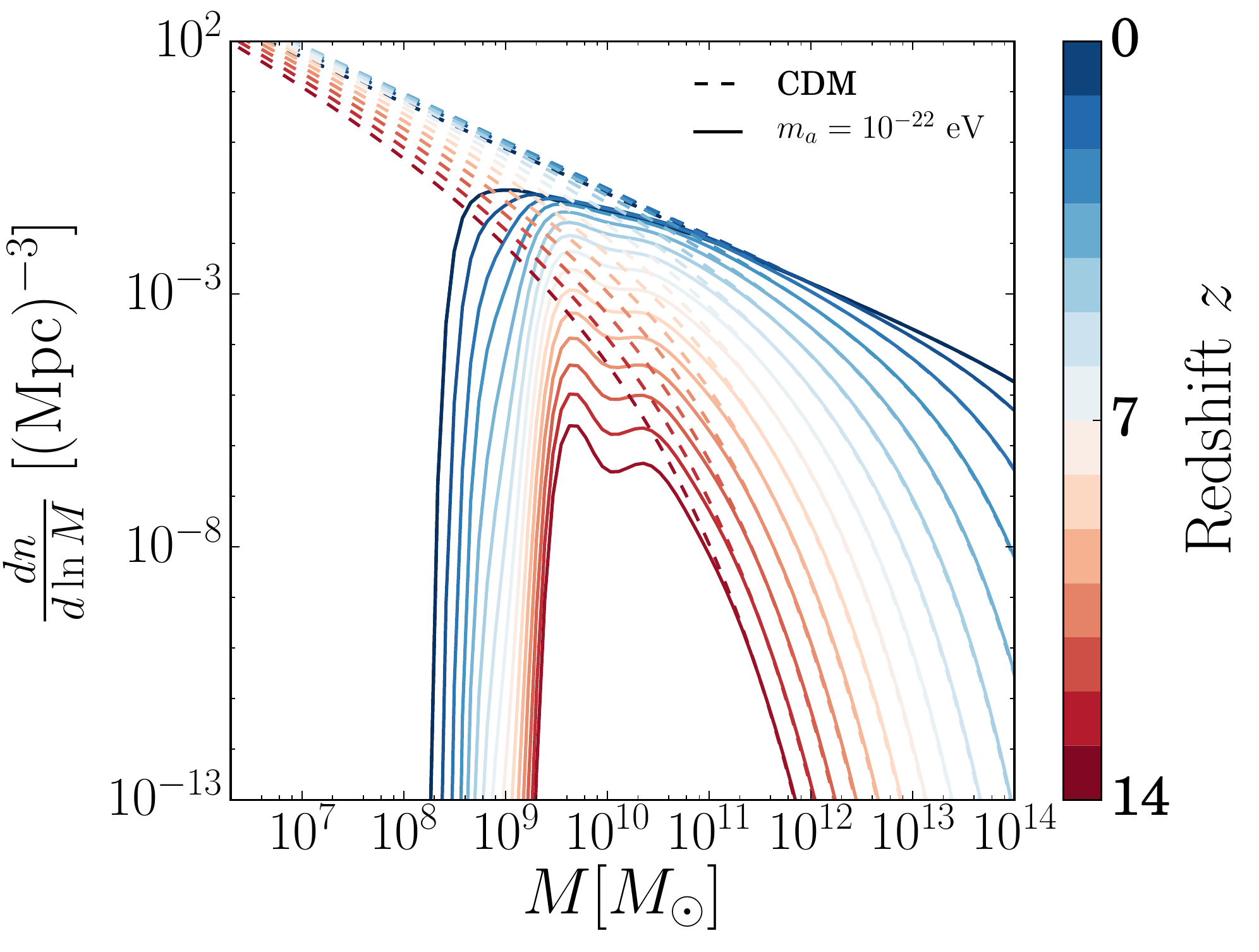}
 \end{array}$
 \end{center}
 \caption{\emph{Right Panel:} Critical overdensity for collapse, $\delta_{\rm crit}(M,z)$ for $m_a=10^{-22}\text{ eV}$ computed from scale-dependent growth using Eq.~\eqref{eqn:delta_crit_ula}, normalizing for the growth in $\Lambda$CDM using $D(z)$. \emph{Left Panel:} Resultant halo mass function, compared to CDM. Modified from Ref.~\cite{2015MNRAS.450..209B}, Figs. 1 and 2.}
\label{fig:delta_hmf}
\end{figure} 

The critical overdensity appears in the HMF in the argument of a Gaussian. Thus, even a modest increase in $\delta_{\rm crit}$ causes a sharp cut-off in the HMF: this is shown in Fig.~\ref{fig:delta_hmf}.\footnote{The fact that the barrier appears in a Gaussian also renders the details of the barrier function, such as the acoustic features and smoothing scheme at masses much below the axion Jeans scale, largely irrelevant for halo statistics.} The cut-off makes physical sense: there are no seed density perturbations on scales below the Jeans scale, and even if there were, growth is so suppressed there that density perturbations cannot collapse into virialized objects. At higher redshifts, when density perturbations are smaller, and the Jeans scale is larger, the effect is more pronounced. We learn that: \emph{ULAs dramatically suppress halo formation compared to CDM at low halo masses and at high redshifts.}\footnote{There is some discussion and debate concerning the location and origin of the HMF cut-off in both WDM (filtering, spurious structure~\cite{2007MNRAS.380...93W}) and CDM (baryonic effects) that I will not go into here. For axions, numerical simulations such as those of Ref.~\cite{2014NatPh..10..496S,2016ApJ...818...89S}, with the addition of hydrodynamics and star formation, are necessary in order to be more precise. For basic, semi-analytic results, the intuitive notion of a cut-off at the Jeans scale provided by scale-dependent growth is sufficient.} 

For the QCD axion, the cut-off in the HMF induced by the Jeans scale is on extremely small scales $M<10^{-9}M_\odot$ (c.f. the standard WIMP, where the smallest halos have mass $M\approx 10^{-6}M_\odot$~\cite{2004MNRAS.353L..23G}). These smallest halos will certainly be tidally disrupted today, but are interesting to study the very first moments of structure formation at $z\approx 60$ in CDM models. Axion miniclusters produced in the unbroken PQ scenario for the QCD axion in the classic window have $M_{\rm mc}\approx 10^{-9}M_\odot$~\cite{1994PhRvD..49.5040K}. Miniclusters of ALPs may be more, or less, massive. Being denser than ordinary halos, axion miniclusters survive to the present day and are relevant to observational searches for minihalos (e.g. Refs.~\cite{2012PhRvD..85l5027B,2010AdAst2010E...9Z,2012PhRvD..86d3519L}).

\subsection{Constraints from High-$z$ and the EOR}
\label{sec:ula_highz_constraints}

There is accumulating data about the high-$z$ Universe. We see a number of very high redshift galaxies with Hubble Ultra Deep Field (HUDF, e.g.~Ref.~\cite{2015ApJ...803...34B}). We also know that the intergalactic medium (IGM) is reionized by star formation. Reionization is known to be essentially complete by $z\sim 6$ (e.g. observation of Gunn-Peterson trough~\cite{1965ApJ...142.1633G} in quasar spectra~\cite{2006AJ....132..117F}). Furthermore, reionization of the IGM produces an optical depth to the CMB, which is constrained by a combination of large angle temperature and polarization correlation functions to be $\tau=0.07$--$0.08\pm 0.02$ (central value depends on dataset combinations in Ref.~\cite{planck_2015_params}).

The suppression of halo formation at high-$z$ by ULAs cannot be too severe, or else it would be inconsistent with these observations, producing too few high-$z$ galaxies to match HUDF and to efficiently reionize the IGM. Getting these things right places a lower bound on $m_a$ if ULAs are to contribute significantly to the DM density. Ref.~\cite{2015MNRAS.450..209B} investigated these bounds, following similar work on WDM in Ref.~\cite{2014MNRAS.442.1597S}.

In order to obtain constraints from the HMF, one needs to relate the halo mass to the UV magnitude of the galaxy, $M_{\rm UV}$. This can be done by abundance matching~\cite{2004ApJ...609...35K,2004MNRAS.353..189V}. The luminosity function, $\phi_{\rm lum} (M_{\rm UV},z)$, is fit and matched to the low-$z$ observations. The integrated (cumulative) luminosity function is then matched by number count to the cumulative halo mass function: $\Phi_{\rm lum}(<M_{\rm UV},z)=n(>M_h,z)$. This chain of relations fixes $M_h(M_{\rm UV})$. Therefore, once the low redshift data are fixed, the high redshift value of $\Phi_{\rm lum}(M_{\rm UV},z)$ can be predicted for a given DM model, and itself compared to observation. The cut-off in the HMF induced by the axion Jeans scale cuts off the $M_h(M_{\rm UV})$ relation at some brightest magnitude, leaving the function $\Phi_{\rm lum}(M_{\rm UV},z)$ with no support at the faint end.

\begin{figure}
\begin{center}
$\begin{array}{@{\hspace{-0.4in}}c@{\hspace{-0.2in}}c}
\includegraphics[scale=0.22]{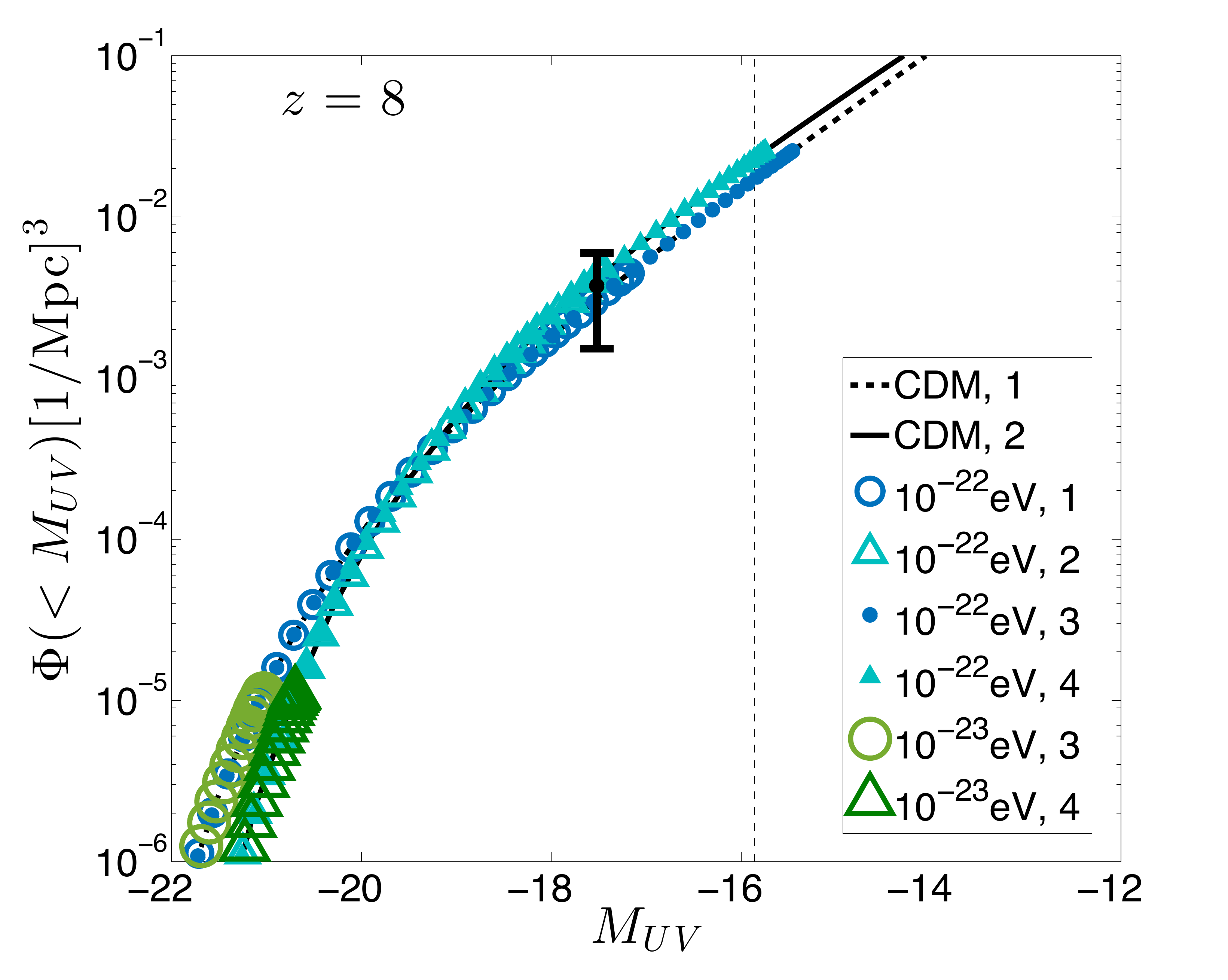} &
\includegraphics[scale=0.22]{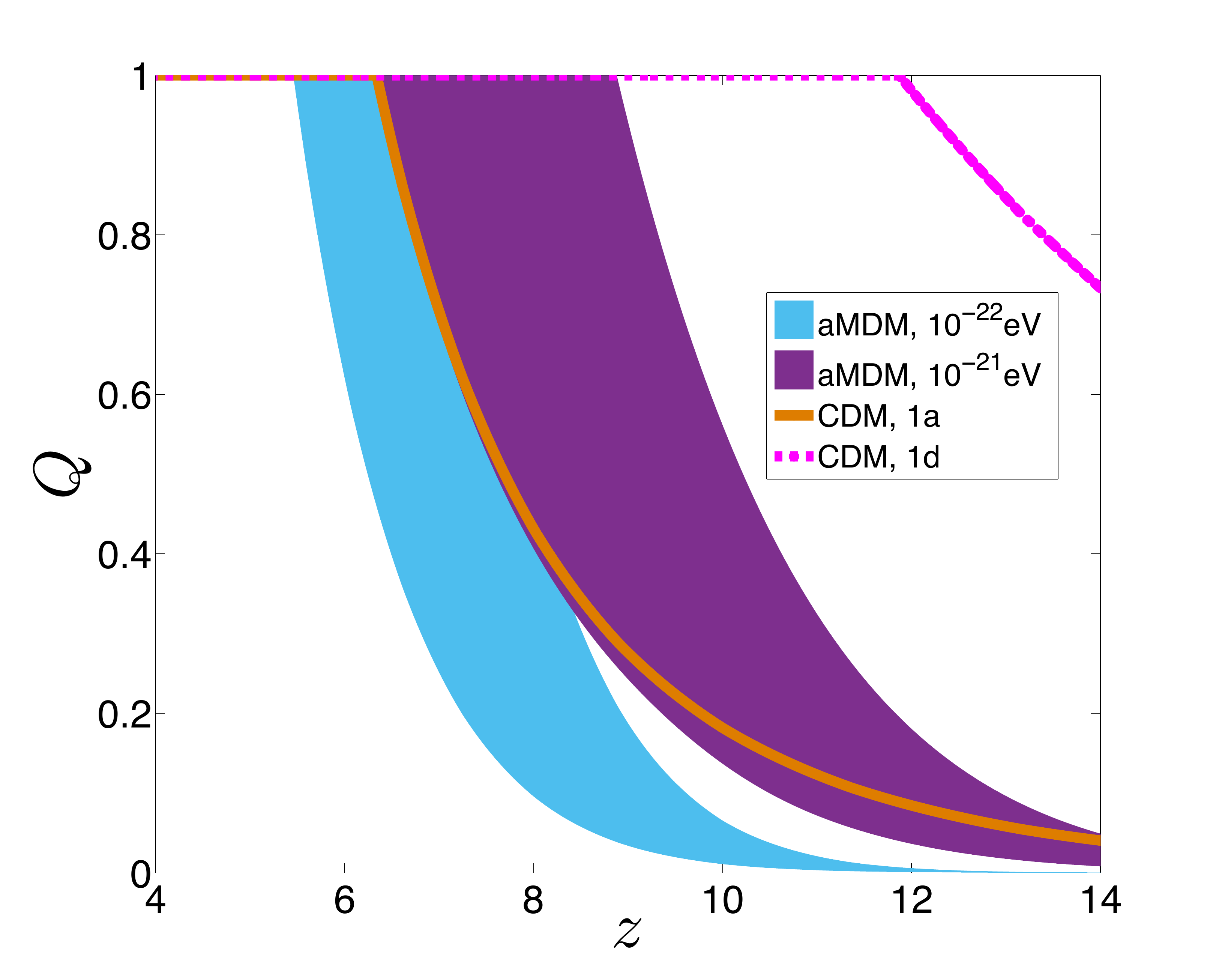}
 \end{array}$
 \end{center}
 \caption{\emph{Left Panel:} Cumulative UV luminosity at $z=8$, here denoted $\Phi$, in axion models using the abundance matching technique. Data: HUDF~\cite{2015ApJ...803...34B}. Dashed line: JWST reach~\cite{2006NewAR..50..113W}. Model numbers are different abundance matching procedures and DM composition. Models 1, 2, ULAs are all the DM. Models 3, 4, ULAs are half of the DM. \emph{Right Panel:} Ionization fraction. Shaded regions cover model uncertainties. Only extreme edges shown for CDM. Reproduced from Ref.~\cite{2015MNRAS.450..209B}, Figs.~4 and 6.}
\label{fig:bozek_eor}
\end{figure}

Fig.~\ref{fig:bozek_eor} (Left Panel) shows the predicted cumulative luminosity function for axion DM at $z=8$. If ULAs are too light, or make up too much of the DM, it is impossible to match the observed HUDF UV luminosity. The model $m_a=10^{-23}\text{ eV}$ with $\Omega_a h^2>0.06$ is ruled out at $>8\sigma$ by HUDF. The model $m_a=10^{-22}\text{ eV}$ with $\Omega_a h^2=0.12$ is consistent with HUDF, but only just: the UV luminosity function cuts off at $M_{\rm UV}\approx -18$, right where the constraint is. This model could be excluded by a JWST measurement of the faint-end luminosity function at $M_{\rm UV}\approx -16$~\cite{2006NewAR..50..113W} if it were found to be consistent with the larger CDM value of $\Phi_{\rm lum}(M_{\rm UV},z)$.

The UV luminosity function can also be used to predict the evolution of the ionization fraction, $Q(z)$ (not to be confused with the quantum potential, also denoted $Q$). This involves a fair amount of astrophysical modelling, as described in e.g. Refs.~\cite{2012MNRAS.423..862K,2015MNRAS.450..209B,2016ApJ...818...89S}. The results are shown in Fig.~\ref{fig:bozek_eor} (Right Panel), with shaded regions showing modelling uncertainty. These results are broadly consistent with the studies of Refs.~\cite{2016ApJ...818...89S,2016JCAP...04..012S}, where the underlying halo mass function was computed from $N$-body simulations with modified initial power spectra. Ref.~\cite{2016JCAP...04..012S} also used different methods to model the reionization field.

The ionization fraction gives the optical depth to redshift, $\tau(z)$, from the integral along the line of sight:
\be
\tau(z) = \int_0^z dz' \frac{(1+z')^2}{H(z')}Q(z')\sigma_T\bar{n}_H(1+\eta_{\rm He}Y/4X)\, ,
\ee 
where $\sigma_T$ is the Thompson optical depth, $\bar{n}_H$ is the mean comoving Hydrogen number density, $Y=(1-X)$ is the Helium fraction, $X$ is the Hydrogen fraction, and $\eta_{\rm He}$ is the ionization state of Helium (see Ref.~\cite{2015MNRAS.450..209B} for references and more details on these parameters). The optical depth to the CMB is $\tau(z_{\rm rec}\approx 1100)$. 

Ref.~\cite{2015MNRAS.450..209B} found that, within the modelling uncertainty, all axion DM models with $m_a\geq 10^{-22}\text{ eV}$ can reproduce a CMB optical depth consistent with observations, while $m_a=10^{-23}\text{ eV}$ cannot (though the tension for the lightest masses is slightly less with the revised, \emph{Planck} 2015, value for the optical depth). Thus the CMB optical depth excludes the lightest ULAs with $m_a\lesssim 10^{-23}\text{ eV}$ from being all of the DM.

There is the opportunity in future to constrain axion DM with $m_a\sim 10^{-22}$--$10^{-21}\text{ eV}$ from the evolution of $Q(z)$. The cut off in the HMF delays the formation of the first galaxies, and thus reionization occurs at lower redshift than in CDM. Once collapse has begun, structure builds up more rapidly for ULAs, and reionization completes in a smaller redshift window. These different reionization histories distinguish ULAs and CDM. For example, the amplitude of the kinetic Sunyaev-Zel'dovich effect~\cite{1970Ap&SS...7....3S} in the CMB is sensitive to the duration of reionization (e.g. Ref.~\cite{1998ApJ...508..435G}). This will be measured in the near future by Advanced ACTPol~\cite{2014JCAP...08..010C} and could distinguish $m_a\lesssim 10^{-21}\text{ eV}$ from CDM~\cite{2015MNRAS.450..209B}.

The bottom line is that high-$z$ constraints currently exclude $m_a=10^{-23}\text{ eV}$ from being all of the DM at high confidence, and $m_a=10^{-22}\text{ eV}$ is right on the edge of acceptability. The bounds are only approximate, as a lot of uncertain astrophysics is involved, but Ref.~\cite{2015MNRAS.450..209B} covered a range of models and the lower limit on $m_a\gtrsim 10^{-22}\text{ eV}$ is reliable by order of magnitude. Similar results were also found by Ref.~\cite{2016ApJ...818...89S}, giving $m_a\geq 1.2\times 10^{-22}\text{ eV}\,(2\sigma)$. \emph{This is the current lower limit on DM particle mass from non-linear clustering.} Future constraints on high-$z$ galaxies, and on the mean redshift and duration of reionization, could improve this limit by some two or more orders of magnitude. A measurement of the large scale 21cm power spectrum could constrain ULA mass as high as $m_a\approx 10^{-18}\text{ eV}$~\cite{2015PhRvD..91l3520M}.

\subsection{Halo Density Profiles}
\label{sec:halo_profiles}

$N$-body simulations of pure CDM indicate that halo density profiles have a universal shape, known as the Navarro-Frenk-White (NFW) profile~\cite{Navarroetal1997}:
\be
\frac{\rho_{\rm NFW}(r)}{\rho_{\rm crit.}}=\frac{\delta_{\rm NFW}}{r/r_s(1+r/r_s)^2}\, ,
\ee
where $\delta_{\rm NFW}$ is a function of the ``halo concentration,'' commonly denoted as $c$, and $r_s$ is the scale radius. The concentration is defined such that the virial radius is $r_{\rm vir}=cr_s$.\footnote{The virial radius is taken to be the radius where the density is 200 times the critical density, and the virial velocity is the circular velocity at this radius. The mass of a halo is often defined as $M_{200}=M(<r_{\rm vir})$. One can use this to derive $\delta_{\rm NFW}(c)$. A typical concentration is $c\sim 10$.} Notice that the NFW halo is a smoothly varying power law, with $\rho\sim r^{-1}$ in the centre: the so-called `cusp.' 

A dwarf galaxy in $\Lambda$CDM with $M\sim 10^{10}M_\odot$ has peak circular velocity on the order of 50~km~s$^{-1}$ at a radius of around 10~kpc. The de Broglie wavelength, $\lambda_{\rm dB}=1/mv$, of a particle inside such galaxy is then
\be
\lambda_{\rm dB}\geq 4\times10^{-2} \left(\frac{m_a}{10^{-22}\text{ eV}} \right)^{-1}\text{ kpc} \, ,
\ee
and for a ULA is non-negligible in terms of the galaxy size. Using that $v\sim M/r$ and $M\sim \rho r^3$, setting $\lambda_{\rm dB}=r$ we find that $\lambda_{\rm dB}\sim m_a^{-1/2}\rho^{-1/4}\sim r_J$ where $r_J$ is the Jeans scale.

Let's work directly with the Jeans scale. Taking $r_J=2\pi/k_J$ and simply scaling Eq.~\eqref{eqn:axion_jeans} to the halo density gives
\be
r_J=94.5 \left(\frac{m_a}{10^{-22}\text{ eV}} \right)^{-1/2} \left(\frac{\rho (r_J)}{\rho_{\rm crit.}}\right)^{-1/4} \left(\frac{\Omega_a h^2}{0.12}\right)^{-1/4} \text{ kpc}\, .
\ee
This is a polynomial equation to be solved for $r_J$. Plugging in a typical overdensity of $10^6$ with $m_a=10^{-22}\text{ eV}$ gives $r_J\sim 3\text{ kpc}$. The ULA Jeans scale inside a dwarf halo can be very large. 

The wavelike effects of ULAs (the de Broglie and Jeans scales) affect the halo density profile, and it cannot be completely described by the CDM result. How is the NFW profile modified by the presence of a ULA and what forms on small scales? Clearly there should be some granularity and a smoothing of the central cusp, each caused by the wave-mechanical uncertainty principle. When the density is smoothed over many Jeans scales, the profile should return to being NFW-like. These effects are observed in simple one-dimensional~\cite{hu2000} and full cosmological~\cite{2014NatPh..10..496S} simulations. Both the core and the granularity~\cite{2014PhRvL.113z1302S} can be understood by considering a certain class of soliton solution~\cite{1969PhRv..187.1767R,1991PhRvL..66.1659S} of the axion equations of motion.\footnote{Technically, these solutions are pseudo-solitons since the field is time-dependent, and they are not absolutely stable. This is a distinct difference between axions, which are real-valued fields, and complex scalar field DM. Complex fields have a conserved $U(1)$ charge and true soliton solutions known as boson stars~\cite{Liddle:1993ha}. See e.g. Ref.~\cite{2013PhRvD..88f7302D}, the Appendix of Ref.~\cite{2015MNRAS.451.2479M}, and references therein, for more discussion.}

We work in the non-relativistic Schr\"{o}dinger picture of Section~\ref{sec:schrodinger}. Stationary wave, constant energy solutions take the form
\be
\psi=\mathcal{X}(r)e^{-i\gamma t}\, ,
\ee
where $\gamma$ is the energy eigenvalue. The system possesses a very useful \emph{scaling symmetry}~\cite{1969PhRv..187.1767R}:
\begin{equation}
(r, \mathcal{X}, \Psi, \gamma, M(<r), \rho) \rightarrow
(r/\lambda, \lambda^2 \mathcal{X}, \lambda^2 \Psi, \lambda^2\gamma,\lambda M(<r), \lambda^4 \rho)\, , \label{eqn:scalingRelation}
\end{equation}
where the scale factor is $\lambda$, $\rho=\mathcal{X}^2$ is the soliton density, and $M(<r)$ is the soliton mass enclosed within radius $r$. Imposing the correct boundary conditions~\cite{2015MNRAS.451.2479M,2006ApJ...645..814G} one can numerically solve the resulting system of ordinary differential equations to find $\mathcal{X}(r)$ and $\gamma$. Thanks to the scaling symmetry, this solution need only be found once. The solution with $\mathcal{X}(0)=1$ gives $\gamma=-0.692$ for the zero node groundstate. The ground state solution for an isolated soliton is reached rapidly by a process of ``gravitational cooling''~\cite{1994PhRvL..72.2516S,2006ApJ...645..814G}. The ground state also provides a good description of the cores in virialised DM halos found in the simulations of Ref.~\cite{2014NatPh..10..496S}.

The groundstate soliton solution possess a single characteristic radius, $r_{\rm sol}$, fixed entirely by the choice of units, which in turn is fixed by the axion mass. The scaling symmetry then uniquely fixes the relationship between the central density, $\rho_{\rm sol}$, and the characteristic radius:
\be
r_{\rm sol}\propto m_a^{-1/2}\rho_{\rm sol}^{-1/4} \, .
\ee
The soliton characteristic radius has the same scaling properties as the Jeans scale! This is no surprise: the scalings are derived on dimensional grounds in the non-relativistic limit. The Jeans scale is found from Eq.~\eqref{eqn:approx_sound_speed}, which as we showed can be derived from perturbation theory on the Schr\"{o}dinger equation via the quantum potential.
\begin{figure}
\begin{center}
$\begin{array}{@{\hspace{-0.1in}}c@{\hspace{+0.5in}}c}
\includegraphics[scale=0.62]{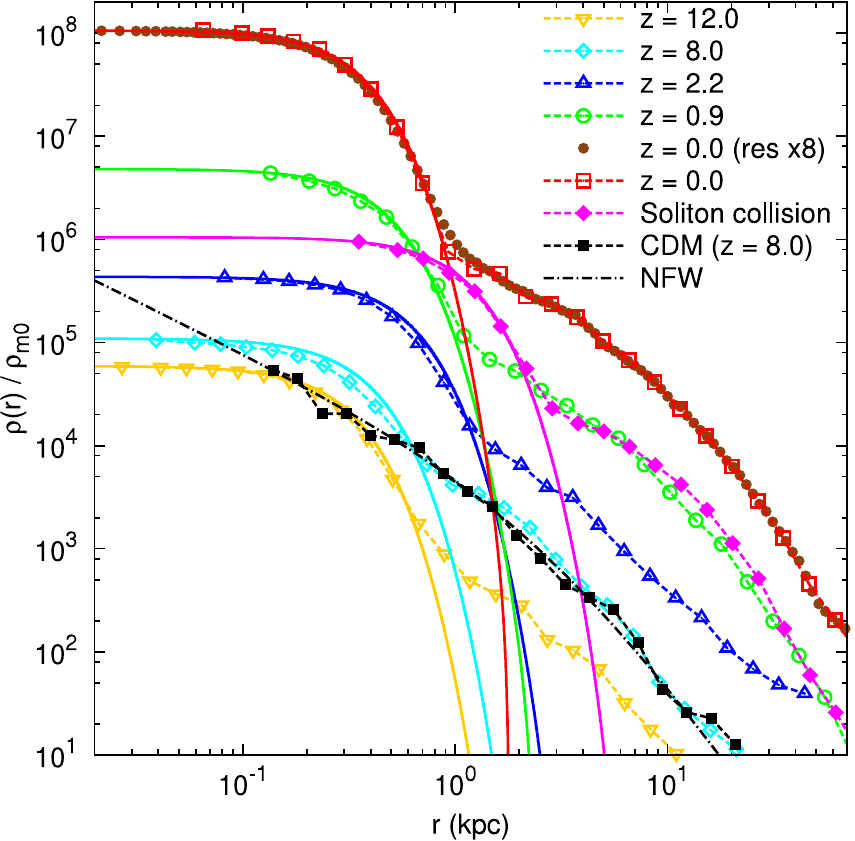} &
\includegraphics[scale=0.26]{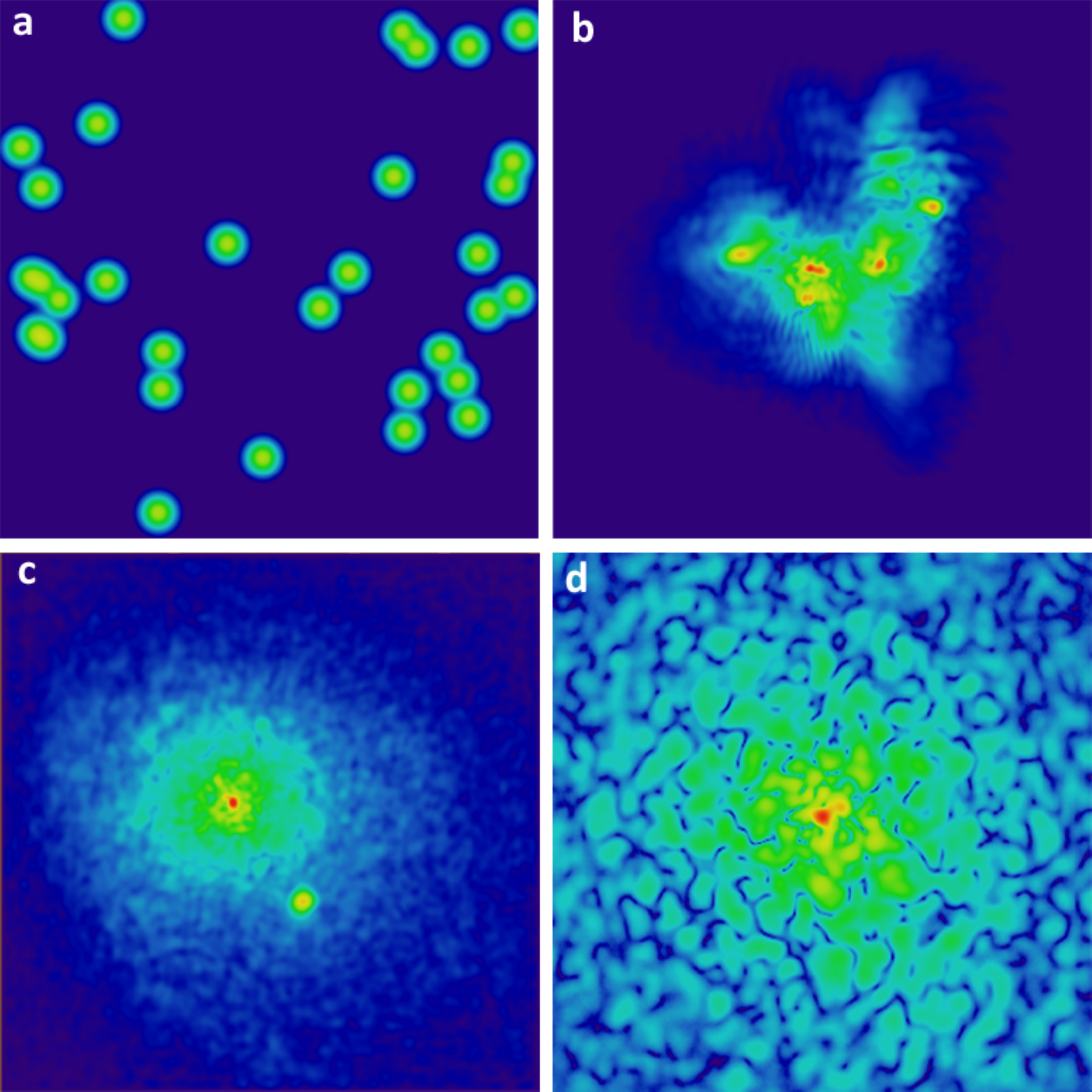}
 \end{array}$
 \end{center}
 \caption{\emph{Left Panel:} Halo density profiles from cosmological simulations of structure formation with a non-relativistic scalar field of mass $m_a=8.1\times 10^{-23}\text{ eV}$ (equivalent to a ULA). There is a central soliton core, transitioning to an NFW profile at large radius, as Eq.~\eqref{eqn:axion_halo_density_profile}. \emph{Right Panel:} Understanding halo formation from soliton collision. The solitons virialize and leave behind a small, dense core, and a granular outer halo: (d) is a close up of (c) detailing this. Reproduced (with permission) from Ref.~\cite{2014PhRvL.113z1302S}. Copyright (2015) by The American Physical Society.}
\label{fig:schive}
\end{figure} 

A good fit to the soliton density profile is provided by:
\be
\rho_{\rm sol}(r)=\frac{\rho_{\rm sol }(0)}{(1+(r/r_{\rm sol})^2)^8}\, ,
\label{eqn:soliton_fit}
\ee
with
\be
r_{\rm sol}=22 \left( \frac{\rho_{\rm sol}(0)}{\rho_{\rm crit}}\right)^{-1/4}\left(\frac{m_a}{10^{-22}\text{ eV}} \right)^{-1/2} \text{ kpc}\, .
\ee
The soliton density has dropped to $\rho_{\rm sol}(0)/2$ at $r_{1/2}\approx 0.3 r_{\rm sol}$, which might be said to be the `core radius.' For a central overdensity of $10^6$ and $m_a=10^{-22}\text{ eV}$ we have $r_{1/2}=0.2\text{ kpc}$, which is smaller than the naive halo Jeans scale, but is of order the de Broglie scale solved for via the circular velocity in the soliton profile~\cite{2015MNRAS.451.2479M}.

A complete model for the axion halo density profile must match the soliton and NFW profiles continuously at some radius. An exact description of the matching is currently lacking (though of course, by order of magnitude it must be at the Jeans/ de Broglie scale), so we can simply parameterize it to occur at $r_\epsilon$ and write
\be
\rho(r)=\Theta (r_\epsilon-r)\rho_{\rm sol}(r)+\Theta (r-r_\epsilon)\rho_{\rm NFW}(r)\, .
\label{eqn:axion_halo_density_profile}
\ee
This profile can be used to compare to galactic rotation curves and stellar kinematical data, either to fix the ULA mass, or to make predictions for a given mass. Similar profiles occur in other models of scalar field DM, such as self-interacting real or complex fields, and can also be used to fit density cores (see Section~\ref{sec:small_scale}) and constrain the parameters of these models~\cite{2013PhRvD..87b1301G,2014PhRvD..90d3517D}.

Fig.~\ref{fig:schive} shows results from numerical simulation of structure formation with a massive scalar field in the non-relativistic regime, taken from Ref.~\cite{2014PhRvL.113z1302S}, and discussed in Section~\ref{sec:schive_simulations}. The left panel shows density profiles taken from a full cosmological simulation at various redshifts, for $m_a=8.1\times 10^{-23}\text{ eV}$~\cite{2014NatPh..10..496S}. The profiles show a central soliton matching to NFW when the density has dropped to $\mathcal{O}(10^{-2})$ of the central density. The soliton profile is well fit by Eq.~\eqref{eqn:soliton_fit}. The right panel shows a numerical experiment of halo formation from collision of multiple solitons. The solitons virialize and leave behind a dense core, with a granular structure in the outer halo on the scale of the core size. The density profile from the soliton collision experiments is also shown in the left panel (arbitrarily normalized to show on the cosmological scale), and also has the same general form as Eq.~\eqref{eqn:axion_halo_density_profile}. The formation of solitons during structure formation with ULAs seems an established numerical fact, but many consequences of this have yet to be fully explored.

\subsection{ULAs and the CDM Small Scale Crises}
\label{sec:small_scale}

The main CDM ``small scale crises'' are~\cite{2013arXiv1306.0913W}:
\begin{itemize}
\item The missing satellites problem~\cite{1999ApJ...524L..19M,1999ApJ...522...82K}: CDM predicts more small Milky Way satellites than are observed.
\item The too-big-to-fail problem~\cite{2011MNRAS.415L..40B}: CDM predicts more massive satellites that should contain stars than are observed.
\item The cusp-core problem~\cite{2008IAUS..244...44W}: many observed low-mass systems contain flat central density profiles, not NFW cusps.
\end{itemize}
All of these problems, and variants of them, are essentially related to the overabundance of structure on small-scales in CDM, which itself is caused by the cold, collisionless, scale-free nature of CDM clustering. 

Methods to address the small-scale problems come in two varieties: baryonic/astrophysical solutions, and dark matter solutions. A recent set of state-of-the-art simulations discussing the baryonic solutions based on feedback from star formation is Ref.~\cite{2015arXiv150202036O}, while a review of the relevant issues if Ref.~\cite{2014Natur.506..171P}.

Dark matter based solutions are interesting, as they attempt to solve the problems by the introduction of a small number of universal parameters. The extent to which these models offer a solution can in principle point to specific values of these parameters. Because of this, we should not only demand solutions to the small-scale crises, but also a complete and consistent cosmological history, which gives the models some predictive power. They also offer us a framework for parameterizing our uncertainty about DM. In the absence of a fundamental theory of DM, as Bayesians we should allow for varying DM properties at the same time as we vary the baryonic physics. Moving away from CDM in this way may allow for a mixed baryon-DM solution with more reasonable priors on astrophysical parameters. Finally, a range of parameters will also be excluded, e.g. providing too few satellites, and independent of offering a solution to the small-scale crises we have learned something new about DM.

So what do DM solutions to the small-scale crises look like? Two popular models are self-interacting (SI)DM~\cite{Spergel:1999mh}, and WDM~\cite{bode2001}. I will only discuss WDM in detail, as it is interesting to contrast with ULAs. For further discussion of SIDM and other interacting models with relation to the small-scale crises and other areas of galaxy formation, see e.g. Refs.~\cite{2011PhRvL.106q1302L,2014MNRAS.444.3684V,2015MNRAS.449.3587S,2015arXiv150605471L}.
\begin{figure}
\begin{center}
\includegraphics[width=0.75\textwidth]{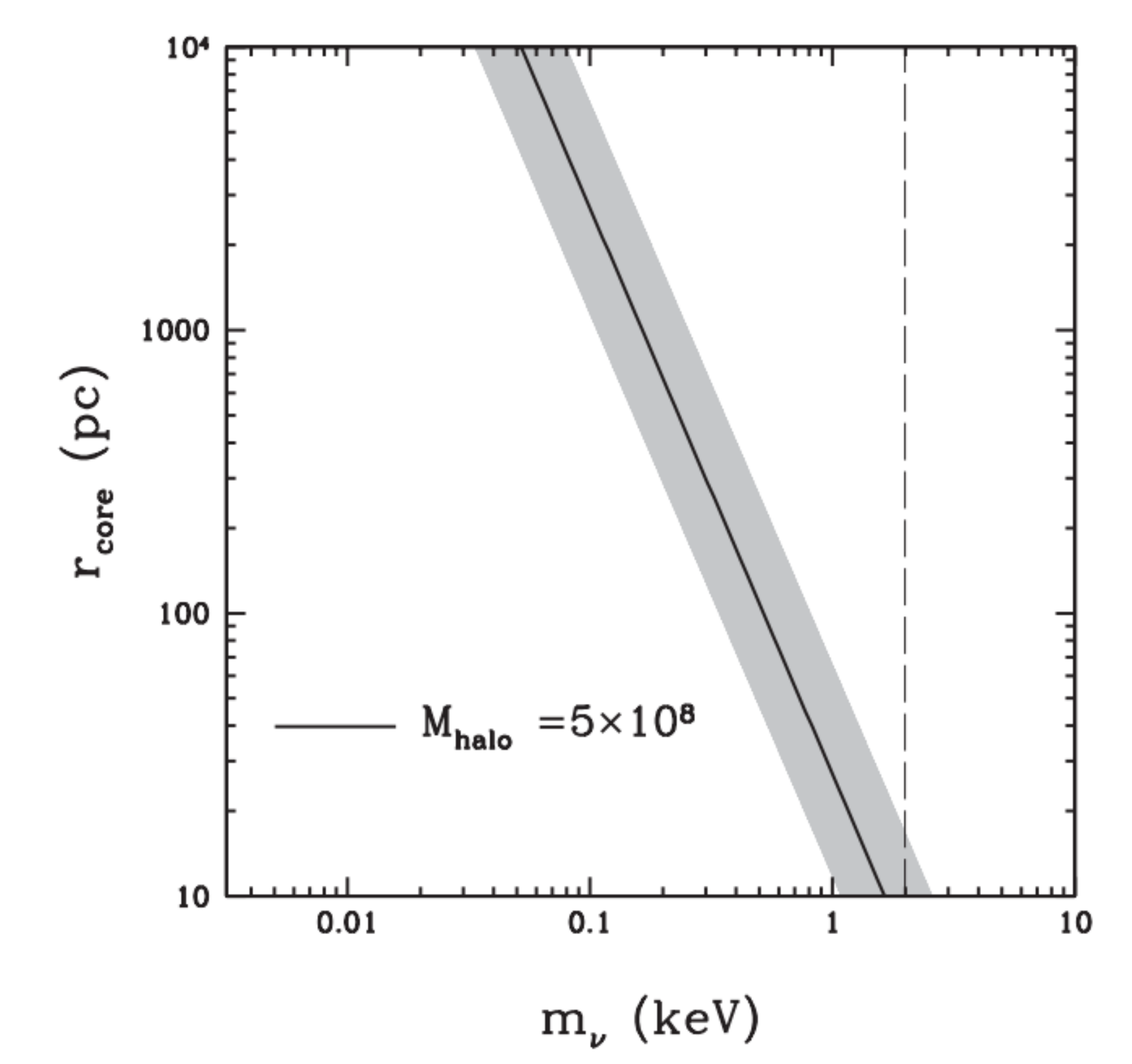} 
 \end{center}
 \caption{Core size in a WDM halo of mass $M=5\times 10^8 M_\odot$ as a function of WDM thermal relic mass, with uncertainties given by the shaded region. A representative constraint of $m_X>2\text{ keV}$ is shown by the vertical dashed line, which leads to small, $\mathcal{O}(10\text{ pc})$ cores and imposes the WDM \emph{Catch 22}. Reproduced (with permission) from Ref.~\cite{2012MNRAS.424.1105M,2013MNRAS.428.3715M}, Fig.~2.}
\label{fig:wdm_catch22}
\end{figure} 

WDM suppresses structure formation by free-streaming and a cut-off in the matter power, as we discussed in Section~\ref{sec:transfers}. This has the ability to address the missing satellites and too-big-to-fail problems for $1.5\text{ keV}\lesssim m_X\lesssim 2.3 \text{ keV}$~\cite{2014MNRAS.439..300L}, while still producing enough satellites and passing constraints on phase space density~\cite{2014PhRvD..89b5017H}. Fermion degeneracy pressure and thermal velocities also allow WDM to form density cores~\cite{1979PhRvL..42..407T}. The core-size-WDM mass relation is plotted in Fig.~\ref{fig:wdm_catch22}, with $r_c\sim m_X^{-1/2}$. Herein lies a problem known as the \emph{Catch 22} of WDM~\cite{2012MNRAS.424.1105M}: core sizes in dwarf galaxies are too small if constraints from satellite abundance and LSS are accounted for. Specifically, the N-body simulations of Ref.~\cite{2012MNRAS.424.1105M} found that masses $m_X\sim$1-2 keV gives a core of size $r_c\sim 10(20)\text{ pc}$ in a dwarf galaxy of mass $10^{10(8)}M_\odot$, far smaller than the $\mathcal{O}(\text{kpc})$ cores required in e.g. Fornax and Sculptor~\cite{Walker&Penarrubia2011}. Ref.~\cite{2013MNRAS.430.2346S} computed the WDM phase space density from N-body simulations and used this to derive the core size expected from free-streaming. A mass $m_X\approx 0.5\text{ keV}$ can provide cores to the Milky Way dSphs, which is too light to be consistent with structure formation.

That an ultralight scalar field, such as an axion, could potentially also resolve the small-scale crises has been known for some time~\cite{1990PhRvL..64.1084P,2000PhRvD..62j3517S,hu2000,2000ApJ...534L.127P}: the Jeans scale suppresses the formation of low mass halos, and at the same time leads to density cores in the form of solitons, as we have already discussed. Here we will address one issue: do ULAs suffer a \emph{Catch 22} like WDM does? The answer, in short, is ``no,'' or more accurately ``not as severely.''

\begin{figure}
\begin{center}
\includegraphics[width=0.75\textwidth]{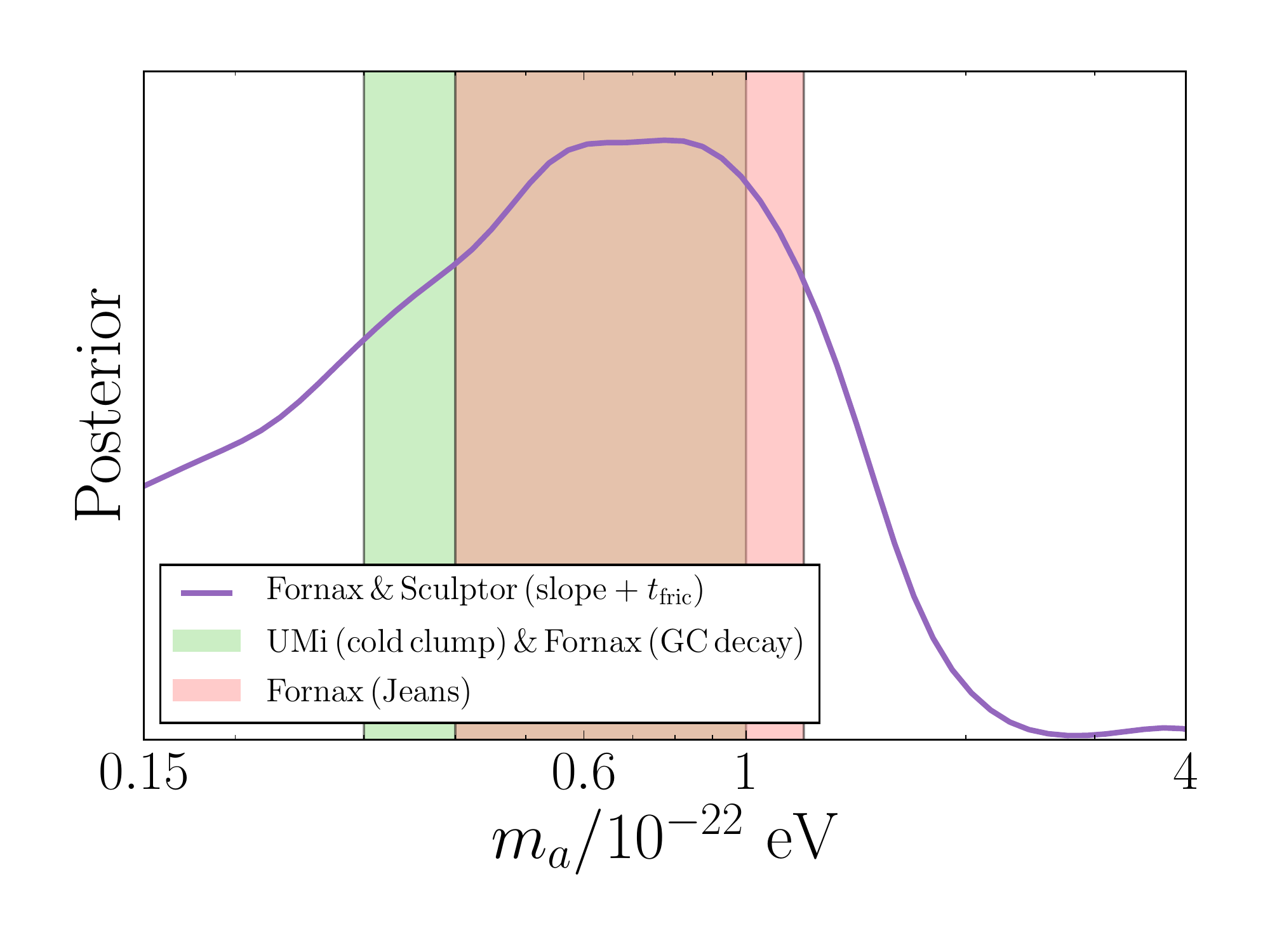} 
 \end{center}
 \caption{One dimensional posterior of ULA mass required to provide soliton cores to Fornax and Sculptor velocity dispersion data~\cite{Walker&Penarrubia2011,2015MNRAS.451.2479M}, including a hard prior $M_{\rm vir}<M_{\rm fric}$~\cite{1992ApJ...389L...9G}. The 95\% C.L. limit is $0.1\times 10^{-22}\text{ eV}<m_a<1.4\times 10^{-22}\text{ eV}$, the upper half of which is consistent with dedicated studies of structure formation and reionization with ULAs~\cite{2015MNRAS.450..209B,2016ApJ...818...89S}. Also shown is the $95\%$ C.L. limit for a Jeans analysis of Fornax~\cite{2014NatPh..10..496S,2016ApJ...818...89S}, and the range required for Ursa Minor (UMi) cold clump longevity and long Fornax globular cluster (GC) orbital decay times~\cite{2012JCAP...02..011L}.}
\label{fig:dwarf_core_ula}
\end{figure}

Fig.~\ref{fig:dwarf_core_ula} shows the one dimensional likelihood for ULA mass from fitting stellar velocity dispersion data of Ref.~\cite{Walker&Penarrubia2011}. This simplified data uses two stellar populations and measures only the slopes of the density profiles within a given radius, in principle allowing an arbitrarily large core outside of this (and hence arbitrarily low axion mass). However, this would allow arbitrarily large dSph mass, while masses $M\gtrsim \mathcal{O}(\text{few})\times 10^{10}M_\odot$ are forbidden by their long dynamical friction time scales~\cite{1992ApJ...389L...9G}.\footnote{I compute the maximum mass for each dSph individually from the formula in Ref.~\cite{1992ApJ...389L...9G} using their co-ordinates~\cite{1998ARA&A..36..435M} and an approximate circular velocity $v_c\approx 200\text{ km s}^{-1}$.} In Fig.~\ref{fig:dwarf_core_ula} the dynamical friction constraint is imposed as a hard prior, supplementing the density profile slope analysis~\cite{Walker&Penarrubia2011} of Ref.~\cite{2015MNRAS.451.2479M}.

Matching the Fornax and Sculptor data with ULAs alone, i.e. with the halo profile Eq.~\eqref{eqn:axion_halo_density_profile}, requires $0.1\times 10^{-22}\text{ eV}<m_a<1.4\times 10^{-22}\text{ eV}$ at 95\% C.L. The best fit using a simplified Jeans analysis on Fornax alone is $m_a=8.1^{+1.6}_{-1.7}\times 10^{-23}\text{ eV}$~\cite{2014NatPh..10..496S} ($1\sigma$ errors). Ref.~\cite{2012JCAP...02..011L} found that a range $0.3\times 10^{-22}\text{ eV}<m_a<1\times 10^{-22}\text{ eV}$ can explain the cold clump longevity in Ursa Minor, and the distribution of globular clusters in Fornax, while respecting some constraints on the maximum dSph mass. All of these limits hint at a mass $m_a\sim 10^{-22}\text{ eV}$ to solve CDM small-scale problems. Recall that this mass is allowed by constraints from halo formation and reionization~\cite{2015MNRAS.450..209B,2016ApJ...818...89S}, reviewed in Section~\ref{sec:ula_highz_constraints}, i.e. ULAs do not suffer from the \emph{Catch 22} like WDM does.

Eq.~\eqref{eqn:mw_ma_half_mode} translates the lower bound on ULA mass from high-$z$ galaxies, $m_a\gtrsim 10^{-22}\text{ eV}$, into an equivalent WDM mass of $m_X\gtrsim 0.8\text{ keV}$, which from Fig.~\ref{fig:wdm_catch22} gives a minuscule core size of $\mathcal{O}(30\text{ pc})$. A harder constraint on $m_X>2\text{ keV}$ implies, by scaling of the half-mode, $m_a>10^{-21}\text{ eV}$. Scaling the core size (from the 1~kpc core in Fornax with $m_a=10^{-22}\text{ eV}$) as $m_a^{-1/2}$ still provides a significant $\mathcal{O}(300\text{ pc})$ core even for this hypothetically stronger constraint. 

Translating bounds from WDM to ULAs using Eq.~\eqref{eqn:mw_ma_half_mode} is good for order-of-magnitude estimates only. The exact constraints from structure formation depend sensitively on the slope of the transfer function and mass function near the cut off (e.g. Ref.~\cite{2016ApJ...818...89S}), which distinguishes WDM and ULAs, such that dedicated studies are necessary. There are tantalizing hints for $m_a=10^{-22}\text{ eV}$ as a solution to the small-scale crises. It is on the edge of current constraints, and of detectability in the EOR. Dedicated studies of this model, including full simulations with star formation and feedback (such as those comparing WDM and CDM including feedback in Ref.~\cite{2015MNRAS.448..792G}), are necessary to explore this further. 


\section{Axions and Accelerated Expansion}

\subsection{Axions and the Cosmological Constant Problem}
\label{sec:dark_energy}

Our discussion in this review began with one of the greatest unsolved problems in modern physics: the cosmological constant (c.c.) problem~\cite{1989RvMP...61....1W}, one of the most notoriously hard problems to solve in high energy physics~\cite{2013arXiv1309.4133B}. One particularly attractive solution to this problem is anthropic tuning, which can be realized by eternal inflation populating a large number of vacua in the string landscape~\cite{2000JHEP...06..006B,2003dmci.confE..26S} (the original idea dates back to Ref.~\cite{1987PhLB..195..177B}). In this picture, four-form fluxes and topologically complex compact spaces with $\mathcal{O}(100)$ or more cycles both play important roles.\footnote{This ``100" is one origin of the famous statement that the string theory landscape contains $10^{500}$ vacua. In this context it arises from demanding that the number of vacua is densely enough distributed near the observed value of the c.c. to make a universe in this region sufficiently likely.} Recall from Section~\ref{sec:string_models} that axions arise from the wrapping of such fluxes on cycles. Furthermore, the canonical axion potential $V(\phi)\propto \cos \phi/f_a$ can provide positive and negative contributions to the vacuum energy, allowing axions to cancel contributions to the c.c. from other sources in a cosmologically dynamical manner. 

The above observations suggest that:
\begin{itemize}
\item Axions may play a central role in the solution of the c.c. problem.
\item The anthropic solution of the c.c. problem in the string landscape provides good motivation for the existence of the axiverse.
\end{itemize}
In this section we will briefly discuss a few ideas relating axions to the c.c. problem.

We begin with the simplest model of axion quintessence. As we already saw in Section~\ref{sec:cosmological_constraints}, ULAs with $m_a\sim H_0\sim 10^{-33}\text{ eV}$ can act as DE, with the axion potential energy providing an effective cosmological constant and driving accelerated expansion as a form of quintessence. Since the axion mass is protected by a shift symmetry and can easily remain so light, the idea of axion and general pNGB~\cite{1988NuPhB.311..253H} quintessence is natural, and has a long history~\cite{1995PhRvL..75.2077F}.\footnote{For a review of DE and quintessence models, see Ref.~\cite{2006IJMPD..15.1753C}.} 
\begin{figure}
\begin{center}
$\begin{array}{@{\hspace{-0.5in}}c@{\hspace{+0.2in}}c}
\includegraphics[scale=0.8]{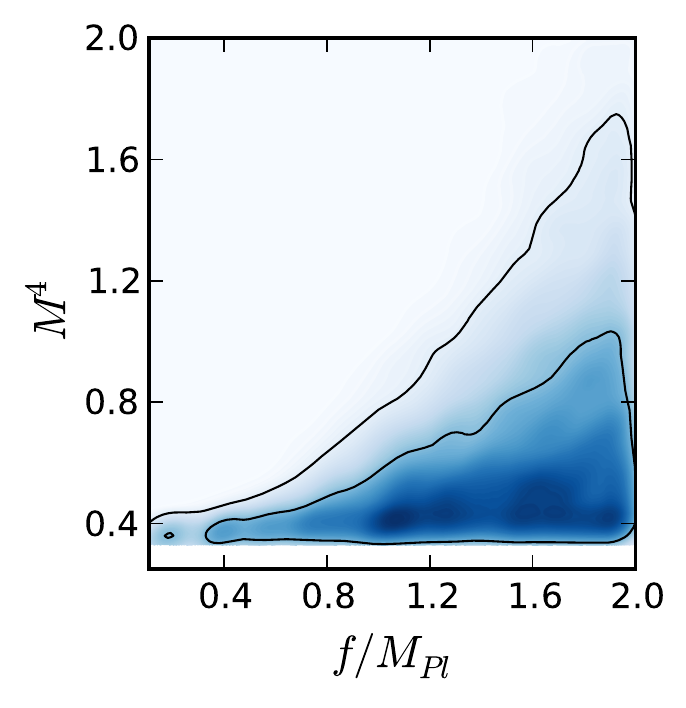} &
\includegraphics[scale=0.8]{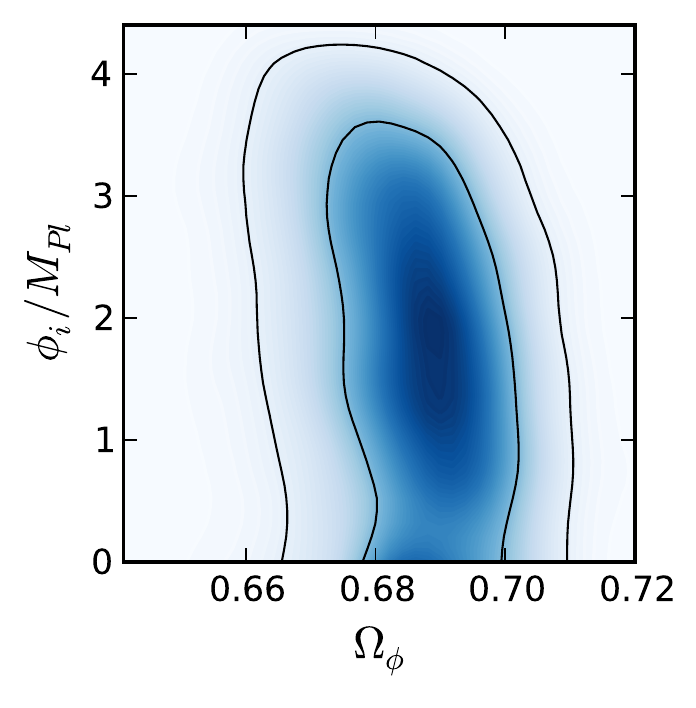}
 \end{array}$
 \end{center}
 \caption{Constraints on axion DE from \emph{Planck}. \emph{Left Panel:} Potential mass scale in units of the critical density, versus decay constant. Note that here $M$ rather than $\Lambda_a$ is used. \emph{Right Panel:} Field displacement versus density fraction. The density fraction is well constrained by the demand that the axion cause accelerated expansion with zero overall vacuum energy. Reproduced (with permission) from Ref.~\cite{2015arXiv150306100S}.}
\label{fig:liddle}
\end{figure} 

The model is specified by the potential
\be
V(\phi)=\Lambda_a^4\left[ 1+\cos\left(\frac{\phi}{f_a}\right) \right]\, ,
\ee
(note the phase shift from our previous definition). The most recent constraints on this scenario using \emph{Planck} data can be found in Ref.~\cite{2015arXiv150306100S} and are summarized in Fig.~\ref{fig:liddle}.  Since the vacuum in this potential has zero energy, the quintessence contribution to the energy budget, $\Omega_\phi$, is controlled by the initial field displacement, $\phi_i$. The value of $\Omega_\phi\approx 0.69$ is well constrained by the requirement of driving accelerated expansion, and just as we saw in Fig.~\ref{fig:contours_combinedLinear} (right panel) large field displacements and decay constants are required to achieve this. There is a degeneracy between the energy density and the decay constant caused by the requirement of keeping the potential roughly flat compared to $H_0$: increasing $\Lambda_a$ requires increasing $f_a$ to retain flatness.

A simple generalization of this quintessence model  goes along the lines of $N$-flation (see Section~\ref{sec:natural_inf_variants}), and was discussed in Ref.~\cite{2006hep.th....7086S}. Taking the string theory-inspired potential in Eq.~\eqref{eqn:v_sinst_string} for $N$ axions of almost degenerate mass, and assuming a fixed decay constant:
\be
f_a=\frac{M_{pl}}{S_{\rm inst.}}\, ,
\ee 
it can be shown that axion quintessence requires
\be
S_{\rm inst.}\sim 200 - 300 \, \text{  and  } N\gtrsim S_{\rm inst.}^2 \, ,
\ee
if the axion contribution to DE is to be non-negligible. 

Alternatively, successful quintessence can occur for sub-Planckian decay constants if the initial displacement $\phi_i/f_a\sim \pi$. This idea was considered in Ref.~\cite{2014PhRvL.113y1302K} for the case of multiple axions. Taking constant $f_a\approx 10^{17}\text{ GeV}$, potential energy scale $\Lambda_a=10^{12}\text{ GeV}e^{-S_{\rm inst}}$ and assuming that the instanton action changes by $\mathcal{O}(10)$ for each axion, then with 24 axions the probability that one axion is close enough to the top of the cosine potential to drive successful quintessence occurs in approximately 1\% of cases. This relatively modest number of axions can achieve successful quintessence with sub-Planckian $f_a$ and minimal fine-tuning. However, the limiting case in this study was the assumption of constant $f_a$, rather than considering the variation of $f_a$ with $S_{\rm inst.}$. The heavier axions in this scenario will be subject to all the phenomenology and constraints discussed elsewhere in this review. In Ref.~\cite{2014PhRvL.113y1302K} it was proposed to avoid unwanted impacts on cosmology by having the heavy axions decay, or evolve in a modified potential.

The models of Refs.~\cite{2015arXiv150306100S,2006hep.th....7086S,2014PhRvL.113y1302K} simply require that axions provide successful quintessence, and assume that the bare c.c. is of an acceptably small value, due to some unknown physical mechanism, or due to anthropics. This is a solution to the ``new c.c.'' or ``why now?'' problem of obtaining small masses and potential energies of order the present critical density. Let us now turn to the role of axions in solving the ``old c.c.'' problem, i.e. the much more taxing problem that
\be
\rho_{\Lambda,{\rm obs.}}\sim 10^{-120}M_{pl}^4\,, \text{ while  } \rho_{\Lambda,{\rm theory}}\sim M_{pl}^4 \, .
\ee 

\begin{figure}
\begin{center}
\includegraphics[scale=1.2]{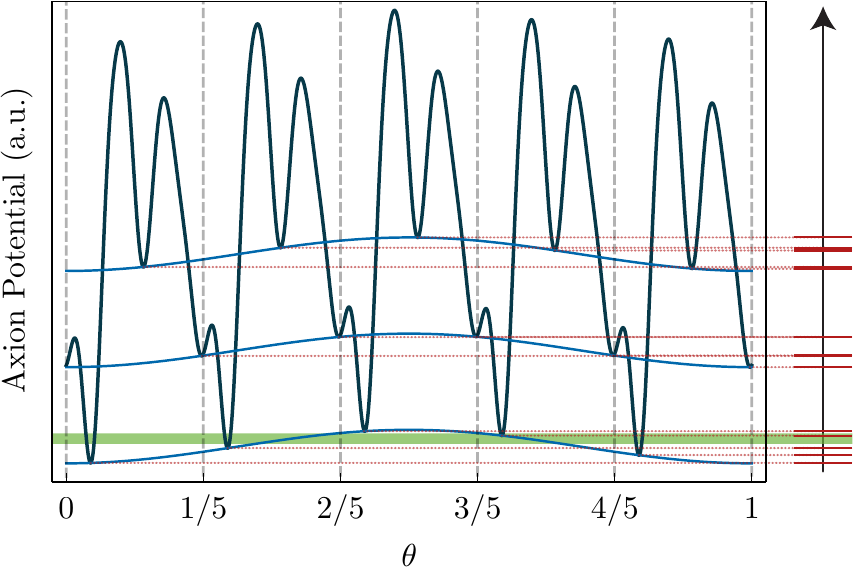}
 \end{center}
 \caption{The axionic band structure of the cosmological constant. A multi-axion theory with sub-leading instanton contributions can give rise to an exponentially large number of vacua, with energy splittings inversely proportional to the number of vacua. This mechanism may provide a solution to the cosmological constant problem. Reproduced, with permission, from Ref.~\cite{2015arXiv151006388B}. Copyright (2016) by The American Physical Society.}
\label{fig:bachlechner}
\end{figure}

Ref.~\cite{2015arXiv151006388B} considered the possibility of using subleading instanton corrections in a multi-axion model to generate a field space with an exponentially large number of vacua. The potential for the $N$ axion fields $\theta_i$ charged under instantons labelled by $j$ with charge $\mathcal{Q}^j_i$ has the form
\be
V(\vec{\theta})=\sum_j\Lambda_j^4\left[ 1-\cos (2\pi \mathcal{Q}^j_i\theta^i+\delta^j)\right]+V_0 \, ,
\ee
where where $\delta^j$ is an arbitrary phase. The leading potential is split into bands of width $\Lambda^4_{\rm sub.}$ by the subleading pieces, with each band containing $\mathcal{N}_{\rm sub.}$ different vacua. This splitting leads to vacua within $\Lambda^4_{\rm sub.}/\mathcal{N}_{\rm sub.}$ of zero, as illustrated in Fig.~\ref{fig:bachlechner}. Therefore, if we take $\Lambda^4_{\rm sub.}\sim M_{pl}^4$ one requires $\mathcal{N}_{\rm sub.}\sim 10^{120}$ distinct vacua to solve the c.c. problem.

For a random matrix model of the instanton charges, Ref.~\cite{2015arXiv151006388B} showed that that expected number of vacua in a theory with $N$ axions obeys the bound
\be
\langle \mathcal{N}_{\rm sub.}^2\rangle \gtrsim \sqrt{2\pi N}\left( \frac{3}{e}\right)^N\, .
\ee
Thus there is an exponentially large number of vacua. An example with 500 axions suffices to obtain the desired factor if $10^{120}$. In this model, the expected mass distribution of the axions was not computed, but the logarithmic distribution of $\Lambda_j$ was invoked. It is thus not clear at this stage what the role of these axions would be in terms of a DM model. Some evidence suggests that this model could incorporate successful axion inflation, a topic to which we now turn.

\subsection{Axion Inflation}
\label{sec:axion_inflation}
\begin{figure}
\begin{center}
\includegraphics[scale=0.6]{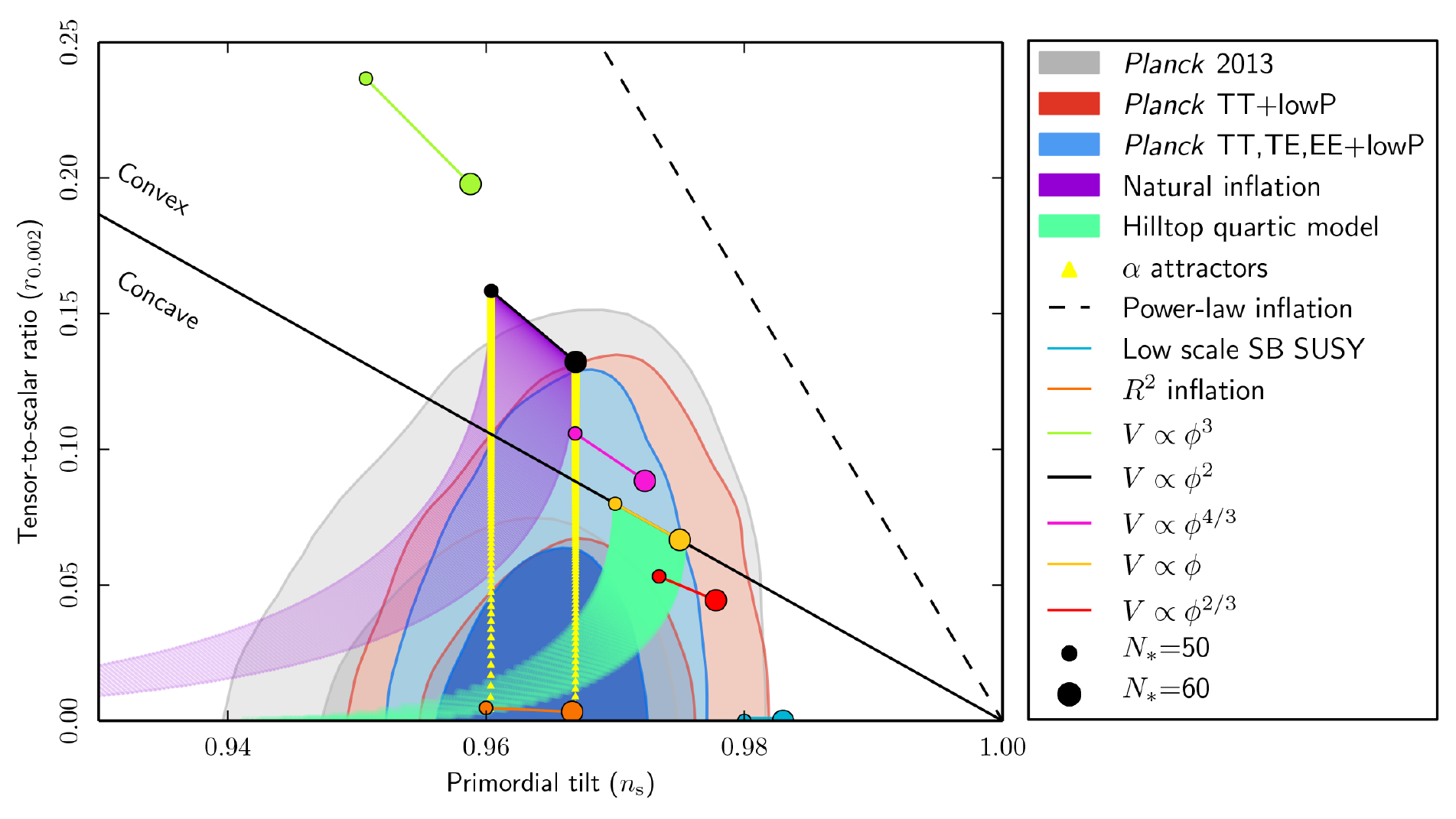}
 \end{center}
 \caption{Constraints on inflationary models from \emph{Planck}~\cite{2015arXiv150202114P}, showing 1 and 2$\sigma$ marginalized confidence regions. Note that the potentials $\sim \phi^{2/3}$, $\sim\phi$, and $\sim\phi^{4/3}$ are the approximate predictions of axion monodromy models if power spectrum oscillations are ignored.}
\label{fig:planck_inflation}
\end{figure} 

In Section~\ref{sec:axion_during_inflation} we discussed the role of stable axion DM fields as spectators during inflation. Here, we discuss the scenario where an unstable axion field itself drives inflation.

Inflation~\cite{1981PhRvD..23..347G,1982PhLB..108..389L,1982PhRvL..48.1220A} is a hypothetical period of accelerated expansion in the early Universe, invoked to explain certain cosmological puzzles relating to initial conditions.\footnote{It is not my purpose here to give a review of inflation, and I defer all detailed calculations and notation. For a general review of inflation, see Ref.~\cite{2010LNP...800....1L}, for inflation in string theory, see Ref.~\cite{2014arXiv1404.2601B}, and for specifics of axion inflation, see Ref.~\cite{2013CQGra..30u4002P}. The state of the art in constraints on inflation can be found in Refs.~\cite{2015arXiv150202114P,2014JCAP...03..039M}, while an exhaustive list of single-field-slow-roll models can be found in Ref.~\cite{2014PDU.....5...75M}.} The simplest inflationary models involve a single, minimally coupled, scalar field (``the inflaton"), driving the expansion by the existence of a potential, $V(\phi)$, on which the field is slowly rolling. Inflation ends when this field reaches the minimum of its potential, oscillates, and decays into radiation: a process known as ``reheating." This reheating must occur in order to produce a hot big bang cosmology and all its successful predictions, from BBN to the CMB.

The inflaton potential must be very flat compared to the other scales in play, namely the Hubble scale. The expansion is driven by the potential, and so $3H_I^2M_{pl}^2\approx V(\phi)$. This defines the inflationary ``slow roll parameters," which depend on the flatness of the potential. The first two slow roll parameters are:
\be
\epsilon_{\rm inf} = \frac{M_{pl}^2}{2}\left(\frac{V'}{V}\right)^2\, , \quad \eta_{\rm inf}=M_{pl}^2\frac{V''}{V}\, ,
\ee
and inflation requires each of these be very much less than unity over a large, relative to $H_I$, field range. Axions are extraordinarily good inflaton candidates because the shift symmetry protects the flatness of the potential from quantum corrections. It is important to note that, because the inflaton must decay, \emph{the axion driving inflation is not a dark matter (or dark energy) axion}. In particular, therefore, the inflaton is not the QCD axion!

The standard view of constraints on inflationary models is shown in Fig.~\ref{fig:planck_inflation}, taken from Ref.~\cite{2015arXiv150202114P}. These simple constraints allow the cosmological initial conditions two degrees of freedom after normalization by $A_s$. These are the tilt, $n_s$, and the tensor-to-scalar ratio, $r_T$. These numbers are determined by the parameters of the inflaton potential. Additional freedom is afforded to the model by the number of e-folds of observable inflation, $N_*$, which takes into account uncertainty about the reheating epoch~\cite{1997PhRvD..56.3258K,2010ARNPS..60...27A,2014PhRvD..89b3522E} and the initial conditions of the inflaton itself~\cite{2001PhRvD..63l3501M,2002JHEP...11..037K}. The constraints shown assume that the primordial power spectra are described by power laws. We will briefly discuss spectra with features later.

\subsubsection{Natural Inflation and Variants}
\label{sec:natural_inf_variants}

So-called ``Natural Inflation"~\cite{1990PhRvL..65.3233F} is the simplest example of inflation with an axion. It simply takes our usual potential
\be
V(\phi)=\Lambda_a^4\left[1\pm \cos \left(\frac{\phi}{f_a}\right)\right]\, .
\label{eqn:natural_inflation_v}
\ee
Natural Inflation is a standard single field slow roll model, giving power law scalar and tensor power spectra.

In its original incarnation, Natural Inflation takes $\Lambda_a\sim m_{\rm GUT}$ and $f_a\sim M_{pl}$. One combination of these parameters is fixed by normalizing $A_s$, and so, including $N_*$, the model has two additional parameters specifying its location on the $(n_s,r_T)$ plane. Thus, in Fig.~\ref{fig:planck_inflation}, Natural Inflation sweeps out a broad region, a portion of which is consistent with the observational constraints. In the limit $f_a\rightarrow\infty$ with $\Lambda_a^2/f_a$ held fixed, Natural Inflation approaches $m^2\phi^2$ ``chaotic'' inflation. Furthermore, we see that Natural Inflation consistent with the observed value of $n_s$ predicts a measurably large value of $r_T\gtrsim 10^{-2}$. This is a reasonable sensitivity to expect for near-future CMB experiments~\cite{2014SPIE.9153E..13R,2013arXiv1309.5381A}, and so Natural Inflation makes testable predictions.\footnote{Up to the usual caveats made by notable detractors.}

The value of the tensor-to-scalar ratio in single field slow roll inflation is closely tied to the field range, $\Delta \phi$, over which the potential is flat, and for which inflation occurs. The ``Lyth bound''~\cite{1997PhRvL..78.1861L} states:
\be
\Delta \phi= 0.46 M_{pl}(r_T/0.07)^{1/2} \, .
\ee
It is generally held that over such large field excursions one loses perturbative control over quantum mechanical corrections to the potential (in particular, those of quantum gravity~\footnote{See also Ref.~\cite{2012JCAP...09..019C}, which suggests that large field inflation in general might be forbidden by entropy bounds in quantum gravity.}), and so achieving large amplitude tensor modes is hard to achieve in a theoretically consistent manner. 

The natural field range in the potential Eq.~\ref{eqn:natural_inflation_v} is $f_a$, and so for Natural Inflation the Lyth bound implies that $f_a$ must be of order the Planck scale. The potential is protected from other corrections by the axion shift symmetry, which is restored in the limit $\Lambda_a\rightarrow 0$, making the the theory technically natural. This is where the ``natural'' in Natural Inflation comes from: the axion potential is flat over scales $\Delta \phi\sim f_a$, and is immune to radiative corrections. ``Standard'' inflation at the GUT scale, with observably large $r_T$, can be achieved with a Planckian decay constant.

As we have already mentioned, however, the weak gravity conjecture~\cite{2007JHEP...06..060A} places some constraints on $f_a\gtrsim M_{pl}$ in theories of quantum gravity, in particular forbidding it in the case of a single canonically normalised axion field. We have also seen that in string theory one finds $f_a< M_{pl}$ in our simple example. One should therefore worry about embedding Natural Inflation in a UV complete theory. The simplest models, which remain quasi-single field and produce power-law initial power spectra, are based on the general idea of ``Assisted Inflation''~\cite{1998PhRvD..58f1301L} (or even more generally, on ``kinetic alignment''~\cite{Bachlechner:2014hsa}).

In Assisted Inflation, one uses the frictional coupling of multiple fields induced by the Hubble expansion to provide extra damping to the collective motion in field space. This slows the collective motion down, effectively flattening the potential of the quasi-single field trajectory. A simple example of Assisted Inflation applied to axion models is ``N-flation''~\cite{2008JCAP...08..003D}. N-flation takes $N$ axions with identical potentials: 
\be
V(\vec{\phi})=\sum_{n=1}^N V_n(\phi_n)\, ,
\ee
where $V_n=\Lambda_n^4\cos (\phi_n/f_n)$ is the familiar cosine potential.\footnote{I drop the higher order instanton corrections discussed in Ref.~\cite{2008JCAP...08..003D}. The radiative stability of N-flation in field theory and in string theory was also established in Ref.~\cite{2008JCAP...08..003D}, and so it fits the maxims of a natural theory.} One now simply applies Pythagoras theorem to the $N$-dimensional field space. 

For simplicity, consider the case of all equal decay constants, $f_n=f_a$, and scales $\Lambda_n=\Lambda_a$. Now displace each field from the origin by an equal amount,\footnote{The equal displacement trajectory is an attractor of Assisted Inflation~\cite{1998PhRvD..58f1301L}. N-flation also takes initial conditions with zero angular momentum in field space. For a discussion of the dynamics with angular motion, see Ref.~\cite{2002PhLB..545...17B}.} $\phi_n=\alpha M_{pl}$, with $\alpha^2<2\pi f_a^2/M_{pl}^2$. The total radial displacement is $\phi_r=\sqrt{N}\alpha M_{pl}$ and the mass of the radial field is $m=\Lambda_a^2/f_a$. It is clear that we can arrange for super-Planckian displacement of $\phi_r$, with $f_a<M_{pl}$ and $\alpha^2\ll 1$, if $N$ is large enough. As in Assisted Inflation, each individual $\phi_n$ feels the friction of all its brothers and sisters, and it is the collective radial motion in field space that acts as the inflaton.

Finally, the Kim-Nilles-Peloso model~\cite{2005JCAP...01..005K} generalizes the multi-axion potential allowing rotations between the fields. This occurs if multiple axions, $i$, each obtain potentials from multiple non-perturbative sources, $j$, but with different strengths, $f_{ij}$. ``Decay constant alignment'' then allows to create a flat-direction on the potential with a large effective value of $f_{a,{\rm eff}}>M_{pl}$ even is each individual $f_{ij}<M_{pl}$, so long as sufficient degeneracy between the decay constants occurs.

\subsubsection{Axion Monodromy}

Axion Monodromy~\cite{2008PhRvD..78j6003S,2010PhRvD..82d6003M}\footnote{For some possible issues in explicit realisations of this model, see e.g. Refs.~\cite{2014JHEP...01..179G,2015arXiv151002005A}.} is another model within the pantheon of UV completions of axion inflation allowing for large field excursions, and thus measurably-large $r_T$. It differs from the examples discussed above, however, in that it \emph{does not produce power-law initial power spectra}, but instead modulates the power law spectra with periodic features.

In string theory, a monodromy occurs when an axion field winds around a particular location in moduli space, like the Riemann sheets of $\log z$ winding around the origin in the complex plane. The monodromy provides an explicit breaking of the periodicity of the axion potential, and lifts it at large field values. The extra potential energy is supplied by the wrapping of branes around compact dimensions. It has been described colloquially as a ``wind up toy.''

Over large field excursions $\Delta \phi\gg f_a$ the potential is on average described as $V\propto \phi^p$ for some $p$, while on small scales the potential is modulated by the usual, instanton-induced, axion cosine. The potential is of the form
\be
V(\phi)=\mu^{4-p}\phi^p+\Lambda_a^4 \left[1-\cos \left(\frac{\phi}{f_a}\right) \right]\, .
\ee
As inflation proceeds along the $\phi$ direction, one has slow roll on the $\phi^p$ piece. From Fig.~\ref{fig:planck_inflation} we see that the predictions of large-field $\phi^p$ models of inflation, with $p=2/3,1,4/3$, motivated by axion monodromy, are consistent with the observations, and predict measurably large tensor modes.

The cosine part of the potential, however, modulates the slow-roll trajectory with oscillations. This leads to an oscillatory power spectrum for the primordial curvature perturbations of the form~\cite{2010JCAP...06..009F}
\be
P_\zeta(k)=A_s\left(\frac{k}{k_0} \right)^{n_s-1+\frac{\delta n_s}{\ln k/k_0}\cos (\phi_k/f_a)} \, ,
\ee
with $\phi_k=\sqrt{\phi_0^2-2\ln (k/k_0)}$, $\phi_0$ the value of the field at horizon crossing of the pivot scale, and $\delta n_s \propto \Lambda_a^4/\mu^3 f_a$ for $p=1$. 

The axion monodromy power spectrum undergoes rapid oscillations in $\log k$, and constraining it properly using CMB data requires special care~(e.g. Refs.~\cite{2014PhRvD..89f3536M,2014PhRvD..89f3537M}). The latest \emph{Planck} data show no statistically significant evidence for the presence of power spectrum oscillations, though there are various low-significance hints~\cite{2015arXiv150202114P}. Axion monodromy also predicts ``resonant non-Gaussianity''~\cite{2010JCAP...06..009F}. Current data cannot reach the sensitivity to confirm hints of oscillations in the power spectrum through resonant non-Gaussianity in the bispectrum, though this may be possible in future.

\section{Gravitational Interactions with Black Holes and Pulsars}

In this section we consider two astrophysical probes of axion DM that arise purely from gravitational interactions, and are quite distinct from any signatures we have considered so far.

\subsection{Black Hole Superradiance}
\label{sec:superradiance}

BHSR is a very general way to search for light bosonic fields. It relies only on their gravitational interaction and assumes nothing about couplings to the standard model or their cosmological energy density.

Massive bosonic fields can form bound states around astrophysical black holes (BHs), just like the energy levels of electrons in the hydrogen atom. Infalling scalar waves extract energy and angular momentum from a spinning Kerr BH and emerge with more energy than they went in with; this is known as the Penrose process~\cite{1969NCimR...1..252P}. Being bosons, the energy levels in the ``gravitational atom'' can be filled exponentially via this superradiant instability (see Ref.~\cite{2015LNP...906.....B} for a review). The boson mass leads to the existence of stable orbits, like the energy levels of an atom. These stable orbits lead to a barrier in the effective potential, and act like the mirror in Press and Teukolsky's ``black hole bomb''~\cite{1972Natur.238..211P,1973ApJ...185..649P}. The energy levels then fill up via the superrandiant instability until they eventually radiate away the extracted energy, for example as gravitational waves. The bosons do not even need to be present initially (i.e. they do not have to be the DM) for this process to occur: superradiance can start from a quantum mechanical fluctuation. It is thus a completely generic feature of massive bosonic fields in astrophysics, and turns astrophysical BHs into sensitive detectors of bosons in the mass range $10^{-20}$ to $10^{-10}\text{ eV}$~\cite{2015PhRvD..91h4011A,axiverse,2011PhRvD..83d4026A,2012PhRvL.109m1102P,2015CQGra..32m4001B}.
\begin{figure}
\begin{center}
$\begin{array}{@{\hspace{-0.8in}}c@{\hspace{+0.2in}}c@{\hspace{-0.5in}}}
\includegraphics[scale=0.25]{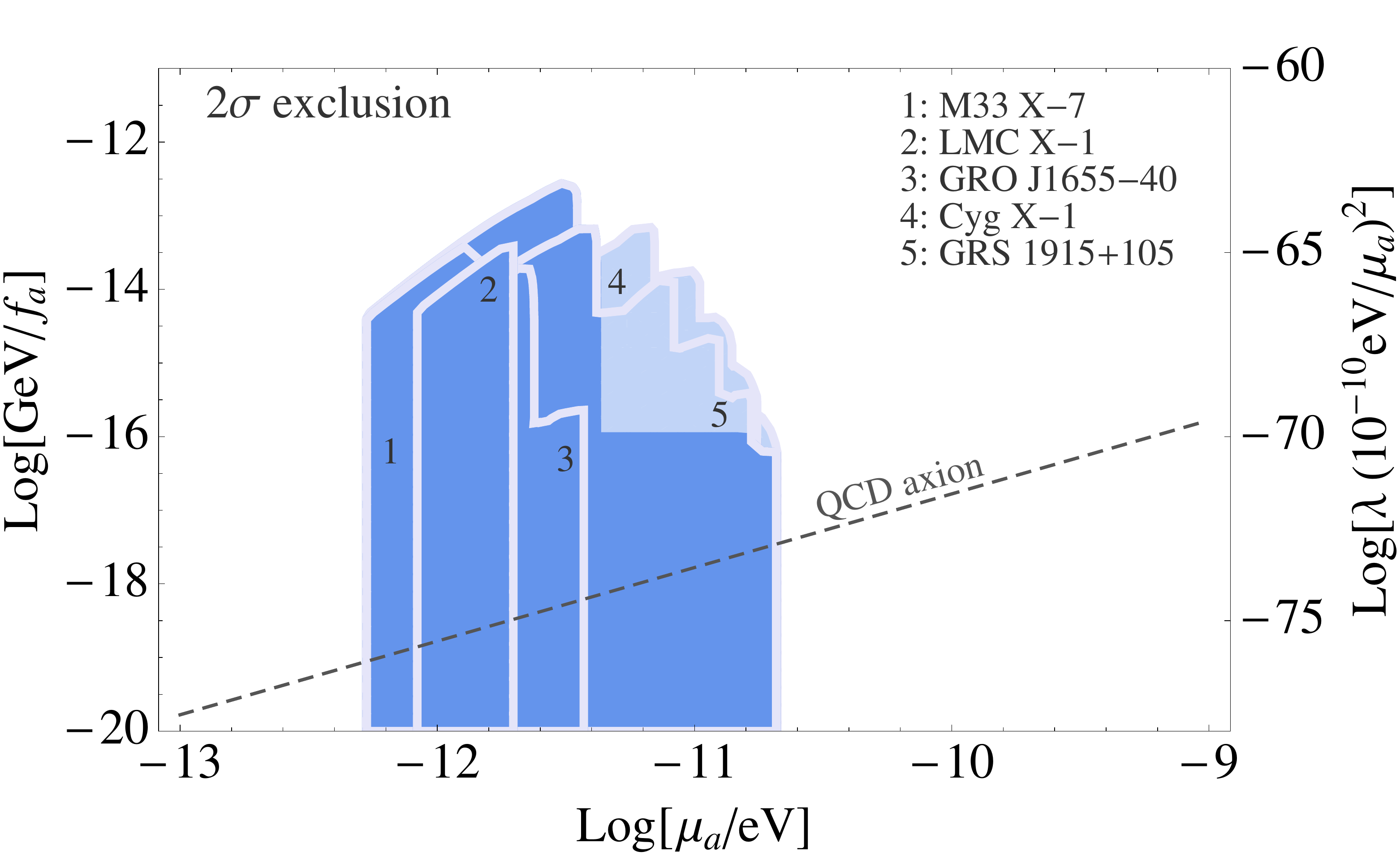} &
\includegraphics[scale=0.4]{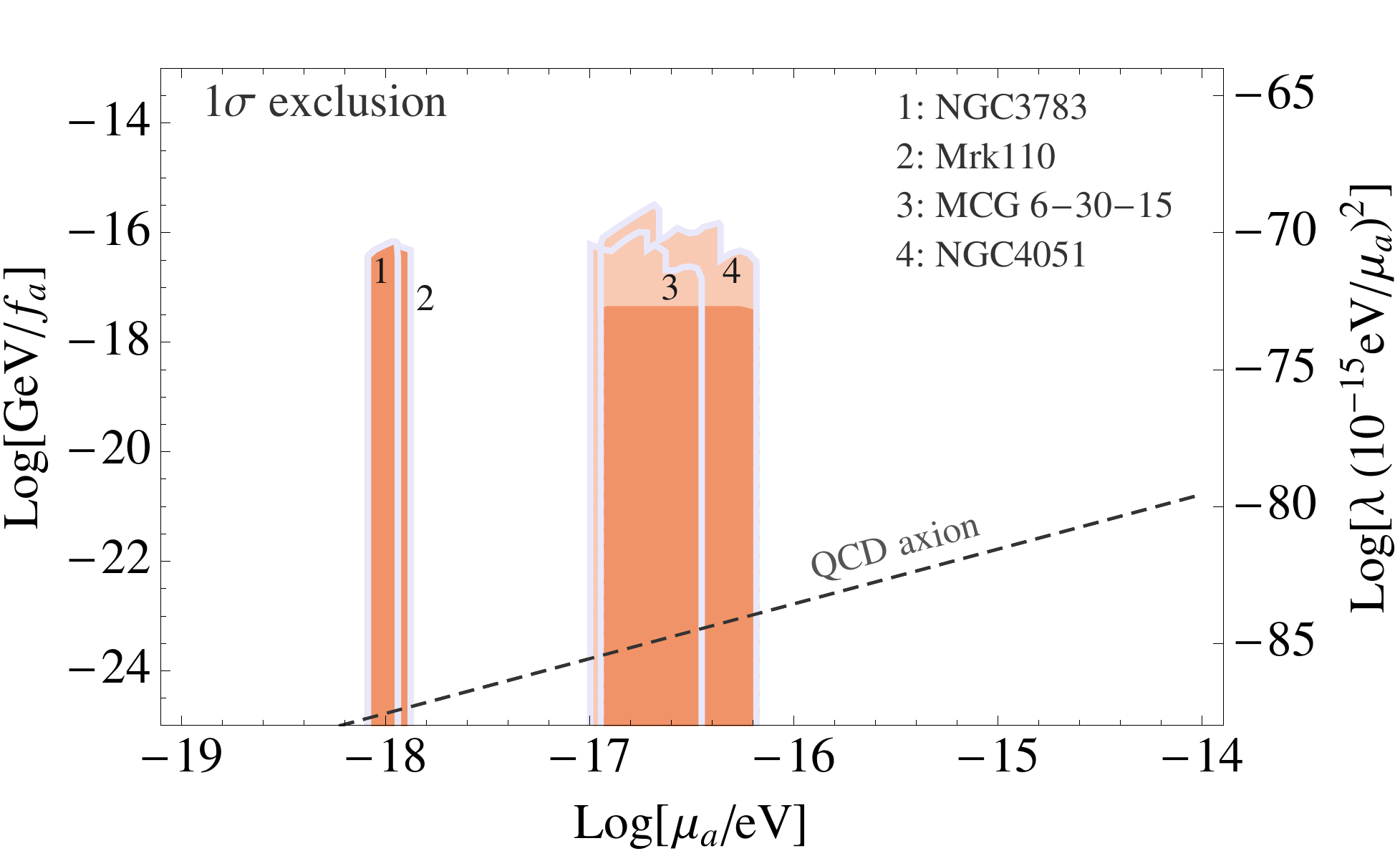}
 \end{array}$
 \end{center}
 \caption{Constraints on axions from BHSR. \emph{Left Panel:} Solar mass black holes, $2\sigma$. \emph{Right Panel:} Supermassive black holes, $1\sigma$. Reproduced (with permission) from Ref.~\cite{2015PhRvD..91h4011A}. Copyright (2015) by The American Physical Society.}
\label{fig:arvanitaki}
\end{figure} 

The instability leads to the spin down of BHs. The spin-down rate is controlled by the effective coupling of the gravitational atom:
\be
\alpha_G=r_G m_a\,, \quad r_G\equiv GM \, ,
\ee
where $M$ is the BH mass. The size of the ``cloud'' formed around the BH is fixed by the orbital velocity $v\sim \alpha_G/\ell$ to be $r_c\sim n^2r_G/\alpha_G^2$ (where $\ell$ is the orbital quantum number and $n$ is the energy level). This is approximately the de Broglie scale for a circular orbit of radius $r_c$, and we observe the link to our previous discussions of density profiles and the Jeans scale.

With $\alpha_G=0.3$ the superradiance time-scale is short ($\sim$ years) for both stellar mass ($M=10M_\odot$) and super-massive ($M=10^7 M_\odot$) BHs, which sets the characteristic axion mass for spin-down. A number of BHs are observed, and their masses and spins have been measured (data are given with citations in Ref.~\cite{2015PhRvD..91h4011A}). Since the spinning BHs would be spun-down in the presence of a light boson, these observations can be used to exclude various axion masses. 

The exclusions are shown in Fig.~\ref{fig:arvanitaki}. Stellar mass BHs exclude a range of masses $6\times 10^{-13}\text{ eV}<m_a<2\times 10^{-11}\text{ eV}$ at $2\sigma$, which for the QCD axion excludes $3\times 10^{17}\text{ GeV}<f_a<1\times 10^{19}\text{ GeV}$. The supermassive BH measurements are more uncertain: there are fewer measurements excluding a narrower range of masses at $1\sigma$ only. The range probed is roughly $10^{-18}\text{ eV}<m_a<10^{-16}\text{ eV}$. Higher precision measurements in future could improve these bounds.

Finally, transitions and annihilations within the axion cloud predict the emission of monochromatic gravitational waves. The detection prospects for such a signal with advanced LIGO~\cite{2010CQGra..27h4006H} and eLISA~\cite{2012CQGra..29l4016A} are discussed in Ref.~\cite{2015PhRvD..91h4011A}. Advanced LIGO has the potential to discover evidence for the QCD axion with $m_a\sim 10^{-10}\text{ eV}$ in the not-too-distant future. Further in the future, eLISA may be sensitive to the lower-frequency emission for ULAs with $m_a\sim 10^{-17}\text{ eV}$, with the possibility to detect $\sim 10$'s of events from axion annihilations out to a radius of $\sim 100\text{ Mpc}$.

\subsection{Pressure Oscillations and Pulsar Timing}

The pressure, $P_a=w_a\rho_a$, in the axion energy momentum tensor undergoes rapid oscillations as $\cos 2m_at$, leading to the $\langle w_a\rangle=0$ DM-like properties of the axion. Local pressure perturbations, $\delta P_a$, also undergo such oscillations. Such pressure oscillations induce oscillations of the gravitational potential, which in turn induce a time-dependent frequency shift and a time delay for any propagating signal. If the DM in the Milky Way is composed of ULAs, then the amplitude of the signal is fixed by the local DM abundance, $\rho_{\rm DM}\approx 0.3\text{ GeV cm}^{-3}$. Ref.~\cite{2014JCAP...02..019K} considered the effect of such oscillations on pulsar timing experiments. 

Consider the energy momentum tensor, Eq.~\eqref{eqn:axion_em_tensor}. The local axion field can be described as
\be
\phi (\vec{x},t)=\phi_0(\vec{x})\cos [mt+\xi (\vec{x})]\, ,
\ee
where $\phi_0$ is the local amplitude and $\xi$ is a local phase. To leading order, the energy density is static, but the pressure oscillates. The local amplitude is fixed by the DM density as:
\be
\phi_0(\vec{x})=\frac{\sqrt{2\rho_{\rm DM}}}{m_a} \, ,
\ee
which in turn fixes the local pressure:
\be
P(\vec{x},t)=-\frac{1}{2} m_a^2\phi_0^2 \cos (2m_at+2\xi)\, .
\ee
The Newtonian potentials, $\Psi$ and $\Phi$, are sourced by the density and the pressure. They have dominant time-independent pieces, and sub-dominant oscillating pieces, found from the Einstein equations.

The oscillating potential induces an oscillating delay in arrival time of pulsar signals, with frequency $2m_a$ and amplitude~\cite{2014JCAP...02..019K}:
\be
\Delta t_\phi = \frac{\pi G_N\rho_{\rm DM}}{m_a^3}\sin \left[m_a D+\xi(\vec{x}_0)-\xi(\vec{x}_p) \right] \, ,
\ee
where $D$ is the distance to the pulsar, $\vec{x}_p$ is the pulsar location, and $\vec{x}_0$ is the position of the Earth. In the variance of this signal the unknown local phases, $\xi$, and the pulsar distance, $D$, drop out. The amplitude of the signal decreases for heavier axions, and has a maximum at a given mass set by the DM density.
\begin{figure}
\begin{center}
\includegraphics[scale=0.5]{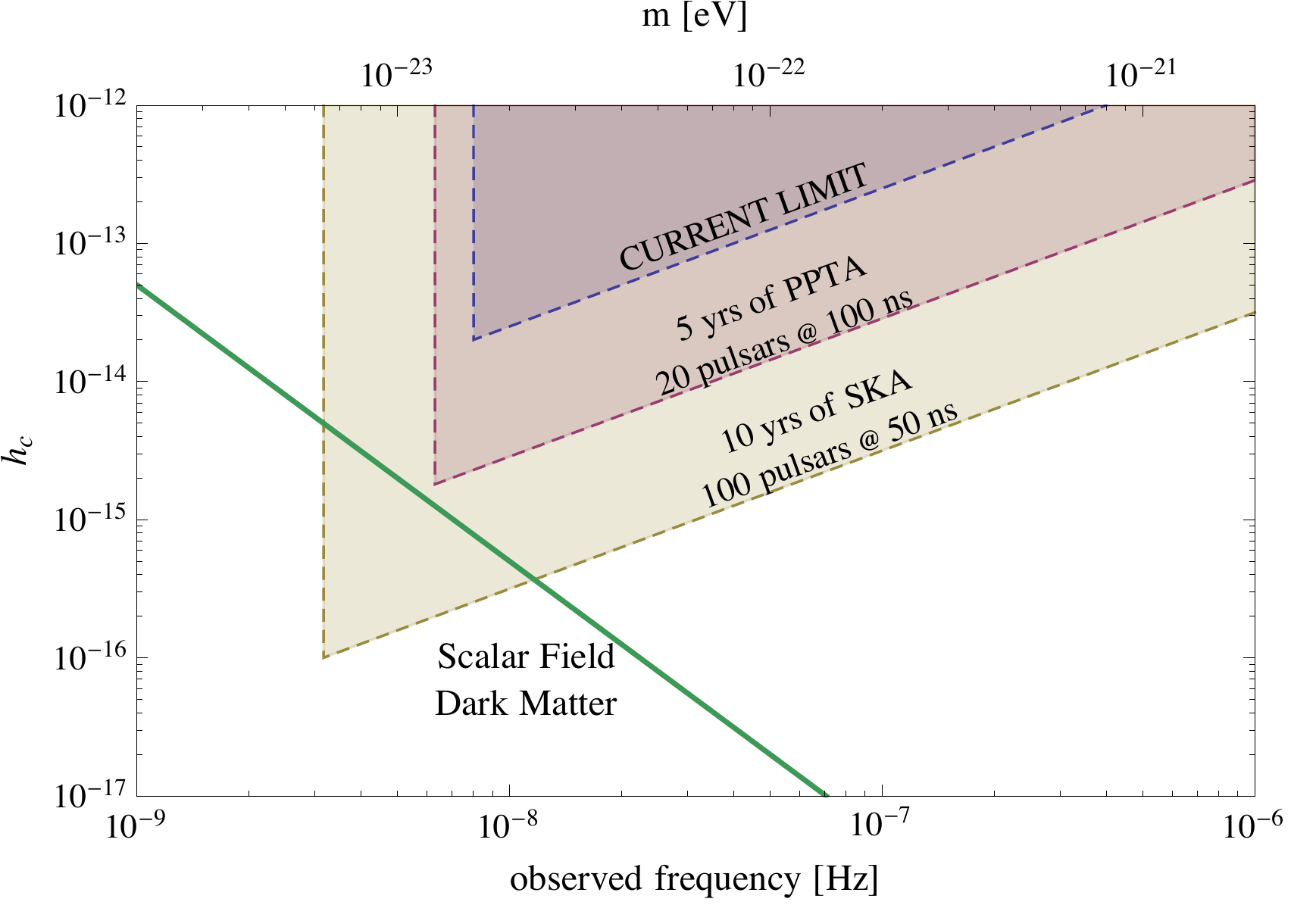}
 \end{center}
 \caption{Current and forecasted constraints on axion/scalar DM from the effect of pressure oscillations on pulsar timing. The green line shows the expected level of signal if axions compose the local DM. SKA will be sensitive to masses $m_a\lesssim 2.3\times 10^{-23}\text{ eV}$. Reproduced (with permission) from Ref.~\cite{2014JCAP...02..019K}.}
\label{fig:pulsar_forecast}
\end{figure}

Ref.~\cite{2014JCAP...02..019K} considered the sensitivity of pulsar timing arrays to this signal by comparing the amplitude $\Delta t_\phi$ to the corresponding time delay from a stochastic gravitational wave background. Fig.~\ref{fig:pulsar_forecast} shows the current constraints from Parkes Pulsar Timing Array (PPTA)~\cite{2006ApJ...653.1571J}, forecasts for a 5 year observation with PPTA, and forecasts for ten years with the Square Kilometre Array (SKA). Current limits do not reach the level of the expected signal from ULAs, however SKA will be sensitive to masses $m_a\lesssim 2.3\times 10^{-23}\text{ eV}$ and DM fractions as low as one percent. This is a powerful probe complementary to the constraints from structure formation discussed in Sections~\ref{sec:cosmological_constraints} and~\ref{sec:galaxy_formation}. 

The best current limits from pulsar timing come from the analysis of Ref.~\cite{2014PhRvD..90f2008P} from the NANOGrav PTA. The limits are an order of magnitude higher than the expected signal at $m_a=10^{-23}\text{ eV}$, consistent with the rougher bounds shown in Fig.~\ref{fig:pulsar_forecast}. Uncertainties in the analysis of PTA data relevant for constraining pressure oscillations include characteristics of the partner in binary pulsars, and modelling of radio wave propagation through the ionized interstellar medium. In the Bayesian analysis of Ref.~\cite{2014PhRvD..90f2008P}, the unknown pulsar parameters were marginalized over, following Ref.~\cite{2013MNRAS.428.1147V}.

As already mentioned, the pulsar timing signal from pressure oscillations depends only on gravitational interactions. Recently, Ref.~\cite{2015arXiv151206165G} considered the pulsar timing signal from interactions between scalar DM and the standard model. For typical coupling strengths, these model-dependent signals are much stronger than the pressure oscillation signal. For $m\lesssim 10^{-22}\text{ eV}$ the PTA limits from interactions can be stronger than e.g. torsion balance or atom interferometry constraints.

\section{Non-Gravitational Interactions}

\begin{figure}
\begin{center}
\includegraphics[scale=0.5]{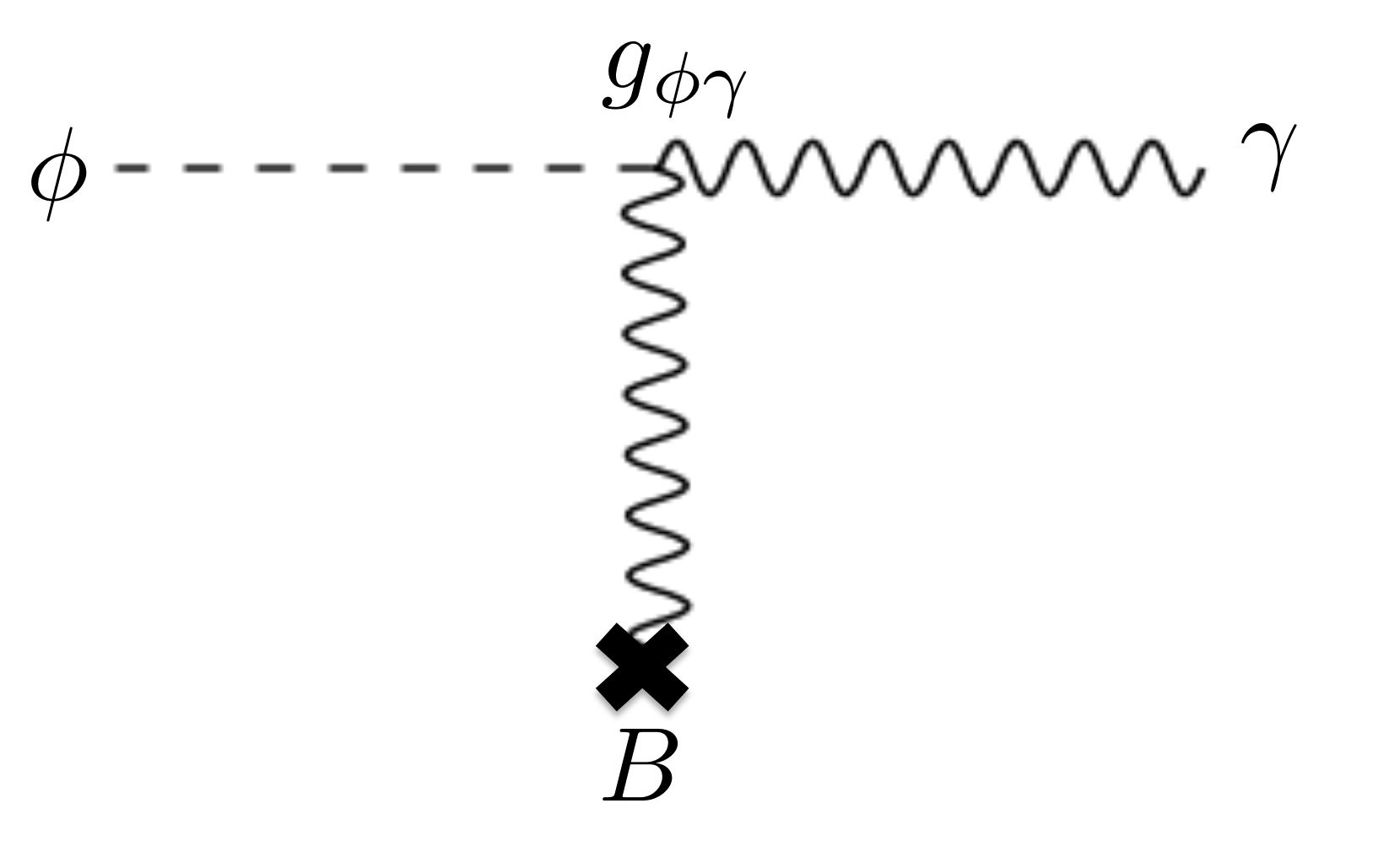}
 \end{center}
 \caption{Axion-photon interaction via the Primakoff process. In the presence of an external magnetic field, $B$, axions can convert into photons, and vice versa. This basic process, arising from the electromagnetic anomaly and expressed in the effective interaction with co-efficient $g_{\phi\gamma}$ in Eq.~\eqref{eqn:interaction_lagrangian}, underpins many constraints on axions and efforts to detect them.}
\label{fig:primakoff_process}
\end{figure}

\begin{figure}
\begin{center}
\includegraphics[scale=0.5]{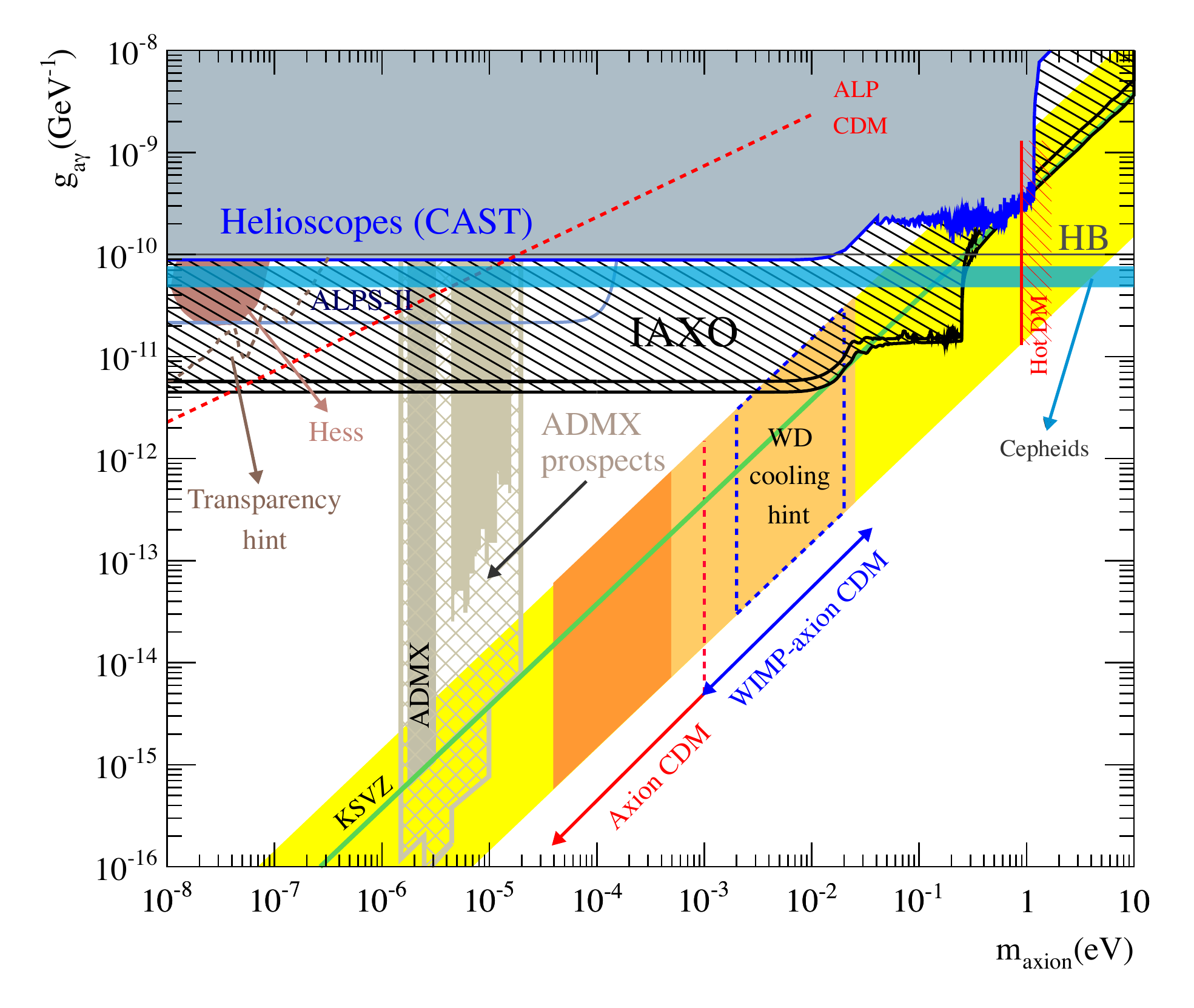}
 \end{center}
 \caption{Summary of constraints on the axion-photon coupling $g_{\phi\gamma}$, Eq.~\eqref{eqn:interaction_lagrangian} (here labelled $g_{a\gamma}$) as a function of axion mass. The line ``ALP CDM'' corresponds to setting $g_{\phi\gamma}=\alpha_{\rm EM}/2\pi f_a$ and requiring $f_a$ to be large enough such that $\Omega_a h^2\approx 0.12$ (c.f. Fig.~\ref{fig:alp_misalignment}). Reproduced (with permission) from Ref.~\cite{2013arXiv1309.7035C}.}
\label{fig:gagamma_review}
\end{figure}

Two classic methods for detecting the QCD axion were proposed by Sikivie in Ref.~\cite{1983PhRvL..51.1415S} and are known as \emph{haloscopes} and \emph{helioscopes}. Another archetypal axion experiment is ``light shining through a wall'' (LSW)~\cite{1987PhRvL..59..759V}. In recent years there has been a flurry of new ideas in axion (and scalar) direct detection (see, for example, Refs.~\cite{2013PhRvD..88c5023G,2015arXiv150608364S}). Some of the most important bounds on axions, in particular establishing the lower limit on $f_a\gtrsim 10^9\text{ GeV}$ for the QCD axion, come from considering stellar processes (e.g. Ref.~\cite{2008LNP...741...51R}). Many bounds on axions from their interactions exploit the two-photon coupling in the presence of magnetic fields (the Primakoff~\cite{1951PhRv...81..899P} process, see Fig.~\ref{fig:primakoff_process}), though we will also discuss the fermion and $G\tilde{G}$ couplings. A recent review of constraints on the axion-photon coupling is given in Ref.~\cite{2013arXiv1309.7035C}, and shown in Fig.~\ref{fig:gagamma_review}. We do not discuss collider signatures of axions in any detail. A recent discussion of existing constraints and future prospects is given in Ref.~\cite{2015JHEP...06..173M}.


\subsection{Stellar Astrophysics}
\label{sec:stellar_astro}

Axion emission is an energy-loss channel for stars and supernovae. The observed properties of stars can be used to limit the existence of such a channel, and the emitted stellar axions can be searched for. The stellar astrophysics limits \emph{apply regardless of whether the axion is DM}, because we are producing axions directly, and not relying on a cosmic population.

The solar luminosity in axions is
\be
L_a=1.85\times 10^{-3}\left(\frac{g_{\phi\gamma}}{10^{10}\text{ GeV}}\right)^2L_\odot\, ,
\ee
where $L_\odot$ is the photon luminosity. The maximum luminosity is at $3\text{ keV}$, and the average is $4.2\text{ keV}$~\cite{2008LNP...741...51R}. Axion production occurs as long as $m_a$ is less than the cental temperature of the sun, $T_\odot \approx 1\text{ keV}$ and leads the sun to consume nuclear fuel faster. A very crude bound can be found by demanding that the axion luminosity is less than the photon luminosity. Equating $g_{\phi\gamma}\sim (\alpha_{\rm EM}/2\pi f_a)$ for the QCD axion gives $f_{\rm QCD}\gtrsim 5\times 10^5\text{ GeV}$. 

The strongest bound on solar axions can be derived from direct searches for them. The helioscope converts solar axions back to photons in a macroscopic $B$ field on earth, and observes the photons in the X-ray. The state-of-the-art helioscope is the CERN Axion Solar Telescope (CAST)~\cite{2005PhRvL..94l1301Z,2007JCAP...04..010A,2014PhRvL.112i1302A}. The 95\% C.L. bounds are:
\begin{align}
g_{\phi\gamma}&<8.8\times 10^{-11}\text{ GeV}^{-1}\quad (m_a\lesssim 0.02\text{ eV})\, , \\
 g_{\phi\gamma}&<3.3\times 10^{-10}\text{ GeV}^{-1}\quad (m_a\lesssim 1.17\text{ eV})\, ,
\end{align}
where the two bounds refer to two different experimental configurations (low mass, vacuum; high mass, $^3$He). The proposed International AXion Observatory (IAXO)~\cite{2013arXiv1302.3273V} could improve the bound on $g_{\phi\gamma}$ by an order of magnitude  (see Fig.~\ref{fig:gagamma_review}).

The ratio of horizontal branch (HB) stars to red giants in galactic globular clusters is altered by axion-photon conversion inside stars, and places a competitive constraint on $g_{\phi\gamma}$ for axions with masses less than the stellar internal temperatures, $T\lesssim 100\text{ keV}$. In Fig.~\ref{fig:gagamma_review}, this is shown as $g_{\phi\gamma}<1\times 10^{-10}\text{ GeV}^{-1}$ (this constraint is also shown in terms of the axion lifetime in Fig.~\ref{fig:axion_decays}.). The most up-to-date constraint using 39 galactic globular clusters and state-of-the-art stellar modelling is that of Ref.~\cite{2014PhRvL.113s1302A}, which gives:
\be
g_{\phi\gamma}<6.6\times 10^{-11}\text{ GeV}^{-1}\quad (95\% C.L.)\quad (m_a\lesssim 100\text{ keV})\, .
\ee

Supernova SN1987a places the strongest limit on $g_{\phi\gamma}$ for low mass axions from the lack of observation of a gamma ray signal coincident with the neutrino burst due to axion-photon conversion within the Milky Way. The most up-to-date limit from Ref.~\cite{2015JCAP...02..006P} is 
\be
g_{\phi\gamma}<5.3\times 10^{-12}\text{ GeV}^{-1}\quad (m_a< 4.4\times 10^{-10}\text{ eV})\, .
\ee
Note that this limit is not shown on Fig.~\ref{fig:gagamma_review}, which does not extend to such low mass axions. SN1987A also places bounds on heavier axions with masses less than the SNe internal temperature, $T\approx 50 \text{ MeV}$, where axion emission leads to additional cooling. An approximate bound is (e.g. Ref~\cite{1995PhRvD..52.1755M}):
\be
g_{\phi\gamma}<10^{-9}\text{ GeV}^{-1}\quad (m_a< 50\text{ MeV})\, .
\ee

Energy loss in globular cluster stars and white dwarfs sets limits on the axion-electron coupling, $g_{\phi e}$. The strongest constraint comes from axion bremsstrahlung in globular cluster red giants~\cite{1995PhRvD..51.1495R}:
\be
g_{\phi e}<3.3\times 10^{-13}\, .
\ee

Finally, the duration of the neutrino burst from SN1987a can be used to constrain the axion-nucleon interaction, $g_{\phi N}$. If axions interact strongly enough with nuclei, then axion emission via nuclear bremsstrahlung, $N+N\rightarrow N+N+\phi$, is a more efficient energy-loss channel than neutrino emission, shortening the observed neutrino burst~\cite{1992PhR...220..229K}.  The theoretical calculation of supernova energy loss involves many uncertainties, but approximate bounds can be obtained. For a KSVZ type axion with no tree-level fermion couplings the bound is~(see Ref.~\cite{2008LNP...741...51R} for discussion)
\be
f_a \gtrsim 4\times 10^8\text{ GeV}\quad (\text{KSVZ})\,.
\ee

\subsection{``Light Shining Through a Wall''}

LSW is based on a very simple idea: shine a laser beam at a wall; apply a magnetic field so that it converts into axions, which travel freely through the wall; on the other side of the wall apply another magnetic field to convert the axions back to observable photons (for a review, see Ref.~\cite{2011ConPh..52..211R}). Just like the stellar astrophysics limits, this is direct axion production and \emph{applies regardless of whether the axion is DM}.

The conversion probability, $P(\gamma\rightarrow\phi)$, for photons of energy $\omega$ into axions in the presence of a coherent magnetic field, $B$, of length $L$ is
\be
P(\gamma\rightarrow\phi)=4\frac{g_{\phi\gamma}^2B^2\omega^2}{m_a^4} \sin^2 \left(\frac{m_a^2L}{4\omega}\right) \, .
\ee
The conversion probability can also be affected by using a medium with a refractive index $n_r\neq 1$, and by use of resonant cavities to enhance conversion and reconversion on either side of the wall.

The constraints from current LSW experiments are not particularly strong compared to astrophysical constraints, and do not appear on the scale of Fig.~\ref{fig:gagamma_review}. The strongest bounds come from the Any Light Particle Search (ALPS) experiment~\cite{2010PhLB..689..149E} and are roughly
\be
g_{\phi\gamma}\lesssim 7\times 10^{-8}\text{ GeV}^{-1}\quad (m_a\lesssim 10^{-3}\text{ eV})\, .
\ee 
The planned experiment ALPS-II~\cite{2013JInst...8.9001B} will improve these limits by more than three orders of magnitude, sensitive to $g_{\phi\gamma}\sim 2\times 10^{-11}\text{ GeV}^{-1}$ over a similar range of masses. The projected reach is shown in Fig.~\ref{fig:gagamma_review} and will be competitive with astrophysical and helioscope limits discussed in Sec.~\ref{sec:stellar_astro}.

\subsection{Vacuum Birefringence and Dichroism}

In the presence of a magnetic field, the Primakoff interaction between axions and photons allows for the vacuum to become birefringent and dichroic~\cite{1983PhRvL..51.1415S}. These effects cause the polarization plane of linearly polarized light to be rotated as it propagates. With no external magnetic field, we simply have birefringence (rotation with no absorption, we consider this effect in a cosmological context in Section~\ref{sec:cosmic_bire}), while in the presence of a magnetic field, there is absorption of one polarization state, i.e. dichroism. The amplitude of the dichroism is given by~\cite{2006PhRvL..96k0406Z}
\be
\varepsilon = \sin 2\theta \left(\frac{B L g_{\phi\gamma}}{4}\right)^2 \left[ \frac{\sin (m_a^2L/4\omega)}{m_a^2L/4\omega}\right]^2 \, ,
\ee
where $\theta$ is the angle between the magnetic field, $B$, and the polarization direction, $L$ is the length of the magnetic region, and $\omega$ is the photon energy. The effect can be enhanced in a Fabrey-Perot cavity by increasing the number of passes the light makes in the cavity. Measuring the dichroism of the vacuum in the presence of a $B$-field can thus be used to place constraints on the existence of axions possessing the two-photon coupling.

Using this technique, in 2006 PVLAS reported evidence for a polarization rotation in the presence in a $B\approx 5\text{ T}$ field of $\alpha=(3.9\pm 0.5)\times 10^{-12} \text{ rad/pass}$ ($3\sigma$ uncertainties). This was interpreted as evidence for an axion with $m_a\approx 1\text{ meV}$ and $g_{\phi\gamma}\approx 10^{-5}\text{ GeV}^{-1}$~\cite{2006PhRvL..96k0406Z}. Although this signal was already in tension with results from helioscopes, considerable interest was generated. The relevant parameter space was later directly excluded by the LSW experiment, GammeV~\cite{2008PhRvL.100h0402C}. Furthermore, reruns of PVLAS at different field strengths~\cite{2008PhRvD..77c2006Z} showed that the signal of Ref.~\cite{2006PhRvL..96k0406Z} was in fact due to instrumental artefacts. Nevertheless, this remains an interesting part of the story of axion constraints.

\subsection{Axion Mediated Forces}
\label{sec:axion_forces}

The couplings $g_{\phi e}$ and $g_{\phi N}$ of Eq.~\eqref{eqn:interaction_lagrangian} cause the axion to mediate spin-dependent forces. Such force exists \emph{independently of whether the axion is DM}. The resulting dipole-dipole interaction in the non-relativistic limit gives rise to the following potential~\cite{Moody:1984ba}:
\be
V(r)=\frac{g_{\phi i}g_{\phi j}}{16\pi M_i M_j}\left[ (\hat{\sigma}_i\cdot\hat{\sigma}_j)\left(\frac{m_a}{r^2}+\frac{1}{r^3}\right)-(\hat{\sigma}_i\cdot \hat{r})(\hat{\sigma}_j\cdot \hat{r})\left(\frac{m_a^2}{r}+\frac{3m_a}{r^2}+\frac{3}{r^3}  \right) \right]e^{-m_ar}\, ,
\ee
where $i,j$ labels the electron or nucleon with mass $M$, $\hat{\sigma}$ is a unit vector in the direction of the spin, and $\hat{r}$ is a unit vector along the line of centres. 

The interaction is of Yukawa-type and its range is suppressed by $e^{-m_ar}$. Even though this force can be long-range for ULAs, they are not subject to standard fifth-force constraints since the macropscopic sources must be spin-polarized. The dipole-dipole interactions between nucleons and electrons are only weakly constrained by current experiments, and the resulting bounds on $g_{\phi e}$ and $g_{\phi N}$ are not as strong as those from stellar astrophysics. They are~\cite{2012PhRvD..86a5001R}
\begin{align}
g_{\phi N}&<0.85\times 10^{-4} \quad (m_a\lesssim 10^{-7}\text{ eV})\, , \\
g_{\phi e}&<3\times 10^{-8} \quad (m_a\lesssim 10^{-6}\text{ eV})\, . \\
\end{align}

If the axion also has scalar interactions of the form $g_s\phi\bar{\psi}\psi$, then monopole-monopole and monopole-dipole potentials are induced~\cite{Moody:1984ba}. For a general ALP, $g_s$ should be very small on symmetry grounds. The limits on the scalar interaction strength for the QCD axion are given by the limits on $d_n$ and by the amount of $CP$ violation in the standard model. Current bounds are weaker than the astrophysical limits and do not reach the level of sensitivity to constrain the QCD axion-induced nucleon-nucelon monopole-dipole and monopole-monopole interactions. However, the proposed method of Ref.~\cite{2014PhRvL.113p1801A} using Nuclear Magnetic Resonance to probe the monopole-dipole interaction could cover a wide range corresponding to the entire classic axion window, $10^9\text{ GeV}\lesssim f_a\lesssim 10^{12}\text{ GeV}$. Despite its tiny value, the scalar coupling of the QCD axion offers a very promising avenue for discovery.

\subsection{Direct Detection of Axion DM}

\subsubsection{Haloscopes and ADMX}
\label{sec:admx}

Let's begin with the classic haloscope experiments~\cite{1983PhRvL..51.1415S}, which search for DM axions using the $g_{\phi\gamma}$ coupling. A haloscope currently in operation is the Axion Dark Matter eXperiment (ADMX)~\cite{2010PhRvL.104d1301A}. 

A DM axion enters a microwave cavity, where it interacts with an applied magnetic field, converting into a photon which is then detected. The cavity geometry is tuned such that this conversion is resonant, enhancing the conversion rate. The power generated in the cavity is
\be
P=g_{\phi\gamma}^2 \frac{V B_0 \rho_a C}{m_a}\text{min }(Q,Q_a)\, , 
\ee
where $\rho_a$ is the local DM density in axions, $V$ is the cavity volume, $B_0$ is the applied magnetic field strength, $Q$ is the quality factor of the cavity, $Q_a$ is the ratio of the halo axion energy to energy spread, and $C$ is a mode dependent form factor for the cavity. For approximate ADMX parameters $V=500 \text{ L}$, $B_0=7\text{ T}$, $Q=10^5$, in the classic QCD axion window with $f_a\approx 10^{12}\text{ GeV}$, the power is $P\approx 10^{-21}\text{ W}$. 

Since ADMX is a DM detector, it also relies on $\rho_a$ being large, and quoted constraints assume that axions in its sensitivity range compose all the DM. Because of the resonant tuning required, ADMX is very precise, but is only able to probe a narrow range in the mass-coupling plane (see Fig.~\ref{fig:gagamma_review}). ADMX is sensitive to axions with $m_a\approx 10^{-6}\text{ eV}$. Current constraints exclude ALPs of this mass more strongly coupled to photons than the QCD axion. In the near future ADMX will able to probe most of the model space (KSVZ and DFSZ) for the QCD axion with $10^{-6}\text{ eV}\lesssim m_a\lesssim 10^{-5}\text{ eV}$, i.e. $f_a\sim 10^{12}\text{ GeV}$. 

\begin{figure}
\begin{center}
\includegraphics[width=0.8\textwidth]{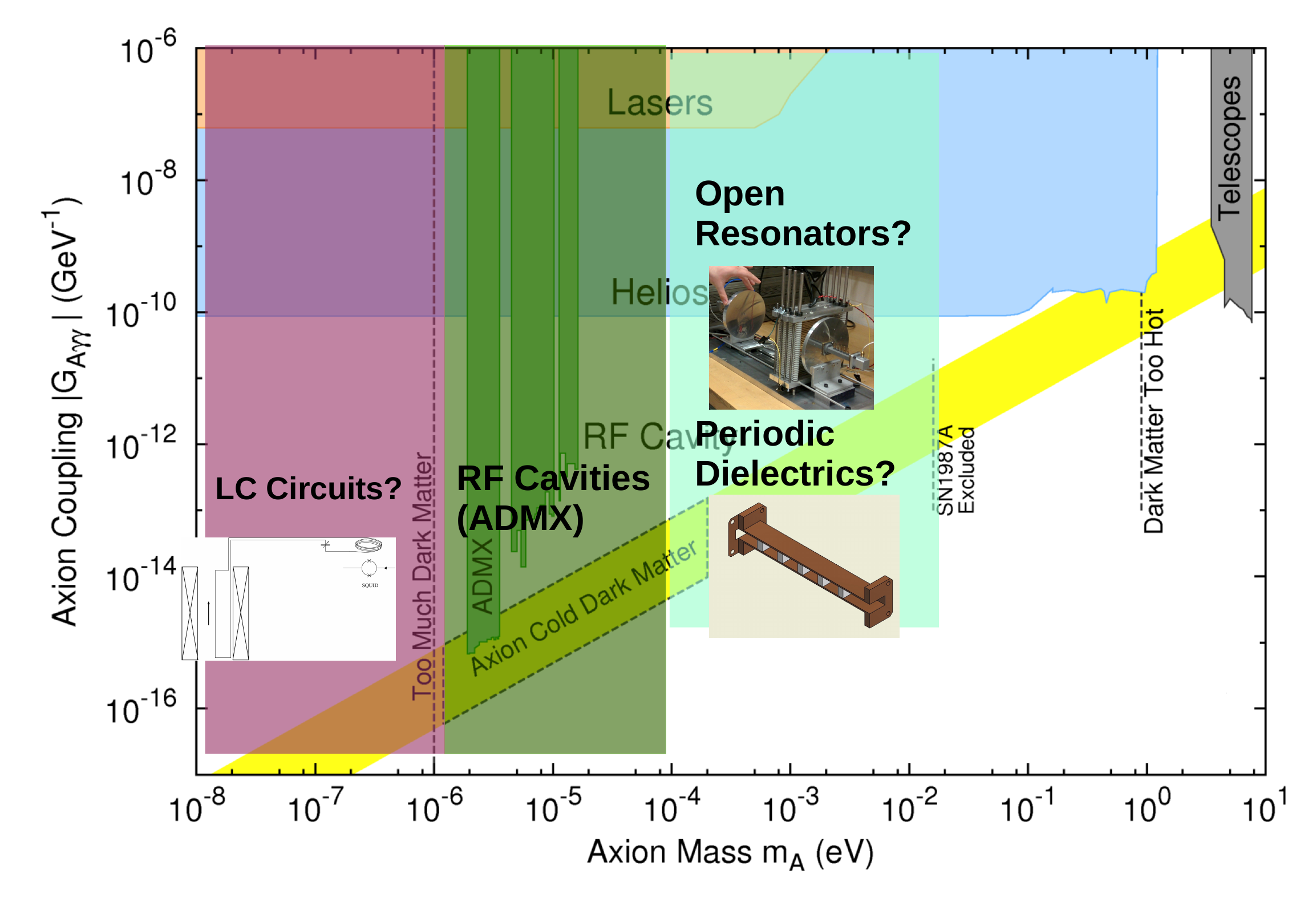}
 \end{center}
 \caption{Future reach of axion DM direct detection experiments using the two-photon coupling. The classic window of the QCD axion can be completely covered. Reproduced, with permission, from ``Resonant Dark Matter Detectors Beyond 10 GHz,'' Gray Rybka, PATRAS10 (2014).}
\label{fig:rybka_new_searches}
\end{figure}

Other upgrades and new proposals for axion DM direct detection experiments in the classic QCD axion window using the two-photon coupling include the use of open resonators (the ORPHEUS experiment)~\cite{2015PhRvD..91a1701R}, LC-circuits~\cite{2014PhRvL.112m1301S} and broadband searches with SQUIDs~\cite{2016arXiv160201086K}. Projections for some of these techniques are shown in Fig.~\ref{fig:rybka_new_searches}, and could cover the mass range $10^{-8}\text{ eV}\lesssim m_a\lesssim 10^{-2}\text{ eV}$ of the QCD axion.

\subsubsection{Nuclear Magnetic Resonance and CASPEr}
\label{sec:casper}

\begin{figure}
\begin{center}
$\begin{array}{@{\hspace{-0.8in}}c@{\hspace{-0.2in}}c}
\includegraphics[scale=0.6]{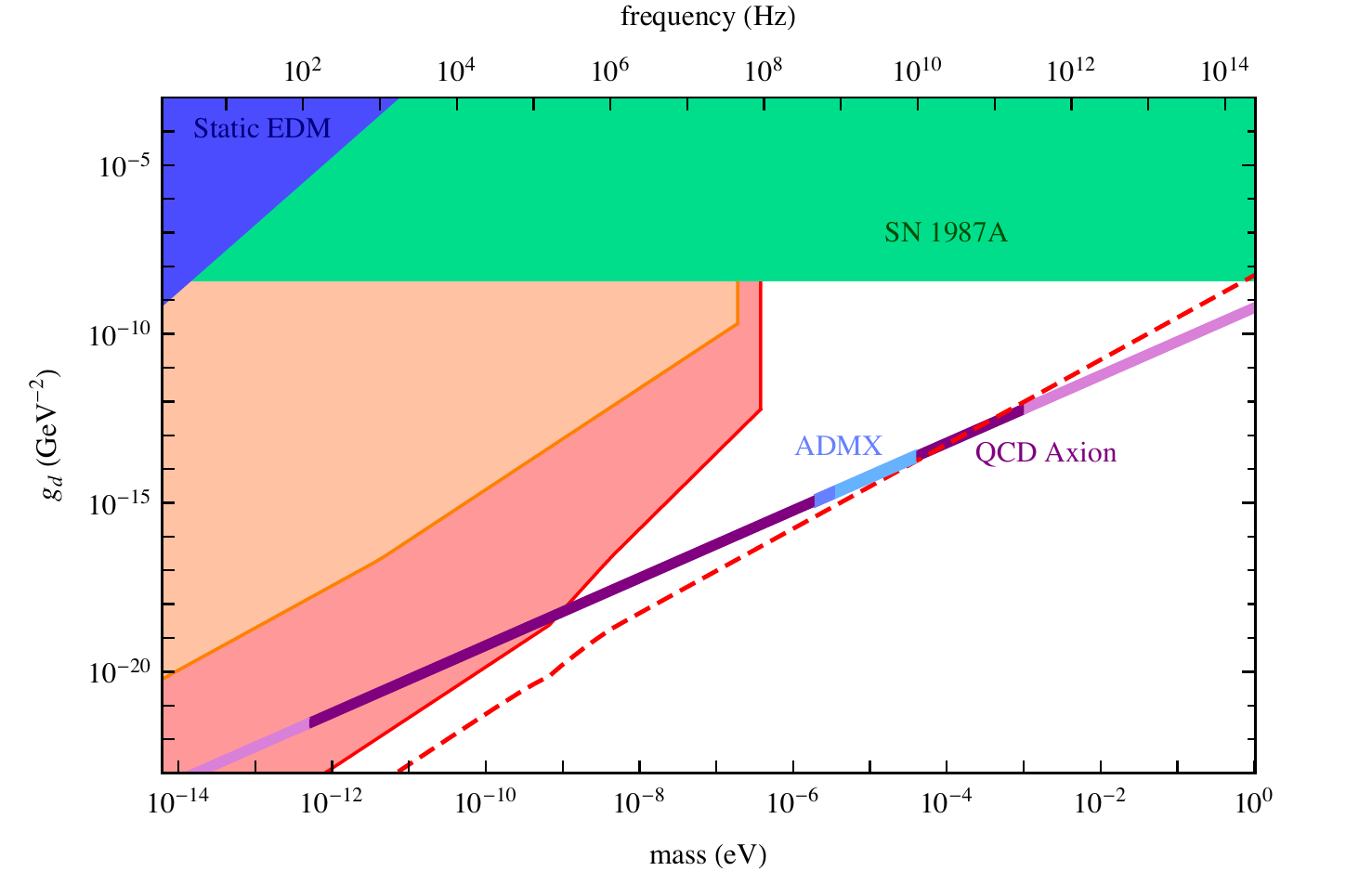} &
\includegraphics[scale=0.6]{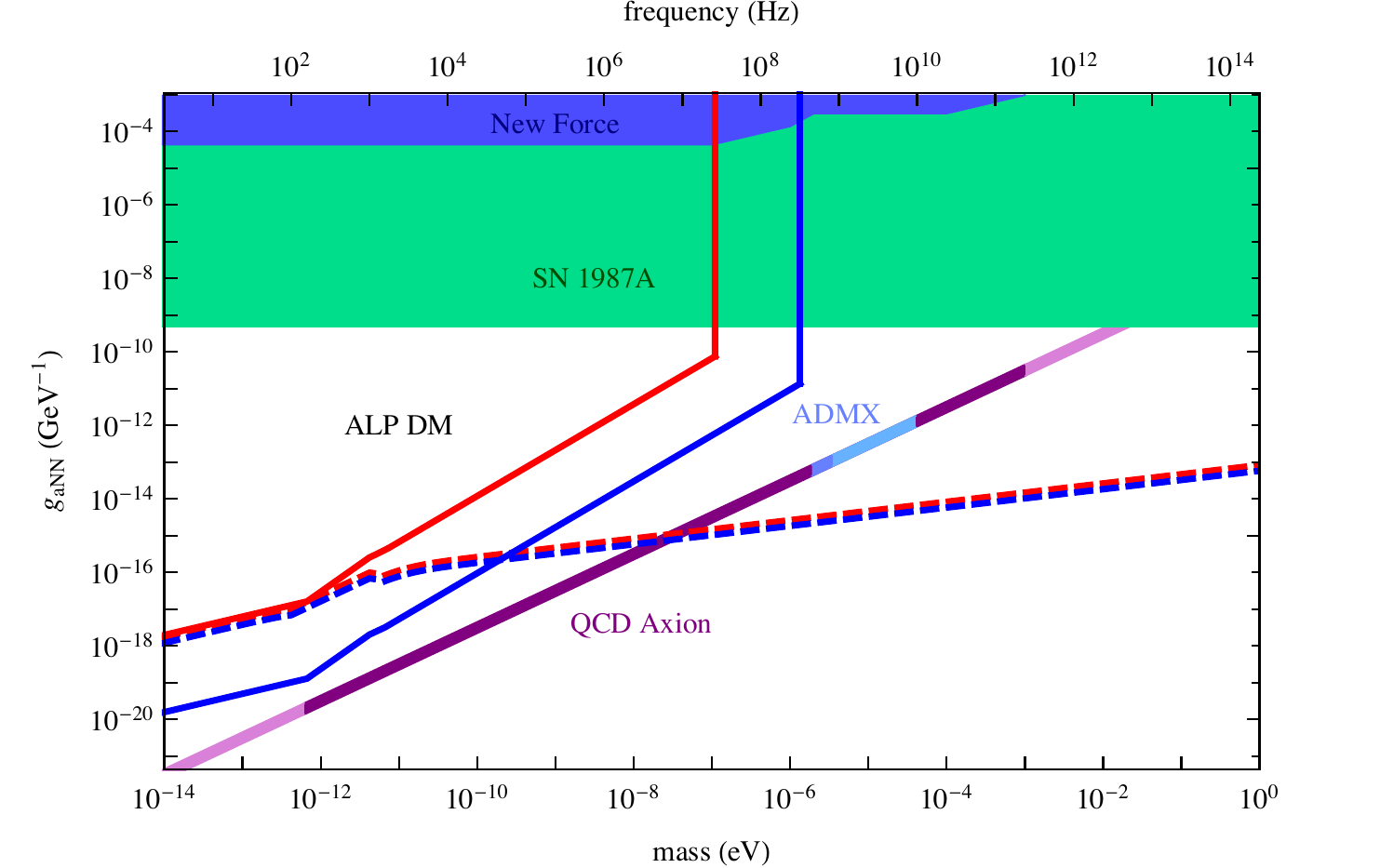}
 \end{array}$
 \end{center}
 \caption{Sensitivity of the CASPEr experiments, assuming the DM is contained exclusively in a single ALP. CASPEr is a resonant experiment and sensitivity assumes a 3 year operation of scanning. \emph{Left Panel:} CASPEr-Electric and the nucleon EDM coupling. Orange shaded: phase 1. Red shaded: phase 2. Dashed red: magnetometer noise limit in phase 2. \emph{Right Panel:} CASPEr-Wind and the axial nucleon moment (note their $g_N$ is our $\tilde{g}_N$). Red: Xe sample. Blue: $^3$He sample. Dashed lines: magnetization noise limits. Reproduced (with permission) from Refs.~\cite{2014PhRvX...4b1030B,2013PhRvD..88c5023G}. Copyright (2014,2013) by The American Physical Society.}
\label{fig:casper_figs}
\end{figure} 

The Cosmic Axion Spin Precession Experiment (CASPEr)~\cite{2014PhRvX...4b1030B}, comes in two varieties. Both strategies are novel, as they do not rely on the ``standard" two-photon coupling. Each CASPEr experiment uses the property that the axion couplings to nucleons are spin dependent. The interactions can be detected by spin-polarizing a sample in an applied magnetic field, and searching for spin-precession using nuclear magnetic resonance techniques. The induced magnetization is resonant at the Larmour frequency of the applied magnetic field, $2\mu_{m} B_{\rm ext}=m_a$ (where $\mu_m$ is the nuclear magnetic dipole moment) and is detected using a SQUID magnetometer. For reasons that will become apparent, we refer to the two distinct experiments as ``CASPEr-Electric"~\cite{2014PhRvX...4b1030B} and ``CASPEr-Wind"~\cite{2013PhRvD..88c5023G}. Just like with ADMX, CASPEr is a DM detector and the sensitivity to axions scales with the DM abundance. CASPEr has not yet been constructed, and we discuss projected sensitivities.

CASPEr-Electric exploits the axion coupling to $(\phi/f_a){\rm Tr}\, G\tilde{G}$, which gives rise to the EDM coupling, $g_d$. CASPEr-Electric thus explores the defining property of the QCD axion. The dipole moment induced by an axion is $d_n=g_d\phi$. Recall that the QCD axion solves the strong-$CP$ problem by setting the \emph{time-average} of the nucleon EDM to zero, as required by experiments constraining the static EDM~\cite{2006PhRvL..97m1801B}. The same oscillations in the axion field that allow it to function as a DM candidate, however, lead to \emph{EDM oscillations}, $d_n\sim 10^{-16}(\phi/f_a)\cos (m_at)\, e\,{\rm cm}$, where $\phi$ is the \emph{local} value of the axion field amplitude. CASPEr-Electric applies an electric field to a spin-polarized sample and detects the precession of the nuclear spins about the $\vec{E}$ field axis caused by the non-zero EDM.  

The projected sensitivity of CASPEr-Electric is shown in Fig.~\ref{fig:casper_figs}, Left Panel. In phase 2 CASPEr-Electric will be able to detect the QCD axion for $f_a\gtrsim 10^{16}\text{ GeV}$, with ultimate limits from magnetization noise able to reach $f_a\gtrsim 3\times 10^{13}\text{ GeV}$. CASPEr-Electric is thus highly complementary to ADMX and astrophysical bounds.

CASPEr-Wind exploits the axion coupling to the axial nuclear current, $g_{\phi N}$, and the induced spin-dependent force. As the earth moves relative to the DM halo of our galaxy, so we experience  a ``DM wind'' of axions. The effective coupling in the nucleon Hamiltonian is $H_N\supset \tilde{g}_{\phi N}m_a\phi \cos (m_a t)\vec{v}\cdot\vec{\sigma}$, where $\vec{\sigma}$ is the nuclear spin, and $\vec{v}$ is the DM wind velocity. The spin-dependent force creates a torque around the direction of the DM wind and leads to spin precession of nuclei without the need for an applied electric field. CASPEr-Wind is thus somewhat simpler to implement than CASPEr-Electric. 

The projected sensitivity of CASPEr-Wind is shown in Fig.~\ref{fig:casper_figs}, Right Panel. While CASPEr-Wind is not sensitive to the QCD axion (except in the noise-limited regime), it is sensitive to the ULA model of Ref.~\cite{2015arXiv151001701K}, and is complementary to cosmological axion searches.

\subsection{Heavy Axions and Axion Decays}
\label{sec:decays}

In this section we consider constraints on axions with masses $m_a\gg 1\text{ eV}$. Note that the constraints summarised in Fig.~\ref{fig:gagamma_review} (and much of the phenomenology discussed elsewhere in this review) typically do not apply to such high masses, as they rely on the coherence of the axion field. The bounds from stellar astrophysics in Section~\ref{sec:stellar_astro} can apply for $m_a$ as large as $1\text{ keV}$. We consider primarily the astrophysical and cosmological consequences of axion decay, but mention some other constraints in passing.

Consider the axion-photon coupling, $g_{\phi\gamma}$, defined in Eq.~\eqref{eqn:interaction_lagrangian}, which we recall has mass-dimension $-1$, and is in general a free parameter for ALP models, with approximate scale $1/f_a$. This coupling allows axions to decay into two photons, with a lifetime:
\be
\tau_{\phi\gamma}=\frac{64\pi}{m_a^3g_{\phi\gamma}^2} \approx 130\text{ s}\left(\frac{\text{GeV}}{m_a}\right)^3\left(\frac{10^{-12}\text{ GeV}^{-1}}{g_{\phi\gamma}}\right)^2\, .
\ee
Consider the KSVZ axion, with the photon coupling fixed by Eq.~\eqref{eqn:photon_dfsz_ksvz}. Taking the age of the Universe to be $\tau_{\rm univ.}\approx 10^{10}\text{ years}$ we find that the QCD axion is stable on the lifetime of the Universe for $f_a\gtrsim 1.9\times 10^6\text{ GeV}$. Thus, the QCD in the allowed range of $f_a$ is stable on the lifetime of the Universe, and hence is a DM candidate. 

\begin{figure}
\begin{center}
\includegraphics[scale=0.6]{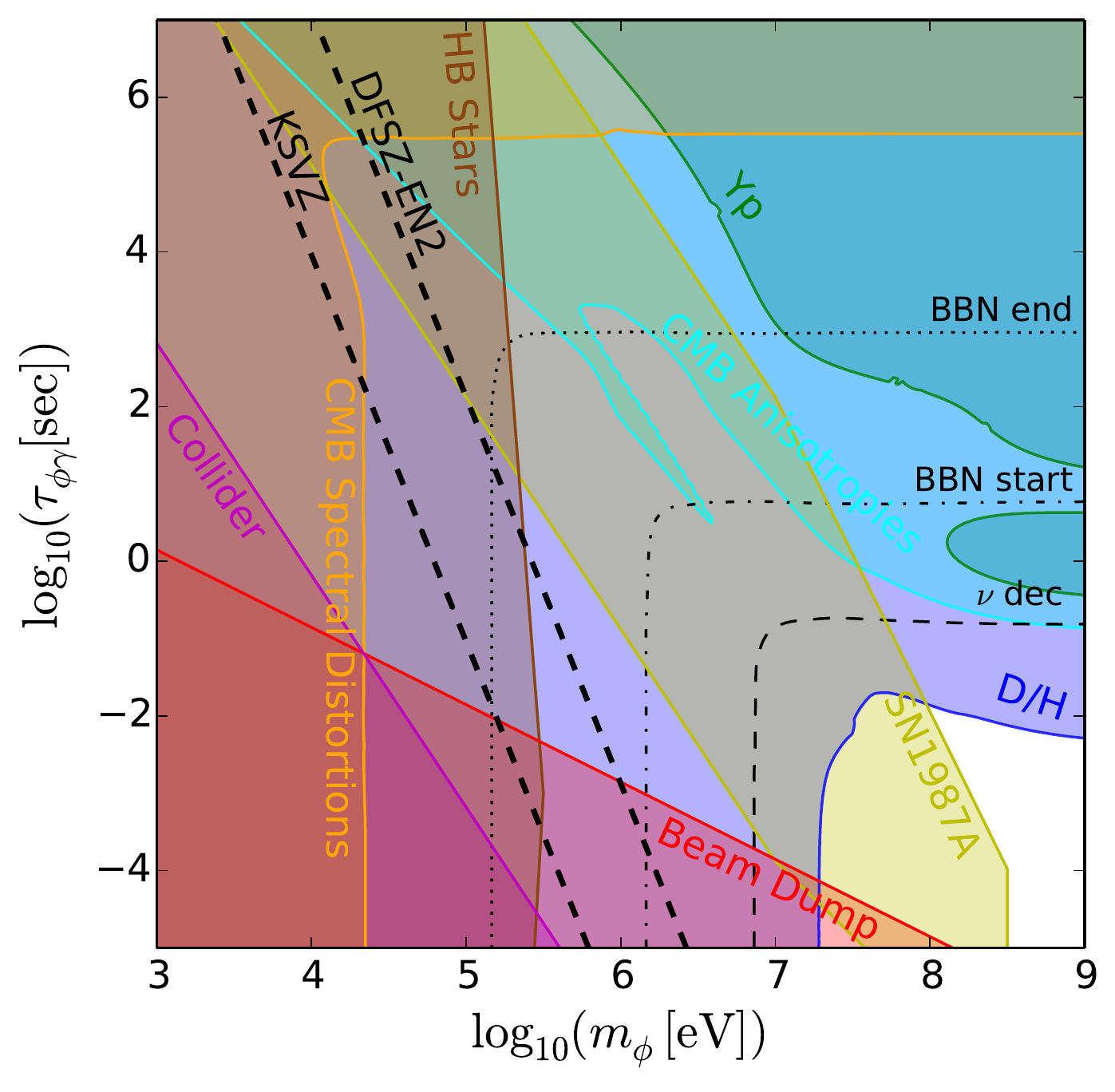} 
 \end{center}
 \caption{Constraints on heavy ALPs from decays, in the mass-lifetime plane. The axion mass is here labelled $m_\phi$. The CMB, D/H, and Yp regions are excluded at 3$\sigma$, the Collider and Beam Dump regions are excluded at 2$\sigma$, and the SN1987a and HB Stars regions are less formal. Reproduced (with permission) from Ref.~\cite{2015PhRvD..92b3010M}.}
\label{fig:axion_decays}
\end{figure}

ALPs, on the other hand, may decay on much shorter time scales. The coupling of ALPs is in general proportional to the mass, since couplings go as $1/f_a$ and $m_a=\Lambda_a^2/f_a$. Thus heavier ALPs can be unstable on cosmological timescales and will decay to standard model particles (or light dark sector particles). The decay of such a population of ALPs injects additional relativistic energy density into the Universe, which is constrained by a number of probes. We will closely follow the recent compilation of constraints in Ref.~\cite{2015PhRvD..92b3010M}, as shown in Fig.~\ref{fig:axion_decays}. Some early constraints on ALPs from decays can be found in Refs.~\cite{1995PhRvD..52.1755M,1997PhRvD..55.7967M}, while further reading can be found in Ref.~\cite{2012JCAP...02..032C} (for general physics and consequences of decaying particles, see Ref.~\cite{1992NuPhB.373..399E}). 

The presence and later decay of ALPs in the early Universe can change the effective number of relativistic species, $N_{\rm eff}$ (Eq.~\ref{eqn:neff_def}), and the baryon-to-photon ratio, $\eta_b\equiv n_b/n_\gamma$, at different times in cosmological history. A lower value of $N_{\rm eff}^{\rm CMB}$ affects the CMB power spectrum, as discussed in Section~\ref{sec:direct_decay}. The baryon ratio at the CMB is well measured, fixind $\eta_b^{\rm CMB}=2.74\times 10^{-8}\Omega_b h^2$. The photon energy density is also fixed by the equally well measured $T_{\rm CMB}$. Therefore ALP decays can actually \emph{reduce} $N_{\rm eff}$ and \emph{increase} $\eta_b$. An ALP decaying between BBN and the CMB reduces $N_{\rm eff}^{\rm CMB}$ if the decay occurs after neutrino decoupling, by heating of the plasma.\footnote{It is interesting to note the opposite effects of different ALPs on $N_{\rm eff}$: decay of a heavy particle to an ALP leads to an increase, while decay of a heavy ALP to photons leads to a decrease. The effects of light and heavy ALPs and moduli could conspire to hide them from our view.} Decay before BBN also reduces $N_{\rm eff}^{\rm BBN}$. On the other hand, if the ALPs are themselves relativistic at BBN, $N_{\rm eff}^{\rm BBN}$ is increased. ALP decay between BBN and the CMB leads to a relative increase $\eta_b^{\rm BBN}$ compared to $\eta_b^{\rm CMB}$. 

Changes of the expansion rate , via $N_{\rm eff}$, and baryon abundance during BBN affect the light element abundances. The standard model predictions of the BBN light element abundances are extremely well verified (with the famous exception of Lithium): see Refs.~\cite{pdg,2004IJTP...43..669M} for reviews. The helium abundance, $Y_p$ and the deuterium-to-hydrogen ratio, $D/H$, place strong constraints on ALPs, both from decays and from the contribution of thermally produced axions with $m_a\lesssim 1\text{ MeV}$ to the radiation density at BBN.

Energy injections at different epochs can also change the shape of the CMB frequency power spectrum, such that it is no longer a perfect black body.  Such effects are known as \emph{CMB spectral distortions}, and are strongly constrained by the COBE-FIRAS measurements (for a review, see Ref.~\cite{2014arXiv1405.6938C}). Early decays of axions heat the plasma leading to distortions of ``$\mu$-type'' (chemical potential) for decays between $10^5\lesssim z\lesssim 10^6$, or ``$y$-type'' (Compton scattering) for decays between $1100\lesssim z\lesssim 10^5$. These effects are computed in e.g. Refs.~\cite{2015PhRvD..92b3010M,2012JCAP...02..032C,1997PhRvD..55.7967M}.

In the life-time range of relevance to cosmological axion decays, the axion-photon coupling also has collider signatures, allowing, for example, single-photon final states in electron-positron colliders. The constraint from LEP~\cite{Abbiendi:2000hh,Heister:2002ut,Abdallah:2003np} is~\cite{2015JHEP...06..173M}
\be
g_{\phi\gamma}<4.5\times10^{-4}\text{ GeV}^{-1} \quad (\text{LEP: }m_a\lesssim \text{ GeV})\, .
\ee
In fact, a stronger bound due to the single photon final state was derived much earlier, using ASP data~\cite{1989PhRvD..39.3207H} in Ref.~\cite{1995PhRvD..52.1755M}: $g_{\phi\gamma}\leq 5.5\times 10^{-4}\text{ GeV}^{-1}$ for $m_a\gg 29\text{ GeV}$. Anomalous decays of heavy quark states lead to similar bounds.

The summary of these constraints is shown in Fig.~\ref{fig:axion_decays}. The DFSZ and KSVZ axion models are excluded for $m_a$ in the keV to MeV range, as are most ALPs  with
\begin{align}
1\text{ keV}\lesssim &m_a \lesssim 1\text{ GeV}\, , \\
10^{-4}\text{ s}\lesssim &\tau_{\phi\gamma} \lesssim 10^6\text{ s}\, .
\end{align}
There is an open window for short-lived, $\tau_{\phi\gamma}<0.01\text{ s}$, heavy, $m_a\gtrsim 1\text{ GeV}$, ALPs that decay early enough and are sufficiently non-relativistic at BBN to not alter the light element abundances.

\subsection{Axion Dark Radiation}
\label{sec:axion_dr}

We discussed in Section~\ref{sec:direct_decay} how a population of relativistic axions can be created by decay of a modulus. The CMB power spectrum and other cosmological observables constrain the simplest consequence of this: the relativistic axion energy density, parameterized by $\Delta N_{\rm eff}$. This population of axions, if coupled to the standard model, can also be probed by axion scattering. 

If the modulus decay that produced the axion DR also reheats the Universe, then the axion energy is $E\sim m_\sigma\sim T_{\gamma}\sqrt{M_{pl}/m_\sigma}\gg T_\gamma$. Because the energy is much higher than the plasma temperature, this gives access to processes that are otherwise kinematically forbidden. This leads to interesting constraints and phenomenology despite the $f_a$-suppressed axion couplings. Ref.~\cite{2013JHEP...10..214C} discussed the phenomenology in detail. 
\begin{figure}
\begin{center}
\includegraphics[width=0.75\textwidth]{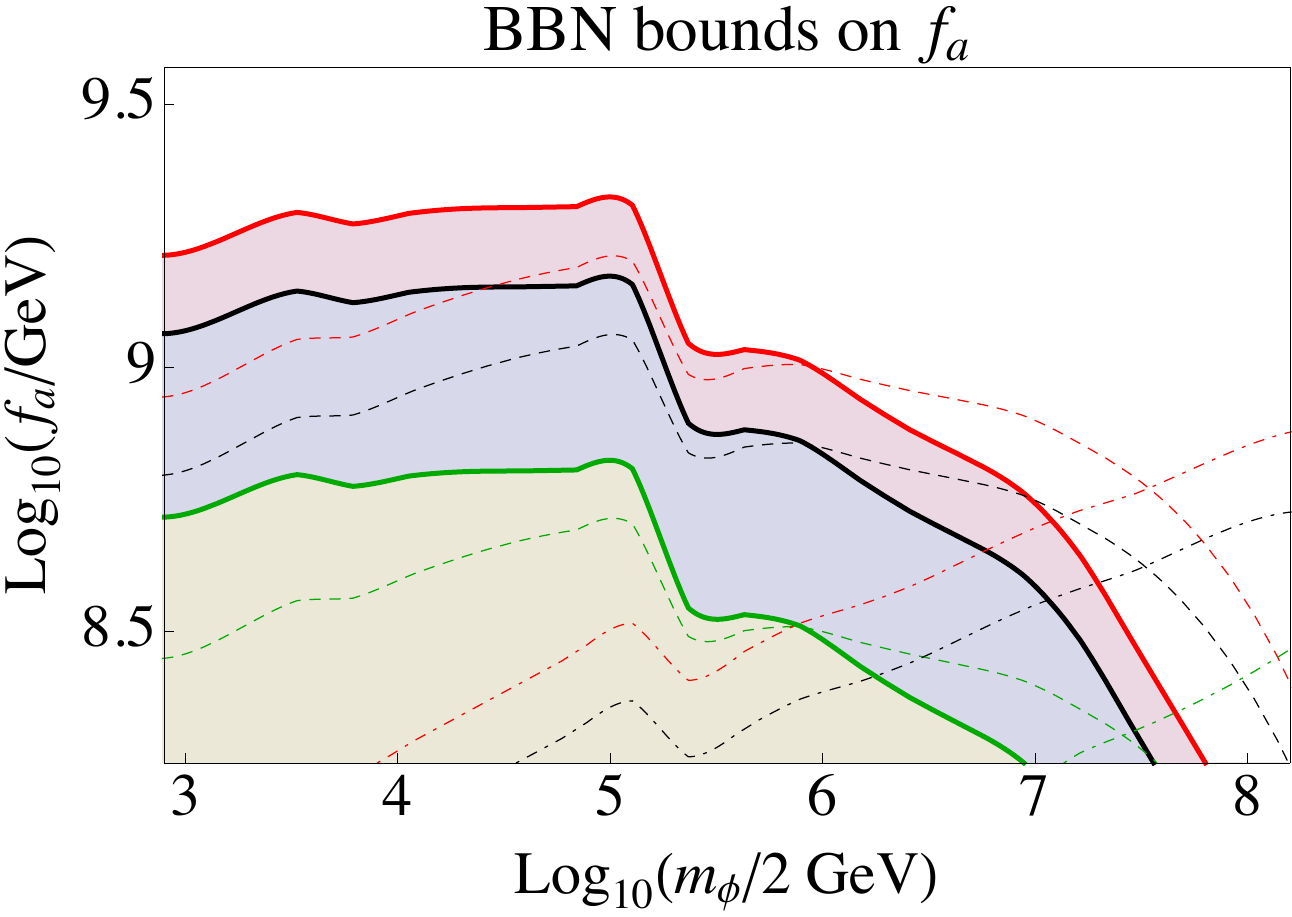} 
 \end{center}
 \caption{Constraints on axion DR from the primordial helium abundance~\cite{2006PhRvD..74j3509J}. The fermion interaction is taken to have strength $c_f=1$, and here $m_\phi$ is the modulus mass. Final states to $b\bar{b}$ (solid), $c\bar{c}$ (dashed) and $s\bar{s}$ (dot-dashed) are considered, with varying amounts of DR, $\Delta N_{\rm eff}=0.1,0.5,1$ (green, black, red; corrected labelling from typo in original). Areas below curves are excluded. Reproduced (with permission) from Ref.~\cite{2013JHEP...10..214C}.}
\label{fig:conlon_marsh_bbn}
\end{figure} 

An axion-fermion coupling of the form $\mathcal{L}_f=c_fm_f\phi\bar{\psi}\gamma^5\psi/f_a$ (this form can be obtained from the axial current interaction in Eq.~\ref{eqn:interaction_lagrangian} by use of the equations of motion) allows for production of heavy fermions via the process $a+\gamma\rightarrow f+\bar{f}$. The secondary decay of the fermions can alter the proton to neutron ratio during BBN, and thus the primordial helium abundance. Each axion scattering process can be mapped onto an ``effective decay process''~\cite{2013JHEP...10..214C} for which constraints can readily be found in the literature (e.g. Ref.~\cite{2006PhRvD..74j3509J}). The constraints are shown in Fig.~\ref{fig:conlon_marsh_bbn}. Taking $c_f=1$, BBN constraints rule out values of $f_a\lesssim 10^9\text{ GeV}$ over a wide range of modulus masses.

Axion DR also has a flux at Earth and, if the axion-photon coupling is non-vanishing, could be detected by helioscopes like CAST. The axion DR flux is distinct from the solar flux in two important ways: firstly, because of its cosmological origin, it is isotropic; secondly, the DR flux is not suppressed by as many powers of $g_{\phi\gamma}$, due to the different production mechanism compared to solar axions. Taking $g_{\phi\gamma}\sim f_a^{-1}$, the DR signal in a heliscope is thus suppressed as only $f_a^{-2}$, compared to the $f_a^{-4}$ suppression for solar axions. For a modulus mass of $m_\sigma=5\times 10^6\text{ GeV}$ and $\Delta N_{\rm eff}\approx 0.6$ the flux is $\Phi_a\approx 1.09\times 10^6\text{ cm}^{-2}\text{s}^{-1}$~\cite{2013JHEP...10..214C}, which is of order the solar QCD axion flux for $f_a=10^{10}\text{ GeV}$. The DR background in this model is thus in reach of IAXO. For these same parameters, the energy spectrum peaks in the keV range, and has a form characteristic of the axion DR background from modulus decay.

\subsection{Axions and Astrophysical Magnetic Fields}

Let's further consider the Primakoff process, but now for the case of ULAs in the presence of astrophysical magnetic fields. Gamma rays from blazars suggest that the cosmic background field exceeds $B\sim 10^{-16}\text{ G}$ in large voids~\cite{2010MNRAS.406L..70T,2010Sci...328...73N}, while it could be large as nG, with Mpc coherence length. Larger magnetic fields are present in clusters of galaxies, with strength $B\sim \mu\text{G}$ and coherence length of order kpc.

\subsubsection{CMB Spectral Distortions}

In the presence of a background magnetic field axion photon mixing occurs and, just like in the case of massive neutrinos, propagation and interaction eigenstates are not the same. Furthermore, plasma effects lead to an effective photon mass:
\be
m_\gamma^2=\omega_p^2(z)-2\omega^2(n_H-1)\, ,
\ee 
where $\omega$ is the photon frequency, and the refractive index of neutral hydrogen is $n_H$. The plasma frequency, $\omega_p$, depends on the free electron density, and is thus a function of redshift determined by recombination and reionization. At $\omega=T_{\rm CMB}$ the photon plasma mass at $z=0$ is $m_\gamma\sim 10^{-14}\text{ eV}$. Resonant axion-photon conversion occurs when $m_\gamma=m_a$. Since for high frequency photons $m^2_\gamma$ passes through zero, resonant conversion can occur for arbitrarily low axion mass, and can occur multiple times as $m_\gamma^2$ changes sign. 

The frequency dependence of the resonant conversion epoch leads to a spectral distortion~\cite{1988PhLB..202..301Y}. COBE-FIRAS~\cite{1994ApJ...420..439M,1996ApJ...473..576F} measured the CMB to be a black body to high precision. This constrains the resonant conversion probability, which in turn leads to a constraint on the product $g_{\phi\gamma}B_0$, where $B_0$ is the spatially averaged magnetic field strength today. 
\begin{figure}
\begin{center}
\includegraphics[width=0.75\textwidth]{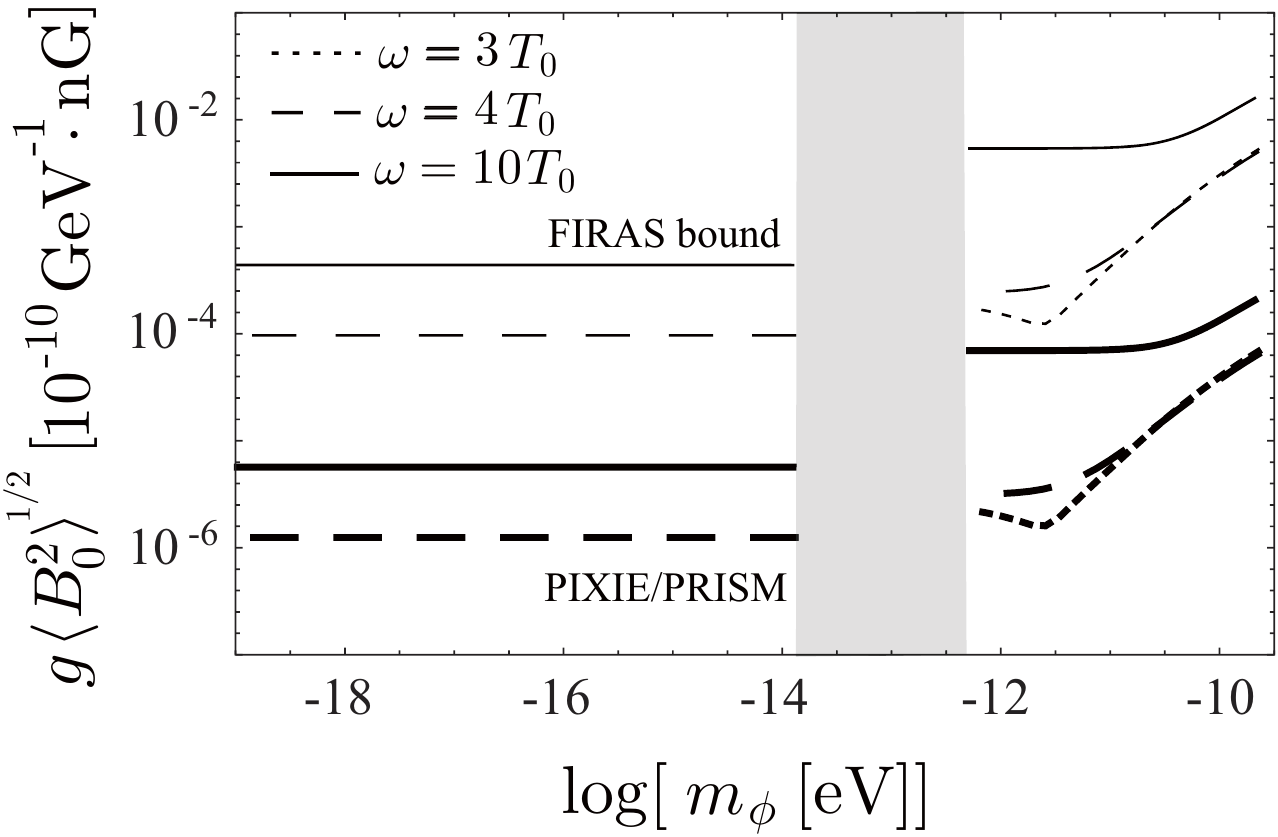} 
 \end{center}
 \caption{Constraints on ULAs from CMB spectral distortions from FIRAS, and projected for PIXIE/PRISM. The axion mass is labelled $m_\phi$ in this plot. The dark band shows masses where multiple resonant conversions effectively exclude such axions entirely, for $g_{\phi\gamma}\neq 0$. Reproduced (with permission) from Ref.~\cite{2013PhRvD..88l5024T}. Copyright (2013) by The American Physical Society.}
\label{fig:tashiro}
\end{figure} 

The constraints have been addressed in detail in Refs.~\cite{2009JCAP...08..001M,2013PhRvD..88l5024T}. Fig.~\ref{fig:tashiro} shows the constraints on ULAs from FIRAS, and projected constraints from a PIXIE~\cite{2011JCAP...07..025K}/PRISM~\cite{2014JCAP...02..006A}-like mission. Multiple resonant conversions occur for $10^{-14}\text{ eV}\lesssim m_a\lesssim 10^{-12}\text{ eV}$, effectively excluding any $g_{\phi\gamma}\neq 0$ for this mass range. While constraints are only on the product $g_{\phi\gamma}B_0$, they are stronger than the product of current upper limits on $g_{\phi\gamma}$ and $B_0$ individually.

\subsubsection{X-ray Production}
\label{sec:x-ray_bg}

As discussed a number of times, axion DR can be produced by the decay of a modulus, and the axion DR energy today is $E_0\sim T_{\rm CMB}\sqrt{M_{pl}/m_\sigma}$. For a modulus mass $m_\sigma\sim 10^6\text{ GeV}$ (suggested by string theory solutions to the EW hierarchy problem) this gives rise to a cosmic axion background (CAB) with energy $E\sim 0.1$ - $1$ keV. The energy density in the CAB is
\be
\rho_{\rm CAB}=1.6\times 10^{60}\text{ erg Mpc}^{-3} \left(\frac{\Delta N_{\rm eff}}{0.57}\right) \, ,
\ee 
Conversion of the CAB to photons in the presence of magnetic fields leads to production of X-rays. 

Clusters of galaxies are permeated by magnetic fields with $B\sim \mu G$ and coherence lengths  $L\sim \text{ kpc}$. Axion-photon conversion in this environment is predicted to lead to excess X-ray emission from clusters~\cite{2013JHEP...10..214C,2013PhRvL.111o1301C}. The X-ray luminosity of a typical Mpc sized cluster is $\mathcal{L}_{\rm cluster}\sim 10^{44}\text{ erg s}^{-1}$. The excess soft X-ray luminosity in Coma is $1.6\times 10^{42}\text{ erg s}^{-1}$~\cite{1996Sci...274.1335L}, which could plausibly be explained with an axion-photon coupling $g_{\phi\gamma}\sim 10^{-14}\text{ GeV}^{-1}$~\cite{2013PhRvL.111o1301C}, depending on the axion mass and the photon plasma mass in the intra-cluster medium. This emission has fixed redshift scalings, since the CAB is cosmological in origin. It is also predicted to correlate with cluster magnetic fields, unlike an annihilating DM signal. 

Ref.~\cite{2015arXiv150605334D} considered X-ray production within galactic magnetic fields. For the strength of coupling required to explain the soft X-ray excess in Coma, conversion within the Milky Way is negligible. Star burst galaxies, with larger magnetic fields, may produce an observable signal, in particular if the inhomogeneous free electron density is accounted for in modelling the emission.

Conversion in cosmological magnetic fields could contribute to an unresolved cosmic X-ray background. This is essentially the inverse of the spectral distortion effect discussed in the previous subsection, with a different energy spectrum. A diffuse cosmic X-ray background in the keV energy range is observed~\cite{2007ApJ...661L.117H}, with diffuse intensity that could be explained by the CAB with $g_{\phi\gamma}\sim 10^{-13}\text{ GeV}^{-1}$, assuming nG strength cosmological magnetic fields~\cite{2013PhRvL.111o1301C}. From Fig.~\ref{fig:tashiro} we see that this explanation for the X-ray background will in addition produce a CMB spectral distortion close to the FIRAS bound, and observable with PIXIE/PRISM.

\subsection{Cosmological Birefringence}
\label{sec:cosmic_bire}

CMB polarization comes in E-modes and B-modes. E-modes are generated from temperature fluctuations at last scattering by the quadrupole anisotropy, and the E spectrum can be predicted from the measurement of the adiabatic temperature fluctuations. B-modes can be generated in three ways: primordially, by tensor fluctuations with relative amplitude $r_T$; by gravitational lensing along the line of sight; and finally by the birefringent effect, rotating of E into B.

\begin{figure}
\begin{center}
\includegraphics[width=0.75\textwidth]{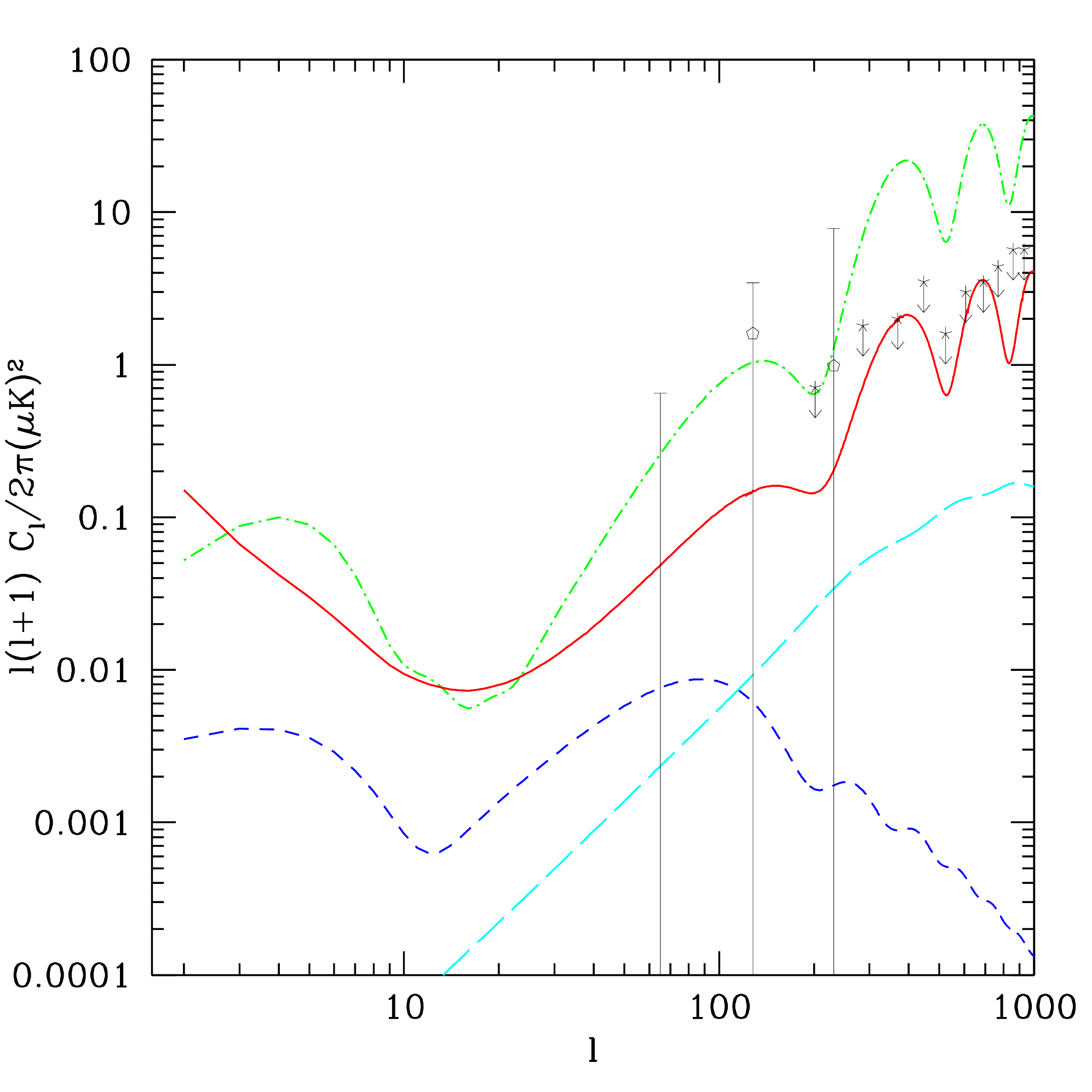} 
 \end{center}
 \caption{CMB B-mode power from birefringence caused by ULAs coupled to magnetic fields with $(H_Ig_{\phi\gamma})^2\approx 0.17$ (red, solid). The large angle signal can mimic tensor modes with $r\sim 0.1$ (blue, short dashed), while the small angle signal contains distinctive BAO from the E-modes (green, dot-dashed) and, for this choice of parameters, dominates over lensing power (cyan, long dashed). Reproduced (with permission) from Ref.~\cite{2009PhRvL.103e1302P} (where the data are described). Copyright (2009) by The American Physical Society.}
\label{fig:pospelov}
\end{figure} 

In the presence of the axion-photon coupling in Eq.~\eqref{eqn:interaction_lagrangian}, the fields satisfying free wave equations are $\vec{D}=\vec{E}+\frac{g_{\phi\gamma}}{2}\phi\vec{B}$ and $\vec{H}=\vec{B}-\frac{g_{\phi\gamma}}{2}\phi\vec{E}$~\cite{1992PhLB..289...67H} (note $\vec{E}$ and $\vec{B}$ are the fields of electromagnetism, and are not the same as E and B mode polarization). Therefore, if the axion field $\phi$ varies in time or space it can cause rotation of the plane of polarization of the CMB~\cite{1990PhRvD..41.1231C}:
\be
\Delta \beta = \frac{g_{\phi\gamma}}{2}\int d\tau \phi' \, ,
\ee
where it is reminded that $\tau$ is conformal time, and primes denote derivatives with respect to this. The integral is performed along the line of sight from the surface of last scattering at $z_{\rm dec}$ to today. When the axion is oscillating, the integral vanishes. Therefore, significant rotation only occurs for ULAs that begin oscillations after photon decoupling. Using $z_{\rm dec}=1020$, $\Omega_m=0.31$, $\Omega_\Lambda=0.69$, $h=0.67$, we find that ULAs with a mass $m_a\lesssim 3H_{\rm dec}=1\times 10^{-28}\text{ eV}$ can cause significant cosmological birefringence.

The uniform misalignment of ULAs in the broken PQ scenario (see Section~\ref{sec:broken_PQ}) leads to a uniform rotation of the plane of CMB polarization. Such a uniform rotation is constrained to be $|\Delta\beta|<1.4\times 10^{-2}$~\cite{2015ApJ...805..107M}. If we assume $\phi (\tau_0)=0$, this gives the approximate constraint $\phi_i g_{\phi\gamma}<2.8\times 10^{-3}$. Taking $g_{\phi\gamma}\sim \alpha_{\rm EM}f_a^{-1}$, CMB polarization rotation imposes a constraint on the $(\phi_i,f_a)$ plane. For ULAs, and using Eq.~\eqref{eqn:full_ula_omega} for $a_{\rm eq}<a_{\rm osc}$, the birefringence constraint is of order the constraint on the DM abundance from temperature anisotropies (Fig.~\ref{fig:contours_combinedLinear}, right panel), assuming $f_a<M_{pl}$ and $m_a\lesssim 1\times 10^{-28}\text{ eV}$ (excluding also the lightest DE like axions where $\phi (\tau_0)\neq 0$). Thus, if a sub-dominant population of such ULAs is detected in LSS in future, e.g. by \emph{Euclid} (Fig.~\ref{fig:modified_euclid}), then this may well be accompanied by birefringence in the CMB.

Anisotropies in the axion field cause anisotropic rotation. This leads to generation of BB anisotropy power from EE, and thus EB cross-correlations, and can be significantly sourced by ULA isocurvature pertrubations: see e.g. Refs.~\cite{1999PhRvL..83.1506L,2015ApJ...805..107M,2009PhRvL.103e1302P,2012PhRvD..86j3529G,2014PhRvD..89j3518Z}. The resulting CMB power spectra are shown in Fig.~\ref{fig:pospelov}. The amplitude of the power spectrum scales as $(H_Ig_{\phi\gamma})^2$. This effect is particularly interesting as it can generate B-modes that dominate over those produced by tensor perturbations. This could source large angle B-mode power in low-scale inflation if $H_Ig_{\phi\gamma}\sim 0.1$. Since the power is generated from the E-modes, there is also oscillating, large amplitude, small-angle B-power in this scenario. This would be present even after de-lensing and is distinct from the tensor mode power, which falls rapidly on small angular scales.  

The most recent constraints on anisotropic birefringence come from the $B$-mode power and 4-point function measured by \textsc{Polarbear}~\cite{2015arXiv150902461P}. These constraints are consistent with zero signal.

\section{Concluding Remarks}

In this review we have presented the vast cornucopia of axion physics. We have considered the motivations and models for axions coming from particle physics and string theory. We have seen how axions can be produced in the early Universe by a variety of mechanisms. Axions can play important roles in all of the unsolved mysteries of cosmology: inflation, dark matter, and dark energy. They also lead to novel phenomena, such as fuzzy dark matter, and dark radiation. Axion couplings to the standard model are fixed by symmetry considerations, and can be computed in specific models. We studied the tailored direct and indirect searches for axions, which are quite different to more ``standard" searches for new particle physics.

I hope, dear reader, that you have come away from this review with a sense for the fascinating progress that has been made in axion physics over the last years and decades. I also hope that you can see the places on the horizon where new opportunities are arising. Let me briefly reiterate some of these:
\begin{itemize}

\item \emph{The dark sector and large scale structure:} Soon, large scale structure measurements will reach the precision to test in detail aspects of standard neutrino physics, such as the neutrino mass, and number of neutrinos. Axions share many degeneracies with the neutrino sector. Misalignment-produced ULAs suppress structure formation on cluster scales; hot axions contribute to dark radiation either via thermal production or via modulus decay. Improved measurements and studies of CMB polarization and gravitational lensing of galaxies could easily discover these effects at the same time as testing neutrino physics. Breaking degeneracies via multiple probes is an important endeavour for both axion and neutrino physics.

\item \emph{Axions with $m_a\sim 10^{-22}\text{ eV}$ and the CDM small-scale crises:} The CDM small-scale crises, if they are indeed crises, can be solved by ULAs. Observational and simulation techniques on these scales are always improving, and axion physics must keep up. There are some simulations on the market, but the field has not been studied in anywhere near as much depth as competing models, such as WDM. The tantalizing prospect to see evidence for axions on these scales, in galactic dynamics and in the epoch of reionization, must not be overlooked, and much work is necessary to exploit this opportunity.

\item \emph{Progress in string theory model building and the axiverse:} A large part of the motivation to study axions comes from their apparent prevalence in string theory. In principle, therefore, constraints on axions can be interpreted as constraints on string theory. There is already a large program of model building in this direction. The focus has largely been on inflation, but extensions to other parts of cosmology are slowly being made. This model building should also be done holistically, with emphasis on the many different facets of axion physics that combine and provide the opportunity to make unique and verifiable predictions.

\item \emph{Novel experiments for axion direct detection:} Axion direct detection has, for many years, focused on the $\vec{E}\cdot\vec{B}$ coupling and the QCD axion. Recent years have seen an upsurge in interest in searching for the other possible axion couplings in terrestrial experiments. These searches are more generally applicable to ALPs, which may only possess a fraction of the couplings allowed by symmetry, for example having no coupling to photons. All direct searches for axions provide vital information to cosmology, not least by limiting the decay constant in specific models, but also by allowing the possibility to actually identify the DM as axion-like by the form of its couplings.

\end{itemize}

This summary is not the end. Axion physics is alive and well, and growing: long may it be so.

\vskip 0.25in
\noindent
{\bf Acknowledgments.}
This review was prepared in part for a lecture presented at the mini-workshop ``Axion Theory and Searches" at IPhT CEA/Saclay. I would like to thank the organisers of the workshop, Marco Cirelli, Bradley Kavanagh and Filippo Sala, for inviting me to lecture, and the other lecturers, Joerg Jaeckel and Pierre Sikivie, for stimulating discussion. I thank all my collaborators on the work presented here: Brandon Bozek, Malcolm Fairbairn, Pedro Ferreira, Daniel Grin, Ren\'{e}e Hlozek, Robert Hogan, Luca Iliesiu, Edward Macaulay, Kavilan Moodley, Ana Pop, Joseph Silk, Hiroyuki Tashiro, Maxime Trebistch, Scott Watson and Rosemary Wyse. Special thanks to the authors of Refs.~\cite{2015PhRvD..91h4011A,2014JCAP...02..019K,2013JHEP...10..214C,2013arXiv1309.7035C,2014PhRvL.113z1302S,2012MNRAS.424.1105M,2015PhRvD..92b3010M,2009PhRvL.103e1302P,2015arXiv150306100S,2014PhRvX...4b1030B,2013PhRvD..88c5023G,2015arXiv151006388B} for permission to reproduce their figures. I thank Jihn Kim and Maxim Pospelov for discussions on the particle theory of axions, Thomas Bachlechner for discussion on the weak gravity conjecture, Sacha Davidson for discussion of BEC's, and Ren\'{e}e Hlozek and Cliff Burgess for reading parts of the manuscript. Finally, I thank the anonymous referee, whose comments helped greatly improve the breadth of this work. This work was supported at Perimeter Institute by the Government of Canada through Industry Canada and by the Province of Ontario through the Ministry of Research and Innovation; and by a Royal Astronomical Society research fellowship, hosted at King's College London.

\appendix

\section{Theta Vacua of Gauge Theories}
\label{appendix:theta_vacua}
\setcounter{equation}{0} 
\renewcommand{\theequation}{A\arabic{equation}}

I will simply state some relevant results to give you a feel for this topic: see the wonderful book by Coleman, Ref.~\cite{1988assy.book.....C}, for the gory details. I follow Coleman's notation and normalisation in this discussion. 

Quantum theory depends on the Euclidean functional integral, with the path integral being dominated by field configurations of finite Euclidean action. These dominant contributions to the semi-classical approximation are known as \emph{instantons}. The action for a gauge field theory with gauge group $G$ (for definiteness, take $G=SU(N)$) and gauge coupling $g_G$ in 4 flat Euclidean dimensions is
 \be
 S=\frac{1}{4 g_G^2}\int d^4 x (F_{\mu\nu},F_{\mu\nu})
 \ee
 A field configuration of finite action must have $F\sim O(1/r^3)$ as $r\rightarrow \infty$ and so the gauge field must be of the form
 \be
 A_\mu = g\partial_\mu g^{-1}+O(1/r^2) \, ,
 \ee 
for some gauge transformation $g(x)$, which is a function mapping $G$ to the variables of Euclidean 4-space. In order not to alter the asymptotic baheviour in $r$ we must have that $g(x)$ maps $G$ to only the angular vairables. That is, the field configurations are defined \emph{up to a mapping of $G$ to the space-time boundary}, which in this case is topologically the three-dimensional hypersphere, $S^3$.
 
How many different mappings are there, and how can we classify them? Firstly, we can always make a gauge transformation by some other element $h$, which is a continuous function, and continuously deform one into another. That means that all \emph{homotopoically equivalent} mappings are equivalent field configurations.\footnote{An important consequence of this is the fact that $U(1)$ gauge theory has no instantons in 3+1 dimensions. $U(1)$ is topologically the circle, $S^1$, which, when wrapped around $S^3$, can be continuously deformed to a single point: the trivial mapping.} We now need to classify the homotopically distinct mappings. 

A theorem~\cite{bott_proof} states that we need only consider the $SU(2)$ subgroups of our group $G$. $SU(2)$ is topologically $S^3$, and so one such mapping is the trivial mapping
\be
g^{(1)}(x) = (x_4+i\vec{x}\cdot\vec{\sigma})/r \, ,
\ee
where $\sigma_{1,2,3}$ are the Pauli matrices. It is then also possible to prove (Coleman does not prove it, and I certainly won't) that all mappings from $S^3$ to $S^3$ are homotopic to a family of mappings
\be
g^{(\nu)}(x)=[g^{(1)}(x)]^\nu \, ,
\ee
where $\nu$ is an integer called the \emph{winding number}. For the simple example of wrapping $U(1)$ round a circle, this is easy to visualise, and $\nu$ labels the representations of $U(1)$ as $e^{i\nu\theta}$, with $\theta$ the angle on $S^1$.

Finally, it is possible to show that the winding number of a field configuration is given by the integral
\be
\nu = \frac{1}{32\pi^2}\int d^4x (F,\tilde{F}) \, ,
\ee
where $\tilde{F}$ is the dual field strength as defined below Eq.~\eqref{eqn:qcd_topological}. The winding number is a topological invariant of the field configuration, providing a finite contribution to the Euclidean action proportional to the integral of Eq.~\eqref{eqn:qcd_topological}. 

The winding number describes the boundary conditions of the gauge fields with $\nu=n$ in some state $|n\rangle$. The vacuum of the theory is given by a superposition of states
\be
|\theta\rangle = \sum_n e^{in\theta}|n\rangle \, .
\ee
such that
\be
\langle\theta|e^{-HT}|\theta\rangle \propto \int [dA]e^{-S}e^{i\nu\theta} \, .
\ee
A $\theta$-vacuum is thus described by a term in the action
\be
S_\theta = \frac{\theta}{32\pi^2}\int d^4x (F,\tilde{F}) \, .
\ee
All the $\theta$-vacua are topologically distinct, and transitions between them are forbidden as they involve discontinuous changes in the gauge field boundary conditions.

By considering a gas of $n$ instantons and $\bar{n}$ anti-instantons, such that $\nu=n-\bar{n}$, Coleman goes on to show that
\be
\langle\theta|e^{-HT}|\theta\rangle \propto \exp [e^{-S_0}\cos\theta]\, ,
\ee
so that the energy of the $\theta$-vacuum is
\be
E(\theta)\propto e^{-S_0}\cos\theta \, ,
\ee
with the one-instanton action
\be
S_0 = \frac{8\pi^2}{g_G^2}\, .
\ee

\section{EFT for Cosmologists}
\label{appendix:eft}
\setcounter{equation}{0} 
\renewcommand{\theequation}{A\arabic{equation}}

This is an extremely heuristic description of EFT. For a rigorous treatment, see e.g. Ref.~\cite{2007ARNPS..57..329B}.

The general notion of EFT is based on the idea that at low energies, $q$, we can replace a ``fundamental'' action, $S$, with an effective action, $S_{\rm eff}(q)$. In the jargon, this is thought of in terms of the Wilsonian picture of the renormalization group: we define an action in the UV at a scale $\Lambda_{\rm UV}$ and then use the renormalization group equations to ``run'' down to $q<\Lambda_{\rm UV}$. This is referred to as ``integrating out'' fields with masses $m>q$. Quantum field theory (e.g. Refs.~\cite{1995iqft.book.....P,2007qft..book.....S}) allows for interactions mediated by virtual particles, and when these particles are integrated out this leads to effective interactions in the low-energy theory that were not present in the UV theory.

\begin{figure}
\begin{center}
\includegraphics[scale=0.5]{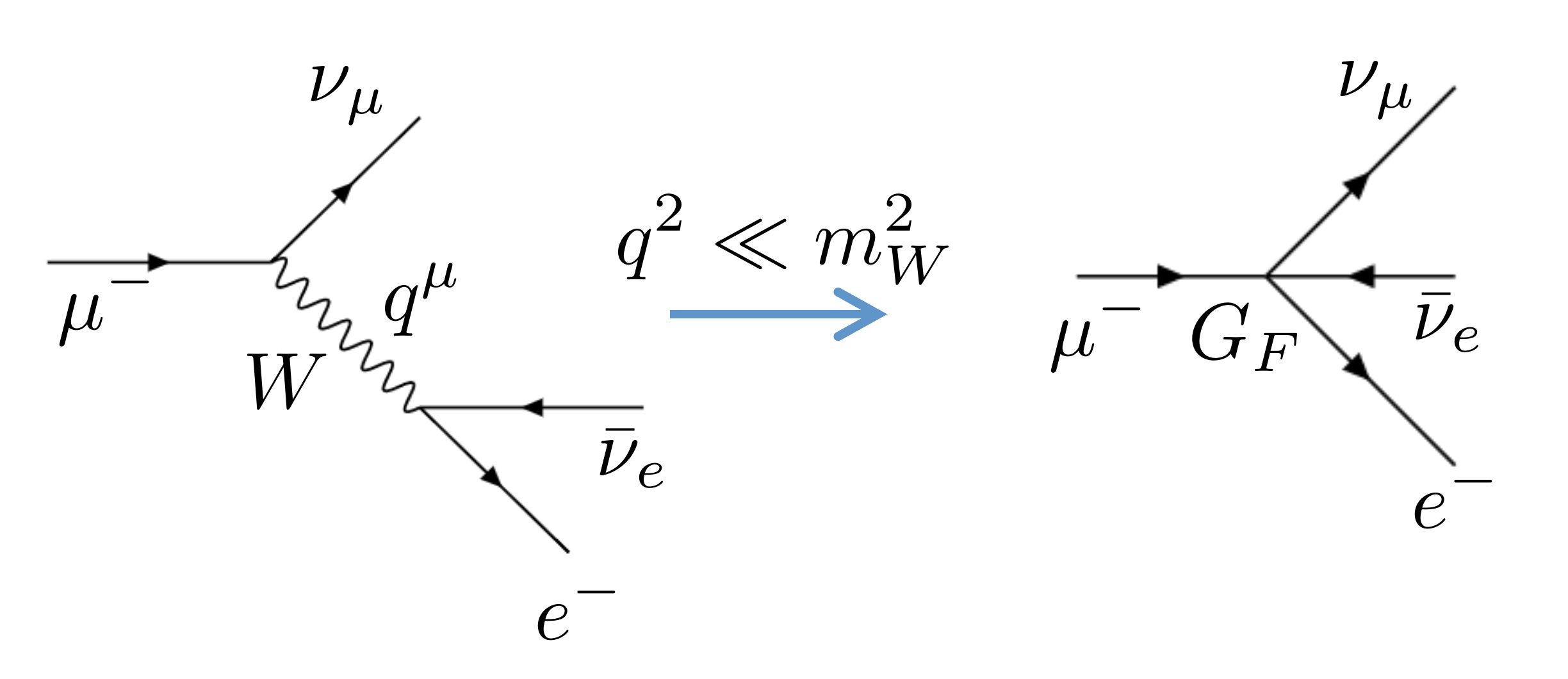}
 \end{center}
 \caption{Muon decay and the Fermi interaction as an example of EFT. The fundamental theory involves exchange of virtual $W$ bosons with momentum $q^\mu$. At low-momentum transfer, $q^2\ll m_W^2$, the interaction can be replaced with the effective 4-fermion interaction proportional to $G_F$.}
\label{fig:fermi_4pt}
\end{figure}

Consider the case of the Fermi interaction, represented in Fig.~\ref{fig:fermi_4pt} for muon decay. In the EW theory we know that, at a fundamental level, charged lepton-neutrino interactions are governed by a term in the action $S\supset ig_2W_\mu\bar{\ell}_i\gamma^\mu\nu_i+h.c.$, where $g_2$ is the EW coupling constant, $\ell_i$ is the charged lepton field, $\nu_i$ its corresponding neutrino, and $W_\mu$ is a charged $W$ boson. This allows for $W^\pm$ particles to mediate muon decay (recall that a similar process involving quarks and the CKM matrix elements mediates nuclear $\beta$-decay, and was the original use of the Fermi interaction). The exchanged 4-momentum is $q^\mu$, and the $W$-boson propagator is proportional to $1/(q^2+m_W^2)$, where $m_W=80.4\text{ GeV}$~\cite{pdg} is the mass of the $W$. At small momentum transfer, $q^2\ll m_W^2$ (corresponding via the uncertainty principle to large distances) the propagator can be replaced by an effective 4-fermion interaction proportional to $g_2^2/m_W^2$. Higher order interactions come suppressed by higher powers of $m_W$. In the low-energy EFT we replace the EW gauge invariant interaction with the Fermi interaction using $G_F=\sqrt{2}g_2^2/8m_W^2$. For muon decay, the low energy theory has a term in the effective action $S_{\rm eff}(q<m_W)\supset G_F(\bar{e}\nu_e)(\bar{\nu}_\mu\mu)+h.c.$

\begin{figure}
\begin{center}
\includegraphics[scale=0.5]{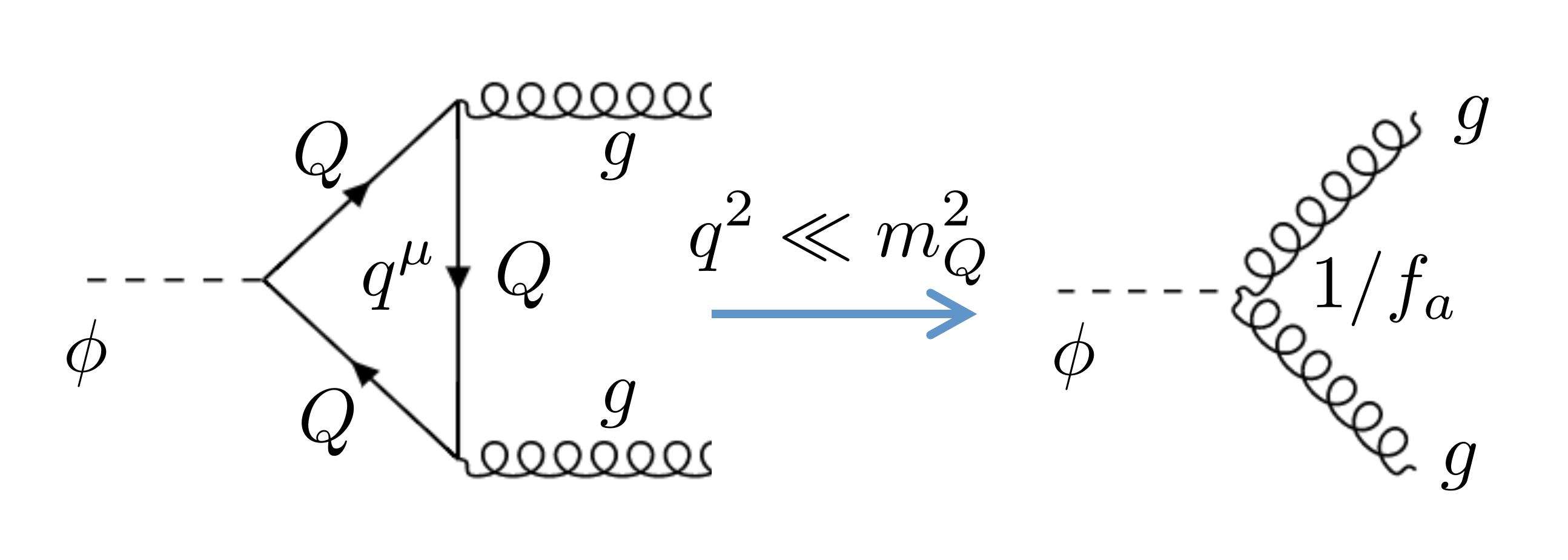}
 \end{center}
 \caption{The colour anomaly in the KSVZ axion model. Heavy quarks, $Q$, run in a loop with momentum $q^\mu$. At low-momentum transfer, $q^2\ll m_Q^2$, the interaction can be replaced with the effective $\phi G\tilde{G}/f_a$ interaction.}
\label{fig:eft_ksvz}
\end{figure}

The situation with axions and the chiral anomaly is more complicated to compute, but is easy to represent in pictures. The case of the KSVZ axion model is shown in Fig.~\ref{fig:eft_ksvz}. The fundamental action contains Yukawa interactions between the axion and the heavy quark fields, $Q$. The $Q$ fields also interact with gluons. Virtual $Q$-particles then induce an effective axion-gluon interaction at loop-level. At low momentum transfer, $q^2\ll m_Q^2$, the heavy quarks can be integrated out and the effective action has a term $S_{\rm eff}(q<m_Q)\supset \phi G\tilde{G}/32 \pi^2 f_a$. This is the dominant term in the expansion in powers of $1/m_Q$. It gives the largest contribution to the explicit breaking of $U(1)_{\rm PQ}$, and thus the axion potential, and also generates the necessary $G\tilde{G}$ interaction required for a solution to the strong-$CP$ problem. EFT can also be applied to light quarks after chiral symmetry breaking. This gives rise to the second term in Eq.~\eqref{eqn:c_phi_gamma}, which gives a contribution to the axion-photon coupling from the colour anomaly.



\section{Friedmann Equations}
\label{appendix:frw}
\setcounter{equation}{0} 
\renewcommand{\theequation}{B\arabic{equation}}

Consider the line element for the flat Friedmann-Robertson-Walker (FRW) Universe:
\be
ds^2=-dt^2+a(t)^2d\vec{x}^2 \, ,
\ee
where $a(t)$ is the cosmic scale factor. The scale factor obeys the Friedmann equations:
\begin{align}
3H^2M_{pl}^2&=\bar{\rho} \, , \nonumber\\
6(\dot{H}+H^2)M_{pl}^2&=\bar{\rho}+3\bar{P} \label{eqn:friedmann} \, , 
\end{align}
where $H=\dot{a}/a$ is the Hubble rate, $\bar{\rho}$ and $\bar{P}$ are the homogeneous background values of the components of the energy momentum tensor as defined in Eqs.~\eqref{eqn:em_tensor_fluid}, and homogeneity and isotropy of the FRW metric demand the vanishing of velocity and anisotropic stress at the background level. The current cosmic time is $t=t_0$, and the current Hubble rate is $H(t_0)\equiv H_0 = 100h\text{ km s}^{-1}\text{ Mpc}^{-1}=2.13h\times 10^{-33}\text{ eV}=hM_H$. Normalising $a(t_0)=1$, the redshift is given by $z=1/a-1$. The scale factor and redshift can both serve as useful time co-ordinates.

Cold (C)DM, baryons and non-relativistic massive-neutrinos have zero pressure, and the energy density in matter scales as $\bar{\rho}_m=\bar{\rho}_{m,0}a^{-3}$. Radiation, including photons and relativistic neutrinos, has pressure $\bar{P}_r=\bar{\rho}_r/3$ and the energy density scales as $\bar{\rho}_r=\bar{\rho}_{r,0}a^{-4}$

The first of Eqs.~\eqref{eqn:friedmann} is commonly known as the Friedmann equation, while the second is known as the Raychaudhuri equation. The Friedmann equation is a first order constraint, and is sufficient to solve the background evolution in the case of a flat or open universe with positive energy density. The Raychaudhuri equation is only necessary to solve for collapsing universes (closed, or an AdS scalar field potential), although there are occasions when it is more numerically stable than the Friedmann equation.

\section{Cosmological Fluids}
\label{appendix:fluid}
\setcounter{equation}{0} 
\renewcommand{\theequation}{C\arabic{equation}}

Useful references for this section include Refs.~\cite{bertschinger1995,Hlozek:2014lca,hu1998b,2004astro.ph..2060H}. The components of the energy momentum tensor can be identified with the energy-density, $\rho$, pressure, $P$, velocity, $v_i$, and anisotropic stress, $\Sigma_{ij}$ of a perfect fluid:
\begin{align}
T^0_{\,\,\,0}&=-\rho\, , \nonumber\\
T^0_{\,\,\,i}&=(\rho+P)v_i \, , \nonumber\\
T^i_{\,\,\,j}&=P\delta^i_{\,\,\,j}+\Sigma^i_{\,\,\,j} \, . \label{eqn:em_tensor_fluid} 
\end{align}
In full General Relativity this decomposition holds for linear perturbations, where $T=\bar{T}+\delta T$, and helps identify the physical meaning of the sources of the Einstein equation. Perturbations are defined such that $\bar{T}$ has the symmetries of the FRW metric. Perturbations in fluid components are defined as $\rho=\bar{\rho}+\delta\rho=\bar{\rho}(1+\delta)$, $P=\bar{P}+\delta P$. Homogeneity and isotropy at the background level imply that $v_i$ and $\Sigma_{ij}$ are (at least) first order. The related variables $\theta$ and $\sigma$ and are defined by
\begin{align}
\theta&=ik^jv_j \, , \\
(\bar{\rho}+\bar{P})\sigma &= -\left(\hat{k}^j\hat{k}_i-\frac{1}{3}\delta^j_{\,\,\,i}\right)\Sigma^i_{\,\,\,j} \, ,
\end{align}
where $\hat{k}$ is a unit vector in Fourier space. 

The continuity equation for the energy density is
\be
\dot{\bar{\rho}}=-3H(1+w)\bar{\rho} \, ,
\ee
where the equation of state is $w=\bar{P}/\bar{\rho}$. Matter and radiation have constant equations of state, $w_m=0$, $w_r=1/3$. The cosmological constant has equation of state $w_\Lambda=-1$. In the general, the equation of state can evolve in time. It's equation of motion is
\be
\dot{w}=-3H(1+w)(w-c_{\rm ad}^2) \, ,
\ee
where the adiabatic (background) sound speed is
\be
c_{\rm ad}^2=\frac{\dot{\bar{P}}}{\dot{\bar{\rho}}}=c_s^2-\frac{w}{\delta}\Gamma \, .
\ee
The sound speed in fluctuations is
\be
c_s^2=\frac{\delta P}{\delta \rho} \, ,
\ee
and $\Gamma$ is the non-adiabatic pressure perturbation.

It is important to note that definitions of ``sound speed'' are not universal, and that the sound speed itself is not gauge invariant. I adopt the definitions above, and apply them in whatever gauge we happen to be working in (synchronous or Newtonian). This is in keeping with the treatment of Ref.~\cite{bertschinger1995}, and is convenient and intuitive for standard cosmological perturbation theory as applied to the post-inflationary universe.

Some authors define the sound speed as the co-efficient in the equation of motion of the gauge invariant ``Mukhanov-Sasaki'' variable, $\nu$. This is common in inflationary theory, and among relativists. For a scalar field, let's denote this particular sound speed $c_\phi^2$. One can prove that $c_\phi^2=1$: i.e. it is the sound speed in the gauge in which $\delta\phi=0$ (flat scalar field slicing). The non-trivial growth and scalar field Jeans scale in this formulation can be understood from the behaviour of the background (anti-)friction terms induced by gauge transformations from, e.g., the Newtonian gauge to the $\delta\phi=0$ gauge~\cite{2015arXiv150106918A}. This is consistent with the time-averaged effective sound-speed we employed in Section~\ref{sec:eff_sound_speed}, and the driven nature of Eqs.~\eqref{eqn:dphi_sync} and Eqs.~\eqref{eqn:dphi_newt} in the oscillating regime~\cite{hu1998b}.

\section{Bayes Theorem and Priors}
\label{appendix:bayes}
\setcounter{equation}{0} 
\renewcommand{\theequation}{D\arabic{equation}}

All cosmologists worth their salt are Bayesians. This happy state of affairs is forced upon us by the unavoidably one-shot nature of observing the cosmos. An introduction to Bayesian methods in cosmology can be found in Ref.~\cite{2003moco.book.....D}, with a more advanced specific treatment in Ref.~\cite{2009bmc..book.....H}.

We are interested in the probability of our theory, specifed by a vector of parameters $\vec{\theta}$, given the data $D$: $P(\vec{\theta}|D)$. What we have access to is the \emph{likelihood}, $\mathcal{L}$, i.e. the probability of the data given the theory: $P(D|\vec{\theta})=\mathcal{L}(D,\vec{\theta})$. Bayes theorem relates these for us:
\be
P(\vec{\theta}|D)=\frac{P(D|\vec{\theta}) P(\vec{\theta})}{P(D)}\, ; \quad {\rm posterior}=\frac{{\rm likelihood\,}\times{\rm \, prior}}{{\rm evidence}}\, .
\ee

The probability of the theory, $P(\vec{\theta})$, is the all-important \emph{prior}. In an MCMC setting, the prior can be thought of as the distribution from which we draw sample theory curves to compare to the data (although it can also be imposed later on top of uniform sampling). The probability of the data, $P(D)$, can be computed as a normalization. It can often be ignored, since we are interested in ratios of probabilities, although it is important for model comparison and Bayesian evidence. 

The likelihood reflects our uncertainty on the data. A very simple assumption is to weight data points individually, and assume Gaussian errors, so that a model has a likelihood as a product of Gaussians given by the distance of the theory curve from each data point. In many real-world examples, the likelihood is much more complicated. For example, the \emph{Planck} likelihood is discussed in Ref.~\cite{2014A&A...571A..15P}.

The prior reflects our degree of belief in a model, and is often where physics can be put in. See Ref.~\cite{2014PhRvD..90j5023M} for an example in dark energy theory, and the formalism for treating information gain over the prior in a Bayesian context.

An ``uniformative'' prior is the Jeffreys prior, which for most practical purposes is flat in log space. It is a suitable prior for unknown energy scales, for example the axion mass and decay constant. The log-flat prior on axion mass is also physically motivated: in string theory the mass scales exponentially with some modulus, $\sigma$, of the compact space: $m_{a,i}\propto e^{-c\sigma_i}$, where $i$ labels the axion species. The moduli are expected to have a uniform distribution in real space (since the scale is set by the compactification volume), leading to a log-flat axion mass distribution. String theory predicitions for the $f_a$ distribution are in general not log-flat, since $f_{a,i}\propto M_{pl}/\sigma_i$~\cite{2006JHEP...06..051S}. The distribution can be calculated from random matrix theory, which selects some preferred scale somewhat below the Planck scale (e.g. Refs.~\cite{acharya2010a,2014JHEP...10..187L}).

The axion initial misalignment angle, on the other hand, is a compact variable, and so the natural prior is a uniform prior. For the QCD axion, holding $f_a$ fixed and using that $\Omega_a h^2\propto \theta_{a,i}^2$ this gives the prior distribution for the relic density (e.g. Ref.~\cite{2006PhRvD..73b3505T}):
\be
P(\Omega_a h^2)\propto \frac{1}{\sqrt{\Omega_a h^2}}\, .
\ee
This fixed prior from theory makes axions uniquely predictive in landscape and multiverse scenarios (e.g. Refs.~\cite{2004hep.ph....8167W,2010JCAP...03..021F,2013PhRvD..88f3503B}). Incorporating additional information such as the prior on $f_a$ for the QCD axion, or on $m_a$ for ALPs, has not yet been fully explored in the literature.

\section{Degeneracies and Sampling with ULAs}
\label{appendix:degeneracies}
\setcounter{equation}{0} 
\renewcommand{\theequation}{E\arabic{equation}}

\begin{figure}
\begin{center}
\includegraphics[width=0.75\textwidth]{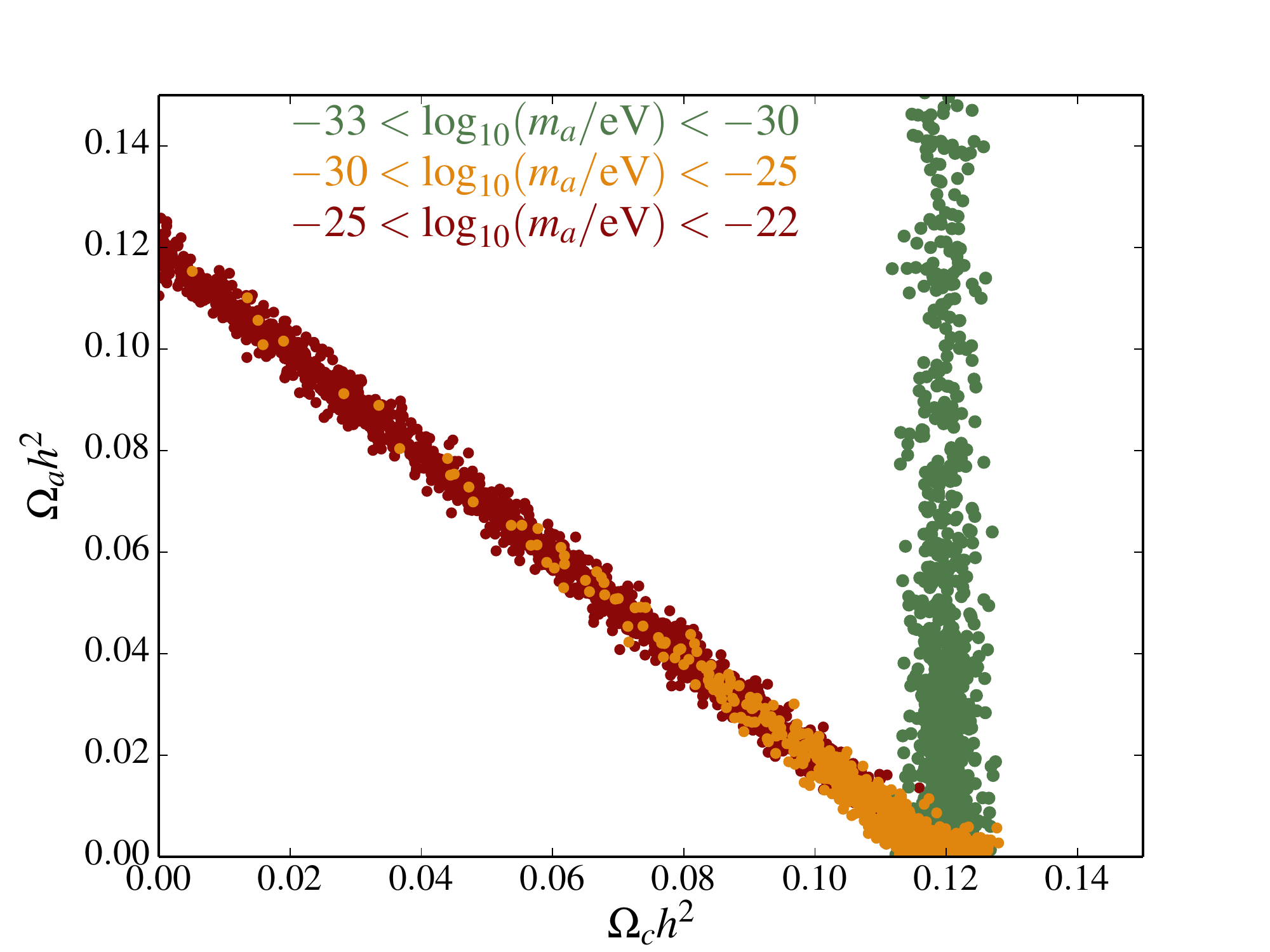}
\caption{Degeneracy of $\Omega_a h^2$ and $\Omega_c h^2$. Sample points for an MCMC chain are shown, binned by axion mass, $m_a$. High axion mass leads to a one-to-one degeneracy, with $\Omega_a h^2+\Omega_c h^2\approx 0.12$. Low mass axions behave as DE, allowing for large $\Omega_a h^2$ and fixing $\Omega_c h^2=0.12$. Intermediate masses are constrained to have $\Omega_a h^2<0.12$. Reproduced (with permission) from Ref.~\cite{Hlozek:2014lca}. Copyright (2015) by The American Physical Society.}
\label{fig:degeneracy_scatter}
\end{center}
\end{figure}

On scales much larger than the Jeans scale, axion DM is degenerate with CDM. For very low mass axions with $m_a\sim H_0$, the axion equation of state is $w_a\approx -1$ even today, and axions are degenerate with the cosmological constant and DE. Our goal is to use precision cosmology to map out the range of axion masses in between, i.e. those masses constrained by cosmology because such axions are neither equivalent to CDM nor DE. This leads to a very challenging degeneracy structure for $\Omega_a h^2$ as a function of $m_a$, which is illustrated in Fig.~\ref{fig:degeneracy_scatter}.

Standard cosmological parameter estimation is carried out using MCMC analysis (the industry standard used by \emph{Planck} is \textsc{cosmomc} \cite{cosmomc}; see e.g. Ref.~\cite{2005MNRAS.356..925D} for a description of the methodology). The chain is begun at some location close to the maximum likelihood, and then randomly (and ergodically) explores this likelihood, with the density of samples reflecting the value of the likelihood. With infinite computing time, the process is guaranteed to explore the entire likelihood. Allowing for a wide prior on $m_a$ makes the convergence of this process very slow, and the chain can get ``stuck'' in particular regions (modes) of the likelihood. For example, we might get stuck in a high-likelihood region with large $m_a$, and $\Omega_a h^2\approx \Omega_c h^2$. What we really want to know is the constraint on $\Omega_a h^2$ at intermediate masses, and what the range of ``intermediate'' really is for a given observable.

Working around this bottleneck requires using different tools to estimate the likelihood than a standard ``out-of-the-box'' MCMC. The method employed in Ref.~\cite{Hlozek:2014lca} used nested sampling with \textsc{multinest} \cite{multinest}, an algorithm designed for multi-modal likelihoods, instead of MCMC. However, it still proved prohibitively expensive to have enough sample points to achieve accurate limits on $\Omega_a h^2$ across the full range of $m_a$ in the two dimensional $(m_a,\Omega_a)$ plane. A two-step procedure was used to overcome this. Three separate mass ranges ran independently. A more coarse global chain was then ran, and the information from this was used to importance-sample the individual chains together on the $(m_a,\Omega_a)$ plane.\footnote{A similar procedure using a `hot' MCMC chain as the global sample could also have been used, but \textsc{multinest} was found to be more efficient. Another alternative would be to use an ensemble sampler, such as \textsc{emcee}~\cite{emcee}.}

The procedure described above was able to deal with the degeneracies between CDM, DE and axions that occur for high and low ULA masses respectively. A separate issue that has yet to be addressed fully is the degeneracy between ULAs and neutrinos at intermediate ULA mass. Cosmology is approaching the required precision to detect the effects of $\sum m_\nu=0.06\text{ eV}$, the minimum consistent with oscillation experiments. It is crucially important to address all possible degeneracies so that a future detection can be considered robust. Ref.~\cite{amendola2005} used a grid-based likelihood, where convergence is not an issue, but only constrained $m_a$ and $m_\nu$ independently. Grids scale poorly for large numbers of parameters, and are unsuitable for precision analysis. Ref.~\cite{marsh2011b} performed a preliminary investigation using a Fisher matrix formalism to perform forecasts. At the level of the study, degeneracies were not too severe: the difference in behaviour between axions and neutrinos during the radiation era breaks the degeneracy in the effect on structure formation. However, Ref.~\cite{marsh2011b} looked at individual ULA masses independently, and did not study the degeneracies as a function of ULA mass. Including ULAs, CDM and neutrinos in a full parameter estimation pipeline will likely require further tricks like those described here to be employed when sampling the likelihood.

In general, when considering degeneracies, it is important to break the effects of axion DM up into two parts: effects on the background expansion, and effects on the perturbations. Axion cosmology coming purely from the misalignment production is a well defined model where all effects on the expansion rate, clustering and initial conditions come packaged together. As we saw in Section~\ref{sec:transfers}, and has been discussed extensively elsewhere in the literature, the axion transfer function is similar to the WDM and neutrino transfer functions. However, these thermal and non-thermal components behave quite differently in their effects on the background expansion, leading to, for example, very different CMB signatures for similar transfer functions. It also might naively appear that any effect on the transfer function can be mimicked by a change in the primordial power. However, the primordial power affects radiation \emph{and} DM, and so its effects show up in the CMB as well as in the matter power spectrum. The DM transfer function will only show up at leading order in the matter power spectrum. Multiple measurements can thus break that possible degeneracy. Similarly, axion effects on the background expansion could be mimicked by some particular model for the DE equation of state or modified gravity (MG). However, the particular physical DE/MG model may have different clustering or early Universe behaviour from the corresponding axion model, allowing the two to be distinguished. 

\section{Sheth-Tormen Halo Mass Function}
\label{appendix:hmf_def}
\setcounter{equation}{0} 
\renewcommand{\theequation}{F\arabic{equation}}

The HMF is given by
\begin{align}
\frac{dn}{d\ln M} &= -\frac{1}{2} \frac{\rho_m}{M} f(\nu) \frac{d \ln \sigma^2}{d \ln M} \, , \\
\nu &\equiv \frac{\delta_{\rm crit}}{\sigma} \, .
\end{align}
For $f(\nu)$ we use the Sheth-Tormen function~\cite{1999MNRAS.308..119S}:
\be
f(\nu)=A \sqrt{\frac{2}{\pi}} \sqrt{q} \nu (1+(\sqrt{q}\nu)^{-2p})\exp \left[  -\frac{q\nu^2}{2} \right] \, ,
\ee
with parameters $\{A=0.3222,p=0.3,q=0.707\}$. This is a semi-analytic result for the HMF derived in ellipsoidal collapse, which fits results from CDM N-body simulations reasonably well. Other fits for $f(\nu)$ can be found by fitting directly to N-body simulations, but the Sheth-Tormen result will do for us.

The variance is defined by smoothing the power spectrum with some window function, $W(k|R)$, of radius $R$ and assigning a mass using the enclosed matter density:
\be
\sigma^2(M,z)=\frac{1}{2 \pi^2}\int_0^\infty \frac{dk}{k}\Delta^2(k,z)W^2(k|R(M)) \, ,
\ee
where $\Delta^2(k,z)=k^3P(k,z)$. A real-space spherical top-hat window function assigns mass unambiguously:
\begin{align}
W(k|R)&=\frac{3}{(kR)^3}(\sin kR-kR\cos kR) \, , \\
M&=\frac{4}{3}\pi\rho_m R^3 \, .
\end{align}

\bibliographystyle{h-physrev3.bst}
\bibliography{axion_review}

\end{document}